%% file: main.tex
\documentclass{article}
\input{admin/preamble}
\input{admin/macros}

\input{admin/glossary}

\usepackage{authblk}

\title{Towards near-term quantum simulation of materials}
\date{\today}
\author{Laura Clinton}
\author{Toby Cubitt}
\author{Brian Flynn}
\author{Filippo Maria Gambetta}
\author{Joel Klassen}
\author{Ashley Montanaro}
\author{Stephen Piddock}
\author{Raul A. Santos}
\author{Evan Sheridan}
\affil{Phasecraft Ltd.}

\begin{document}

\maketitle

\begin{abstract}
  Simulation of materials is considered one of the most promising applications of quantum computers.
  On near-term hardware, the crucial constraint on simulations of real material models is the quantum circuit depth (and, to a lesser extent, the number of qubits required).
  The core of many quantum simulation algorithms, including time-dynamics and the variational quantum eigensolver under the Hamiltonian variational ansatz, is a layer of unitary evolutions by each local term in the Hamiltonian: a single Trotter step in time-dynamics simulation, or a single layer of VQE.

  In this work we develop a new quantum algorithm design for materials modelling where this depth is {\it independent} of the system's size.
  To achieve this, we take advantage of the locality of materials Hamiltonians in the Wannier single-particle electron basis and construct a tailored, multi-layer fermionic encoding that keeps the weight of Pauli operators appearing in the Hamiltonian independent of the system's size.
  We analyse the circuit and measurement cost of this approach and present a compiler that is able to produce quantum circuit instructions for our method starting from density functional theory data, thus bridging from the physics description of a material to the quantum circuit needed to simulate it.
  The quantum circuits produced by our compiler are automatically optimised at multiple levels, not only at the circuit level but also incorporating optimisations derived from the physics of the specific target material.
  We present detailed numerical results for different materials spanning a wide structural and technological range.
  Our results demonstrate a reduction of many orders of magnitude in circuit depth over standard prior methods that do not consider the structure of the Hamiltonian.

  For example, for Strontium Vanadate (SrVO\textsubscript{3}), our results improve the resource requirements over standard quantum simulation algorithms from 864 to 180 qubits for a $3\times3\times3$ lattice, and the circuit depth of a single Trotter or variational layer from $7.5\times 10^8$ to depth $884$.
  Although these circuit depths are still beyond current hardware, our results show that realistic materials simulation may be feasible on quantum computers without necessarily requiring fully scalable, fault-tolerant quantum computers, providing quantum algorithm design incorporates deeper understanding of the specific target materials and applications.
\end{abstract}

\tableofcontents

\input{Introduction}

\input{Design_strategy}

\input{Fermion_description} 

\input{Qubit_Hamiltonian} 

\input{Algorithms} 

\input{Compiler_Description}

\input{Results} 

\input{Outlook}

\appendix
\input{Appendix}

\clearpage
\printbibliography

\end{document}

%% file: admin/preamble.tex
\usepackage[a4paper, total={6in, 9in}]{geometry}
\usepackage[utf8]{inputenc}
\usepackage[T2A]{fontenc}
\usepackage{tikz}

\usepackage[
    backend=bibtex,
    natbib = true,
    style = phys,
    isbn = false,
    hyperref = true,
    sorting = none,
    giveninits = true,
    maxbibnames = 9,
    biblabel = brackets,
    eprint = true
]{biblatex}

\bibliography{references}

\usepackage{amsmath}
\usepackage{braket}

\usepackage{listings}
\usepackage{amssymb}
\usepackage[ampersand]{easylist}
\usepackage{enumitem}
\usepackage{hyperref}
\usepackage{subcaption}
\usepackage{comment}
\usepackage{makecell} 
\usepackage{enumitem} 

\usepackage{chngcntr} 
\counterwithin{figure}{section}
\counterwithin{table}{section}

\usepackage{glossaries-extra}
\setabbreviationstyle[acronym]{long-short}
\glssetcategoryattribute{acronym}{nohyperfirst}{true}

\usepackage{mathtools}
\usepackage{longtable}
\usepackage{multirow}

\DeclarePairedDelimiter{\ceil}{\lceil}{\rceil}
\newcommand{\ttt}[1]{\texttt{#1}}

\hypersetup{
  colorlinks = true, 
  linkcolor={blue!50!black}, 
  citecolor=red, 
  urlcolor=blue, 
  linktoc=page,
}
\usepackage[capitalise, nameinlink]{cleveref}
\usepackage{booktabs} 


%% file: admin/macros.tex
\newcommand{\bm}[1]{\textbf{#1}}

\definecolor{asparagus}{rgb}{0.53, 0.66, 0.42}



%% file: admin/glossary.tex
\newacronym{jw}{JW}{Jordan Wigner}
\newacronym{vqe}{VQE}{Variational Quantum Eigensolver}
\newacronym{tds}{TDS}{Time Dynamics Simulation}
\newacronym{qc}{QC}{quantum computer}
\newacronym{nisq}{NISQ}{noisy intermediate-scale quantum}
\newacronym{dft}{DFT}{density functional theory}
\newacronym{dof}{DoF}{degrees of freedom}

\newacronym{srv}{SrVO$_3$}{strontium vanadate}
\newacronym{gaas}{GaAs}{gallium arsenide}
\newacronym{h3s}{H$_3$S}{hydrogen disulfide}
\newacronym{li2}{Li$_2$CuO$_2$}{lithium copper oxide}
\newacronym{mlwf}{MLWF}{maximally localized Wannier function}
\newacronym{qma}{QMA}{Quantum  Merlin Arthur}
\newacronym{bqp}{BQP}{Bounded-error Quantum Polynomial time}

%% file: Introduction.tex
\section{Introduction}

The race to demonstrate useful applications for near term quantum computers has begun in earnest, with quantum simulation being one of the leading candidates \cite{georgescu14,bharti2021noisy}.
Accurate simulations of complex materials yield valuable insight into their behaviour.
This understanding serves to predict macroscopic properties and facilitates the rational design of materials with novel characteristics.
The capability to understand and design characteristics in chemicals and materials is crucial for scientific, industrial, and commercial purposes, evidenced by the central role of classical simulation in guiding innovation in the multi-billion dollar chemical industry \cite{icca_rep, Chem}.
There are various challenges involved in the rational design of properties in novel materials, encompassing vastly different length and time scales; notably, fundamentally inefficient descriptions of electron--electron interactions hinder the ability to make predictions in the strong-coupling regime, where many relevant technological applications are expected to appear \cite{quant_materials}.
\par

A \gls{qc} can simulate these correlated processes natively, by decomposing the quantum evolution into a sequence of elementary operations (i.e.\ a quantum circuit), applied to a specified quantum state.
The state obtained from this procedure is then queried by measuring relevant quantities.
Crucially, the advantage of this approach over direct classical simulation of the state vector appears for large enough systems (in terms of qubits and quantum circuit complexity), where the exponential growth of the Hilbert space outpaces state-of-the-art supercomputer capabilities~\cite{arute19}.
\par
Two main challenges pervade the \gls{nisq} era of \glspl{qc}.
First, the number of physical qubits is restricted,
although various hardware providers project that this will increase dramatically in the next decade \cite{IBM,Google,IonQ}.
The second challenge is to obtain gate fidelities sufficient to facilitate reliable circuits.
Taken together, both capabilities are critical to the production of large quantum circuits, where error mitigation techniques can handle the noisy outcomes to generate meaningful signals beyond classical capabilities.
Further developments will be required for quantum error correction and fault tolerance to become feasible, which themselves will allow for arbitrarily deep circuits.
\par

These limitations on \glspl{qc} constrain the types of algorithm that can in principle be implemented -- and algorithms targeting near-term hardware should use all available strategies to minimize the required number of qubits and required circuit depth in order to meet these limitations.
Materials' simulation is well suited to this domain.
Although the number of electrons in a large piece of material is of the order of Avogadro's number, the regularity of the lattice restricts the behaviour of electrons, allowing us to concentrate the important degrees of freedom into a relevant {\it active space}, wherein the dominant mechanisms of interest lie.
In doing so, we may minimise the number of qubits required to perform an accurate simulation.
The periodic structure of materials also usually offers a great deal of symmetries, that can be leveraged to generate a compact representation of the Hamiltonian, lowering the number of interactions and ultimately the cost of implementing a circuit based on that Hamiltonian. Symmetries can also be used to mitigate errors in the measured signals, as already demonstrated experimentally \cite{stanisic2021}.

Materials' systems also enjoy some useful properties.
Band theory -- namely the description of materials in terms of single-particle physics -- is a well defined limit underlying the success of \gls{dft}. This limit provides a natural starting point for quantum state preparation, where one can initialize the system in the correct symmetry sector with an efficient quantum circuit.
Moreover, using the single-particle state as a starting point has already been shown as a useful technique for  error mitigation~\cite{montanaro2021_FLF}.
Here, by training on the data obtained in a non-interacting instance, a map between the data obtained in the \gls{qc} and the exact values can be inferred, and used to correct extracted data in the instances where classical simulations are not feasible.
Similar error mitigation approaches are plausible for chemical systems, but have not yet been explored.

\begin{table}[t]
    \centering
    \input{Figs_Results/materials_results/intro_table}

    \caption{
    Summary of resources needed to implement a single \gls{vqe} layer that simulates the Hamiltonian of a material, without accounting for initial state preparation.
    The properties of the materials are listed in \cref{tab:mat_props}.
    \textit{Baseline} estimates are based on standard methods available in the literature -- namely employing the Jordan-Wigner transform in the Bloch basis -- without considering the structure of the Hamiltonian (see \cref{app:previous_method}).
    For further details, see \cref{sec:res_materials}.
    }
    \label{table:estimates_intro}

\end{table}

The description of electronic systems in digital quantum computers also presents particular challenges.
While the most important components describing the physics of materials are electrons, most digital quantum computers operate with two-level systems, i.e. qubits.
In order to properly account for fermion statistics and the Pauli principle, an algebraic mapping between fermions and qubits is needed. The usual mapping -- the \gls{jw} transform -- can increase the cost of a computation by a multiplicative factor which scales according to the size of the system, outweighing the benefits yielded by preparing a local fermion Hamiltonian.
Further, crystalline solids possess at least two natural bases for the single-particle electrons, the band (Bloch) basis, which represents electrons in momentum space, and the Wannier basis, which represents electrons in real space.
Each single-particle basis affects the final cost of implementing a circuit differently, so we must establish stringent principles upon which we may choose between them.

In this work we take advantage of the interplay between single-particle bases, locality, symmetries, fermionic encodings, fermionic swap networks, and measurement strategies in order to develop novel, efficient algorithms for simulating materials' systems.
In doing so, we demonstrate a speed up of multiple orders of magnitude over standard methods, in a cost model assuming all-to-all hardware connectivity and cost 1 for each 2-qubit gate.
Selected results appear in \cref{table:estimates_intro}, where we compare the circuit depth obtained by our methods with a standard, generic method that does not exploit the structure of the Hamiltonian (see  \cref{app:previous_method} for details).
The materials analysed here represent a selection of systems whose behaviours are dominated by distinct underlying mechanisms: they span a minimal but wide structural, chemical, and technological range. \Gls{srv} is a strongly correlated material that serves as a benchmark for post-DFT methods \cite{Sheridan2019}, \gls{gaas} is a fairly well-understood material with many technological applications. Likewise, Si is the cornerstone material used in modern electronics \cite{si_sc} and is also important in many other applications, such as solar technologies \cite{si_sc}.   Recently, \gls{h3s} has been found to host a high superconducting transition temperature at high pressures \cite{Drozdov2015}. Finally, \gls{li2} is a material used in advanced lithium-ion battery technology \cite{Jing2017Li2CuO2}.

We present a unified resource estimate for quantum algorithms, namely \gls{vqe},
and \gls{tds}, where a layer represents a single Trotter step or single Hamiltonian Variational Ansatz VQE layer in the overall evolution.
The remainder of this text describes the numerous strategies employed to achieve these resource estimates.

\subsection{Comparison with earlier works}

Qubit and gate resources required for Trotterized Hamiltonian simulation algorithms of local Hamiltonians have recently been investigated by Kanno {\it et al.} \cite{Kanno22}. Here, effective Hamiltonians of several unit cells of materials have been constructed starting from a classical description that accounts for the important chemistry of the active space \cite{Imada10}. The resources to implement a single Trotter step are investigated on devices with nearest-neighbor connectivity in terms of CNOT and arbitrary single-qubit gates.
They use a \gls{jw} transform to encode the fermionic modes, and fermionic swaps \cite{kivlichan18, OGorman19} to deal with the large operator weight of the encoded Pauli operators. This leads to a scaling of the gate count that is $O(N_{\rm cells}^2)$ for a Hamiltonian defined in $N_{\rm cells}$ unit cells.

In comparison, our approach attains $O(N_{\rm cells})$ scaling of the number of gates, as the intercell interactions are implemented through ancillas in the compact encoding \cite{derby2021compact}. This incurs a qubit overhead proportional to the number of unit cells.
Importantly, considering that the main problem of current QCs is the presence of gate errors, our approach allows us to achieve a layer depth for single Trotter step that is independent of the size of the system,  in stark contrast with the $O(N_{\rm cells}^{2/3})$ depth using \gls{jw} (in a cubic system with nearest neighbour interactions).

Delgado et al.\ recently gave a detailed resource analysis of quantum algorithms for determining properties of battery materials, such as equilibrium voltages and thermal stability~\cite{delgado22}. They use a first quantisation approach with the plane wave basis and compute the cost of the quantum phase estimation algorithm. Considering one unit cell of the material Li$_2$FeSiO$_4$ with 156 electrons, these authors find a Toffoli gate cost of between $10^{11}$ and $10^{15}$ for quantum phase estimation, depending on the number of plane waves and level of accuracy required.

Counting the overall number of Toffoli or T gates is an appropriate approach to estimate complexity in the fault-tolerant regime, as this quantity directly determines the (very significant) overhead required for fault-tolerance. For near-term quantum computers, depending on the architecture, quantum circuit depth can be more appropriate, for several reasons. First, quantum computations are limited by decoherence, which sets an upper bound on the overall running time, as measured by circuit depth. Second, as errors can be seen as spreading out across a quantum circuit within a ``lightcone'', lower-depth circuits lead to improved localisation of errors. Third, as the circuit depth determines the running time, a lower-depth circuit executes more quickly.

Several other works have produced quantum algorithmic resource costs tailored for the fault tolerant era in molecular systems \cite{PRXQuantum.2.030305,Su2021,kim2022} and the interacting electron gas (Jellium) \cite{Babbush2018,Kivlichan2020improvedfault,Su2021nearlytight,McArdle2022}.

%% file: Figs_Results/materials_results/intro_table.tex
\begin{tabular}{lllllr}
\toprule
         &                                                                              &                   & Bands & Qubits &   Depth \\
Material & Applications & Method &       &        &         \\
\midrule
\multirow{2}{*}{GaAs} & \multirow{2}{*}{\makecell{Semiconductors \cite{GaAs_semi}, transistors \cite{shur1987gaas}, \\solar cells \cite{Jun2013}, spintronics \cite{Okamoto2014}}} & This work &     4 &   1120 & 7.9E+03 \\
         &                                                                              & Baseline estimate &    18 &   4500 & 1.2E+11 \\
\cline{1-6}
\cline{2-6}
\multirow{2}{*}{H$_3$S} & \multirow{2}{*}{Superconductors \cite{Drozdov2015}} & This work &     7 &   1870 & 3.7E+04 \\
         &                                                                              & Baseline estimate &     6 &   1500 & 4.0E+09 \\
\cline{1-6}
\cline{2-6}
\multirow{2}{*}{Li$_2$CuO$_2$} & \multirow{2}{*}{High-capacity battery cathode \cite{Jing2017Li2CuO2}} & This work &    11 &   1024 & 8.4E+03 \\
         &                                                                              & Baseline estimate &    11 &    990 & 1.1E+09 \\
\cline{1-6}
\cline{2-6}
\multirow{2}{*}{Si} & \multirow{2}{*}{Semiconductors \cite{si_semi}, solar cells \cite{si_sc}} & This work &     4 &   1120 & 8.6E+03 \\
         &                                                                              & Baseline estimate &     3 &    750 & 4.8E+08 \\
\cline{1-6}
\cline{2-6}
\multirow{2}{*}{SrVO$_3$} & \multirow{2}{*}{Solar cells \cite{srvo3_solar_cell}, batteries \cite{svo_anode, svo_cathode}} & This work &     3 &    180 & 8.8E+02 \\
         &                                                                              & Baseline estimate &    16 &    864 & 7.5E+08 \\
\bottomrule
\end{tabular}

%% file: Design_strategy.tex
\section{Design strategy}

The \gls{nisq} era is characterised by \glspl{qc} operating without fault tolerance, so the depth of implementable quantum circuits is fixed by the error level present in the available device. 
Therefore the construction of compact circuits for simulation is crucial, as it can enable meaningful results (i.e., circuits where the accumulated error can be mitigated), as opposed to the random noise otherwise likely.
Such constructions rely on two critical components: the physical instance being simulated, and an efficient decomposition of the physical information into layers of quantum gates. Our design strategy tackles these aspects in tandem. 
\par
 
The first step is to identify the relevant \gls{dof} of the phenomena under investigation. This is not a sharp (or even well defined) procedure, but instead depends on the nature of the question being asked. For a given material, for example, studying electric transport at low temperatures involves different physical processes than the melting behaviour at high temperature. At a high level, this approach consists of choosing an {\it active space}, commonly discussed in chemistry and materials science \cite{jensen2016book}. This active space can be seen as a distillation of the relevant \gls{dof} at a certain energy scale. Once the relevant \gls{dof} in the active space have been identified, their dynamics are constructed: these dynamics are governed by an \emph{effective Hamiltonian} which describes their interactions. 
\par 

Once this effective Hamiltonian has been obtained, a map between the physical and the logical \gls{dof} is required. Abstractly, this procedure maps interactions between the original \gls{dof} to qubit operations. In particular for fermions, the interplay between the structure of the Hamiltonian interactions and the fermionic encoding plays an important role in the ability to create compact circuits. At the end of this step a collection of Pauli operators is derived, comprising the qubit Hamiltonian $H_Q$. 

Following this, the protocol implementing all the terms in the qubit Hamiltonian is computed. Here, the general approach that we use has the same structure as Trotterization of the evolution operator ${\rm exp}(iH_Q t)$. The structure of this step is indicative of the cost of finding a ground state via a \gls{vqe} approach (in particular under the Hamiltonian variational ansatz~\cite{wecker15}), or \gls{tds}. This produces the circuit for a single Trotter step (or a single layer of \gls{vqe}). Finally, we determine the measurement protocol that produces the minimum measurement overhead.
Clearly, the decisions at each stage will have an effect on final cost of implementing all the qubit operators present in $H_Q$ through a quantum circuit. Hence we adopt a multi-tiered strategy for minimizing the cost of the quantum circuit, that we describe below in the context of materials simulations.
\par 

For the physics-based construction of the Hamiltonian of a material we adopt the Born-Oppenheimer approximation \cite{Cederbaum_BO, Solyom2_book, Ashcroft_book}, and concentrate on the quantum description of the electron \gls{dof}, including the nuclei as a classical background potential. While this approach is general enough to be used in chemistry and materials science, we note that including the quantum mechanical \gls{dof} of the nuclei is also possible within this framework.
The existence of the periodic ionic potential is a distinctive feature of materials, which sets them apart from molecules. We use \gls{dft} for a low level exploration of the active space of materials, defined as an energy window around the Fermi level. Using this window containing the relevant \gls{dof}, we construct an effective Hamiltonian by classically computing its matrix elements.\footnote{The problem of properly computing these matrix elements is ambiguous, for at least two reasons. An unavoidable problem that appears once an active space is used is that the electrons outside the active space renormalise the interactions that the electrons in the active space feel with the nucleus. To fully characterise that renormalisation, the solution of the many-body interacting problem has to be found, which is what we are trying to do in the first place. The second reason is that any realisation of \gls{dft} is an approximation in itself, as the exchange correlation functional is unknown. Both problems are known in the community, and are handled in a plethora of different ways see, e.g., \cite{Georges2004,Haule2015,Imada10,Kanno22} }

We study two natural single-particle bases for the electrons: the Bloch basis, and the Wannier basis \cite{Solyom2_book, Ashcroft_book}. The bands kept in the active space become modes in the unit cell, and the size of the material determines the number of unit cells. Due to the locality in real space achieved by the Wannier basis, a bespoke selection of bands allows us to construct a local Hamiltonian in real space, where the Coulomb interactions are localised, and the hopping range of electrons between unit cells does not scale with the system size. This local Hamiltonian defines a {\it motif} that can be used to tile a system of any size without increasing the depth, i.e. the number of layers, each containing many quantum gates. 

To leverage the locality of the obtained fermion Hamiltonians, we introduce a novel fermionic encoding that uses the local structure of Coulomb interactions and hopping terms by hybridizing two existing encodings: the \gls{jw} transform within a unit cell (where the majority of the electron-electron interactions are present), and the compact encoding \cite{derby2021compact} between unit cells, where fewer interactions have to be considered, following from the locality of the fermion Hamiltonian. 
This comes at the cost of introducing further ancillary qubits. 
In order to deal with the existence of large weight operators along the \gls{jw} line, we introduce an algorithm based on the use of fermionic swap operations (fswaps)~\cite{kivlichan18}. 
These operations can bring operators closer together along the \gls{jw} line and minimise their weight, by relabelling the fermionic modes.
This construction can be expanded to span structures other than single unit cells, depending on the connectivity graph of the Hamiltonian in question.
\par 

We invoke a cost model where all-to-all qubit interactions are available, arbitrary 2-qubit gates  have cost 1 each, and 1-qubit gates are free (i.e. negligible). 
We hereby perform an in-depth analysis of the cost of implementing the most general terms allowed by symmetry, and the cost of performing fswaps to bring modes into an adjacent ordering within the \gls{jw} string. 
Finally, we analyse the cost of executing a full \gls{vqe} layer of the Hamiltonian  $H_Q = \sum_k H_k$ given by $\prod_k e^{i\theta_k H_k}$, where $\set{\theta_k}$ is a set of variational parameters. 
Here $H_Q$ is obtained from a fermion representation in the Wannier bases, with different fermionic encodings. 
Additionally, we calculate the classical measurement overhead, i.e., we determine how many times a given circuit must be repeated to estimate an observable of interest, according to a series of measurement strategies.
\par 

\begin{figure}[ht]
 \centering
 \includegraphics[scale=0.5]{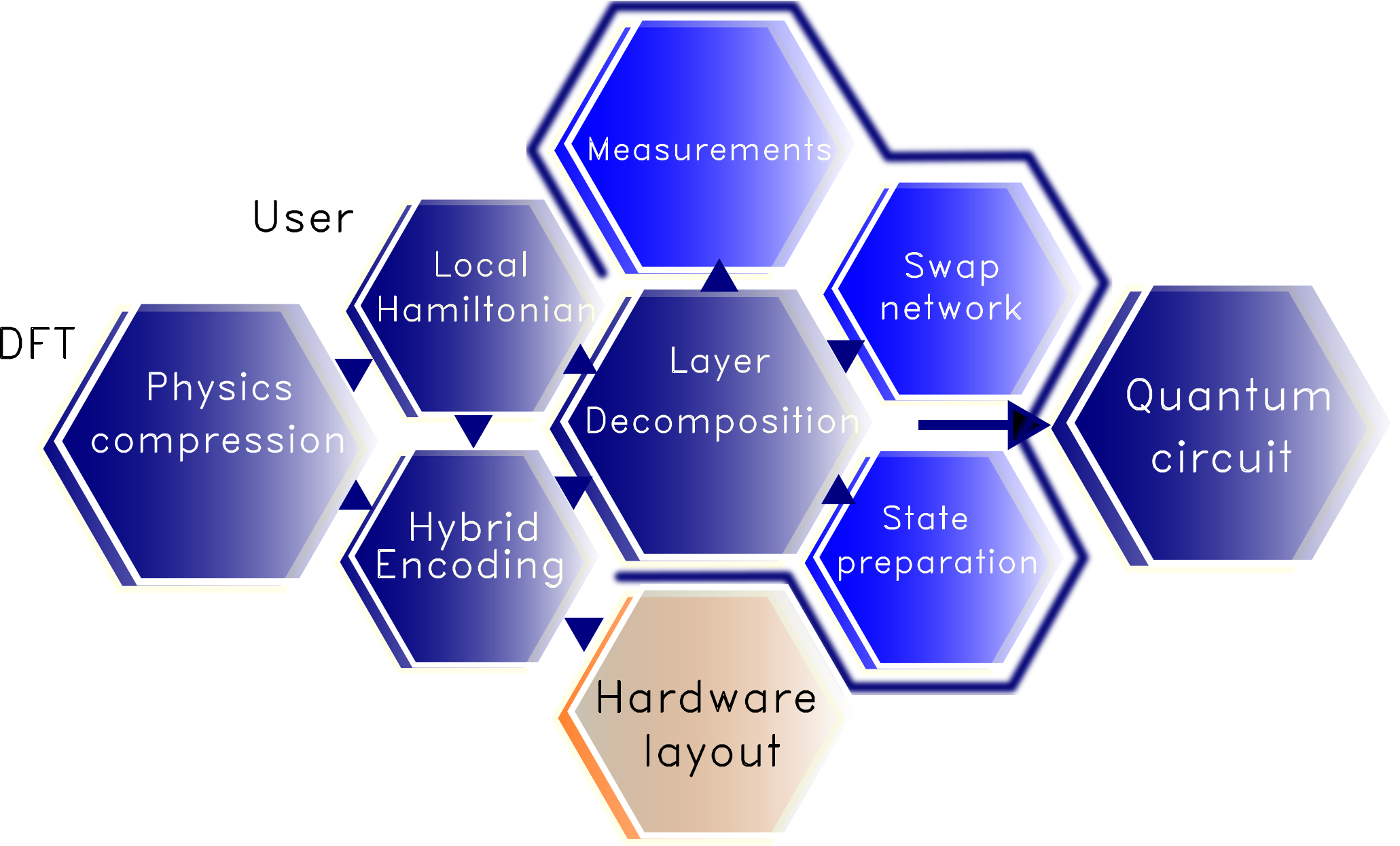}
 \caption{Summary of the strategy developed in this work to minimise circuit depth in the simulation of materials. Starting from a low level calculation based on \gls{dft}, we perform a compression of the physical information into relevant degrees of freedom. The locality of the interactions and the hardware layout determine the structure of the hybrid encoding. In order to minimise the circuit depth, the layer decomposition module determines the appropriate fswap networks, state preparation layers, and a measurement protocol. These elements constitute the quantum circuit that implements a layer of either \gls{vqe} or \gls{tds}.}
  \label{fig:Strategy}
 \end{figure}

We develop a tool to perform the necessary decomposition into layers of simultaneously implementable terms from a given Hamiltonian. 
This allows us to study different materials and model Hamiltonians, to understand their cost complexity, and to find a full decomposition into quantum circuits. The summary of the design strategy to optimise over circuit cost is shown in \cref{fig:Strategy}.

A self-contained exposition of the physics behind the construction of Hamiltonians is presented in \cref{sec:Physics}. The role of symmetries, Wannier and Bloch functions, and efficient techniques to construct the matrix elements are discussed there.

In \cref{sec:qubit_rep} we introduce a hybrid fermionic encoding, and discuss its use in the context of materials' simulation, where it represents an efficient fermion-to-qubit mapping.
In \cref{sec:vqe_alg} we first concentrate on the quantum algorithm (\gls{vqe}) itself and then discuss the decomposition of operators in terms of gates, initial state preparation, time evolution according to the material's Hamiltonian, and measurement protocols.
Combining these ideas, in \cref{sec:compiler} we discuss the design of our circuit compiler, which we go on to use in \cref{sec:results} to analyse the cost of running a single layer of \gls{vqe} or a single Trotter step for \gls{tds}, in examples of increasing complexity.

%% file: Fermion_description.tex
\section{Effective description of the Hamiltonian}\label{sec:Physics}

The full simulation of a physical system comprises infinitely many DoF, which makes it infeasible. This has never been a problem in domains where the relevant energy scale of the problem is restricted to a finite range. In this situation, the DoF at that scale are the ones that mostly contribute to the physical phenomena in question. For everyday applications, where most of the processes are controlled by the behaviour of the electron DoF in atoms,  
the Hamiltonian\footnote{The full Hamiltonian includes the lattice ions. The mass of the ions is much larger 
than the mass of the electrons, so a good approximation is to consider the ions frozen. The lattice of frozen ions then acts as an external potential on the electrons. This approach, known as the Born-Oppenheimer approximation \cite{Cederbaum_BO}, has found success outside typical everyday experimental phenomena, from the prediction of the optimal structural configuration of the rare earth hydrides used in high pressure room temperature superconductors \cite{SCs_at_room_temp} to understanding the role of Li-ion migration in conventional batteries \cite{PARK20107904}.}
\begin{align}\label{eq:Ham_everyday}
 H&=\sum_\sigma\int d{\bm r}\left[\frac{\hbar^2}{2m}|\nabla\hat\psi_\sigma(\bm r)|^2+\tilde{U}(\bm{r})\hat\psi^\dagger_\sigma(\bm r)\hat\psi_\sigma(\bm r) \right] \nonumber\\
 &+\frac{1}{2}\sum_{\sigma,\sigma'}\int d\bm{r} \int d\bm{r}'\hat\psi^\dagger_\sigma(\bm{r})\hat\psi^\dagger_{\sigma'}(\bm{r}')V(|\bm{r}-\bm{r}'|)\hat\psi_{\sigma'}(\bm{r}')\hat\psi_{\sigma}(\bm{r})
\end{align}
describes all the possible non-relativistic physical systems in the absence of external magnetic fields. Here, $\hat\psi^\dagger_\sigma(\bm r)$ $(\hat\psi_\sigma(\bm r))$ is an operator that creates (destroys) an electron at position $\bm r$ of spin $\sigma$. For the sake of notational simplicity, in what follows we will omit the hat when denoting operators. In \cref{eq:Ham_everyday}, $V(|\bm{r}-\bm{r}'|)$ is the distance-dependent repulsive potential between electrons. To derive explicit formulas, in what follows we will consider the screened Coulomb potential $V(|\bm{r}-\bm{r}'|) = q_e(4\pi\epsilon_0)^{-1} e^{-\mu|\bm{r}-\bm{r}'|}/|\bm{r}-\bm{r}'|$, with $\mu$ being the inverse screening length, but our results hold for any positive definite, central, and spin-independent potential. The constants $\hbar$, $m$, $q_e$ and $\epsilon_0$, are Planck's constant, the electron mass, electron charge, and the vacuum permittivity of space respectively.

The abundant phenomena we observe in nature day-to-day is due, in part, to the structure of the potential $\tilde{U}(\bm{r})$, which characterises the Coulomb potential produced by the positively charged nucleus of the atoms in the system.

In materials, the external potential created by the ions in the lattice heavily influences the electrons. Assuming a block of material is invariant under lattice translations $\bm R_n$, the external potential satisfies $\tilde{U}({\bm r}+{\bm R_n})=\tilde{U}({\bm r})$. A usual way of parameterising it is
\begin{equation}\label{external_pot}
 \tilde{U}(\bm r)=\frac{q_e}{4\pi\epsilon_0}\sum_I\frac{Z_I}{|\bm r- \bm r_I|},
\end{equation}
where $Z_I$ is the charge of the ions and $\bm r_I$ is their position.

Starting from this scenario, in this section we discuss how the reduction of the Hamiltonian in \cref{eq:Ham_everyday} (which from now on we assume to represent a block of material, and thus lattice periodic) is performed, leading to a Hamiltonian over finitely many degrees of freedom and with an interaction structure that makes it amenable to simulation using shorter quantum circuits. As quantum simulation brings different communities together, we present a self contained discussion, revising familiar concepts to condensed matter physicists and materials scientists, but which may be not completely familiar to other communities.

\subsection{General characteristics of fermion Hamiltonians}
\label{sec:general_hamiltonians}
\subsubsection{Structure of two and four fermion integrals}
In this section we examine the general properties of the two- and four-fermion integrals occurring in the Hamiltonian of \cref{eq:Ham_everyday}.
We first expand the electron operator $ {\psi}_{\sigma}(\bm{r}) $ in a basis of single-particle wavefunctions $\set{\phi_{\lambda}(\bm{r})} $ as 
\begin{equation}
	{\psi}_{\sigma}(\bm{r}) = \sum_{\lambda} \phi_{\lambda}(\bm{r}) c_{\lambda,\sigma},
\end{equation}
where $ \lambda $ represents the collection of all the particles' quantum numbers but the spin\footnote{In systems with strong spin-orbit coupling, a more general single-particle spinor wavefunction $\phi_{\lambda,\sigma}(\bm{r})$ is possible. We do not consider this case here.}, and $ c_{\lambda,\sigma} $ ($ c^{\dagger}_{\lambda,\sigma} $) is the annihilation (creation) operator for a fermion in the state $ (\lambda,\sigma) $. In terms of the latter, \cref{eq:Ham_everyday} becomes
\begin{equation}\label{eq:H_general}
	H = \sum_{\sigma}\sum_{\lambda_1,\lambda_2} t_{\lambda_1\lambda_2} c^{\dagger}_{\lambda_1,\sigma} c_{\lambda_2,\sigma} + \sum_{\sigma,\sigma'} \sum_{\lambda_1,\lambda_2,\lambda_3,\lambda_4} V_{\lambda_1\lambda_2\lambda_3\lambda_4} c^{\dagger}_{\lambda_1,\sigma} c^{\dagger}_{\lambda_2,\sigma'} c_{\lambda_3,\sigma'} c_{\lambda_4,\sigma}.
\end{equation} 
Here, the \emph{hopping matrix} is defined as 
\begin{equation}\label{eq:HM}
	t_{\lambda_1\lambda_2} = \int d\bm{r}\ \phi^*_{\lambda_1}(\bm{r}) \left[-\frac{\hbar^2\nabla^2}{2m}+
\tilde{U}(\bm{r})\right] \phi_{\lambda_2}(\bm{r}),
\end{equation}
while the \emph{Coulomb tensor} is 
\begin{equation}\label{eq:CT}
	V_{\lambda_1\lambda_2\lambda_3\lambda_4} = \frac{1}{2}\int d\bm{r} \int d\bm{r}'\ \phi^*_{\lambda_1}(\bm{r})\phi^*_{\lambda_2}(\bm{r}') V(|\bm{r}-\bm{r}'|)\phi_{\lambda_3}(\bm{r}')\phi_{\lambda_4}(\bm{r}).
\end{equation}
In particular, both the hopping matrix and the Coulomb tensor are Hermitian, i.e., $	t_{\lambda_1\lambda_2} = t^{*}_{\lambda_2\lambda_1} $ and $ V_{\lambda_1\lambda_2\lambda_3\lambda_4} = V^{*}_{\lambda_4\lambda_3\lambda_2\lambda_1}  $, where $a^*$ denotes the complex conjugate of $a$.  From \cref{eq:CT} it immediately follows that the Coulomb tensor obeys the index-swap symmetry $ V_{\lambda_1\lambda_2\lambda_3\lambda_4} = V_{\lambda_2\lambda_1\lambda_4\lambda_3}  $. 

\subsubsection{Cauchy-Schwarz inequality for the Coulomb tensor}
\label{sec:CS_ineq}
Exploiting the fact that $ V(|\bm{r}-\bm{r}'|) $ is a real positive definite function, one can rewrite \cref{eq:CT} in terms of an inner product. The latter can be defined in two possible ways. The first one is
\begin{equation}
\label{eq:CT_inner_prod_1}
	V_{\lambda_1\lambda_2\lambda_3\lambda_4} \equiv \langle \rho_{\lambda_1\lambda_2}, \rho_{\lambda_4\lambda_3}   \rangle_1 = \frac{1}{2}\int d\bm{r} \int d\bm{r}'\ V(|\bm{r}-\bm{r}'|) \rho^*_{\lambda_1\lambda_2}(\bm{r}, \bm{r}') \rho_{\lambda_4\lambda_3}(\bm{r}, \bm{r}'),
\end{equation}
where $ \rho_{\lambda_i\lambda_j}(\bm{r}, \bm{r}') \equiv \phi_{\lambda_i}(\bm{r}) \phi_{\lambda_j}(\bm{r}')$. Hence, the following inequality between the elements of the Coulomb tensor follows from the Cauchy-Schwarz inequality applied to \cref{eq:CT_inner_prod_1}
\begin{equation}
\label{eq:CT_Cauchy-Schwarz_1}
	|V_{\lambda_1\lambda_2\lambda_3\lambda_4}|^2 \leq |V_{\lambda_1\lambda_2\lambda_2\lambda_1} V_{\lambda_4\lambda_3\lambda_3\lambda_4}|.
\end{equation}
On the other hand, another well-defined inner product can be introduced as
\begin{equation}
\label{eq:CT_inner_prod_2}
	V_{\lambda_1\lambda_2\lambda_3\lambda_4} \equiv \langle \rho'_{\lambda_1\lambda_4}, \rho'_{\lambda_2\lambda_3}   \rangle_2 = \frac{1}{2}\int d\bm{r} \int d\bm{r}'\ V(|\bm{r}-\bm{r}'|) \rho^{\prime *}_{\lambda_1\lambda_4}(\bm{r}) \rho'_{\lambda_3\lambda_2}(\bm{r}'),
\end{equation}
where $ \rho'_{\lambda_i\lambda_j}(\bm{r}) \equiv \phi_{\lambda_i}(\bm{r}) \phi^*_{\lambda_j}(\bm{r})$. Similarly to the previous case, the Cauchy-Schwarz inequality associated with this inner product implies the following relation between the Coulomb tensor elements
\begin{equation}
\label{eq:CT_Cauchy-Schwarz_2}
	|V_{\lambda_1\lambda_2\lambda_3\lambda_4}|^2 \leq |V_{\lambda_1\lambda_4\lambda_1\lambda_4} V_{\lambda_3\lambda_2\lambda_3\lambda_2}|.
\end{equation}
\cref{eq:CT_Cauchy-Schwarz_1} and \cref{eq:CT_Cauchy-Schwarz_2} can be exploited to obtain bounds on the Coulomb tensor coefficients, allowing one to truncate the elements smaller than a given threshold without having to directly compute them. This is usually very useful in reducing the classical computation needed to determine a quantum Hamiltonian.

\subsection{Momentum-space single-particle bases}

In this and the following sections we will introduce some of the most common single-particle bases to study condensed matter systems. 
As we will be discussing different bases for the same Hamiltonian, to avoid confusion, especially when these Hamiltonians are mapped into qubit operators, we will explicitly add a superscript to a Hamiltonian in a particular basis, each of which will be defined below. We will have:
\begin{itemize}
    \item $H^P$: Hamiltonian \cref{eq:Ham_everyday} in the plane wave single-particle electron basis. The second quantized creation (annihilation) operators of momentum $k$ and spin $\sigma$ in this context are denoted by $c^\dagger_{k,\sigma}$ $(c_{k,\sigma})$. Choosing a lattice of discrete translations, the total momentum $k$ can always be decomposed in the lattice momentum $\bm{k}$ and reciprocal lattice vector $\bm{G}$ as $k=\bm{k}+\bm {G}$.
    \item $H^B$: Hamiltonian \cref{eq:Ham_everyday} in the Bloch-wave single-particle electron basis. The creation (annihilation) operators are $f^\dagger_{\bm{k},n,\sigma}$ ($f_{\bm{k},n,\sigma}$), with $\bm{k}$ the lattice momentum, $n$ the band index and $\sigma$ the spin.
    \item $H^W$: Hamiltonian \cref{eq:Ham_everyday} in the Wannier single-particle electron basis. The creation (annihilation) operators of band $n$ and spin $\sigma$ are $w^\dagger_{\bm{R},n,\sigma}$ ($w_{\bm{R},n,\sigma}$), where $\bm{R}$ is the lattice vector.
\end{itemize}
All single-particle basis operators (called generically $A_j$) satisfy the equal-time anti-commutation relations $\{A_i,A^\dagger_j\}=\delta_{ij}$.

\begin{figure}[ht]
 \centering
 \includegraphics[scale=0.3]{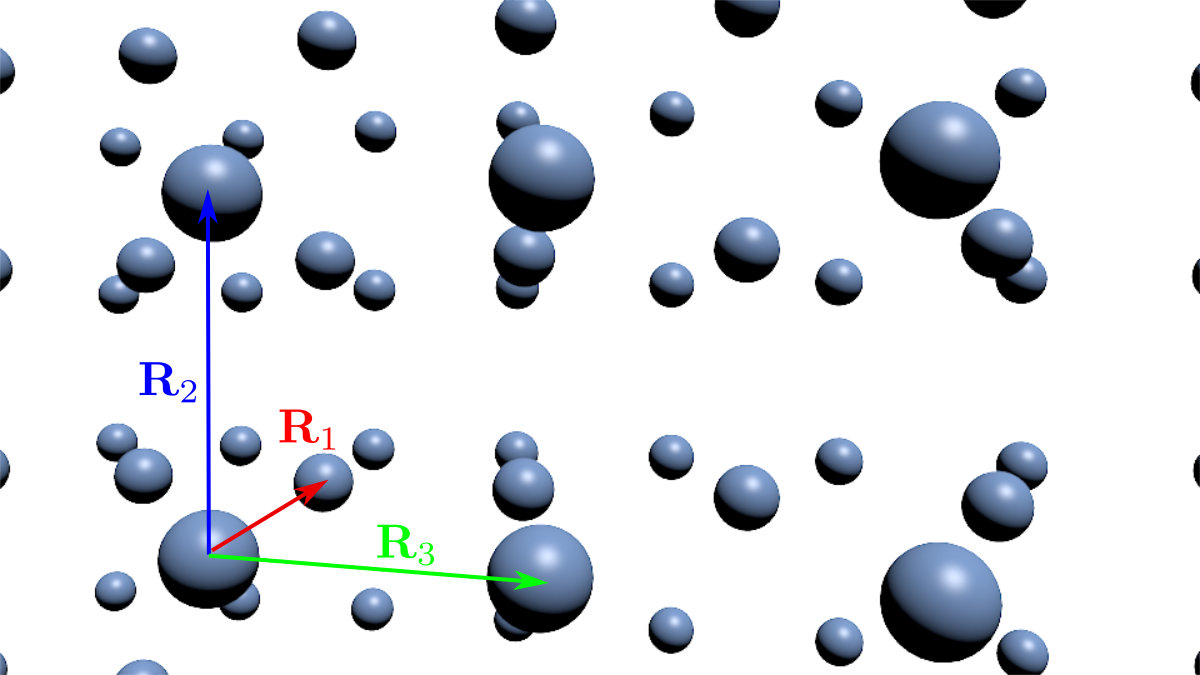}
 \caption{Simple Orthorhombic Bravais lattice, with its lattice vectors.}
  \label{fig:lattice}
 \end{figure}

We begin with momentum-space bases, which fully exploit the translational invariance of crystalline solids. For more details see e.g., Refs.~\cite{Ashcroft_book, Solyom1_book}.
In a material, atoms are arranged in a periodic structure (see \cref{fig:lattice}) which is spanned by the lattice vectors $\bm{R}_a$, $a=1,2,3$. The lattice points correspond to 
\begin{equation}
\label{eq:R_lattice}
 \bm{R}= n_1\bm{R}_1+n_2\bm{R}_2+n_3\bm{R}_3,
\end{equation}
where $n_a\in \mathbb{Z}$. The lattice vector $\bm R_a$ has length $R_a$.  Translations $T_{\bm{R}}$ along these lattice vectors  leave the Hamiltonian $H$ invariant (assuming periodic boundary conditions). Consequently, we can block-diagonalize the Hamiltonian, 
and each block will correspond to a different eigenvalue of the translation operator. The Bloch Theorem allows us to find the simultaneous eigenfunctions of $T_{\bm{R}}$ and 
$H$ \cite{Ashcroft_book, Solyom1_book}. 

The translation operator $T_{\bm R}$ forms an Abelian group, satisfying $T_{\bm R}T_{\bm R'}=T_{{\bm R}+{\bm R'}}$, with $T_{\bm 0}=1$. As $T$ should be represented by a unitary 
operator, in its diagonal basis it acts on the single-particle wavefunctions as 
\begin{equation}\label{bloch_trans}
 T_{\bm R}\phi(\bm r)\equiv\phi({\bm r}+{\bm R})=e^{i{\bm k}\cdot{\bm R}}\phi({\bm r}),
\end{equation}
where the vector ${\bm k}$ is called the crystal momentum\footnote{Note that the crystal momentum does {\bf not} coincide with the momentum of the particle. The latter can be obtained from its group 
velocity according to ${\bm v}_n({\bm k})=\frac{1}{\hbar}\nabla_{\bm k}E_n({\bm k}),$ where $E_n({\bm k})$ is the energy of $n-$th band.}.
In a periodic system with linear size $L_a=N_aR_a$ in each lattice vector direction, the periodic boundary conditions (Born-von Karman boundary conditions) 
$\psi({\bm r}+N_a{\bm R_a})=\psi({\bm r})$ imply the quantization of the crystal momentum as
\begin{equation}\label{allowed_k}
 {\bm k}=\frac{n_1}{N_1}{\bm b_1}+\frac{n_2}{N_2}{\bm b_2}+\frac{n_3}{N_3}{\bm b_3},
\end{equation}
where the reciprocal lattice vectors ${\bm b_j}$ satisfy ${\bm b_i}\cdot{\bm R_j}=2\pi\delta_{ij}$
and $n_a\in [0, N_a - 1]$. The eigenstates of the translation operator $T_{\bm{R}}$ can then be labelled by the triplet $n_1, n_2, n_3$, corresponding to a total of $N=N_1N_2N_3$ states.
The total volume of the crystal is $V_c=N \Omega$, with $\Omega=|{\bm R}_1\cdot({\bm R}_2\times {\bm R}_3)|$ the volume of the unit cell. The relation between direct and reciprocal lattice is shown in \cref{fig:Direct_and_rep_lattice}. 

Using \cref{bloch_trans}, we can define
\begin{equation}
 \label{eq:Bloch_functions}
 \phi_{\bm k}({\bm r})=e^{i{\bm k}\cdot{\bm r}}[e^{-i{\bm k}\cdot{\bm r}}\phi_{\bm k}({\bm r})]=e^{i{\bm k}\cdot{\bm r}}u_{\bm k}({\bm r}),
\end{equation}
with $u_{\bm k}({\bm r}+{\bm R})=u_{\bm k}({\bm r})$ a lattice periodic function. The single electron wavefunction $\phi_{\bm k}({\bm r})=e^{i{\bm k}\cdot{\bm r}}u_{\bm k}({\bm r})$ is called a Bloch wave \cite{Ashcroft_book, Solyom2_book}. Since $u_{\bm k}({\bm r})$ is a lattice periodic function, it may be useful to expand it in Fourier series as 
\begin{equation}
    u_{\bm{k}}(\bm{r}) = \sum_{\bm{G}} u_{\bm{k}, \bm{G}} e^{i\bm{G}\cdot \bm{r}},
\end{equation}
where ${\bm G}$ is a reciprocal lattice vector 
${\bm G}=m_1{\bm b_1}+m_2{\bm b_2}+m_3{\bm b_3}$, with $m_i \in \mathbb {Z}$,
and to write the Bloch wave as
\begin{equation}
    \label{eq:Bloch_wave_Fourier}
    \phi_{\bm{k}}(\bm{r}) = \sum_{\bm{G}} u_{\bm{k},\bm{G}} e^{i(\bm{k}+\bm{G})\cdot \bm{r}}.
\end{equation}

\subsubsection{Plane wave basis}

\begin{figure}[t!]
 \includegraphics[width=\linewidth]{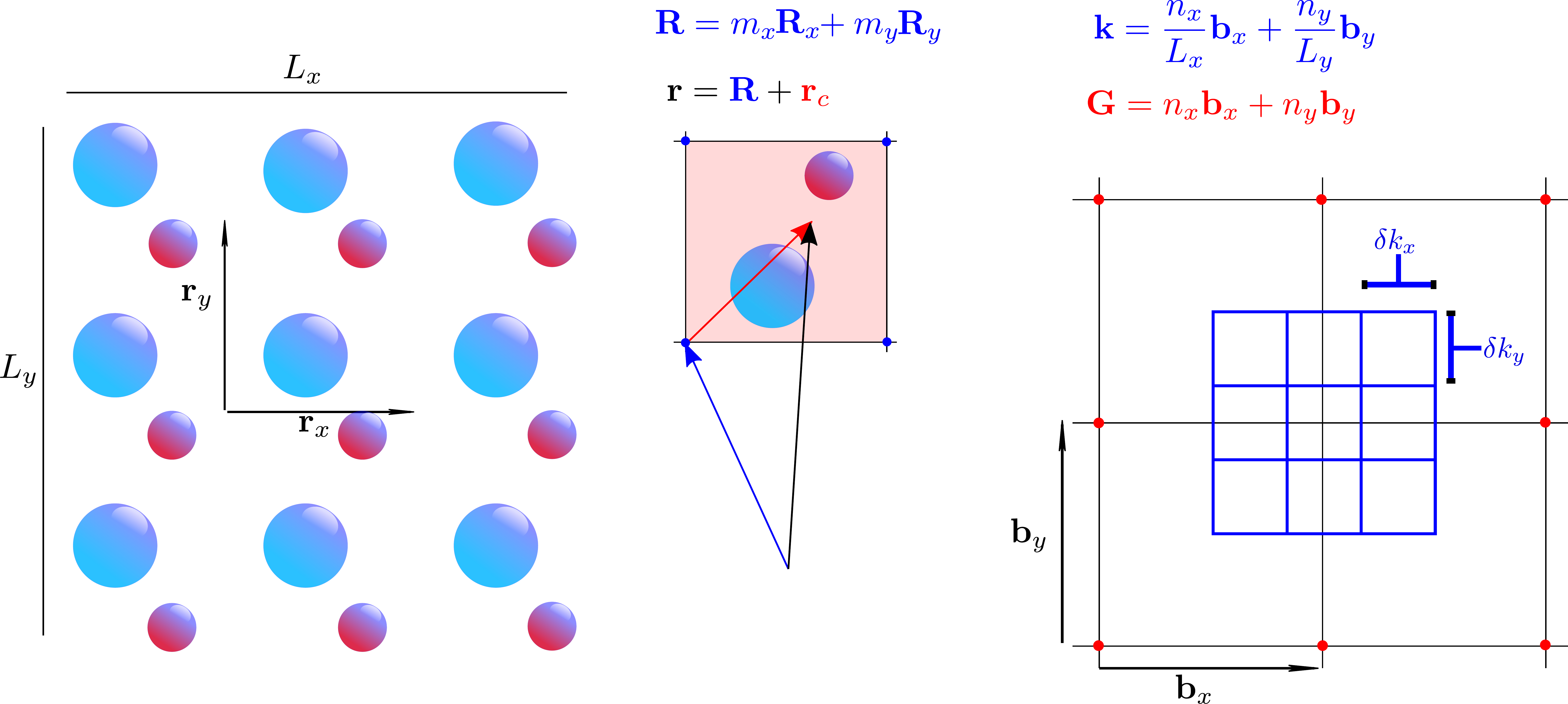}
 \caption{Direct and reciprocal lattice of a 2D square Bravais lattice.
 Left: Square Bravais lattice with two atoms per unit cell. The lattice vectors are $\bm r_x$ and $\bm r_y$ and the size of the system is $L_x$ and $L_y$. Middle: Every position $\bm r$ in the material can be decomposed in the position of the cell $\bm R$ and a position inside the cell $\bm r_c$. Right: The reciprocal lattice (black grid) is unbounded. The finer $k$-mesh (blue lattice) represents the different lattice momentum states $\bm k$. The system size determines the number of $k$ points and does not affect the reciprocal lattice.}
 \label{fig:Direct_and_rep_lattice}
\end{figure}

A particularly simple choice for the functions $u_{\bm{k}}(\bm{r})$ is $u_{\bm{k}}(\bm{r}) = N^{-1}\sum_{R} \delta(\bm{r}-\bm{R})$, with $\delta(\bm{R})$ the Dirac delta function. This choice implies that all the Fourier coefficients $u_{\bm{k}, \bm{G}}$ in  \cref{eq:Bloch_wave_Fourier} are set to 1 and, therefore, it corresponds to  expanding the Bloch wave $\phi_{\bm{k}}(\bm{r})$ in the plane wave basis $\set{e^{i(\bm{k}+\bm{G})\cdot\bm{r}}}$. An advantage of this basis is that plane waves for different momenta are orthogonal. The electron operator takes the form
\begin{equation}
    \psi_\sigma(\bm{r}) = \frac{1}{\sqrt{V_c}}\sum_{{\bm k}, \bm{G}}e^{i({\bm k}+\bm{G})\cdot{\bm r}} c_{\bm{k} + \bm{G}, \sigma},
\end{equation}
where $c_{{\bm k}+{\bm G},\sigma}$ ($c^\dagger_{{\bm k}+{\bm G},\sigma}$) is the annihilation (creation) operator of an electron with momentum $ {\bm k}+{\bm G} $ and spin $\sigma$. In the plane wave basis, the Hamiltonian of \cref{eq:H_general} becomes
\begin{equation}\label{Ham_recip_lattice}
H^{P}=\sum_{\bm k,\bm G,\bm G',\sigma}h_{\bm{k},\bm{G}-\bm{G}'}c^\dagger_{ \bm k+ \bm G,\sigma}c_{\bm k+\bm G',\sigma}
+\sum_{\substack{\bm k,\bm k',p\\ \bm G,\bm G',\sigma,\sigma'}} V_{p}c^\dagger_{\bm k+\bm G + p,\sigma}c^\dagger_{\bm k' +\bm G' - p,\sigma'}c_{\bm k' + \bm G',\sigma'}c_{\bm k+\bm G,\sigma},
\end{equation}
where $h_{\bm{k},\bm{G}-\bm{G}'}=\left[\frac{|\hbar(\bm k+\bm G)|^2}{2m}\delta_{\bm G,\bm G'}+U_{\bm G-\bm G'}\right]$, $V_p = 1/(2V_c) \int d\bm{r} e^{-i p \cdot \bm{r}} V(|\bm{r}|)$, and $p=\bm p +\bm P$ is the total momentum, with $\bm p = \bm k + \bm k'$ and $\bm P = \bm G + \bm G'$. $U_{\bm G}$ is 
the Fourier component of the external lattice potential at reciprocal lattice vector $\bm G$
\begin{equation}
\tilde{U}(\bm r)=\sum_{\bm G}e^{i{\bm G} \cdot \bm r}U_{\bm G}\quad\rightarrow\quad U_{\bm G}=\frac{1}{\Omega}\int_{uc}d{\bm r}e^{-i{\bm G} \cdot \bm r}\tilde{U}(\bm r),
\end{equation}
where the integral is over the unit cell. Using \cref{external_pot}, we find
\begin{equation}
 U_{\bm G}=\frac{q_e}{\epsilon_0\Omega}\sum_{a}Z_a\frac{e^{i\bm r_a\cdot \bm{G}}}{|\bm G|^2},\quad \mbox{for $\bm G\neq0$}\quad (U_0=0),
\end{equation}
where the sum runs over the positions $\bm r_a$ of the atoms in the unit cell (see \cref{fig:Direct_and_rep_lattice}). 

\subsubsection{Bloch wave basis}

Going back to the Hamiltonian of \cref{Ham_recip_lattice}, in the non-interacting limit we see that the lattice momentum $\bm k$ enters as a parameter,
\begin{equation}\label{Ham_non_int}
H_0=\sum_{\bm k,\bm G,\bm G',\sigma}\left[\frac{|\hbar(\bm k+\bm G)|^2}{2m}\delta_{\bm G,\bm G'}+U_{\bm G-\bm G'}\right]c^\dagger_{ \bm k+ \bm G,\sigma}c_{\bm k+\bm G',\sigma}.
\end{equation}
This implies that we can decompose the Hamiltonian in different crystal momentum blocks as $  H=\sum_{\bm k} H_0(\bm k) $ and solve an independent Schr\"odinger equation for each of them,
\begin{equation}
\label{eq:momentum_schro_eq}
H_0(\bm k)|\Psi_n(\bm k)\rangle= \epsilon_n(\bm k)|\Psi_n(\bm k)\rangle,
\end{equation}
with $|\Psi_n(\bm k)\rangle$ being a two-component spinor state.
It is useful to define a particular zone of $\bm k$ values called the Brillouin zone, which corresponds to the Wigner-Seltz cell construction in reciprocal space, i.e.,
the locus of points $\bm k$ in the reciprocal space which is closer to $\bm G=0$. 
The eigenvalues $\epsilon_n(\bm k)$ define the energy bands of the system.

The expansion of the non-interacting Hamiltonian in the basis defined by the momentum block eigenstates of \cref{eq:momentum_schro_eq}, can be obtained by diagonalising $h_{\bm{k}, \bm{G}-\bm{G}'}$ in \cref{Ham_recip_lattice}, 

\begin{equation}
 \label{eq:band_fermions}
 H_0({\bm k})\equiv \sum_{\bm G, \bm G',\sigma} h_{\bm{k},\bm G-\bm G'}c_{\bm k,\bm G,\sigma}^\dagger c_{\bm k,\bm G',\sigma}=\sum_{n, \sigma}\epsilon_n(\bm k)f_{\bm k, n,\sigma}^\dagger f_{\bm k, n,\sigma},
\end{equation}
where $f_{\bm k, n, \sigma}=\sum_{\bm G}S_{n,\bm G}(\bm k, \sigma)c_{\bm k+ \bm G, \sigma}$ ($f^\dagger_{\bm k, n, \sigma}$) is the band fermion annihilation (creation) operator. Here, $S_{n,\bm G}(\bm k, \sigma)$ is the unitary matrix
that diagonalises $h$, i.e., $h_{\bm k, \bm G- \bm G'}=\sum_n(S^\dagger)_{\bm G, n}(\bm k, \sigma)\epsilon_n(\bm k)S_{n,\bm G'}(\bm k, \sigma)$. The index $n$ here denotes the band
and takes the same number of values $|n|$ as the reciprocal lattice vectors $\bm G$, i.e., $|n|=8G_{\rm max}^3$ in $D=3$ dimensions.

For a system with $\nu_{\rm el}$ electrons per unit cell (i.e., corresponding to a total of $\nu_{\rm el}\times N$ electrons), the system will have $\nu_{\rm el}$ occupied bands, as each band can accommodate $N$
states, which is the number of different lattice momentum values in the Brillouin zone (note that $\nu_{\rm el}$ can be a rational number, in which case there are $\lfloor\nu_{\rm el} \rfloor$ fully occupied bands
and the last band $\lceil \nu_{\rm el} \rceil$ is partially occupied).

In real space, the non-interacting Hamiltonian corresponding to each momentum block is $H_0(\bm{k}) = \frac{\hbar^2}{2m}(\bm{k} - i\nabla)^2 + \tilde{U}(\bm{r}) $. The spin components of its eigenstates coincide with the periodic functions $u_{\bm k, n, \sigma}(\bm r)$ introduced in \cref{eq:Bloch_functions}, i.e.,
\begin{equation}
\label{eq:real_space_eigeneq}
H_0(\bm k)u_{\bm k, n, \sigma}(\bm r)= \epsilon_n(\bm k)u_{\bm k, n, \sigma}(\bm r),  
\end{equation}
with the boundary condition $u_{\bm k+\bm G,n,\sigma}(\bm r) =e^{-i\bm G\cdot\bm r}u_{\bm k,n,\sigma}(\bm r)$. Note that the functions $u_{\bm k,n,\sigma}(\bm r)$ are defined within the unit cell via $u_{\bm k,n,\sigma}(\bm r)=u_{\bm k,n,\sigma}(\bm r_c+\bm R)=u_{\bm k,n,\sigma}(\bm r_c)$, where $\bm{R}$ is a lattice vector and $\bm r_c$ is a vector with domain in the unit cell. By expanding $u_{\bm{k},n,\sigma}(\bm{r})$ in \cref{eq:real_space_eigeneq} in Fourier series one can verify that
\begin{equation}
 u_{\bm k,n,\sigma}(\bm r)=\sum_{\bm G}e^{i\bm G\cdot \bm r}S_{n,\bm G}(\bm k,\sigma).
\end{equation}

In the language of \cref{sec:general_hamiltonians}, what we have done so far corresponds to expanding the electron operator on a Bloch wave basis (also called band fermion basis) $\set{\phi_{\bm{k},n,\sigma}(\bm{r})}$, with $\phi_{\bm{k},n,\sigma}(\bm{r}) = e^{i\bm{k}\cdot\bm{r}}u_{\bm{k},n,\sigma}(\bm{r})/\sqrt{V_c}$. 
In this basis, the full Hamiltonian is 
\begin{equation}\label{Ham_band_basis}
H^{B}=\sum_{\bm k, n, \sigma}\epsilon_n(\bm k) f^\dagger_{\bm k,n, \sigma}f_{\bm k,n, \sigma}
+ \sum_{\sigma, \sigma'} \sum_{\substack{n_1,n_2,n_3,n_4\\ \bm k, \bm q, \bm k'}} V_{n_1n_2n_3n_4}^{(\bm k,\bm k',\bm q)}f^\dagger_{\bm k + \bm q,n_1,\sigma}f^\dagger_{\bm k' - \bm q, n_2, \sigma'}f_{\bm k',n_3, \sigma'}f_{\bm k,n_4,\sigma},
\end{equation}
with Coulomb tensor coefficients
\begin{equation}
 V_{n_1n_2n_3n_4}^{(\bm k,\bm k',\bm q)}=\sum_{\bm G,\bm K,\bm G'} V_{\bm{q}+\bm{K}} S_{n_1,\bm G + \bm K}(\bm k + \bm q, \sigma){S}_{n_2,\bm G'-\bm K}(\bm k' - \bm q, \sigma')S^*_{n_3,\bm G'}(\bm k', \sigma')S^*_{n_4, \bm G}(\bm k, \sigma).
\end{equation}
and $V_{\bm q+ \bm K}$ defined after \cref{Ham_recip_lattice}.

\subsection{Real-space single-particle basis: Wannier functions}
\label{sec:Wannier_functions}

In \cref{Ham_band_basis}, the quadratic part of the Hamiltonian is diagonal, but the electron-electron interaction is highly non-local. {On the other hand, in a real-space coordinate basis (such as the one obtained by discretising the position operator $\bm{r}$ on a real-space grid),  the electron-electron interaction is diagonal but non-local, while the kinetic term is not diagonal.} 
{To reduce the number of the relevant coefficients entering the Hamiltonian, one strategy is to look for a representation where both the hopping matrix and Coulomb tensor are not diagonal with respect to the single-particle basis, but as local (in real space) as possible.}
One convenient way to achieve this goal is to consider Wannier functions as the single-particle basis. The fermion annihilation operators associated with the latter are defined in \cite{Marzari12} as 
\begin{equation}
\label{eq:Wannier_operators}
 w_{\bm{R},n, \sigma}=\sum_{\bm{k},m}e^{i\bm{k}\cdot\bm{R}}U^{*}_{mn}(\bm{k})f_{\bm{k},m, \sigma}\rightarrow f_{\bm{k},m, \sigma}=\frac{1}{N}\sum_{\bm{R},n}e^{-i\bm{k}\cdot\bm{R}}U_{mn}(\bm{k})w_{\bm{R},n, \sigma},
\end{equation}
where $U_{mn}(\bm k)$ is a unitary transformation representing the gauge freedom in the definition of the Bloch waves.
In this basis, the Hamiltonian of \cref{eq:H_general} becomes 
\begin{align}\label{Ham_wannier}
H^{W}&=\sum_{\sigma}\sum_{\substack{m,n\\\bm{R}_1,\bm{R}_2}}T(\bm{R}_1-\bm{R}_2)_{mn}w_{\bm{R}_1,m,\sigma}^{\dagger}w_{\bm{R}_2,n, \sigma}\nonumber\\
&+\sum_{\sigma, \sigma'}\sum_{\substack{s,l,m,n\\\bm{R}_{1},\bm{R}_{2},\bm{R}_3,\bm{R}_4}}
\tilde{V}^{(\bm{R}_{1}\bm{R}_{2},\bm{R}_3,\bm{R}_4)}
_{slmn}w_{\bm{R}_{1},s, \sigma}^{\dagger}w_{\bm{R}_{2},l, \sigma'}^{\dagger}w_{\bm{R}_3,m, \sigma'} w_{\bm{R}_4,n, \sigma},
\end{align}
with the matrix elements 
\begin{align}
 T(\bm R)_{mn}&=\frac{1}{N^2}\sum_{\bm{k}}e^{i\bm{k}\cdot \bm{R}}[U(\bm{k})\epsilon(\bm{k})U^{\dagger}(\bm{k})]_{nm},\label{eq:Wannier_T}\\
 \tilde{V}_{slmn}^{(\bm R_1,\bm R_2,\bm R_3,\bm R_4)}&=\frac{1}{N^4}\sum_{\substack{n_{1}n_{2}n_{3}n_{4}\\\bm{k},\bm{q},\bm{k}'}}
 V_{n_{1}n_{2}n_{3}n_{4}}^{(\bm{k},\bm{k}',\bm{q})}U^{*}_{ n_{1}s}(\bm{k} + \bm{q})U^{*}_{n_{2} l}(\bm{k} - \bm{q})U_{ n_{3} m}(\bm{k}')U_{ n_{4} n}(\bm{k})\nonumber\\&\times
 e^{i\bm{k}\cdot(\bm{R}_{1}-\bm{R}_4)}
 e^{i\bm{q}\cdot(\bm{R}_{1}-\bm{R}_{2})}
 e^{i\bm{k}'\cdot(\bm{R}_2-\bm{R}_3)}.
\end{align}

The Wannier functions corresponding to the operators in \cref{eq:Wannier_operators} are
\begin{equation}
\mathcal{W}^{\bm{R}}_{s, \sigma}(\bm{r})=\mathcal{W}^{\bm{0}}_{s, \sigma}(\bm{r}-\bm{R})=\sum_{\bm{k},n}e^{-i\bm{k}\cdot\bm{R}} U_{ns}(\bm{k}) u_{\bm{k},n,\sigma}(\bm{r})e^{i\bm{k}\cdot\bm{r}}.
\end{equation}
In terms of the latter, the matrix elements of the hopping matrix and the Coulomb tensor can expressed as
\begin{subequations}
\label{eq:Wannier_T_V}
    \begin{align}
    T(\bm{R}_1 -\bm{R}_2)_{mn} &= \int d\bm{r} \mathcal{W}^{\bm{R}_1*}_{m,\sigma}(\bm{r}) \left[-\frac{\hbar^2\nabla^2}{2m}+
    \tilde{U}(\bm{r})\right]\mathcal{W}^{\bm{R}_2}_{n,\sigma}(\bm{r}), \\
    \tilde{V}_{slmn}^{(\bm R_1,\bm R_2,\bm R_3,\bm R_4)}&=\frac{1}{2}
    \int d\bm{r} \int d\bm{r}'\mathcal{W}_{s, \sigma}^{\bm{R}_{1}*}(\bm{r})\mathcal{W}_{l, \sigma'}^{\bm{R}_2*}(\bm{r}')V(|\bm{r}-\bm{r}'|)\mathcal{W}_{m, \sigma'}^{\bm{R}_3}(\bm{r}')\mathcal{W}_{n, \sigma}^{\bm{R}_4}(\bm{r}).\label{eq:Wannier_CT}
    \end{align}
\end{subequations}

The discrete translational invariance of the lattice allows us to rewrite the coefficients above as
\begin{subequations}
\label{eq:Wannier_T_V_core}
    \begin{align}
    T(\bm{R})_{mn} &= \int d\bm{r} \mathcal{W}^{\bm{R}}_{m,\sigma}(\bm{r}) \left[-\frac{\hbar^2\nabla^2}{2m}+
    \tilde{U}(\bm{r})\right]\mathcal{W}^{\bm{0}}_{n,\sigma}(\bm{r}), \\
    \tilde{V}_{slmn}^{(\bm 0,\bm R_2,\bm R_3,\bm R_4)}&=\frac{1}{2}
    \int d\bm{r} \int d\bm{r}'\mathcal{W}_{s, \sigma}^{\bm{0}}(\bm{r})\mathcal{W}_{l, \sigma'}^{\bm{R}_2}(\bm{r}')V(|\bm{r}-\bm{r}'|)\mathcal{W}_{m, \sigma'}^{\bm{R}_3}(\bm{r}')\mathcal{W}_{n, \sigma}^{\bm{R}_4}(\bm{r}),\label{eq:Wannier_CT_core}
    \end{align}
\end{subequations}
with $\bm{R} = \bm{R}_1 - \bm{R}_2$.

Since the Coulomb tensor coefficients $\tilde{V}_{slmn}^{(\bm 0,\bm R_2,\bm R_3,\bm R_4)}$ involves integrals over the real space, if the Wannier functions $\mathcal{W}_{s, \sigma}^{\bm R}(\bm{r})$ are localized around $\bm R$, then the coefficients will decay fast for distant cells in the lattice. From the definition of the Wannier functions we have $ \mathcal{W}^{\bm{0}}_{s, \sigma}(\bm{r}) \equiv \sum_{\bm{k}}v_{\bm{k},{s}, \sigma}(\bm{r})e^{i\bm{k}\cdot\bm{r}}$, where $v_{\bm{k},{s}, \sigma}(\bm r) = \sum_{n} U_{ns}(\bm{k}) u_{\bm{k},n,\sigma}(\bm{r}) $ are quasi-Bloch functions. This relation tells us that the quasi-Bloch functions and the Wannier functions are related by a Fourier transform.  As discussed in \cref{app:MLWFs}, we can then use the analytical form of the quasi-Bloch functions $v_{\bm{k},{s},\sigma}(\bm r)$ as a function of the crystal momentum $\bm{k}$ to show that \glspl{mlwf} can be obtained if the following conditions are satisfied \cite{Marzari12,Cloizeaux1,Cloizeaux2,Nenciu1983}:

\begin{enumerate}
 \item The system has a vanishing Chern number. This condition is automatically satisfied in systems with time-reversal symmetry, as in this case where the Chern number is zero. Note that systems without time-reversal symmetry can still have a vanishing Chern number.
 \item An energy gap exists between the bands in the active space (see \cref{sec:active_space} below) and the rest. Note that a system satisfying this condition does not necessarily represent an insulator, as the Fermi energy can lie within an active space which is separated from the rest of the bands.
\end{enumerate}

In this case, both the non-interacting and the electron-electron interaction terms of the Hamiltonian in the Wannier basis of \cref{Ham_wannier} contain only local terms. Moreover, \glspl{mlwf} are always real~\cite{Brouder2007}. This fact, combined with the hermiticity of the hopping matrix $T(\bm{R})_{mn} = T(-\bm{R})_{nm}$ (see \cref{eq:HM_hermiticity} below), implies
\begin{equation}
    \label{eq:Wannier_hopping_matrix_hermiticity}
    T(\bm{R})_{mn} = T(-\bm{R})_{nm}.
\end{equation}
This identity restricts the type of quadratic terms that can appear in the Hamiltonian. As we will discuss in \cref{sec:symmetries} below, symmetry properties of the physical system further constrain the form of the Hamiltonian coefficients.

\subsubsection{Material real-space motif}
\label{sec:motif}
\begin{figure}[ht]
\centering
 \includegraphics[width=0.4\textwidth]{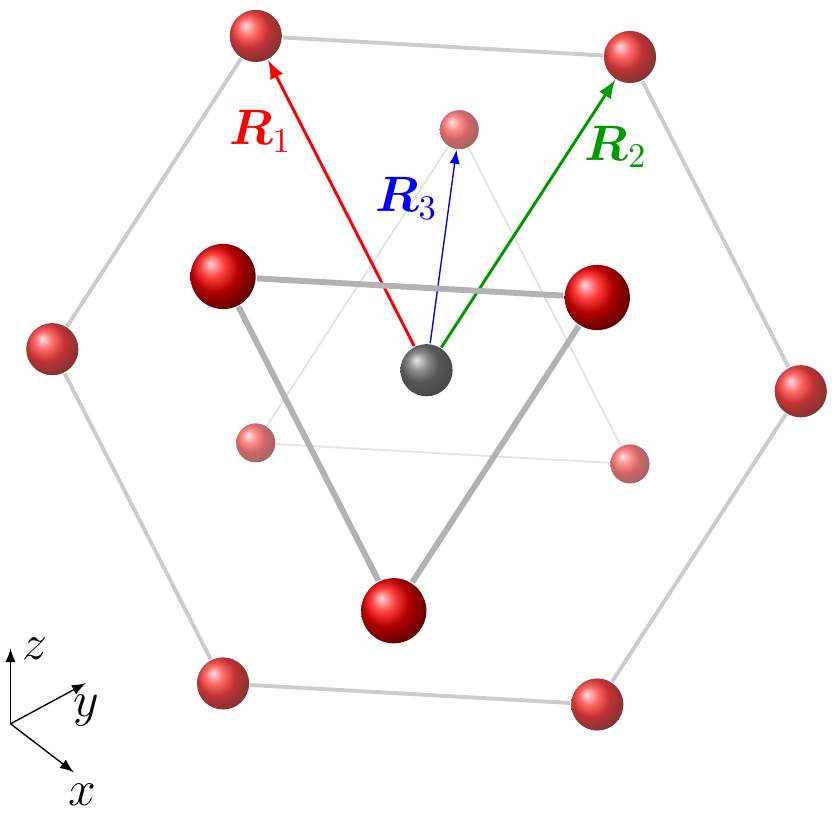} \hspace{1cm}
 \includegraphics[width=0.4\textwidth]{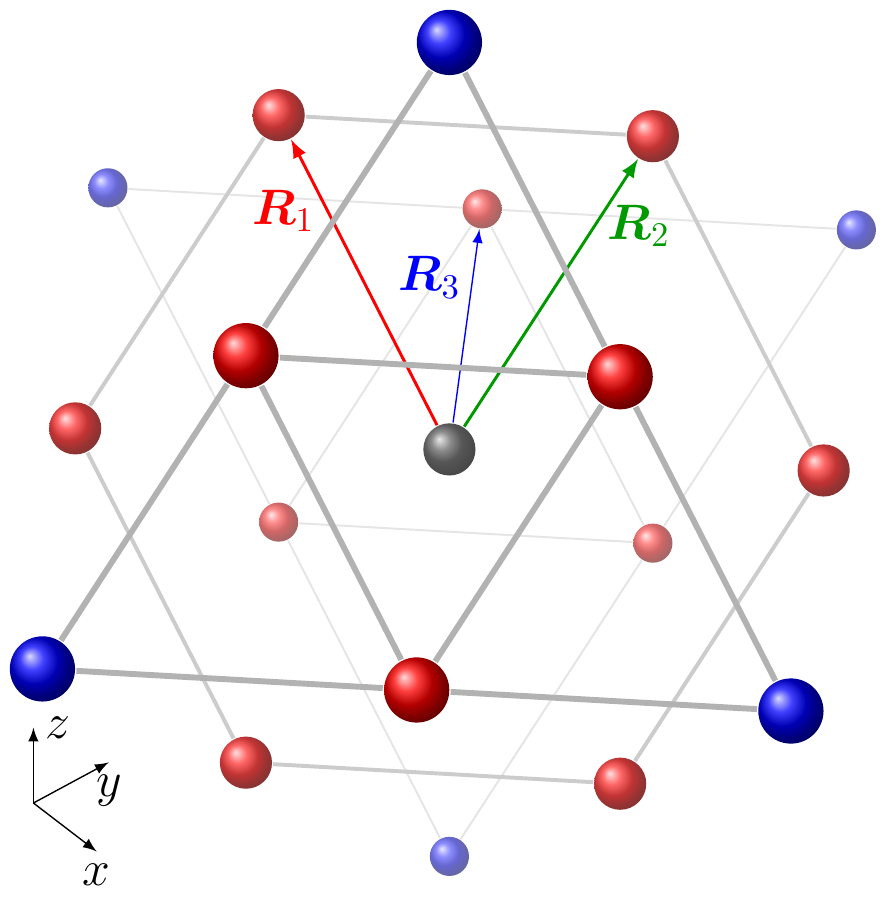} 
 \caption{Motif of order (left) $n=1$ and (right) $n=2$ for silicon. The motif of order $1$ contains the central cell (black) and its $12$ nearest neighbours (red), while the motif of order $2$ includes also the $6$ next-nearest neighbours (blue). The gray links connect the nearest and next-nearest neighbouring cells belonging to the same (111) crystallographic plane, with thinner lines denoting planes with larger distance with respect to the reader along the direction perpendicular to the page. $\bm{R}_1$, $\bm{R}_2$, $\bm{R}_3$ denote the silicon lattice vectors. }
 \label{fig:Si_motif}
\end{figure}

The key advantage of using a localised single-particle basis is that the magnitude of the hopping matrix and Coulomb tensor coefficients of \cref{eq:Wannier_T_V_core} decreases quickly as a function of the distance between the unit cells involved in the integrals. A natural approximation is thus to consider only those coefficients involving cells which are reciprocal nearest neighbours of order $n$, with the latter depending on the material and on the degree of accuracy required for the Hamiltonian coefficients. In particular, we say that two unit cells $A$ and $B$, identified by the lattice vectors $\bm{R}_A$ and $\bm{R}_B$, are nearest neighbors of order $n$ if $|\bm{R}_A - \bm{R}_B| = d_n$, with $|...|$ denoting the Euclidean norm. Here, $ 0 = d_0 < d_1 < d_2 < ... < d_n$ is a ascending sequence of distances corresponding to nearest neighbors of order $0$, $1$, $2$, and $n$, respectively. Within this approximation, we do not need to evaluate the hopping matrix and the Coulomb tensor coefficients over the whole material lattice but we can rather focus on a minimal set of cells which includes a central cell and its nearest neighbors up to order $n$. All the other coefficients can then be obtained by exploiting the discrete translational invariance of the lattice. For this reason, we will name such a minimal set of unit cells a \emph{motif} of order $n$. As an example, in \cref{fig:Si_motif} we show the motifs of order $n=1$ and $n=2$ for silicon. Note that the motif is formed by the material unit cells and not by its individual atoms. We denote by $\mathcal{N}^n_{\mathcal{G}'}$ the set of all the unit cells of a material lattice $\mathcal{G}$ which are nearest neighbors of order $\leq n$ with respect to all the unit cells of a sublattice $\mathcal{G}'\subseteq\mathcal{G}$, i.e., $\mathcal{N}^n_{\mathcal{G}'} \equiv \set{A \in \mathcal{G} | |\bm{R}_A-\bm{R}_B| \leq d_n, \forall B\in\mathcal{G}'}$. The motif of order $n$ centered on the unit cell $O$ coincides with $\mathcal{N}^n_{O}$ and contains $N_{\mathrm{cells/motif}} = \mathrm{dim}(\mathcal{N}^n_{O})$ cells of the material lattice. Its Hamiltonian can be obtained from \cref{Ham_wannier} by restricting the various summations over the lattice vectors only to those $\bm{R}_i$ corresponding to the unit cells forming the motif. Moreover, thanks to the discrete translational invariance of the lattice, only those coefficients involving the central cell of the motif at least once should be included in the motif Hamiltonian. See \cref{app:sec:motif_H} for more details on
the explicit form of the latter. 

As we will see in \cref{sec:compile_unit_cell}, the presence of a motif can be exploited to implement highly parallelisable quantum circuits whose depth is independent from the total number of motifs forming the lattice of the simulated material. This fact leverages the capabilities of a hybrid fermion to qubit encoding \cite{derby2021compact, derby2021compactalt}, which will be discussed in \cref{sec:qubit_rep}. In view of this step, we notice that the sites of an arbitrary lattice can be labeled by the triplets of integers $(n_1,n_2,n_3)$ introduced in \cref{eq:R_lattice}. Therefore, any lattice can be visualized on the Cartesian grid defined by those triplets. This fact establishes a natural mapping between a (real-space) motif of order $n$ and a \emph{Cartesian motif} consisting of $N_C = L_x \times L_y \times L_z$ sites, with $L_i = \max_{(n_1, n_2, n_3)\in\mathcal{N}^n_O}(n_i) - \min_{(n_1, n_2, n_3)\in\mathcal{N}^n_O}(n_i) + 1$

\begin{figure}[ht]
\centering
 \includegraphics[width=0.4\textwidth]{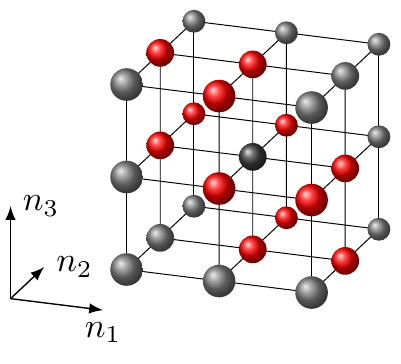} \hspace{1cm}
 \includegraphics[width=0.4\textwidth]{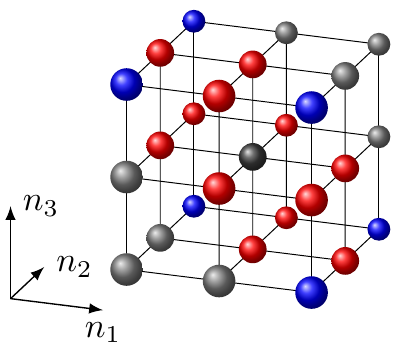} 
 \caption{Cartesian motifs corresponding to the motif of order (left) $n=1$ and (right) $n=2$ for silicon. As in \cref{fig:Si_motif}, the central unit cell is in black, and real-space nearest neighbour and next-nearest neighbour cells are shown in red and blue, respectively. Grey spheres represent the additional $N_D$ sites of the Cartesian motif which do not correspond to any cell of the motifs shown in \cref{fig:Si_motif}. Note that, in general, nearest neighbour cells of order $n$ in the motif can be nearest neighbours of order $n' \neq n$ in the Cartesian motif.}
 \label{fig:Si_cartesian_motif}
\end{figure}

Since, in general, $ N_C \neq N_{\mathrm{cells/motif}}$, the Cartesian motif contains $ N_D = N_C - N_{\mathrm{cells/motif}} $ additional cells. The latter do not enter the motif Hamiltonian but should be taken into account anyway during the encoding stage described in \cref{sec:qubit_rep}. In \cref{fig:Si_cartesian_motif}, we show the Cartesian motifs corresponding to the motifs of order $n=1$ and $n=2$ for silicon. Here, grey spheres denote the additional $N_D$ sites forming the Cartesian motif. Note that distances between unit cells ought to be computed in real space and that, as a result of the mapping, two units cells which are (real-space) nearest neighbours of order $n$ may be nearest neighbours of order $n'\neq n$ on the Cartesian grid. 

\subsection{General constraints and symmetry properties of the fermion integrals}
\label{sec:symmetries}

The single-particle wavefunctions determine the symmetry properties of the second-quantized Hamiltonian. For generic real space wavefunctions, the symmetry properties are discussed below. After that, we discuss in this section  two important generic symmetries in the Bloch and Wannier basis respectively, inversion (\cref{sec:inversion_sym}) and time reversal symmetry (\cref{sec:time_rev_sym}).

\subsubsection{General constraints and symmetry properties of the fermion integrals for general real single-particle wavefunctions}
\label{sec:fermion_integral_symmetries}

The hopping matrix and Coulomb tensor of a general many-body system satisfy the following identities (see \cref{eq:HM}, \cref{eq:CT}):
\begin{align}
    t_{\lambda_1 \lambda_2} &= t_{\lambda_2 \lambda_1}^{*} \label{eq:HM_hermiticity} \quad \text{(hermiticity)},\\
    V_{\lambda_1\lambda_2 \lambda_3 \lambda_4} &= V_{\lambda_2\lambda_1 \lambda_4 \lambda_3} \quad \text{(swap symmetry)},\\
    V_{\lambda_1\lambda_2 \lambda_3 \lambda_4} &= V^{*}_{\lambda_4\lambda_3\lambda_2\lambda_1} \quad \text{(hermiticity)},\\
    V_{\lambda_1\lambda_2 \lambda_3 \lambda_4} &= V^{*}_{\lambda_3\lambda_4\lambda_1\lambda_2} \quad \text{(hermiticity + swap)}.
\end{align}

For single-particle bases with real wavefunctions $ \set{\phi_{\lambda}(\bm{r})} $ (see \cref{sec:Wannier_functions}) the above relations simplify to
\begin{align}
    t_{\lambda_1\lambda_2} &= t_{\lambda_2\lambda_1}, \label{eq:t_hermitian_sym_real} \\
    V_{\lambda_1\lambda_2\lambda_3\lambda_4} & = V_{\lambda_4\lambda_3\lambda_2\lambda_1} = V_{\lambda_2\lambda_1\lambda_4\lambda_3} = V_{\lambda_4\lambda_2\lambda_3\lambda_1} = V_{\lambda_1\lambda_3\lambda_2\lambda_4} \nonumber \\
	& =
	V_{\lambda_3\lambda_4\lambda_1\lambda_2} = V_{\lambda_3\lambda_1\lambda_4\lambda_2} = V_{\lambda_2\lambda_4\lambda_1\lambda_3}.\label{eq:CT_equivalent_configs}
\end{align}
In particular, the latter set of equivalences has been obtained by combining the hermiticity and swap symmetry of the Coulomb tensor with the additional symmetry $V_{\lambda_1\lambda_2\lambda_3\lambda_4} = V_{\lambda_4\lambda_2\lambda_3\lambda_1} =  V_{\lambda_1\lambda_3\lambda_2\lambda_4}$ arising from \cref{eq:CT} for real wavefunctions and it can be exploited to significantly reduce the number of independent Coulomb tensor coefficients one has to directly compute.

After the single-particle basis has been picked and thus the second quantized Hamiltonian is fixed, it will have a form like \cref{eq:H_general}. To pass into a qubit Hamiltonian, as we will discuss in \cref{sec:qubit_rep}, it is useful to group together the single-particle quantum numbers and the spin in a single label $ \xi_i = (\lambda_i, \sigma_i) $. 
Introducing the spin-dependent hopping matrix $ \mathcal{T}_{\xi_1 \xi_2} $ and Coulomb tensor  $ \mathcal{V}_{\xi_1 \xi_2 \xi_3 \xi_4} $, we can re-write \cref{eq:H_general} as
\begin{equation}
	H = \sum_{\xi_1,\xi_2} \mathcal{T}_{\xi_1\xi_2} c^{\dagger}_{\xi_1}  c_{\xi_2} +  \sum_{\xi_1,\xi_2,\xi_3,\xi_4} \mathcal{V}_{\xi_1 \xi_2 \xi_3 \xi_4} c^{\dagger}_{\xi_1} c^{\dagger}_{\xi_2} c_{\xi_3} c_{\xi_4}.
\end{equation}

Here the \emph{spinful} hopping matrix is (with $\xi_i=(\lambda_i,\sigma_i)$)
\begin{equation}
	\mathcal{T}_{\xi_1\xi_2} = \begin{cases}
		t_{\lambda_1\lambda_2} & \text{if } \sigma_1 = \sigma_2  \\
		0 & \text{otherwise},
	\end{cases}
\end{equation}
while the \emph{spinful} Coulomb tensor is
\begin{equation}
	\label{eq:CT_antisymmetric}
	\mathcal{V}_{\xi_1 \xi_2 \xi_3 \xi_4}  = 
	\begin{cases}
	    V^s_{\lambda_1 \lambda_2 \lambda_3 \lambda_4} & \text{for } \sigma_1 = \sigma_2 = \sigma_3 = \sigma_4,\\
		\frac{1}{2}V_{\lambda_1 \lambda_2 \lambda_3 \lambda_4} & \text{if } \sigma_1 = \sigma_4 \text{ and } \sigma_2 = \sigma_3\ (\sigma_1 \neq \sigma_2),\\
		-\frac{1}{2}V_{\lambda_2 \lambda_1 \lambda_3 \lambda_4} & \text{if } \sigma_1 = \sigma_3 \text{ and } \sigma_2 = \sigma_4\ (\sigma_1 \neq \sigma_2)\\
		0 & \text{otherwise},
	\end{cases}
\end{equation}
with $V^s_{\lambda_1 \lambda_2 \lambda_3 \lambda_4} = (V_{\lambda_1 \lambda_2 \lambda_3 \lambda_4} - V_{\lambda_2 \lambda_1 \lambda_3 \lambda_4})/2$.

The spinful Coulomb tensor $ \mathcal{V}_{\xi_1\xi_2\xi_3\xi_4} $ is hermitian, $ \mathcal{V}_{\xi_1\xi_2\xi_3\xi_4} = \mathcal{V}^{*}_{\xi_4\xi_3\xi_2\xi_1} $, and antisymmetric
under exchange of the first or last pair of indices
\begin{equation}\label{eq:CT_anti-symmetry}
	\mathcal{V}_{\xi_1 \xi_2 \xi_3 \xi_4} = - \mathcal{V}_{\xi_2 \xi_1 \xi_3 \xi_4} = -\mathcal{V}_{\xi_1 \xi_2 \xi_4 \xi_3} = \mathcal{V}_{\xi_2 \xi_1 \xi_4 \xi_3}.
\end{equation}
Note that if $\sigma_1 = \sigma_4 $ and $\sigma_2 = \sigma_3$, $ \mathcal{V}_{\xi_1\xi_2\xi_3\xi_4} $ obeys the same identities in \cref{eq:CT_equivalent_configs} as $ V_{\lambda_1\lambda_2\lambda_3\lambda_4} $. In this latter case, by exploiting \cref{eq:CT_equivalent_configs} and \cref{eq:CT_anti-symmetry}, one can show that $ \mathcal{V}_{\xi_1\xi_2\xi_3\xi_4} $ also satisfies the following Jacobi identity:
\begin{equation}\label{eq:CT_Jacobi_identity}
	\mathcal{V}_{\xi_1\xi_2\xi_3\xi_4} + \mathcal{V}_{\xi_1\xi_3\xi_4\xi_2}  + \mathcal{V}_{\xi_1\xi_4\xi_2\xi_3} = 0 \quad \text{if $\sigma_1 = \sigma_4 $ and $\sigma_2 = \sigma_3$}.
\end{equation}

\subsubsection{Inversion symmetry}\label{sec:inversion_sym}

Inversion symmetry $\mathcal{I}$ transforms space and momentum variables according to $\bm{r} \rightarrow -\bm{r}$ and $\bm{k} \rightarrow -\bm{k} $, respectively. In a crystal with inversion symmetry the external potential satisfies $\tilde{U}(\bm{r}) = \tilde{U}(-\bm{r})$.
\paragraph{Bloch basis.}
Under inversion, a Bloch wavefunction $\phi_{\bm{k},n,\sigma}(\bm{r})$ transforms as
\begin{equation}
    \phi_{\bm{k},n,\sigma}(\bm{r}) \rightarrow \mathcal{I} \phi_{\bm{k},n,\sigma}(\bm{r}) = \phi_{\bm{k},n,\sigma}(-\bm{r}) = \phi_{-\bm{k},n,\sigma}(\bm{r}).
\end{equation}
Recalling that in the Bloch basis the hopping matrix is $h_{mn}(\bm{k}, \bm{k}') = \epsilon_m(\bm{k})\delta_{\bm{k},\bm{k}'}\delta_{mn}$, with \begin{equation}
    \epsilon_m(\bm{k}) = \int d\bm{r} \phi^*_{\bm{k},m,\sigma}(\bm{r})\left[-\frac{\hbar^2\nabla^2}{2m}+
\tilde{U}(\bm{r})\right] \phi_{\bm{k},m,\sigma}(\bm{r}),
\end{equation}
in a crystal with inversion symmetry one finds
\begin{equation}
    \label{eq:bands_inv_sym}
    \epsilon_m(\bm{k})=\epsilon_m(-\bm{k}).
\end{equation}
Since the electron-electron interaction potential $V(|\bm{r}-\bm{r}'|)$ is always invariant under inversion, one also obtains
\begin{equation}
    V_{n_1n_2n_3n_4}^{(\bm{k},\bm{k}',\bm{q})} = V_{n_1n_2n_3n_4}^{(-\bm{k},-\bm{k}',-\bm{q})}.
\end{equation}

\paragraph{Wannier basis.}
Under inversion with respect to $\bm{R}=\bm{0}$, a Wannier function $\mathcal{W}^{\bm{R}}_{m,\sigma}(\bm{r})$ transforms as
\begin{equation}
    \mathcal{W}^{\bm{R}}_{m,\sigma}(\bm{r}) \rightarrow \mathcal{I} \mathcal{W}^{\bm{R}}_{m,\sigma}(\bm{r}) = \mathcal{W}^{-\bm{R}}_{m,\sigma}(-\bm{r}).
\end{equation}

First, we note that if the unitary transformation entering the definition of the Wannier basis in \cref{eq:Wannier_operators} is trivial, i.e., $U_{mn}(\bm{k}) = \delta_{mn}$, \cref{eq:Wannier_T} reduces to
\begin{equation}
    T(\bm{R})_{mn}=\frac{1}{N^2} \sum_{\bm{k}} e^{i\bm{k}\cdot\bm{R}}\epsilon_n(\bm{k})\delta_{mn}.
\end{equation}
If the crystal has inversion symmetry, from \cref{eq:bands_inv_sym} we have $\epsilon_n(\bm{k}) = \epsilon_n(-\bm{k})$ and, therefore, 
\begin{equation}
    T(\bm{R})_{mn} = T(-\bm{R})_{mn} = T(\bm{R})_{nm}, \quad\forall m,n,
\end{equation}
where in the last step we used \cref{eq:Wannier_hopping_matrix_hermiticity}. Unfortunately, for \glspl{mlwf} the unitary matrix $U_{mn}(\bm{k})$ is usually more complicated and the relation above does not hold in general.

A more general identity can be obtained if the Wannier functions transform  under inversion as $\mathcal{I}\mathcal{W}^{\bm{R}}_{m,\sigma}(\bm{r}) = \mathcal{W}^{-\bm{R}}_{m,\sigma}(-\bm{r}) =  \sum_{m'}P^{\pi}_{mm'}\mathcal{W}^{-\bm{R}}_{m',\sigma}(\bm{r})$, $\forall m, \bm{R}$, with $P^{\pi}$ the generalized permutation matrix corresponding to a permutation $\pi$ acting on the orbital indices, with $P^{\pi}_{mm'}=\eta_m = \pm 1$ for $m'=\pi(m)$ and $P^{\pi}_{mm'}=0$ otherwise.  In simpler terms, the latter condition implies that under inversion a given Wannier function with orbital index $m$ and centred at $\bm{R}$ transforms into another Wannier function with orbital index $m'$ (which can be different from $m$) centred at $-\bm{R}$. Since usually Wannier functions retain the main features of the corresponding atomic orbitals, this a quite common situation. For Wannier functions centred at the centre of the unit cell one usually has $P^{\pi}_{mm'}=\pm\delta_{mm'}$. For instance, the action of the inversion symmetry on a $p_x$-type orbital centred at $\bm{R}=\bm{0}$ is $\mathcal{I}p^{\bm{0}}_x = -p^{\bm{0}}_x$.
In this case, in a crystal with inversion symmetry, one obtains 
\begin{equation}
    T(\bm{R})_{mn} = \sum_{m',n'} P^{\pi}_{mm'}P^{\pi}_{nn'} T(\bm{R})_{m'n'} = \eta_m \eta_n T(\bm{R})_{\pi(m)\pi(n)},
\end{equation}
from which one can see that
\begin{equation}
    T(\bm{R})_{mn} = 0 \quad \text{if } (\pi(m),\pi(n))=(m,n) \text{ and } \eta_m\eta_n = -1.
\end{equation}
The identity above can be generalized by noting that for $\bm{R} = \bm{0}$ we have $T(\bm{0})_{mn} = T(\bm{0})_{nm}$ and hence the two orbital configurations $(m,n)$ and $(n,m)$ are equivalent in the computation of the hopping matrix. We can then introduce the set of the orbital configurations equivalent to $(m,n)$ as $[(m,n)]^{\bm{0}}=\set{(m,n), (n,m)}$ and $[(m,n)]^{\bm{R}}=\set{(m,n)}$ for $\bm{R}\neq0$. Using this fact, we finally obtain
\begin{equation}
    T(\bm{R})_{mn} = 0 \quad \text{if } (\pi(m),\pi(n))\in{[(m,n)]^{\bm{R}}} \text{ and }\eta_m\eta_n = -1.
\end{equation}

On the other hand, the electron-electron interaction potential $V(|\bm{r}-\bm{r}'|)$ is always invariant under inversion. In cases with Wannier functions transforming as $\mathcal{I}\mathcal{W}^{\bm{R}}_{m,\sigma}(\bm{r}) = \mathcal{W}^{-\bm{R}}_{m,\sigma}(-\bm{r}) =  \sum_{m'}P^{\pi}_{mm'}\mathcal{W}^{-\bm{R}}_{m',\sigma}(\bm{r})$ under inversion, one obtains 
\begin{equation}
    \tilde{V}^{(\bm{R}_1,\bm{R}_2,\bm{R}_3,\bm{R}_4)}_{s,l,m,n} = 0 \quad \text{if } (\pi(s),\pi(l),\pi(m),\pi(n))\in{[(s,l,m,n)]^{\bm{R}_1\bm{R}_2\bm{R}_3\bm{R}_4}} \text{ and }\eta_s\eta_l\eta_m\eta_n = -1.
\end{equation}
Here, $[(s,l,m,n)]^{\bm{R}_1\bm{R}_2\bm{R}_3\bm{R}_4}$ is the set of equivalent orbital configurations according to \cref{eq:CT_equivalent_configs}. See \cref{app:sec:CT_pipeline_CS} for details. The equation above is particularly useful since it allows one to determine a number of coefficients of the Coulomb tensor which are identically zero without having to compute them  explicitly.

A general discussion on discrete crystal symmetries and the structure of matrix elements is presented in \cref{sec:crystal_sym}.

\subsubsection{Time-reversal symmetry}\label{sec:time_rev_sym}

Under time-reversal symmetry $\mathcal{T}$ momentum and spin variables transform according to $\bm{k} \rightarrow -\bm{k}$ and $\bm{s} \rightarrow - \bm{s}$, with $\bm{s} = \boldsymbol{\sigma}/2$ and $\boldsymbol{\sigma} = (\sigma_x, \sigma_y, \sigma_z)$. Time-reversal symmetry is represented by an anti-unitary operator. One common choice is $\mathcal{T}=e^{-i\pi\sigma_y}K$, with $K$ denoting the complex conjugation operator and the unitary operator $e^{-i\pi\sigma_y}$ performing a $\pi-$rotation of the spin around the $y-$axis. If the crystal possesses time-reversal symmetry, the external potential $\tilde{U}(\bm{r})$ is spin-independent and the electronic bands are doubly degenerate. This is always the case for a non-relativistic system (e.g., with no spin-orbit coupling) and in the absence of an external magnetic field. Since both these assumptions have been made in \cref{sec:Physics}, the general Hamiltonians of \cref{eq:Ham_everyday} and \cref{eq:H_general} already take into account the consequences of time-reversal invariance. To identify the latter in an explicit way, in this section we will examine a generalization of the Hamiltonians considered in \cref{sec:Physics} with a spin-dependent external potential $\tilde{U}_{\sigma_1\sigma_2}(\bm{r})$. In this case, the quadratic part of \cref{eq:H_general} should be modified as follows,
\begin{equation}
    H_0 = \sum_{\sigma_1,\sigma_2}\sum_{\lambda_1,\lambda_2} t^{\sigma_1,\sigma_2}_{\lambda_1\lambda_2} c^{\dagger}_{\lambda_1,\sigma_1}c_{\lambda_2,\sigma_2},
\end{equation}
with
\begin{equation}
    t^{\sigma_1,\sigma_2}_{\lambda_1\lambda_2} = \int d\bm{r} \ \phi^{*}_{\lambda_1, \sigma_1}(\bm{r})\left[-\frac{\hbar^2\nabla^2}{2m}+\tilde{U}_{\sigma_1\sigma_2}(\bm{r)}\right]\phi_{\lambda_2, \sigma_2}(\bm{r}).
\end{equation}

\paragraph{Bloch basis.}
Under time-reversal a Bloch wavefunction $\phi_{\bm{k},n,\sigma}(\bm{r})$ transforms as
\begin{equation}
    \phi_{\bm{k},n,\uparrow}(\bm{r}) \rightarrow \phi_{-\bm{k},n,\downarrow}(\bm{r}) = \phi_{\bm{k},n,\downarrow}(-\bm{r}) \quad \text{and} \quad \phi_{\bm{k},n,\downarrow}(\bm{r}) \rightarrow - \phi_{-\bm{k},n,\uparrow}(\bm{r}) = - \phi_{\bm{k},n,\uparrow}(-\bm{r}).
\end{equation}
If the external potential $\tilde{U}(\bm{r})$ is spin-independent, the hopping matrix coefficients $h^{\sigma,\sigma'}_{mn}(\bm{k}, \bm{k}') = \epsilon_m(\bm{k})\delta_{\bm{k,\bm{k}'}}\delta_{m,n}\delta_{\sigma,\sigma'}$ satisfy the identity
\begin{equation}
    \epsilon_m(\bm{k})=\epsilon_m(-\bm{k}),
\end{equation}
while for the Coulomb tensor coefficients one finds
\begin{equation}
    V_{n_1n_2n_3n_4}^{(\bm{k},\bm{k}',\bm{q})} = V_{n_1n_2n_3n_4}^{(-\bm{k},-\bm{k}',-\bm{q})}.
\end{equation}
Note that the above equations are the same as the ones we obtained for a system with inversion symmetry. 

On the other hand, in the presence of a spin-dependent but time-reversal invariant external potential $\tilde{U}_{\sigma_1\sigma_2}(\bm{r}) = \tilde{U}_{\bar{\sigma}_1\bar{\sigma}_2}(\bm{r})$ (where $\bar{\sigma}_i=-\sigma_i$), one finds $h^{\sigma,\sigma'}_{mn}(\bm{k},\bm{k}') = \epsilon^{\sigma,\sigma'}_m(\bm{k})\delta_{\bm{k},\bm{k}'}\delta_{m,n}$, with $\epsilon^{\sigma,\sigma'}_m(\bm{k}) \in \mathbb{R}$ and 
\begin{equation}
    \epsilon^{\uparrow, \uparrow}_m(\bm{k}) = \epsilon^{\downarrow, \downarrow}_m(-\bm{k}) \quad \text{and} \quad \epsilon^{\uparrow, \downarrow}_m(\bm{k}) = - \epsilon^{\downarrow, \uparrow}_m(-\bm{k}) = - \epsilon^{\uparrow, \downarrow}_m(-\bm{k}).
\end{equation}
If, in addition, the crystal also possesses inversion symmetry, the above equation becomes
\begin{equation}
    \epsilon^{\uparrow, \uparrow}_m(\bm{k}) = \epsilon^{\downarrow, \downarrow}_m(\bm{k}) \quad \text{and} \quad \epsilon^{\uparrow, \downarrow}_m(\bm{k}) = - \epsilon^{\uparrow, \downarrow}_m(\bm{k}) = 0.
\end{equation}

\paragraph{Wannier basis.}
Under time-reversal a Wannier function $\mathcal{W}^{\bm{R}}_{n,\sigma}(\bm{r})$ transforms as
\begin{equation}
    \mathcal{W}^{\bm{R}}_{n,\uparrow}(\bm{r}) \rightarrow \mathcal{W}^{\bm{R}}_{n,\downarrow}(\bm{r}) \quad \text{and} \quad \mathcal{W}^{\bm{R}}_{n,\downarrow}(\bm{r}) \rightarrow - \mathcal{W}^{\bm{R}}_{n,\uparrow}(\bm{r}).
\end{equation}
If the external potential $\tilde{U}(\bm{r})$ is spin-independent, time-reversal invariance does not lead to any further relation for the hopping matrix and Coulomb tensor coefficients. 

On the other hand, in the presence of a spin-dependent and time-reversal invariant external potential $\tilde{U}_{\sigma_1\sigma_2}(\bm{r}) = \tilde{U}_{\bar{\sigma}_1\bar{\sigma}_2}(\bm{r})$, one obtains
\begin{equation}
    T(\bm{R})^{\uparrow,\uparrow}_{mn} = T(\bm{R})^{\downarrow,\downarrow}_{mn} \quad \text{and} \quad T(\bm{R})^{\uparrow,\downarrow}_{mn} = - T(\bm{R})^{\downarrow,\uparrow}_{mn},
\end{equation}
where we have used the reality of the matrices $T(\bm{R})$, that follows from the reality of the Wannier functions.
Finally, from the last identity and the hermiticity of the Hamiltonian it follows that $T(\bm{0})^{\uparrow,\downarrow}_{mm} = 0 $.

\subsection{Crystal symmetries}\label{sec:crystal_sym}

A Hamiltonian $H$ invariant under a symmetry group $G$ (e.g., crystal symmetries, inversion, discrete rotations, etc), satisfies particular {\it selection rules} that determine the type of operators allowed in the Hamiltonian. In this section we discuss the general strategy to determine those rules for a general discrete group. Leveraging those constraints allows us to reduce the number of classical computations (computations of fermion integrals) and, more importantly, reduces the number of gates needed to implement the Hamiltonian by constraining the allowed operators in $H$.

Let us consider the electron-electron interaction term in \cref{eq:H_general}. The single-particle wavefunctions used to span the Hilbert space form a basis that can be chosen to transform under a definite representation of the symmetry of the Hamiltonian. In this case, a symmetry operation $g\in G$, with representation $D_{\lambda_i\lambda_i'}^{(\mu)}(g)$ (which can always be chosen to be some irreducible representation $\mu$), satisfies
\begin{equation}
  \sum_{\lambda_1',\lambda_2',\lambda_3',\lambda_4'}D_{\lambda_1\lambda_1'}^{\dagger(\mu_{1})}(g)D_{\lambda_2\lambda_2'}^{\dagger(\mu_{2})}(g)D_{\lambda_3\lambda_3'}^{(\mu_{3})}(g)D_{\lambda_4\lambda_4'}^{(\mu_{4})}(g)V_{\lambda_1'\lambda_2'\lambda_3'\lambda_4'}	=V_{\lambda_1\lambda_2\lambda_3\lambda_4}.
\end{equation}

For simplicity we assume that the symmetries discussed here are unitary. 

The equation above means in particular that the tensor product of different representations involved in a particular element of the interaction tensor should contain the trivial representation, that does not transform under the symmetry. In general, the tensor product of representations can be decomposed into the direct sum of irreducible representations (irreps) as~\cite{Hamermesh_book} 
\begin{equation}
    D^{(\mu_{1}\times\mu_{2})}(g)=\bigoplus_{\nu}c_{\nu}(\mu_{1},\mu_{2})D^{(\nu)}(g),
\end{equation}
where $c_\nu(\mu_1,\mu_2)$ is the number of times that the irrep $\nu$
appears in the tensor product of irreps $\mu_1$ and $\mu_2$.
Using this equation repeatedly and taking the trace, i.e., $\chi^{(\mu)}(g)={\rm Tr}(D^{(\mu)}(g))$ with $\chi^{(\mu)}$ the character of $D^{(\mu)}$, we find the relation between the characters  $\chi^{*(\mu_{1}\times\mu_{2})}(g)\chi^{(\mu_{3}\times\mu_{4})}(g)	=\sum_{\nu,\nu',\rho}c_{\nu}(\mu_{1},\mu_{2})c_{\nu'}(\mu_{3},\mu_{4})\chi^{*(\nu)}(g)\chi^{(\nu')}(g)$. In this product
the number of times that the trivial representation $e$ appears can be computed using the orthogonality relation of the characters
$c_{e}(\nu,\nu')	=\frac{1}{|G|}\sum_{g}\chi^{*(\nu)}(g)\chi^{(\nu')}(g)=\delta_{\nu\nu'} $, where $|G|$ is the dimension of $G$ \cite{Hamermesh_book}.

For a given matrix element of a Hamiltonian to have a chance of being invariant under the symmetry, the representations involved in that matrix element should be such that their tensor product contains a copy of the trivial representation. This means that the tensor product of the irreps corresponding to non-zero coefficients of the Coulomb tensor can be decomposed as $D^{\dagger(\mu_{1}\times\mu_{2})}(g)D^{(\mu_{3}\times\mu_{4})}(g)=\bigoplus_{\nu}c_{\nu}(\mu_{1},\mu_{2})c_{\nu}(\mu_{3},\mu_{4})D^{(e)}(g)$.

Starting from the definition of the Coulomb tensor
\begin{equation}\label{Coulomb_sym}
    V_{\lambda_1\lambda_2\lambda_3\lambda_4}^{\mu_{1}\mu_{2}\mu_{3}\mu_{4}}=\int d\bm{r}d\bm{r}'\phi_{\lambda_1}^{*(\mu_{1})}(\bm{r})\phi_{\lambda_2}^{*(\mu_{2})}(\bm{r}')V(|\bm{r}-\bm{r}'|)\phi_{\lambda_3}^{(\mu_{3})}(\bm{r}')\phi_{\lambda_4}^{(\mu_{4})}(\bm{r}),
\end{equation}
where with a slight abuse of notation we use the label and the representation under which the single-particle wavefunction $\phi_{\lambda_i}^{(\mu_i)}(\bm{r})$ transforms as indices for the tensor, we can decompose it into irreps to find
\begin{equation}
\label{eq:V_irreps_decomposition}
V_{\lambda_1\lambda_2\lambda_3\lambda_4}^{\mu_{1}\mu_{2}\mu_{3}\mu_{4}}=\sum_{\nu}\sum_{s,s'}C_{\lambda_1\lambda_2,s}^{*\mu_{1}\mu_{2},\nu}C_{\lambda_3\lambda_4,s'}^{\mu_{3}\mu_{4},\nu}\int d\bm{r}d\bm{r}'\Psi_{s}^{*(\nu)}(\bm{r},\bm{r}')V(|\bm{r}-\bm{r}'|)\Psi_{s'}^{(\nu)}(\bm{r}',\bm{r}),
\end{equation}
with $\set{\Psi_{s'}^{(\nu)}}$ the set of the basis functions of $D^{(\nu)}(g)$.
The three-leg tensor $C_{\lambda_1\lambda_2,s}^{\mu_{1}\mu_{2},\nu}$ (a Clebsch-Gordan coefficient of the group $G$) transforms the tensor product of basis in the $\mu_1$ and $\mu_2$ irreps  into a new basis transforming in the $\nu$ irrep. They can be computed via the relation \cite{Hamermesh_book}
\begin{equation}
    \frac{1}{|G|}\sum_{g}D_{\lambda_1\lambda_2}^{(\mu_{1})}(g)D_{\lambda_3\lambda_4}^{(\mu_{2})}(g)D_{ss'}^{(\nu)}(g)=\left(\frac{1}{\sqrt{n_{\nu}}}C_{\lambda_1\lambda_3,s}^{\mu_{1}\mu_{2},\nu}\right)\left(\frac{1}{\sqrt{n_{\nu}}}C_{\lambda_2\lambda_4,s'}^{\mu_{1}\mu_{2},\nu}\right)^{*},
\end{equation}
where $n_{\nu}$ is the dimension of the irrep $\nu$. Fixing $\lambda_1=\lambda_2,\lambda_3=\lambda_4$, and $s=s'$, we have
\begin{equation}\label{Clebsch_gordan}
\frac{1}{|G|}\sum_{g}D_{\lambda_1\lambda_1}^{(\mu_{1})}(g)D_{\lambda_3\lambda_3}^{(\mu_{2})}(g)D_{ss}^{(\nu)}(g)=\left|\frac{1}{\sqrt{n_{\nu}}}C_{\lambda_1\lambda_3,s}^{\mu_{1}\mu_{2},\nu}\right|^{2}.
\end{equation}
The equation above allows us to look for the set of labels where the Clebsch-Gordan coefficient does not vanish. Together with \cref{eq:V_irreps_decomposition}, this determines the selection rules, and can be used to find the set of allowed labels (i.e., the ones corresponding to non-vanishing terms) in the Coulomb tensor. 

\paragraph{Example: $\mathbb{Z}_{m}$ group.}

\begin{figure}[ht]
\centering
 \includegraphics[width=0.4\linewidth]{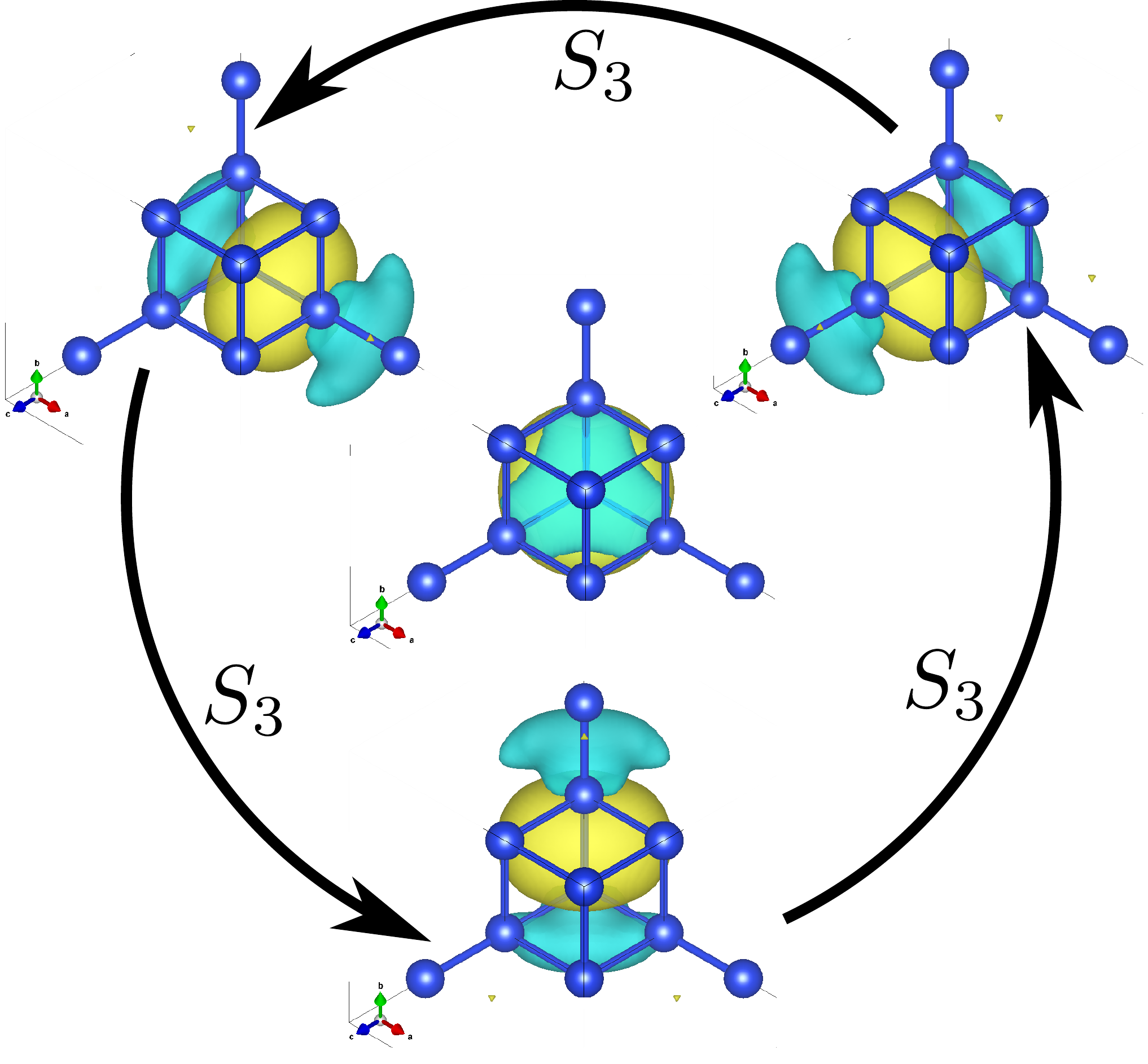}
 \caption{Four valence Wannier functions of Silicon. Each of these is aligned with respect to the axes 1 through 4 of \cref{eq:axes_sil}. A $\mathbb{Z}_3$ rotation around one of the axis maps the Wannier functions into themselves. The transformation matrix together with the invariance of the Hamiltonian determines the selection rules.}
 \label{fig:SiWFs}
\end{figure}

The different irreps of the cyclic group are all one dimensional and they are parameterised by
$D^{(\mu)}(g^{n})=e^{i\frac{2\pi}{m}\mu n}$, with $\mu=0,\dots, m-1$ and $n=0,\dots, m-1$. \cref{Clebsch_gordan} becomes
\begin{equation}\label{sel_Zm}
\frac{1}{m}\sum_{n=0}^{m-1}e^{i\frac{2\pi}{m}(\mu_{1}+\mu_{2}+\nu)n}	=\delta_{\mu_{1}+\mu_{2}+\nu, 0}=\left|C^{\mu_{1}\mu_{2},\nu}\right|^{2},
\end{equation}
from where we find the selection rule $\mu_{1}+\mu_{2} + \mu_{3}+\mu_{4}=0$.
Let us apply this to one of the many symmetries of Silicon. The lattice vectors (in \AA) are $a=(-2.7, 0, 2.7), b=(0, 2.7, 2.7) $ and $c=(-2.7, 2.7, 0)$. Starting with four valence WFs, each one aligned with one of the four axis of a Silicon tetrahedron (see \cref{fig:SiWFs}),
\begin{eqnarray}\nonumber\label{eq:axes_sil}
W_{1}^{Si}	:a+b-3c\rightarrow\mbox{axis 1},&\quad 
W_{2}^{Si}	:-3a+b+c\rightarrow\mbox{axis 2},\\
W_{3}^{Si}	:a-3b+c\rightarrow\mbox{axis 3},&\quad
W_{4}^{Si}	:a+b+c\rightarrow\mbox{axis 4,}
\end{eqnarray}
a $\mathbb{Z}_{3}$ rotation around each of these axes is a symmetry of the crystal (we consider only these symmetries for simplicity of exposition, recalling that the full space group of Silicon is m3m). We call $S_{j}$ the operator associated with a rotation by $2\pi/3$ around an axis $j$. In \cref{fig:SiWFs} the transformation generated by $S_{4}$ is shown. These transformations permute the WFs in the basis $[W_{1}^{Si},W_{2}^{Si},W_{3}^{Si},W_{4}^{Si}]^{T}$. Focusing on $S_4$, we can find a basis where the action of the symmetry is diagonal, i.e., $S_4\tilde{W}_aS_4^\dagger=\omega^a\tilde{W}_a$, with $\omega=e^{\frac{2\pi i}{3}}$. This basis consists of 
$\tilde{W}_4=W_4^{Si}$ and
\begin{equation}
\begin{bmatrix}\tilde{W}_{0}\\
\tilde{W}_{1}\\
\tilde{W}_{2}
\end{bmatrix}=\frac{1}{\sqrt{3}}\begin{bmatrix}1 & 1 & 1\\
1 & \omega & \omega^{2}\\
1 & \omega^{2} & \omega
\end{bmatrix}\begin{bmatrix}W_{1}^{Si}\\
W_{2}^{Si}\\
W_{3}^{Si}
\end{bmatrix}.
\end{equation}
 
The selection rule of \cref{sel_Zm} applies now for the  symmetry $\mathbb{Z}_3$ and is nothing more than the condition that the overall phase that the Coulomb term in the $\tilde{W}$ basis acquires under $S_4$ is zero, fixing the product of functions in \cref{Coulomb_sym} to have the form $\tilde{W}_a\tilde{W}_b\tilde{W}_c\tilde{W}_d$ with $a+b+c+d =0$ mod $3$.

\paragraph{Example: Octahedral metal centre.}

As a further example of how symmetry could help in reducing the number of Coulomb tensor coefficients one has to compute, we now consider a cluster consisting of a transition metal atom (e.g., Manganese (Mn)) surrounded by six Oxygen (O) atoms. This structure is typical of many transition metal oxides and perovskites. 

In our analysis, we employ renormalized Hydrogen-like atomic orbitals. The main reason for that is that they are simpler than Wannier functions and they feature the same symmetry properties. They are defined as

\begin{equation}
	\psi_{nlm}(Z_\mathrm{eff}; \bm{r}) = R_{nl}(Z_\mathrm{eff}; r) X_l^m(\bm{r}),
\end{equation}
with $ n, l, m $ the standard principal, angular, and magnetic quantum numbers, $ Z_\mathrm{eff} $ the effective nuclear charge~\cite{Clementi63}, $X_l^m(\bm{r})$ cubic spherical harmonics, and
\begin{equation}
	R_{nl}(Z_\mathrm{eff};r) = \sqrt{\left(\frac{ Z_{\mathrm{eff}}}{n}\right)^3 \frac{(n-l-1)!}{2n(n+l)!}} e^{-\frac{Z_\mathrm{eff}r}{2n}}\left(\frac{Z_\mathrm{eff}r}{n}\right)^l L^{2l+1}_{n-l-1}\left(\frac{Z_\mathrm{eff}r}{n}\right)
\end{equation}
the radial wavefunction.
Here, lengths are measured in units of $ a_0/2 $, with $ a_0 $ the Born radius, and $ L^{2l+1}_{n-l+1}(z) $ is the generalized Laguerre polynomial of degree $ n-l+1 $.

In what follows we will assume that the largest contribution to the Coulomb tensor is due to the 5 d orbitals centred at the transitional metal, $ d_{xy}, d_{xz}, d_{xz}, d_{x^2-y^2}, d_{z^2} $, whose corresponding wavefunctions are denoted as $ \mathcal{W}_a(\bm{r}) $ with $ a={1,...,5} $, respectively. We are interested in calculating the Coulomb tensor coefficients in the central unit cell
\begin{equation}
	V_{abcd} = \int d\bm{r} d\bm{r}' \mathcal{W}_a(\bm{r}) \mathcal{W}_b(\bm{r}') V(|\bm{r}-\bm{r}'|) \mathcal{W}_c(\bm{r})  \mathcal{W}_d(\bm{r}'). 
\end{equation}
In doing that, symmetry properties can be exploited to determine a priori which elements of $ V $ are vanishing. Here, we take into account the reflection symmetry along the $ x $, $ y $, and $ z $ axes, and $\pi/2$ rotations around the $z-$axis. The corresponding operators are denoted by $ \mathcal{I}_x $, $ \mathcal{I}_y $, $ \mathcal{I}_z $, and $ \mathcal{I}_R $, respectively. Their action on the orbitals $W_a(\bm{r})$ is $ \mathcal{I_\mu} \mathcal{W}_a(\bm{r}) = \sum_{a'} P^{\pi_\mu}_{aa'} \mathcal{W}_{a'}(\bm{r}) $, with $ \mu=\set{x,y,z,R} $ and $P^{\pi_\mu}$ a generalized permutation matrix corresponding to the permutation of the wavefunction indices $\pi_\mu$, with $P^{\pi_\mu}_{aa'}=\eta^{\mu}_a=\pm 1$ if $a'=\pi_\mu(a)$ and $P^{\pi_\mu}_{aa'}=0  $ otherwise. Exploiting the fact that $ V(|\bm{r}-\bm{r}'|) $ is invariant under the operations associated with $\mathcal{I}_{\mu}$, we can identify which elements of the Coulomb tensor are zero. In particular, $ V_{abcd} = 0 $ if $ (\pi_\mu(a),\pi_\mu(b),\pi_\mu(c),\pi_\mu(d)) \in [(i,j,k,l)] $ and $ \eta^{\mu}_a \eta^{\mu}_b \eta^{\mu}_c \eta^{\mu}_d = -1$. Here, $[(i,j,k,l)]$ is the set of all the configurations equivalent to $(a,b,c,d) $ according to \cref{eq:CT_equivalent_configs}. See \cref{app:sec:CT_pipeline_CS} for additional details about the definition of $[(i,j,k,l)]$.

\cref{tab:TM_elements} shows that symmetries may allow us to reduce significantly the number of Coulomb tensor coefficients we need to compute. For the particular example we have illustrated in this section, the number of the required coefficients passes from $325$ (original case with no symmetry exploited), to $157$ if inversion symmetry is taken into account and to $129$ if both inversion and $\pi/2$ rotational symmetry around the $z-$axis are considered, respectively.

\begin{table}
\centering
\begin{tabular}{ccc}
\toprule
No symmetry & Inversion & Inversion + Rotation \\
\midrule
325 & 157 & 129 \\
\bottomrule
\end{tabular}
\caption{Number of independent Coulomb tensor coefficients to be computed for a transition metal cluster with 5 d orbitals without taking into account any symmetry, by exploiting the inversion symmetry only, and by considering both the inversion and rotational symmetries, respectively. Note that the table reports the minimal number of coefficients required to determine the whole Coulomb tensor via the general relations given in \cref{eq:CT_equivalent_configs}, i.e., the number of the independent coefficients.}
\label{tab:TM_elements}
\end{table}

\subsection{Using DFT to choose degrees of freedom}
\label{sec:DFT}
As we discussed in previous sections, in order to obtain the Bloch and Wannier functions in an actual material it is necessary to obtain the eigenstates of the electronic gas moving in the ionic potential $\tilde{U}(\bm{r})$, which is completely determined by just knowing the position of the ions. This requires solving a system of $N$ Schrodinger equations for each possible value of the wavevector $\bm{k}$ (see \cref{eq:real_space_eigeneq}), a task which is classically unfeasible for most common materials. However, this procedure is generally unnecessarily complex as there are many electrons that are strongly bound to the ions (core electrons) which, if the processes involved are not very energetic, will not participate in the chemistry of the system. Therefore, the properties of a material can be effectively determined by studying the motion of the outermost electrons in a modified ionic potential (pseudo-potential) which combines the original ionic potential with the screening effects of the core electrons. An efficient way of dealing with this approach, which has been developed to maturity in the last century, is \glsxtrfull{dft} \cite{Jones1989, Martin_book}. 
In this section, we discuss how to obtain Bloch and Wannier functions within the framework of \gls{dft} and how to select the relevant degrees of freedom for the description of the system.

\gls{dft} is a highly efficient, accurate and flexible
method for simulating atomic systems. It has enjoyed decades of success for
simulating the ground state properties of numerous quantum systems at the atomic
scale across the entire periodic table for translationally invariant systems \cite{Hafner2006,Marzari2021}. 

In this work, we use \gls{dft} to generate single-particle Kohn-Sham states (described below) which
then are used to select an active space of bands where the relevant processes occur. Once this active space has been chosen, we generate Maximally localised Wannier functions (\glspl{mlwf}) for the construction of second quantised many-body Hamiltonians. 

\gls{dft} formally states that there exists an exact mapping
between the external potential of a many-body system and its ground state density
$n_0(\mathbf{r})$, whereby the nondegenerate ground-state wavefunction is a
unique functional of the ground-state density, i.e.,
\begin{equation}
  \Psi_{0} (\mathbf{r_1}, \mathbf{r}_2, \hdots, \mathbf{r}_N) = \Psi_0[n_0(\mathbf{r})].
  \label{eq:gs_dens}
\end{equation}

An important ground-state property is its energy, and Hohenberg and Kohn
additionally proved that it is minimal if the electron density is the
ground-state electron density \cite{hkI},
\begin{equation}
  E[n_{0}(\mathbf{r})] \leq E[n(\mathbf{r})],
  \label{eq:dft_variational}
\end{equation}
where $n(\mathbf{r})$ is the density of the system. Kohn and Sham subsquently showed how to
map the fully interacting many-body problem onto a single-particle problem \cite{KSI},
i.e.,
\begin{equation}
  \left[ \frac{\hbar^{2}}{2m} \nabla^{2} + \tilde{U}_{\text{eff}}(\mathbf{r})\right] \phi_{i}(\mathbf{r}) = \epsilon_{i} \phi_{i}(\mathbf{r}), 
  \label{eq:kohn_sham}
\end{equation}
where $\tilde{U}_{\text{eff}}(\mathbf{r})$ is the effective Kohn-Sham potential,
$\epsilon_i$ are the single-particle Kohn-Sham eigenvalues and, $\set{
\phi_i(\mathbf{r})}$ are the Kohn-Sham states. The latter are fictitious orbitals (i.e., they are not formally related to any physical electron state) such that they must obey the following constraint
\begin{equation}
  n_{0}(\mathbf{r}) = \sum_{i=1}^{\text{occ}}|\phi_i(\mathbf{r})|^{2}, 
  \label{eq:kohn_sham_density}
\end{equation}
where $n_{0}(\mathbf{r})$ is the ground state density and the sum runs over all occupied electron states. This constraint fixes the electron occupation in the system.

While \gls{dft} is in principle an \emph{ab initio} method, i.e., it requires only
the lattice structure of the system, practically it necessitates the choice of
\emph{(i)} a basis for the fictitious single-particle Kohn-Sham states
$\{\phi_i(\mathbf{r})\}$ and \emph{(ii)} a parameterized Kohn-Sham exchange-correlation functional. In this work, for the \gls{dft} exploration we exclusively consider plane-wave basis sets, as implemented in Quantum Espresso \cite{QE_2009, Giannozzi_2017}, which are
subsequently transformed into real-space \glspl{mlwf} using Wannier90 \cite{w90}. We note that it
is also possible to solve the Kohn-Sham equation exclusively in the real space basis, using
the so-called ``all-electron'' methods \cite{dft_comp}. We now summarise some of the main features of employing
a plane-wave basis code and truncation of the Hilbert space using \glspl{mlwf}. 

\subsubsection{Kohn-Sham eigenstates in the plane-wave basis}
\label{sec:Kohn-Sham}
\begin{figure}[t!] \centering
\includegraphics[scale=0.3]{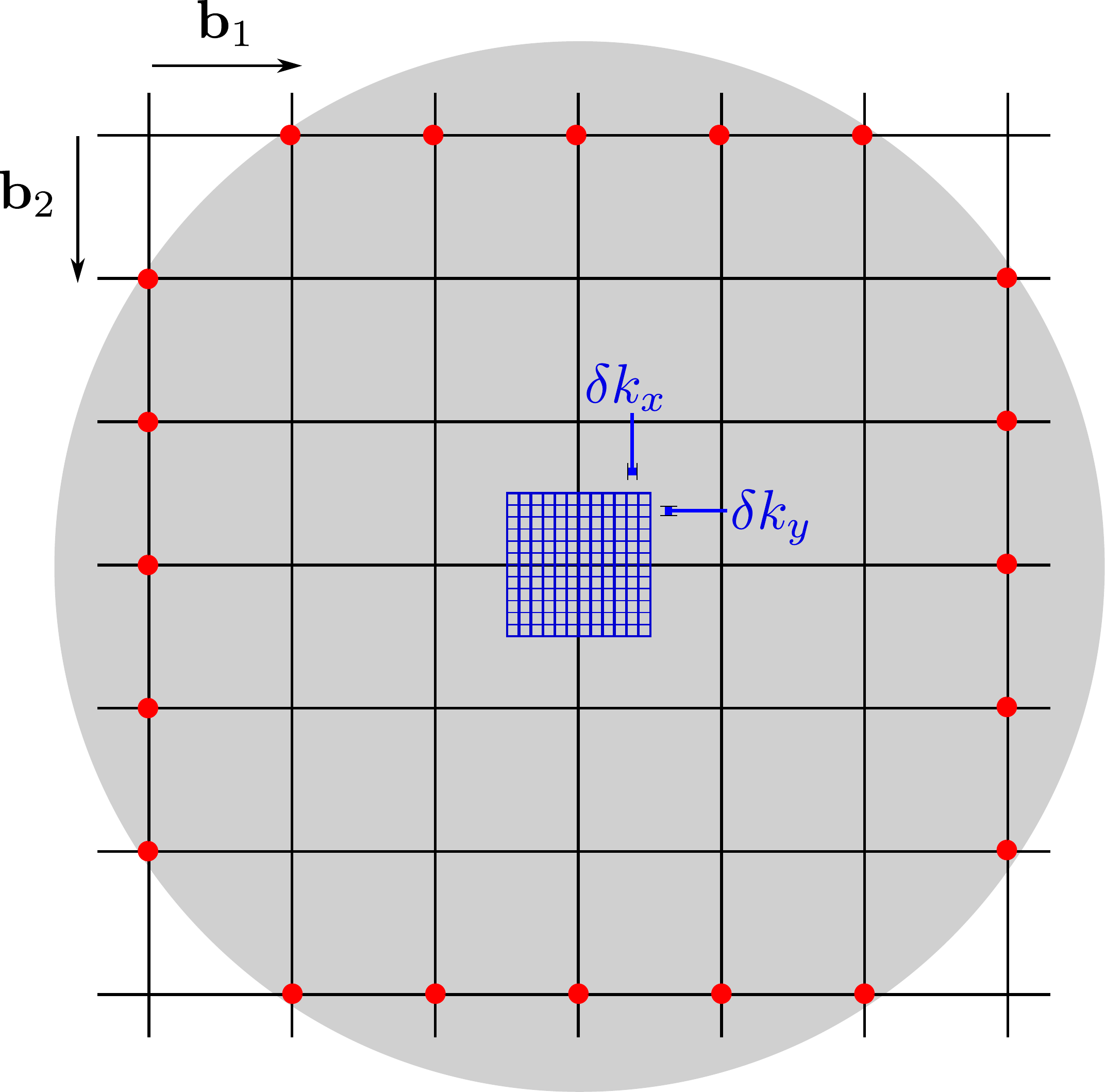}
 \caption{Reciprocal lattice of a 2D square Bravais lattice. The black grid
corresponds to the reciprocal lattice $\bm G=n_1\bm b_1+n_2\bm b_2$, and it is
unbounded $(n_a\in \mathbb{Z})$. The blue lattice represents the different
lattice momentum states $\bm k$. Increasing the system size leads to a finer
blue lattice, without affecting the black lattice.  The red dots represent the
maximum value of $\bm G$ inside a cut-off region, which is depicted by the grey circle.}
 \label{fig:Rep_lattice}
\end{figure}

Bloch's theorem in periodic systems
can be applied to the solution of the full Hamiltonian of any system that can
be written in the plane-wave basis. Similarly, so can the Kohn-Sham equations. 
In the plane-wave basis-set the Hilbert space is truncated by considering only a
finite number of reciprocal lattice vectors $\bm G$. This is motivated by the
fact that the kinetic energy is an unbounded operator that depends on the length of the reciprocal lattice vector, so a
maximum energy sets a maximum value for the norm of $\bm G$ as $\frac{\hbar}{2m}|\bm G_{\rm max}|^2=E_{\rm max}$. After
diagonalization of the single-particle Hamiltonian, only a finite number
of bands below and above the Fermi energy are retained. Moreover, as a
consequence of Bloch's theorem, there is a natural partition of the degrees of
freedom using the reciprocal lattice vectors (see \cref{fig:Rep_lattice}).
While the size of the system under study defines the number of inequivalent
${\bm k}$ points as seen in \cref{allowed_k}, the number of different
reciprocal lattice vectors is unbounded. 

The following cutoff,
\begin{equation}
\label{cut_off}
\frac{\hbar^2}{2m}|\bm G|^2\leq E_{\rm cut}, 
\end{equation}
defines the total number of plane-waves used in the \gls{dft} calculation. Fewer plane waves can be 
used  by replacing the core level Kohn-Sham states by an effective potential, famously known as the
pseudopotential \cite{Martin_book}. It is this crucial observation, i.e., that the core
electrons do not participate in low energy, chemically relevent, excitations
like their valence counterparts, that enables the success of the plane-wave
method. Otherwise, the valence electron wavefunctions require impractically
large number of Fourier components to remain orthogonal to all states within the
core region, which are chemically inert. Thus, ``freezing'' the core states into
an overall effective potential lifts this constraint, and allows the valence
electron wavefunctions to be efficiently represented with far fewer Fourier components,
without any nodes inside the core regions. Practically, this means that smaller matrices have to
be diagonalized when solving \cref{eq:kohn_sham}, due to the reduced basis size.
Pseudopotentials for atoms corresponding to the ions in the lattice are constructed by solving for all eigenvalues of
their atomic wavefunction, fitting them to pseudo-wavefunctions and generating the corresponding atomic pseudopotential by solving all-electron atomic calculations for a single atom and then inverting the corresponding radial Schrodinger equation to find the effective pseudpotential. The constructed
pseudopotential must reproduce the atomic properties of the element, i.e.,
the scattering properties of the ionic potential, and agree with its true
wavefunction outside of a cut-off radius away from the core. Moreover, it should
be transferrable to a variety of chemical enviroments. For the materials
used in this work, we use pseudopotentials from the pregenerated ONCVPSP library \cite{ONCVPSP}. 

As a result, the actual value for the kinetic energy cut-off $E_{\rm cut}$ is
usually chosen such that it corresponds to the minimum energy where the
convergence of the total ground state energy does not change with respect to a
specific tolerance, typically 1 meV per atom. In a material like silicon, a value used in
\gls{dft} calculations \cite{Castep} is $E_{\mathrm{cut}}\sim 200$ eV. Using \cref{cut_off}
this translates into $|\bm{G}_{\rm max}|=\frac{2\pi}{a}N_{\rm max}$ with $N_{\rm
max}\sim 5$. Depending on the pseudopotential used and the material under
consideration, the total ground-state energy converges for larger cut-off
energies. A cut off energy of 800 eV implies a doubling of the value of $N_{\rm max}$ from 5 to 10. From the perspective of the Kohn-Sham approach to \gls{dft}, in
reciprocal space, the Kohn-Sham equations in second quantization are given by (see \cref{eq:real_space_eigeneq})
\begin{equation}
    \label{eq:KS_equation_momentum}
    H^{\mathrm{KS}}(\bm{k})\phi_{\bm{k},n}(\bm{r}) = \epsilon_{\bm{k},n}\phi_{\bm{k},n}(\bm{r}),
\end{equation}
with
\begin{equation}
\label{Ham_KS}
H^{\mathrm{KS}}(\bm{k})=\sum_{\bm G,\bm G'}\left[\frac{\hbar^2|\bm k+\bm
G|^2}{2m}\delta_{\bm G,\bm G'}+U^{\rm eff}_{\bm G-\bm G'}\right]f^\dagger_{ \bm
k+ \bm G}f_{\bm k+\bm G'},
\end{equation}
where the effective potential $U^{\rm eff}_{\bm G-\bm G'}$ also includes the contribution due to the exchange-correlation functional.

\begin{figure}[h!]
 \centering
 \includegraphics[scale=0.65]{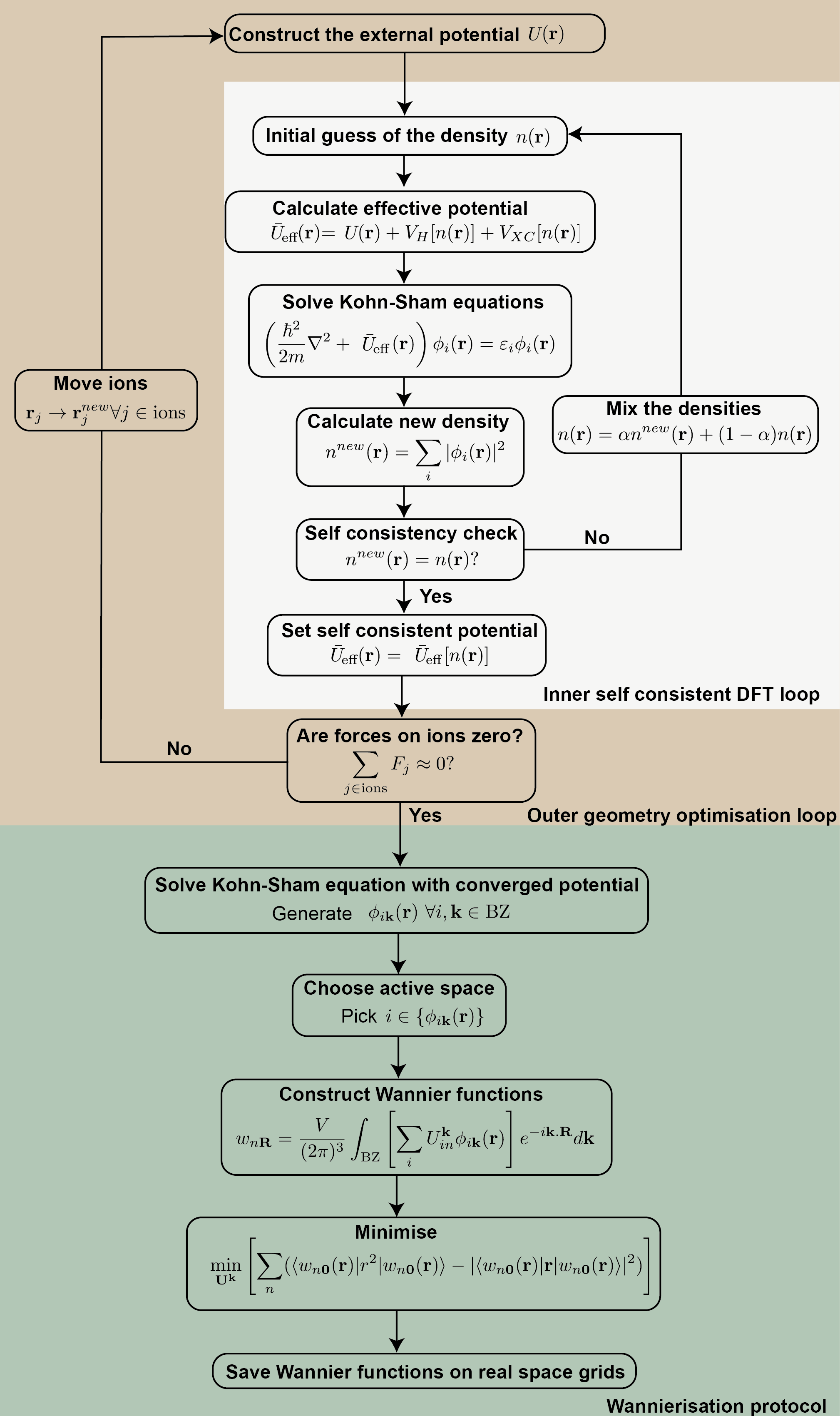}
 \caption{Workflow (in yellow) of a typical Density Functional Theory calculation highlighting the inner electronic self-consistency loop and outer structural optimisation loop. We supplement this procedure with a Wannierisation protocol (in green) illustrating the steps required to transform from a plane-wave basis set to a maximally localised one. }
  \label{fig:dft_workflow}
 \end{figure}

Additionally, before the optimal truncation of the active space occurs, the force on the ions created by the electronic charge of the electrons must be minimised. So far, in the spirit of the Born-Oppenheimer approximation, we have neglected the dynamics of the ions \cite{Cederbaum_BO}. In particular, their positions $\set{\bm{R}_I}$ enter the Kohn-Sham equations of \cref{eq:kohn_sham} as parameters, while their motion occurs on potential energy surfaces which are determined by the eigenvalues $\epsilon_i(\set{\bm{R}_I})$ of the electronic problem. At equilibrium, denoting by $\epsilon_0(\set{\bm{R}_I})$ the ground-state energy of the electronic system, the minimization of the force acting on ion $I$ requires that
\begin{equation}
  \label{eq:forces}
  F_{I} = \frac{\partial \epsilon_0(\set{\bm{R}_I})}{\partial \mathbf{R}_{I}} < \delta \quad \forall I, 
\end{equation}
i.e., the ions' equilibrium positions are obtained from the minimization of $\epsilon_0(\set{\bm{R}})$, which is a function of $3N$ variables, and $\delta$ is the threshold value on the force all ions must satisfy. In the upper part of \cref{fig:dft_workflow} we illustrate the typical workflow of a self-consistent \gls{dft} calculation, from 
calculating the external potential until self consistency is achieved in the electronic density and geometry.  

\subsubsection{Truncation into an active space}
\label{sec:active_space}

\begin{figure}[ht]
 \includegraphics[width=\linewidth]{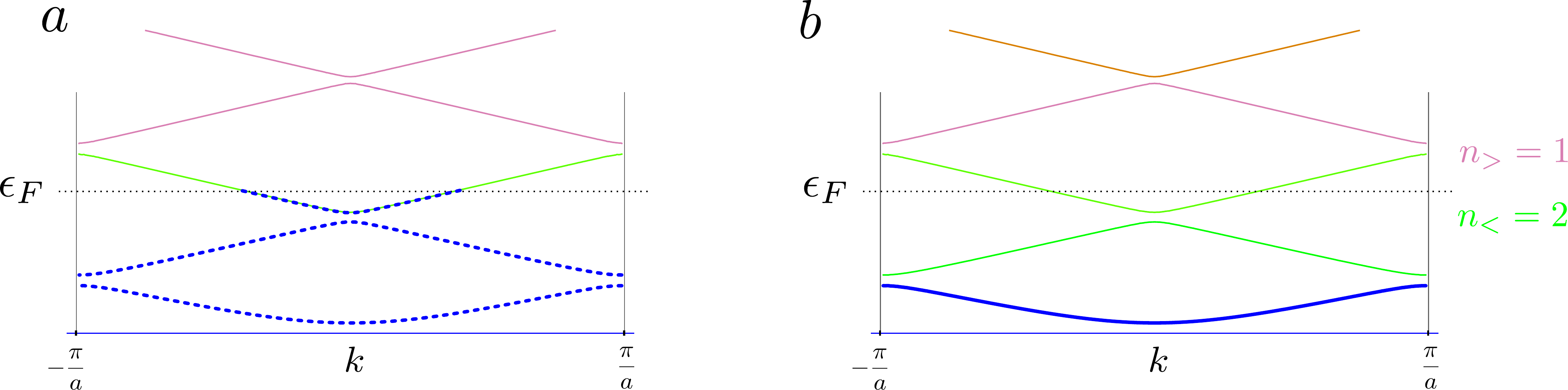}
 \caption{\textbf{(a)} The number of electrons in the system defines the Fermi energy $\epsilon_F$. In the non-interacting picture, the fermions fill the lowest single-particle energy levels, shown here as bands
 in the Brillouin zone, for a one dimensional system. The blue dots represent those energy levels occupied by fermions. In the figure, the lowest two bands are fully occupied, while the third
 band  is partially occupied. \textbf{(b)} Instead of considering all the bands, we can define an active space, where the electrons will reorganize due to interaction effects. This active space is represented 
 here by the states in the shaded area, around the Fermi level $\epsilon_F$. In this example, the number of bands below (or crossing) the Fermi level is $n_<=2$, while the number of bands above the Fermi level is $n_>=1$.}
 \label{fig:bands}
\end{figure}

To further reduce the dimension of the Hilbert space, we can truncate the Kohn-Sham eigenstates
into a minimal representation within an active space of chemical interest. One
such definition, as illustrated in \cref{fig:bands}, is to consider just the
states around the Fermi level. Fixing the number of bands below the last occupied band (including the latter) to be $n_<$ and the number of bands above it to be $n_>$, the dimension of the reduced Hilbert space
$\mathcal{H}_{\rm red}$ for a three dimensional material scales as
\begin{equation}
 {\rm Dim}(\mathcal H_{\rm red})\sim O(e^{(n_>+n_<)N_1N_2N_3\mathcal{F}(x)}),
\end{equation}
with $x=\frac{n_<-1+\nu_{\rm el}-\lfloor\nu_{\rm el}\rfloor}{n_{>}+n_{<}}$ and
$\mathcal{F}(x)=-x\ln x -(1-x)\ln (1-x)$. In addition to knowing the dimension of the Hilbert 
Space, the orbital character of the individual quantum states is required for 
the chemical interpretation of the possible physical processes that can occur between these states. Once the relevant Kohn-Sham orbitals have been selected, this represents the basis of the fermion operator in the active space. In contrast, to generate Wannier functions, further classical computation has to be performed.

A crucial step to generating \glspl{mlwf} is to provide an initial set of sensible
projectors that reflect the orbital character of the Kohn-Sham plane-wave
eigenstates. To achieve this, a local projection operator $\mathbf{\hat{P}}_i^\mathcal{N}$ is
used, which projects onto the subspace $\mathcal{N} = \{ \alpha, l, m \}$, where $\alpha$
is the principal quantum number, $l$ is the azimuthal quantum number and $m$ the
magnetic spin quantum number centered at ion $I$ \footnote{These are the quantum numbers associated with the eigenstates of the angular momentum operator, and here are used as a basis for interpretability of the Kohn-Sham orbitals in therms of atomic or molecular orbitals}. Assuming a paramagnetic spin system,
the local projection is given by, 
\begin{equation}
  \mathbf{\hat{P}}_I^{\mathcal{N}} = | Y_I^{\mathcal{N}} \rangle \langle  Y_I^{\mathcal{N}} |,
\end{equation}
where $Y_I^{\mathcal{N}}$ is the spherical harmonic centred at the centre of ion $I$ and
in the subspace $\mathcal{N}$. Therefore, for the Kohn-Sham eigenpair $(\epsilon_{\mathbf{k},n},
\phi_{\mathbf{k},n})$, the projected weight $p_{\mathbf{k},n}^{\mathcal{N}}$ is defined as,

\begin{equation}
    \label{eq:fatband}
    p_{\mathbf{k},n}^{\mathcal{N}} = \langle \phi_{\mathbf{k},n} | \mathbf{\hat{P}}^\mathcal{N} | \phi_{\mathbf{k},n} \rangle = | \langle \phi_{\mathbf{k},n} | Y_I^{\mathcal{N}} \rangle|^{2}. 
\end{equation}

Subsequently, each Kohn-Sham eigenpair generates a set of weighted projections
$\{(\epsilon_{\mathbf{k},n}, \phi_{\mathbf{k},n}, p_{\mathbf{k},n}^{\mathcal{N}})\}$ at each ion site $I$ for the
chosen subspace $\mathcal{N}$. Typically, for a given subspace, this weight is overlaid
at each point in the band-structure as a colour gradient, and highlights the
orbital character of all the Kohn-Sham eigenpairs in the chosen subspace. For example, $\mathcal{N}=\{2,2,0\}$ determines
the $3d_{z^2}$ subspace, given by the spherical harmonic $Y_{I}^{2,2,0}$
centred at ion $I$. In Quantum Espresso, the axes of the spherical harmonics are
orientated so that they aligns with the Cartesian axes. 

Finally, \glspl{mlwf} are then generated with the Wannier90 code \cite{w90}. The lower half of \cref{fig:dft_workflow} summarises 
the protocol for producing \glspl{mlwf} after the \gls{dft} calculation has been run, which takes as input the Kohn-Sham eigenstates and 
outputs Wannier functions on a real space grid. 

So far we have discussed the construction of effective Hamiltonians starting from a material and reducing it to generate a Hamiltonian with the same general characteristics as the starting system (with the same symmetries, and of the same size in terms of unit cells). Another approach is to appeal to effective descriptions of physical systems, where a portion of the system is considered in a different footing than the rest. If one subsystem is small compared with the other, it is possible to replace the larger portion by an effective description in terms of a bath. This procedure is actually exact in the limit of lattices with infinite connectivity \cite{PhysRevLett.62.324}. At the end of this procedure, a Hamiltonian that looks formally like \cref{eq:H_general} is obtained. The tools that we develop in the sections below work equally well for these systems.

\subsection{Summary} 

Electrons in solids can behave in completely unexpected ways, depending on the ion composition and the interactions between electrons. Using a classically cheap zeroth order description based on \gls{dft}, it is possible to isolate the relevant degrees of freedom that participate in a given phenomenon. 
From this description, we can construct a distilled Hamiltonian that contains the most important interactions and hopping terms within modes in the active space. 
We have also shown that a further compression is possible due to the structure of materials, where the thermodynamic number of degrees of freedom is encapsulated in a separation between bath and impurity modes in embedded approaches. Both strategies ultimately generate a Hamiltonian consisting of a restricted set of modes. 
This effective Hamiltonian is constrained by the symmetries of the system, and the same symmetries can be used to construct the Hamiltonian, reducing the classical cost of computation, but also limiting the possible interaction terms, thus reducing the overall complexity of the quantum circuits that implement the interactions. 
The relevant physics of the system is encoded in this compression.

In the following sections, we discuss in detail how to create a quantum circuit that implements the different terms of an effective Hamiltonian, with the goal of performing VQE or TDS. To achieve that, it is crucial to have an efficient way of representing fermionic degrees of freedom in terms of qubits. 

%% file: Qubit_Hamiltonian.tex
\section{Qubit representation}
\label{sec:qubit_rep}

In order to represent a fermionic system on a \gls{qc}, a mapping must be specified between the fermionic Hilbert space, and the multi-qubit Hilbert space of the \gls{qc}.
Such a mapping is most conveniently specified by a correspondence between fermionic operators and qubit operators. 
There are many design schemes available for such mappings \cite{derby2021compact,derby2021compactalt,bravyi2002fermionic,ball2005fermions,verstraete2005mapping,steudtner2018fermion,chen2018exact,setia2019superfast,jiang2019majorana}, with significant room for variation in the details of their implementation. The most commonly used mapping is the \gls{jw} transform, which maps fermionic creation ($c_i^{\dagger}$) and annihilation ($c_i$) operators to string-like qubit operators:
\begin{equation}
c_i^{\dagger} \leftrightarrow \left( \prod_{j<i} Z_j \right)\frac{(X_i+i Y_i)}{2} \;,\; c_i \leftrightarrow \left( \prod_{j<i} Z_j \right)\frac{(X_i-i Y_i)}{2}.
\end{equation}

The choice of mapping can have important consequences for the circuit depth and qubit requirements of \gls{tds} and \gls{vqe}.
Furthermore the way in which the mapping choice influences these costs will depend strongly on the structure of the given Hamiltonian, as well as the available hardware connectivity. 
Thus it is not obvious what the correct choice of mapping should be in general.

It is generally best to use a mapping which specifically maps the interactions (understood as both electron-electron interactions and hopping terms) present in the Hamiltonian to low-weight operators (i.e., operators that act non-trivially in just a small subset of the qubits, without scaling with the size of the system). 
The \gls{jw} transform is not well equipped to do this in general. An example of a mapping which is better suited to this, and that we will make use of in this work is the Compact Encoding~\cite{derby2021compact}. 

Unfortunately, in cases where there is a high degree of interaction between modes in the Hamiltonian, it is simply not possible to map all interactions to low-weight operators, regardless of the choice of mapping.
In lieu of low-weight representations, a fswap network protocol may be employed, wherein fermionic modes are dynamically re-ordered throughout the algorithm, such that each interaction admits a low-weight representation at some point in the protocol. 
Such an fswap network amortizes the cost of performing high-weight interactions, at the expense of having to actively re-order the fermionic modes in the mapping.
This amortization can be very powerful. Indeed, in the case where we want to implement all-to-all quadratic interactions it can be shown -- under weak algorithmic assumptions -- that fswap network methods in conjunction with the \gls{jw} mapping can yield essentially optimal circuit depths (see \cref{app:jw}).
 More details about the fswap network protocol will be discussed in \cref{sec:time_evolution_by_terms} and \cref{sec:swap_network_details}.

To date, fswap networks of this kind have been employed exclusively in conjunction with the \gls{jw} transform~\cite{kivlichan18,cade20,hagge2021}. 
However, in principle, they may be used in conjunction with any fermion-to-qubit mapping, as the act of reordering fermionic modes admits a representation purely in terms of the fermionic algebra. 
Furthermore, in the case where a subset of modes have a high degree of interactivity, fswap network protocols may be applied to this subset in isolation. This allows us to leverage the optimality of the fswap network protocol for all-to-all interactions, restricted to this subset where it is relevant.   
This suggests that a hybrid strategy may be ideal, wherein clusters of highly interacting modes are handled by an fswap network protocol, while any sparse connectivity is handled by a specific choice of mapping.

With this in mind, and for the purposes of comparison, we focus our attention on two basis choices for the material Hamiltonian: the Bloch basis Hamiltonian (\cref{Ham_band_basis}) and the Wannier basis Hamiltonian (\cref{Ham_wannier}). 

\subsection{Bloch basis mapping}
The Bloch basis Hamiltonian is given by (see \cref{Ham_band_basis}):
\begin{equation}\label{eq:bandbasis}
H^{B}=\sum_{\bm k, n, \sigma}\epsilon_n(\bm k) f^\dagger_{\bm k,n, \sigma}f_{\bm k,n, \sigma}
+ \sum_{\sigma, \sigma'} \sum_{\substack{n_1,n_2,n_3,n_4\\ \bm k, \bm q, \bm k'}} V_{n_1n_2n_3n_4}^{(\bm k,\bm k',\bm q)}f^\dagger_{\bm k + \bm q,n_1,\sigma}f^\dagger_{\bm k' - \bm q, n_2, \sigma'}f_{\bm k',n_3, \sigma'}f_{\bm k,n_4,\sigma}.
\end{equation}
The quartic interactions in the Bloch basis Hamiltonian have no specific local structure -- there are effectively interactions between every mode. 
This suggests that the best choice of mapping is the \gls{jw} transform, in conjunction with an fswap network protocol.

\subsection{Wannier basis mapping}
\label{sec:Wannier_basis_mapping}
The Wannier basis Hamiltonian (see \cref{Ham_wannier}) is given by:
\begin{align}
H^{W}&=\sum_{\sigma}\sum_{\substack{m,n\\\bm{R}_1,\bm{R}_2}}T(\bm{R}_1-\bm{R}_2)_{mn}w_{\bm{R}_1,m,, \sigma}^{\dagger}w_{\bm{R}_2,n, \sigma}\nonumber\\
&+\sum_{\sigma, \sigma'}\sum_{\substack{s,l,m,n\\\bm{R}_{1},\bm{R}_{2},\bm{R}_3,\bm{R}_4}}
\tilde{V}^{(\bm{R}_{1}\bm{R}_{2},\bm{R}_3,\bm{R}_4)}
_{slmn}w_{\bm{R}_{1},s, \sigma}^{\dagger}w_{\bm{R}_{2},l, \sigma'}^{\dagger}w_{\bm{R}_3,m, \sigma'} w_{\bm{R}_4,n, \sigma}. \label{eq:wannierbasis}
\end{align}

\begin{figure}
\center
\includegraphics[scale=0.7]{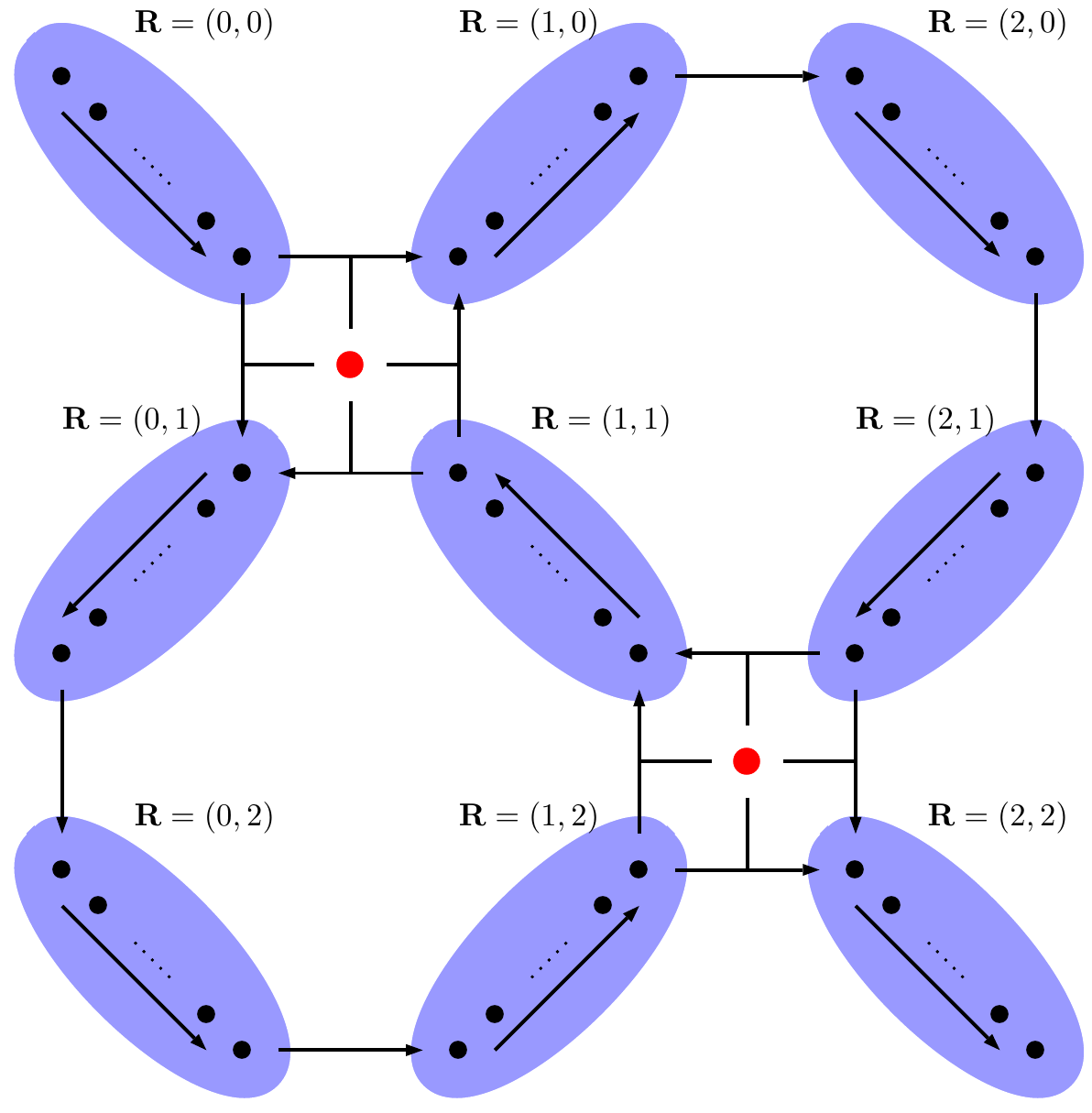}
\caption{Schematic for the hybrid fermion to qubit mapping for a 2D material. Black dots correspond to both fermionic modes and their corresponding data qubit. Red dots correspond to ancillary face qubits operating in the same fashion as in the square compact encoding \cite{derby2021compact}. Blue ellipses surround all fermionic modes assigned to a given site $R$. These modes are arranged in a line, with all interactions within the line taking the same form as in a \gls{jw} transform with an identical linear ordering. Interactions between modes at the ends of neighbouring lines take the form of interactions between neighbouring modes in the compact encoding. Physically, each blue ellipse contains modes associated with the orbitals kept in a unit cell.}
\label{fig:hybrid}
\end{figure}

In contrast to the Bloch basis, the Wannier basis Hamiltonian has some local structure: interactions between modes indexed by ${\bf R}_1$ and ${\bf R}_2$ are suppressed as $|{\bf R}_1-{\bf R}_2|$ increases. However, in the case where $|{\bf R}_1-{\bf R}_2|$ is small, for example nearest neighbour or on-site, interactions are strong, and in general all-to-all with respect to the orbital index. In this case we make use of a hybrid strategy. First, as discussed in \cref{sec:motif}, we label each site index $\bm{R}$ by the corresponding triplet of integers $(n_1, n_2, n_3)$ (defined in \cref{eq:R_lattice}) on a Cartesian grid. Then, we map the system to an expanded compact encoding illustrated in \cref{fig:hybrid} for a 2D material.
In this mapping, all modes which share a common site index ${\bf R}$ are associated with a collection of qubits laid out in a \gls{jw} style string, and each string is connected to nearest-neighbouring strings using the compact encoding design. This encoding is most concisely expressed in terms of ``edge'' ($E_{ij}$) and ``vertex'' ($V_i$) operators, which are defined as:
\begin{equation}
E_{jk} := -i \gamma_j \gamma_k,  \quad V_{j} := -i \gamma_j \bar{\gamma}_j,
\end{equation}
with Majorana operators
\begin{equation}
\gamma_j := w_j+w_j^\dagger, \quad \bar{\gamma}_j := (w_j - w_j^\dagger)/i,
\end{equation}
where $j$ (and $k$) is a multi-index over the site index $\bf R$, mode index $m$, and spin index $\sigma$. The edge and vertex operators are hermitian, they anti-commute when they share an index, and commute otherwise. Furthermore, the edge operators satisfy an important composition relation:
\begin{align}\label{eq:edge_comp}
E_{ik} = i E_{ij} E_{jk}.
\end{align}

The edge and vertex operators can be combined to synthesize any fermionic terms with an even number of creation and/or annihilation operators, i.e., all observables that preserve parity super-selection -- a fundamental requirement of any realistic Hamiltonian. For reference we include these decompositions for quadratic terms:
\begin{align}
w_i^\dagger w_i = (1-V_i)/2, \;&\; w_i^\dagger w_j = \frac{i}{4} (1-V_i)(1+V_j)E_{ij}, \\
w_i w_j = \frac{i}{4}(1+ V_i)(1+ V_j)E_{ij}, \;&\; w_i^\dagger w_j^\dagger = \frac{i}{4} (1-V_i)(1-V_j)E_{ij}, \\
w_i^\dagger w_j + w_j^\dagger w_i &= \frac{-i}{2} \left(E_{ij} V_j + V_i E_{ij} \right).
\end{align}

Quartic terms may be constructed from quadratic terms; however, in this case, there is a freedom in the choice of decomposition into edge operators thanks to \cref{eq:edge_comp}. This freedom may be used to choose a decomposition with the smallest qubit representation. Given that the fermionic encoding ultimately maps products of Majorana operators (Majorana monomials) to Pauli operators, it is most convenient to first decompose the fermionic Hamiltonian into the operator basis of Majorana monomials
\begin{align}
\label{eq:majorana_hamiltonain}
H_M := \sum_{b \in \{0,1\}^{2M}} \alpha_b \prod_j \gamma_j^{b_{2j}} \bar{\gamma}_j^{b_{2j+1}} \;,\; |b| \in \{2,4\},
\end{align}
before proceeding with mapping it to a qubit Hamiltonian by applying the encoding to each Majorana monomial. Here, $M$ is the total number of complex fermion modes and is given by $M = N_{\mathrm{modes/cell}} N_{\mathrm{cells}}$, with $N_{\mathrm{modes/cell}}$ and $N_{\mathrm{cells}}$ the number of modes per unit cells and unit cells in \cref{eq:wannierbasis}, respectively.

\begin{figure}
\center
\includegraphics[scale=0.7]{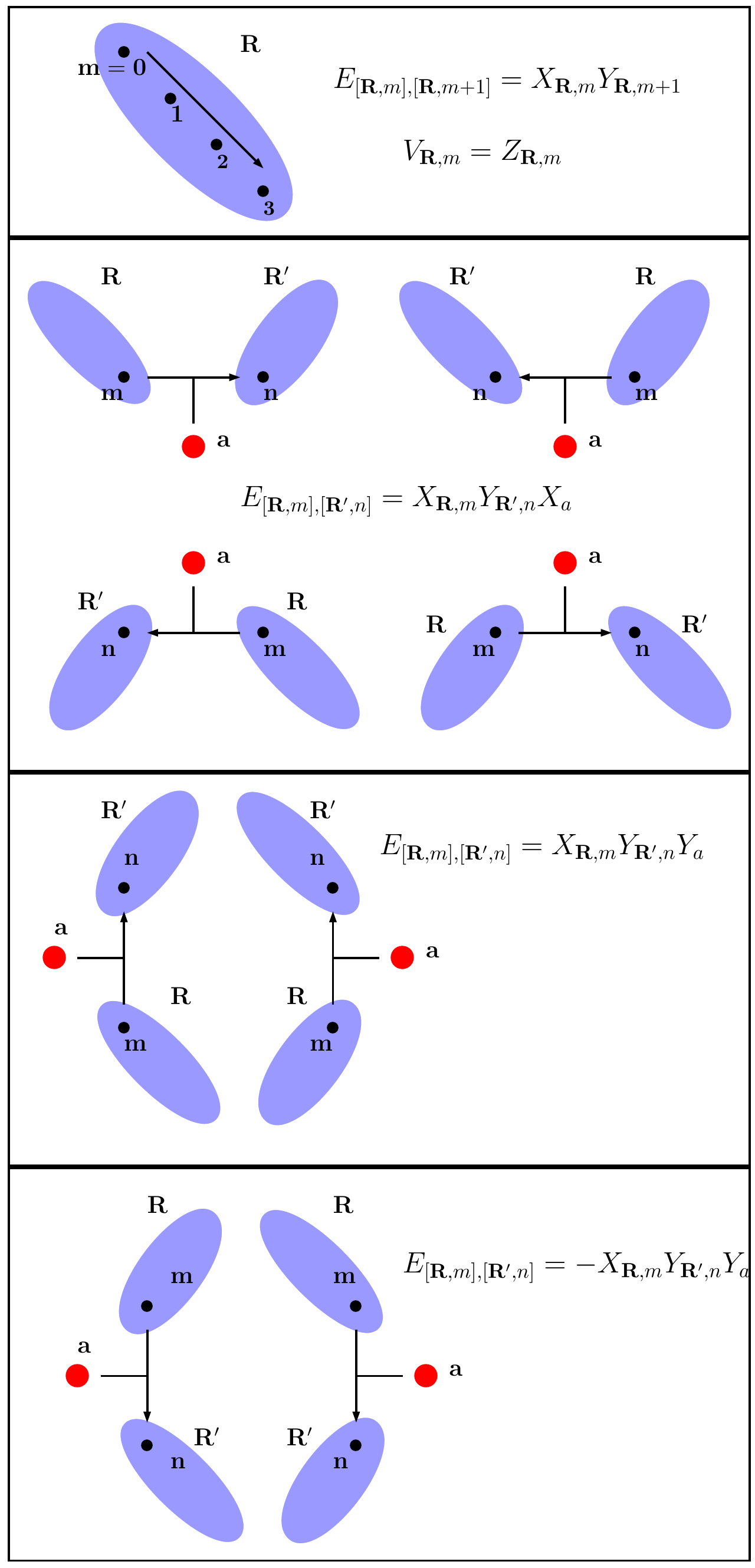}
\caption{Form of the edge ($E_{ij}$) and vertex ($V_i$) operators associated with the diagram in \cref{fig:hybrid}.}\label{fig:hybrid_details}
\end{figure}

The precise details of how the hybrid mapping specifies the edge and vertex operators are given in \cref{fig:hybrid_details}. In the mapping, only certain edge operators are specified. Any other edge operators must be constructed using the composition relation in \cref{eq:edge_comp}. Thus interactions between modes distant from one another on the graph geometry will decompose into products of edge operators, yielding string-like Pauli representations similar to \gls{jw} strings. It is here where the fswap network protocol plays an important role.


This representation takes advantage of both the benefits of fswap network protocols in the context of all-to-all connectivity, which can be naturally applied to these \gls{jw} style strings, and the benefits of the local structure manifest between sites, through the compact encoding. The efficacy of this approach will depend strongly on how localized  the Hamiltonian interactions turn out to be for the particular material.

\subsubsection{Stabilizers}

\begin{figure}
\center
\includegraphics[scale=0.7]{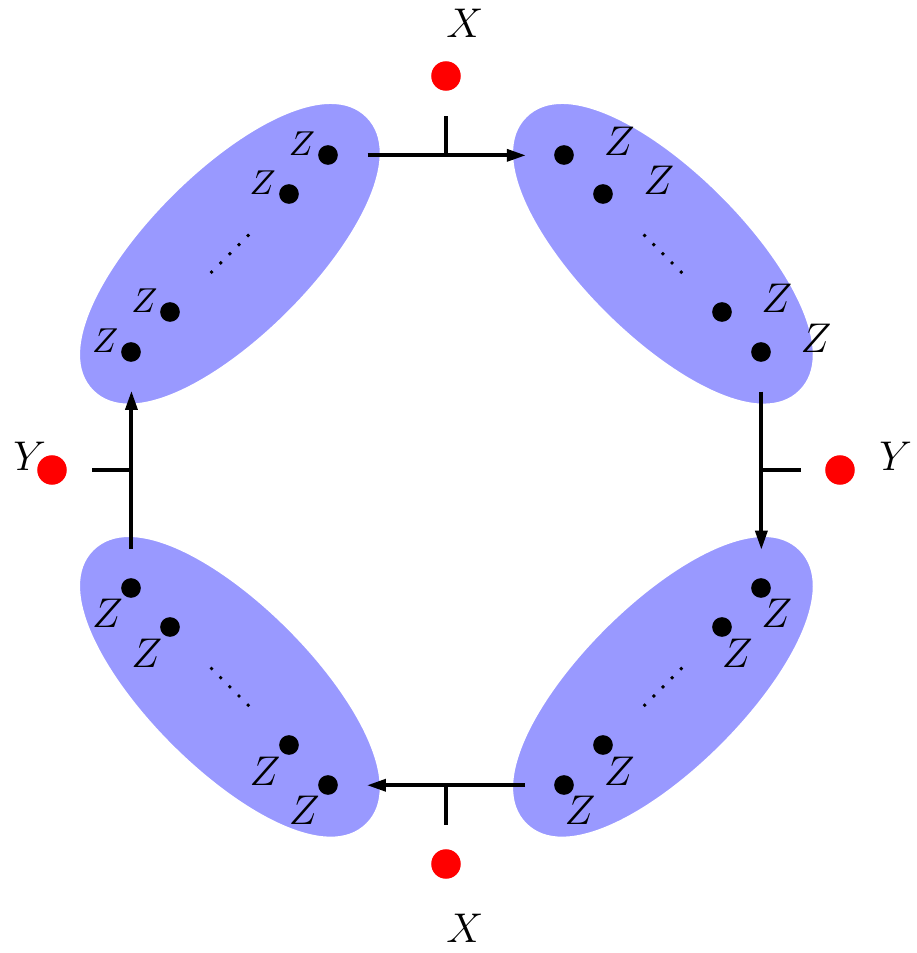}
\caption{Stabilizer of the hybrid encoding.}\label{fig:stabilizer}
\end{figure}

Unlike the \gls{jw} transform, the hybrid encoding represents fermionic states in a subspace of the multi-qubit Hilbert space. This subspace is best described as the code space of a stabilizer code \cite{nielsen_chuang_2010}. The generators of the stabilizer code are given by the ordered product of loops of edge operators around the octagonal faces in \cref{fig:hybrid}, i.e., the faces with no ancillary qubit. An instance of one these generators is illustrated in \cref{fig:stabilizer}. 

The fact that the fermionic system lives in a code space means that state preparation incurs an additional overhead compared to the \gls{jw} transform. This is discussed further in  \cref{sec:state_prep}. However one benefit these stabilizers do supply is error detection. We will not discuss that in detail in this paper. For some examples of this, see \cite{derby2021compact}.

\subsubsection{3D Layout}\label{sec:3dLayout}

For the simulation of the 3D bulk of materials there are two approaches one may take.

\begin{figure}
\center
\includegraphics[scale=0.5]{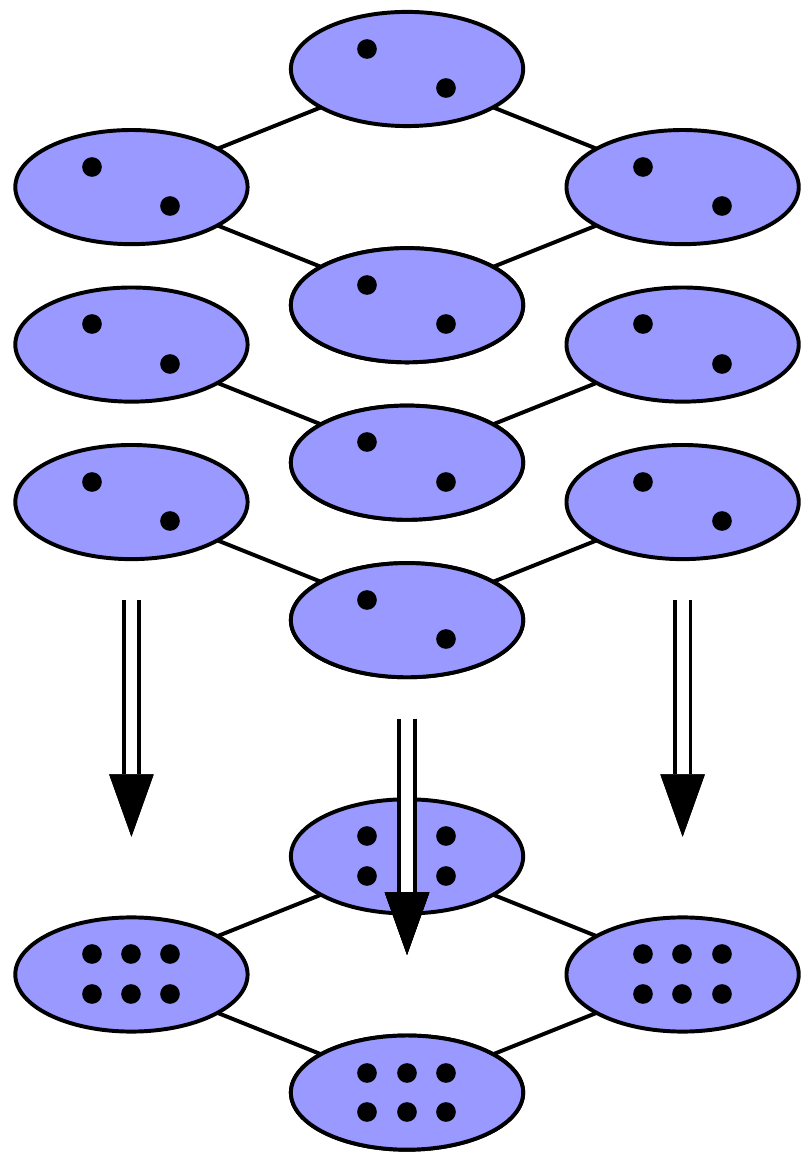}
\caption{Collapsing a 3D model to a 2D model}\label{fig:3dto2d}
\end{figure}

One approach is to collapse the 3D lattice of sites into a 2D lattice that matches the QC layout, with each site containing $N_{\mathrm{modes/cell}}*L_i$ modes, as illustrated in \cref{fig:3dto2d}, where $L_i$ is the side length you collapse. If the original 3D Hamiltonian has a nearest neighbour interaction structure on the Cartesian motif, then the collapsed 2D counterpart will also have a nearest neighbour structure. The downside of this approach is that it increases the depth of the fswap network protocol, since the number of modes on an individual site has increased. Additionally, it does not leverage the full sparsity of the Hamiltonian. The upside is that the qubits on superconducting devices are typically confined to a planar layout, and so this approach lends itself well to such devices. 

\begin{figure}
\center
\includegraphics[scale=0.7]{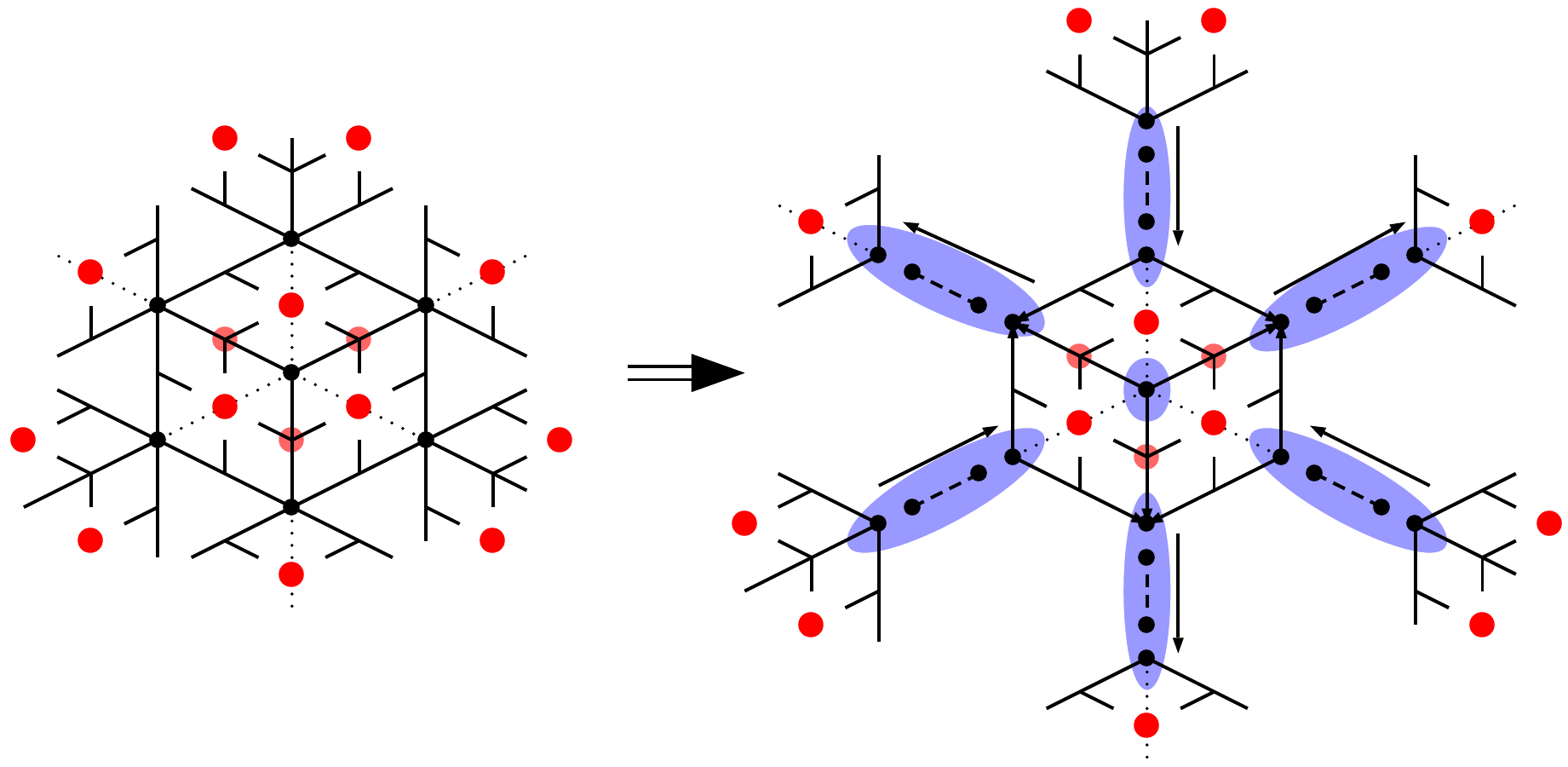}
\caption{(left) 3D compact encoding converted to (right) 3D hybrid encoding. Black dots correspond to both fermionic modes and their corresponding data qubits, and red dots correspond to ancillary qubits. Red dots are positioned on the faces of the cubes.}
\label{fig:3dlayout}
\end{figure}

\begin{figure}
\center
\includegraphics[scale=0.7]{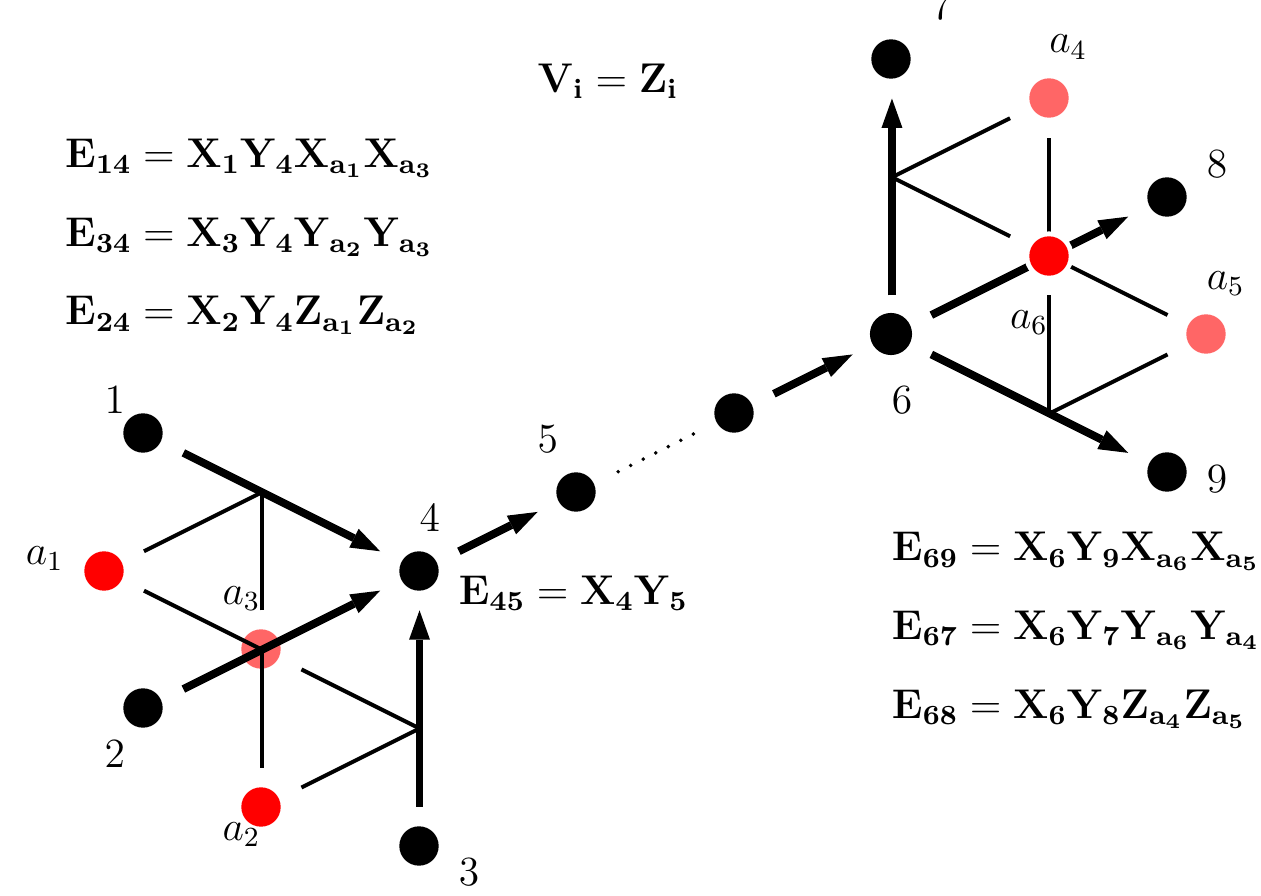}
\caption{The edge and vertex operators of the 3D hybrid encoding.}
\label{fig:3dDetails}
\end{figure}

The second approach is to employ a 3D generalization of the 2D compact encoding, as described in \cite{derby2021compactalt} to construct a 3D generalization of the hybrid encoding -- illustrated in \cref{fig:3dlayout}. The basic principle remains the same, however in this case the edge operators connecting the different sites are weight 4 instead of weight 3. The details of this construction are given in \cref{fig:3dDetails}.

After constructing the map between fermion and qubit operators, we have all the preliminary ingredients to study the complexity of performing a quantum algorithm. In the next section we discuss the implementation of \gls{vqe} and \gls{tds} algorithms.

%% file: Algorithms.tex
\section{VQE and TDS algorithms}\label{sec:vqe_alg}

In this section we will find bounds on the complexity of implementing the \gls{vqe} \cite{bharti2021noisy, peruzzo14} and \gls{tds} algorithms for materials' Hamiltonians.
We begin by reviewing these algorithms before describing the details of how we implement and cost them for the systems of interest to us. We will use a simple model where all-to-all qubit interactions are allowed, arbitrary 2-qubit gates are cost 1 each, and 1-qubit gates are free. Our goal is to minimise the overall circuit depth.

Note that one could also aim to minimise the total gate count. Which metric is most appropriate will depend on the hardware platform being used and whether one is considering a near-term or fault-tolerant model. Assuming that the hardware platform allows gates to be implemented in parallel, minimising the quantum circuit depth can reduce the effect of decoherence as well as the wall-clock running time. Further, by a light-cone argument, quantum circuits with low depth experience less spreading out of local errors. In any case, our low-depth circuits are also efficient in terms of gate count.

\gls{vqe} is a method which aims to produce the ground state of a quantum Hamiltonian, $H$, by optimising over trial quantum circuits picked from some family (``ansatz''), based upon the advance knowledge that such ansatz states should be able to represent the ground state effectively, and/or may be efficiently implementable on quantum hardware~\cite{peruzzo14,bharti2021noisy}. 
Circuits from the ansatz have parameters which are optimised using a classical optimisation routine. This routine aims to minimise the energy with respect to $H$ of the state produced by the quantum circuit. 
By the variational principle, if the ground state can be represented within the ansatz and is unique, minimising the energy output from parameterised circuit will lead to the circuit which prepares ground state. 
In practice, the optimisation routine may find a local minimum rather than the global minimum. Once an approximation to the ground state has been produced, measurements can be performed to determine properties of interest.
    
Here we will use the Hamiltonian variational ansatz~\cite{wecker15} within the \gls{vqe} framework. This ansatz may be used to find the ground state of a Hamiltonian $H_M = \sum_k h_k$, where the terms $h_k$ correspond to the nonzero Majorana monomials in \cref{eq:majorana_hamiltonain}. We assume that we can write $H_M = H_A + H_B$, where we have an efficient quantum circuit for preparing the ground state of $H_A$, and that the time-evolution operations $e^{i t h_k}$ can be implemented efficiently for all $k$ and arbitrary times $t$ (for example, $h_k$ may act non-trivially on only a small number of qubits). Then we perform the following steps:
\begin{enumerate}
    \item Prepare the ground state of $H_A$.
    \item For each layer $l$, implement the operation
    \[ \prod_k e^{it_{lk} h_k}, \]
    for some parameters $t_{lk}$, to produce a state $\ket{\psi}$. Note that we will allow the ordering of the product to be arbitrary below, which can allow for more efficient algorithms.
    \item Measure the energy of $\ket{\psi}$ with respect to $H_M$.
    \item Optimise over the parameters $t_{lk}$ to find the ground state (or a good approximation).
\end{enumerate}
Here we will compute the complexity of step 1, and of one layer of step 2.
We will also determine the number of measurements required to measure the energy of the ground state. To gain a full understanding of the complexity of \gls{vqe}, it is also necessary to understand how many layers are required, and how efficient the optimisation process is, which we will not consider here; see~\cite{cade20,mineh21} for detailed numerical analyses of these points in the case of the Fermi-Hubbard model.

By contrast to \gls{vqe}, the \gls{tds} approach corresponds to approximately implementing the unitary operation $e^{-itH_M}$ for some $t$. The standard method for executing this operation is by Trotterisation, wherein for example $e^{itH_M}$ is approximated by a product of short time steps $(\prod_k e^{i \delta t h_k})^L$, with $\delta t = t/L $. Simulating time-dynamics of a quantum system is theoretically more straightforward than finding a ground state (\gls{bqp}-complete in the worst case, rather than \gls{qma}-complete \cite{watrous09}), yet may be more challenging in practice for near-term quantum computers, as the approximation may demand that the number of Trotter steps $L$ be large. If an algorithm based on Trotterisation is used, \gls{tds} is very similar to implementing step 2 of the \gls{vqe} algorithm. However, note that we have assumed in \gls{vqe} that the time-evolution steps for each $h_k$ can occur in arbitrary order, whereas for some more sophisticated Trotterisation methods we may want to fix a particular order for \gls{tds}. For the purposes of this analysis we assume the Trotterisation does not require a particular order on the sequence of terms -- which is true in the case of first order Trotterisation. The only caveat is that we group together commuting terms in order to be able to compute improved upper bounds on the Trotter error. As in \gls{vqe}, our analysis is for a single short time step. The circuit depth would then need to be multiplied by the desired number of time steps.  

We now discuss how each of the above steps is implemented. The algorithm is based on a set of basic operations, and we begin by calculating their quantum circuit complexity.


\subsection{Gate decompositions of operations}
\label{sec:gate_complexity}

There are three types of operations used in our algorithm: time-evolution according to Majorana operators, fermionic swaps, and Givens rotations.

\begin{figure}[t]
    \centering
    \includegraphics{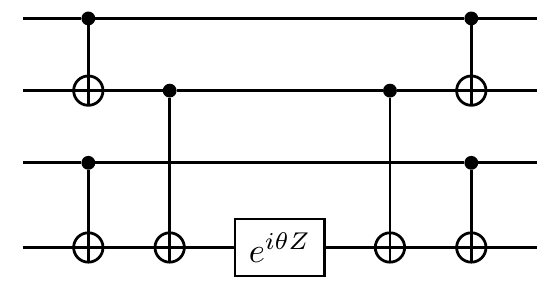}
    \caption{The quantum circuit for implementing $e^{i \theta Z^{\otimes k}}$ in the case $k=4$ in terms of CNOTs and single qubit rotations. The three middle gates can be combined into one 2-qubit operation, given total cost (2-qubit gate depth) 3 in our model.}
    \label{fig:paritycircuit}
\end{figure}

\textbf{Time-evolution.} Whichever fermionic encoding is used, each term of the Hamiltonian will ultimately be represented on the quantum computer as a string of Pauli operators. As 1-qubit gates are free and all Pauli operators are equivalent up to unitary conjugation, implementing a term reduces to implementing the operation $e^{i\theta Z^{\otimes k}}$, acting on $k \ge 1$ qubits, for arbitrary $\theta$. This can be done in depth $2\lceil \log_2 k \rceil - 1$ via a circuit which uses a binary tree of CNOT operations to put the parity of the input state in the last qubit; performs a Z rotation on that qubit; and then performs the CNOT operations in reverse to uncompute the parity. We save depth 1 by combining the last two CNOTs and the Z rotation in one 2-qubit gate. See  \cref{fig:paritycircuit} for an example for the case $k=4$.

Note that we may have multiple commuting terms acting on the same qubits, which can lead to efficiency savings by implementing time evolution according to these simultaneously. However, we do not consider this in our calculations.

\textbf{Fermionic swaps.} In the JW transform, fermionic swaps across adjacent modes are 2-qubit gates with cost 1. In the compact encoding, across most pairs of adjacent modes, the same holds. The exception is fermionic swaps across different material sites, where an ancilla qubit is involved. Thus it is most convenient to express the fermionic swap operator in terms of the fermionic algebra as
\begin{equation}
\textrm{FSWAP}_{ij} = \exp\left(i\frac{\pi}{4} V_i\right) \exp\left(i\frac{\pi}{4} V_j\right)\exp \left( \frac{\pi}{4}E_{ij}V_j\right) \exp \left( \frac{\pi}{4}V_i E_{ij}\right),
\label{eqn:fswap_decomposition}
\end{equation}
The circuit depth may be computed by decomposing each of the terms in terms of Pauli matrices. However, here we can get an efficient decomposition for weight 3 edge operators by combining operators as follows. The first two terms are single-qubit unitaries and hence free in our model. Written in terms of Pauli operators, the remaining terms are of the form
\begin{equation}
\exp\left(\pm i\frac{\pi}{4}(X_1 X_2 + Y_1 Y_2)P_A \right),
\end{equation}
where $P_A$ is a single qubit Pauli operator. We can diagonalise $P_A$ on the ancilla with a local unitary, and can diagonalise $X_1 X_2 + Y_1 Y_2$ with a 2-qubit gate $G$, which turns out to be a Hadamard gate acting only on the odd parity subspace. We get
\begin{equation}
    G(X_1 X_2 + Y_1 Y_2)G^\dag = I - \frac{1}{2}(Z_1 + Z_2).
\end{equation}
Therefore, we can implement the fermionic swap by a circuit of 2-qubit depth 4 (first apply $G^\dag_{12}$ to switch to the correct basis for these qubits; then a gate across qubit 1 and the ancilla; then across qubit 2 and the ancilla; and then $G_{12}$).

%
%
However, for the 3D encoding the edge operators are weight 4, and so we will not be using this decomposition in those cases, but rather the naive \gls{tds} decomposition. 

\textbf{Givens rotations.}
A Givens rotation is a unitary operation which mixes pairs of fermionic operators in the following way \cite{Jiang2018}:
\begin{equation}
\left(\begin{array}{c}
\mathcal{G}_{ij}(\theta, \phi) c_i^\dagger \mathcal{G}_{ij}^\dagger(\theta, \phi) \\
\mathcal{G}_{ij}(\theta, \phi) c_j^\dagger \mathcal{G}_{ij}^\dagger(\theta, \phi)
\end{array} \right) = \left(\begin{array}{cc}
\cos(\theta) & -e^{i \phi}\sin(\theta) \\
\sin(\theta) & e^{i \phi}\cos(\theta)
\end{array} \right)\left( \begin{array}{c}
 c_i^\dagger\\c_j^\dagger
\end{array}\right).
\end{equation}


Givens rotations are again most easily defined in terms of fermionic edge and  vertex operators, which may then be translated into qubit operations using the chosen mapping:
\begin{align}
\mathcal{G}_{ij}(\theta, \phi) &= \exp\left(-i\frac{\phi}{2} V_j\right) \exp \left(i \frac{\theta}{2}[E_{ij}- V_iV_j E_{ij}]\right),
\end{align}

In the JW transform, when acting on adjacent modes this is a 2-qubit operation. In the compact encoding, it is usually a 2-qubit operation, except when acting across sites, when it is an operation of the form
\[ e^{-i\frac{\phi}{2}Z_j} e^{i\frac{\theta}{2}(X_i Y_j- Y_i X_j)P_A}, \]
in 2D, and
\[ e^{-i\frac{\phi}{2}Z_j} e^{i\frac{\theta}{2}(X_i Y_j- Y_i X_j)P_A P_B}, \]
in 3D, where $P_A$ and $P_B$ are single qubit Paulis on face qubits, specified by the orientation of the edge.
By a similar argument to the fermionic swap operation (also see Appendix A of~\cite{cade20}), the 3-qubit operation can be implemented in 2-qubit gate depth 4. The 4-qubit operator can be implemented in 2-qubit gate depth 6.

As pointed out in the discussion at the beginning of this section, when using the Hamiltonian variational ansatz within the \gls{vqe} framework usually one splits the Hamiltonian as $H_M =  H_A + H_B$ and needs to prepare the ground state of $H_A$. Below we discuss how this can be done in the hybrid encoding.


\input{state_prep}


\subsection{Time-evolution according to terms in materials Hamiltonians}
\label{sec:time_evolution_by_terms}

For both the \gls{vqe} and \gls{tds} algorithms, we need to implement time-evolution according to the quadratic and the quartic parts of the materials Hamiltonians $H^B$, $H^W$.
To do this efficiently, we will use a protocol  based on fswap networks~\cite{kivlichan18}. These networks use layers of fswap gates to rearrange the fermionic ordering and to enable terms to be implemented efficiently -- for example, by moving modes to be adjacent in the JW transform. Here we apply fswap networks to more general fermionic encodings than the JW transform, and the notion of efficiency we will use is to apply operations across modes which are adjacent in terms of the encoding graph. That is, we will swap modes (using  fswap operations on adjacent modes) with the intent of bringing modes across which we wish to perform some operation closer together.

Theoretical constructions of fswap networks are known which implement all quadratic~\cite{kivlichan18} or quartic~\cite{Kanno22, microsoft_patent} terms efficiently, in the sense that they reduce the quantum circuit depth by a factor scaling like the number of modes. That is, all quadratic terms on $M$ modes can be implemented in quantum circuit depth $O(M)$, and all quartic terms can be implemented in quantum circuit depth $O(M^3)$. However, here we will want to apply fswap networks to specific materials' Hamiltonians that do not include all terms. We are therefore led to an algorithmic approach to produce an efficient fswap network protocol for a given Hamiltonian. We consider protocols that alternate between layers of the following form:
\begin{enumerate}
    \item Fswap gates across modes that are adjacent with respect to the graph of the fermionic encoding that we are using. For example, in the JW transform, fswaps across qubits of the form $(i,i+1)$ would be allowed.
    \item Time-evolution by all terms that are efficiently implementable given the current permutation of the graph of the fermionic encoding\footnote{Note that following this structure gives a ``greedy'' protocol where we implement all terms that are available at each step. One could also use a more incremental strategy where only some of these terms are implemented before the next layer of fswaps.}. We consider a term to be efficiently implementable if there is a split of the modes on which it acts into pairs such that all pairs are adjacent within the encoding graph. For the fermionic encodings we use, such terms correspond to low-weight Pauli operators.
\end{enumerate}
The aim is then to find a protocol using a small number of layers (corresponding to a good choice of positions for fswap gates), as well as implementing the interactions within each layer efficiently. We can achieve both of these using computational techniques, which we summarise here, with a more detailed description in \cref{sec:compiler}.

To find a good sequence of fswap gates, we use a greedy protocol. At each layer, we look at the set $\mathcal{T}$ of interactions $t$ which have not yet been implemented, and define a distance function which measures the difficulty of implementing these interactions. Here we focus on $\ell_p$ distance functions of the form
\begin{equation}
    \label{eqn:swap_net_full_cost}
    D = \left(\sum_{t \in \mathcal{T}} d(t)^p\right)^{1/p}, 
\end{equation}
where $p > 0$ and $d(t)$ is the ``distance'' of a term $t$. This is defined as the minimum, over all splits of the modes into pairs, of the distance within the encoding graph of those modes. This is closely related to the number of fswaps required to bring these modes together. Empirically, we found that choosing $p = 0.5$ seems to produce good results. We believe that this is because choosing $p<1$ puts greater weight on bringing terms with low distance closer together.

We consider swapping each possible adjacent pair of modes in the interaction graph, and compute $D$ for each choice. 
If there exists a pair of modes which reduces $D$ upon being swapped, we then fswap the pair and mark it as used. We repeat this process until all modes have been used, or there is no choice of modes to fswap that reduces $D$. A potential issue with this approach is that even in the first step, there may be no choice of pairs to fswap that reduces $D$. We can handle this by adding a fallback step where an fswap operation is chosen that at least reduces $d(t)$ for some term $t \in \mathcal{T}$.

To implement all terms efficiently in step 2 above, we express this problem in terms of graph colouring. We define a graph whose vertices are terms that should be implemented in the current layer, and where two vertices are connected if the corresponding terms can be implemented simultaneously. Here we take the simple view that two terms can be implemented simultaneously if they act on disjoint sets of qubits. Then the minimal number of colours required to colour this graph such that no two adjacent vertices have the same colour is the same as the minimal number of sublayers required to implement all the terms.
As graph colouring is an NP-complete problem, we do not expect to be able to find the exact minimal number of sublayers for large numbers of terms and many qubits.
However, we can use graph colouring heuristics to find an upper bound on the minimal number of layers efficiently. Here we use a greedy colouring algorithm with the DSATUR heuristic implemented in the NetworkX package \cite{networkx08}.

As an additional optimisation, as the first step of our protocol we implement the fswap network of~\cite{kivlichan18}, which enables all quadratic terms to be implemented using $M$ layers of fswaps, for a system with $M$ modes. During this process, we can also implement some other terms, if they happen to become efficiently available. As we expect the overall complexity to be significantly greater than $M$, this is a lower-order cost that can reduce the number of terms used substantially.

Finally, we need to decompose the operations we apply in terms of elementary quantum gates. For this we use the costing procedure described in Section \ref{sec:gate_complexity}.

We remark that recent work by Lao and Browne~\cite{lao21}, on quantum circuit compilation for efficiently simulating time-dynamics of 2-local qubit Hamiltonians, follows a similar strategy of decomposing the overall quantum simulation in terms of alternating layers of swaps and implementation of time-evolution operations, implemented in an arbitrary order. Their work also aims to find ``good'' swaps that minimise a distance measure. As here we are instead simulating fermionic Hamiltonians, this leads to a different (though conceptually related) notion of distance. In particular, in our setting we need to handle quartic terms, whereas the 2-local terms in Lao and Browne's work correspond to quadratic terms. Another point of difference is that these authors use a distance measure corresponding to reducing the distance of the ``closest'' term, rather than our distance measure that aims to track distance more globally, and hence to bring many terms closer together at once; also, here we introduce the use of a Steiner tree to reduce the size of the graph being considered, and use the ``chain'' swap network of~\cite{kivlichan18} as a subroutine. See \cref{sec:fswap_network} below for details.


\subsection{Measurements}
\label{sec:measurements}

The final step of a variational quantum algorithm is to measure the energy of the quantum state produced, with the overarching goal of minimising this energy. 
One may also wish to measure some other operator to extract a physical property of the state. In this section we discuss how measurement can be achieved efficiently, in the rather general setting where one wishes to measure an arbitrary set of quadratic or quartic terms, expressed as Majorana operators. 

We will focus on operators expressed in the JW transform. This also allows us to handle the case of the hybrid encoding with nearest-neighbour interactions between sites, because we can split terms into 4 groups (in 2D) or 6 groups (in 3D) that are only connected to nearest neighbours and locally look like Majorana operators in the Jordan-Wigner transform, with the exception of an ancilla qubit, which is always measured in the same basis. However, terms acting on next-nearest neighbours and beyond do not necessarily have this property.

Then a quadratic fermion term corresponds to Pauli strings of the form $AZZ\dots ZB$, where $A,B \in \{X,Y\}$, and the quartic case is either a product of two such strings on disjoint sets of qubits, or the product of a quadratic string and a $Z$ operator elsewhere. In the two-string case, we can assume that we only need to measure terms containing an even number of $X$'s (or $Y$'s), since our Hamiltonian has time reversal symmetry.

A naive measurement strategy would measure each term in sequence, using a number of measurements equal to the number of terms in the Hamiltonian, which can be at worst $\Theta(M^4)$ measurements for a quartic Hamiltonian on $M$ modes. Our goal here will be to do better by measuring multiple terms at a time. Minimising the number of rounds may not always give the strategy requiring the minimal number of measurements to achieve a certain level of accuracy, as this also depends on the variance of each measurement~\cite{crawford21}; however, this approach is a reasonable starting point. We also look for the measurements to be implemented using straightforward (ideally constant-depth) quantum circuits. To this end, we consider three types of measurement strategies:

\begin{itemize}
	\item \textbf{(QWC)} Measuring qubitwise commuting terms simultaneously. These are Pauli terms which commute when restricted to individual qubits. This family of measurement strategies is easy to implement by measuring each qubit in the correct X/Y/Z basis, so requires only additional single-qubit gates. In the literature, this concept is sometimes called measuring in a tensor product basis (TPB) \cite{kandala2017,Verteletskyi2020}.

	\item \textbf{(NC)} Measuring a family of \emph{non-crossing} terms simultaneously. We say that a pair of distinct quadratic Majorana operators acting on modes $i\le j$ and $k\le l$ is non-crossing if either:
	\begin{enumerate}
	    \item $j < k$, or $l < i$, or $i<k\le l<j$, or $k<i\le j<l$;
	    \item or $i=k$, $j=l$, and the endpoints of the two operators are picked from the set $\{XX,YY\}$, or the set $\{XY,YX\}$. 
	\end{enumerate}
    A set $T$ of Majorana operators is non-crossing if there exists a set $S$ of non-crossing quadratic Majorana operators such that all operators in $T$ are equal to a product of terms from $S$. 	
	A set of Majorana operators can be measured simultaneously in a simple way if they are non-crossing. This is because $XX$, $YY$, and $ZZ$ commute and hence can be measured simultaneously (by a simple local transformation, corresponding to transforming to the Bell basis); the same is true for the set $\{XY,YX,ZZ\}$. So if all endpoints of all operators are either both contained within the $Z$-string of another Majorana operator, or avoid that operator completely, we can measure the terms simultaneously.
	
	An NC measurement protocol on $M$ modes gives rise to a non-crossing matching on the complete graph with $M$ vertices (see \cref{fig:matchingoncircle} below for some examples), where we apply a 2-qubit unitary to the endpoints of each edge in the matching, and then measure in the computational basis. Such a protocol allows us to measure all the corresponding quadratic terms (and hence their products) simultaneously.
	
    Explicitly, the protocol is as follows: to measure a set $S$ of non-crossing quadratic Majorana operators, for each operator $O \in S$, measure the pair of qubits at the endpoints of $O$ either in the basis with respect to which $\{XX,YY,ZZ\}$ are diagonal, or the basis in which $\{XY,YX,ZZ\}$ are diagonal, depending on the endpoints. If $O$ has just one endpoint, measure it in the $Z$ basis. Any remaining qubits are measured in the $Z$ basis.
	
	This is a well-defined protocol, because by the non-crossing constraint, each qubit can be the endpoint of at most two distinct operators, which completely overlap and are jointly measurable. To show that it allows all of the operators in $S$ to be measured, observe that the endpoints of each operator are measured in the correct basis, and the qubits between each endpoint are all measured in a basis that allows $Z Z \dots Z$ to be measured. This is because each qubit is either measured in the $Z$ basis directly, if it is not the endpoint of a quadratic operator; or is one of a pair of qubits measured in an entangled basis allowing $ZZ$ to be measured (by the non-crossing constraint).
	
	\item \textbf{(COM)} Measuring a family of commuting operators simultaneously. Note that non-crossing operators are always also commuting, but the converse is not true. This is the most general approach we will consider, so it will require correspondingly fewer measurement rounds. However, the quantum circuits required to simultaneously diagonalise a set of measurement operators may be relatively difficult to implement (requiring depth $\Theta(M)$). 
\end{itemize}

Finding an efficient measurement procedure based on one of the above strategies corresponds to decomposing a set of Majorana operators into groups, such that each group only contains operators which are qubitwise commuting, non-crossing, or commuting (respectively). The number of groups corresponds to the number of measurement settings. 
Since in the commuting and qubitwise commuting cases a group of operators can be simultaneously measured if and only if all pairwise combinations can be, this is then a graph colouring problem and can be solved computationally for any given set of operators. 
The non-crossing case is similar if we first choose a decomposition of each term into a particular product of quadratic terms.
Each vertex corresponds to a  term, and two vertices are connected by an edge when they are incompatible with respect to one of the above strategies (i.e., act incompatibly on the same qubit, cross, or anticommute). Compatible terms can be measured in the same round. The number of colours required then corresponds to the number of measurement rounds.
Often in the literature the complement graph (where edges correspond to compatible operators) is considered in which case the graph colouring problem is equivalent to the problem of finding a minimum clique cover, as discussed for example in \cite{Verteletskyi2020}.

We remark that, as mentioned above, when using the hybrid encoding it does not seem straightforward to apply the non-crossing condition for next-nearest-neighbour terms. These can be handled separately and measured using a QWC strategy.

\subsubsection{Previous work}

Each of the above families of measurement strategies has been studied previously. 

Qubitwise commuting strategies were used in \cite{kandala2017}, and were studied using algorithms for solving graph colouring and minimum clique cover in \cite{Verteletskyi2020}.
We are not aware of any lower bounds in the literature for strategies of this form.

Commuting strategies have been studied as a graph colouring problem \cite{jena2019pauli} and as a minimum clique cover problem \cite{yen2020measuring,gokhale2019minimizing}. But the strongest results are in \cite{Bonet-Monroig2020}, where $\Theta(M^2)$ bounds are shown.
We will discuss these bounds in detail and adapt the upper bound to the non-crossing setting.

Non-crossing measurements have been studied in different contexts (and with different terminology). Cade et al.\ used the concept of non-crossing measurements for the special case of the Fermi-Hubbard model~\cite{cade20}. This enabled energies to be measured using only 5 computational basis measurements. Hamamura and Imamichi~\cite{hamamura20} considered measuring general sets of Pauli operators given the ability to measure pairs of qubits in an entangled basis such as the Bell basis. They give an algorithm based on a greedy approach for finding pairs of qubits that are suitable for applying this method, and carried out numerical tests and experiments on quantum hardware to validate their method.

By restricting to measuring Majorana operators, here we obtain the advantage that the non-crossing condition defines joint measurability, via a simple (constant-depth) measurement strategy, without needing to fix the measurement in advance. This allows the problem of finding an efficient non-crossing strategy to be efficiently reduced to graph colouring, for which any desired approximate or exact algorithm can be used. This is because the inclusion of any term in a group uniquely specifies how the measurement of that term should be performed. By contrast, using more general Hamiltonians or measurement strategies, there does not seem to be such a unique specification and one seems to need to resort to a strategy such as trying each measurement operator in turn.

Next we will obtain analytical upper and lower bounds for each type of strategy in the case where arbitrary quartic terms are allowed in the Hamiltonian. These bounds highlight the differences between the strategies and show their worst-case behaviour. The bounds are presented in Table~\ref{tab:measurementbounds}.

\begin{table}
	\centering
	\begin{tabular}{c|c|c}
		Strategy & Lower bound & Upper bound \\
		\hline
		\textbf{QWC} & $\frac{M^4}{16}$ & $\frac{M^4}{3}$ \\
		\textbf{NC} & $\frac{2M^2}{3}$ & $\frac{7M^2}{3}$ \\
		\textbf{COM} & $\frac{2M^2}{3}$ & $\frac{5M^2}{3}$ \\
	\end{tabular}
	\caption{Summary of the lower and upper bounds on the number of measurement rounds required.}
	\label{tab:measurementbounds}
\end{table}

\subsubsection{Analytical lower bounds}

\textbf{(QWC)} We produce a set of $\Omega(M^4)$ quartic terms, each pair of which are not qubitwise commuting. Assume for simplicity that $M$ is a multiple of 4. 

We consider the set of quartic interactions acting on modes $(i,j,k,l)$ such that 
\[1 \le i \le  M/4, \qquad M/4+1 \le j < k \le 3M/4, \qquad 3M/4+1\le l \le M.\]
There are $\frac{M}{4} \binom{M/2}{2} \frac{M}{4}=M^4/2^7-O(M^3)$ quadruples of this form, and so there are at least $8\times M^4/2^7 = M^4/16$ corresponding quartic interactions.

We claim that any pair of interactions $P_1,P_2$ from this set are not qubitwise commuting.
Consider a pair of interactions $P_1,P_2$ from this set acting on $(i_1,j_1,k_1,l_1)$ and $(i_2,j_2,k_2,l_2)$ respectively. 
If $(i_1,j_1,k_1,l_1)=(i_2,j_2,k_2,l_2)$, then they must act differently at one of $i_1,j_1,k_1,l_1$ in order to be distinct.
So instead consider the case where $(i_1,j_1,k_1,l_1)\neq (i_2,j_2,k_2,l_2)$. Suppose $i_1<i_2$, then at qubit $i_2$, $P_1$ acts as $Z$ but $P_2$ acts as $X$ or $Y$. The same argument applies if $k_1<k_2$.
And similarly, if $j_1<j_2$ (or $l_1<l_2$), then at qubit $j_1$ ($l_1$), $P_1$ acts as $X$ or $Y$, but $P_2$ acts as $Z$.

%

\textbf{(COM)} In \cite{Bonet-Monroig2020}, it is shown that a maximal set of commuting quartic terms is of size at most $\binom{M}{2}$ (in the large $M$ limit). Since we wish to measure $8\binom{M}{4}$ terms in total, any strategy will require at least 
\[8\binom{M}{4}\big/ \binom{M}{2}=\frac{2(M-3)(M-2)}{3}=\frac{2M^2}{3}-O(M),\]
measurement rounds. This result also bounds the NC strategy as all non-crossing operators commute.

\subsubsection{Analytical upper bounds}
\label{sec:upper_bounds}
\textbf{(QWC)} There is a trivial upper bound of $8\binom{M}{4}=\frac{M^4}{3}+O(M^3)$ from the strategy of measuring each term in turn. 

We can do marginally better by splitting $\{1,\dots, M\}$ in half and doing the measurements that act only on $\{1,\dots, M/2\}$ in parallel with the ones that act only on $\{M/2+1,\dots ,M\}$. There are $8\binom{M/2}{4}\approx \frac{1}{16} \frac{M^4}{3}$ terms that act only on $\{1,\dots, M/2\}$, so the total number of measurements is $\frac{15}{16} \frac{M^4}{3}$.

We can iterate this procedure, by first measuring the $\frac{14}{16}$ fraction of the terms that act on both sides of $M/2$. Then, of the $\frac{1}{16}$ fraction of terms that act within the first half, do the $\frac{14}{16}$ fraction of terms acting across $M/4$ and so on. This gives a total number of measurements that asymptotically approaches:
\[\left[\frac{14}{16} +\frac{1}{16}\left(\frac{14}{16}+\frac{1}{16}\dots \right)\right]\frac{M^4}{3}=\frac{14}{16}\frac{1}{1-\frac{1}{16}}\frac{M^4}{3}=\frac{14}{15}\frac{M^4}{3}.\]

\textbf{(COM)} An upper bound is provided in \cite{Bonet-Monroig2020} for measuring all quartic interactions, including those that do not respect time reversal symmetry. Here we describe how the same method can be used to measure all the interactions we are interested in. Then we will show how to adapt this method into a NC strategy.

The abstract combinatorial problem that needs to be solved is the following. We say a list of disjoint pairs of a set is a \emph{matching}. We need to construct a list of matchings of $\{1,\dots, M\}$ such that each quadruple $\{a,b,c,d\}$ appears in a matching as the union of two pairs (i.e., there is a matching that contains either $\{a,b\},\{c,d\}$ or $\{a,c\},\{b,d\}$ or $\{a,d\},\{b,c\}$). We say that a set of matchings with this property \emph{covers all quadruples}. Let $mq(M)$ denote the minimal number of matchings required to cover all quadruples in the set $\{1,\dots ,M\}$.
To get all quartic fermionic terms as in \cite{Bonet-Monroig2020}, we can use this list of matchings as follows.
Two quadratic Majorana terms $\gamma_i\gamma_j$ and $\gamma_k\gamma_l$ commute if and only if $i,j,k,l \in \{1, \dots 2M\}$ are all distinct. A collection of commuting quadratic Majorana terms can be labelled by matchings of $\{1, \dots,  2M\}$.
A list of matchings that covers all quadruples therefore corresponds to a list of measurement settings that measures all quartic Majorana operators.
The total number of measurements required is $mq(2M)$.
If we are interested only in terms that respect time-reversal symmetry, we can instead use a list of matchings of $\{1,\dots,M\}$ that covers all quadruples. For each matching we do two measurements: (i) for each pair in the matching we measure the corresponding pair of qubits in the $\{XX,YY,ZZ\}$ basis, (ii) for each pair in the matching we measure the corresponding pair of qubits in the $\{XY,YX,ZZ\}$ basis. This results in a total of $2mq(M)$ measurements.
Each matching contains $\lfloor M/2 \rfloor$ pairs and so $\binom{\lfloor M/2\rfloor}{2}$ quadruples are covered by each matching. Since there are $\binom{M}{4}$ quadruples in total, we have a lower bound of $mq(M)\ge\binom{M}{4}/\binom{\lfloor M/2\rfloor}{2}\approx M^2/3 $.
It is shown in \cite{Bonet-Monroig2020} that, when $M$ is a power of 2, $mq(M)\le 5M^2/6$. This therefore corresponds to an upper bound of
\[2mq(M) \le \frac{5M^2}{3}.\]

Before moving on to discuss non-crossing strategies, we will first discuss the algorithm of \cite{Bonet-Monroig2020} in more detail so that we can later adapt it into a non-crossing strategy.
There are two important subroutines for creating matchings that contain certain pairs (which can be thought of as methods for measuring quadratic interactions). The first is a list of matchings to get all pairs of a set $\{1,\dots,M\}$ which is equivalent to finding an edge colouring of the complete graph. 
Baranyai's theorem \cite{baranyai} says that this takes $M-1$ colours if $M$ is even and $M$ colours if $M$ is odd.
The second subroutine is a list of matchings to get all pairs of the form $\{x,y\}$ where $x \in \{1,\dots ,M/2\}, y \in \{M/2+1,\dots,M\}$. This can be done in $M/2$ matchings and corresponds to an edge colouring of the complete bipartite graph. 

We now describe the algorithm when $M$ is a power of $2$. Split the set $\{1,\dots, M\}$ into blocks of size $2^n$ labelled by $B^n_m=\{m2^n < i \le (m+1)2^n\}$.
For each $n \in \{1,\dots, \log_2(M)\}$ do the following:
\begin{itemize}
	\item Divide the set $B^n_m$ in half (using the same notation, $B^n_m$ splits into $B^n_m =B^{n-1}_{2m} \cup B^{n-1}_{2m+1}$).
	Generate a list of $2^{n-1}-1$ matchings of $B^{n-1}_{2m}$ to get all pairs in $B^{n-1}_{2m}$. Extend each of these matchings with each  of the $2^{n-1}-1$ matchings of $B^{n-1}_{2m+1}$ that covers all pairs in $B^{n-1}_{2m+1}$. This gives a total of $< 4^{n-1}$ matchings that covers all quadruples in $B^n_m$ that consist of a pair in $B^{n-1}_{2m}$ and a pair in $B^{n-1}_{2m+1}$. Do this in parallel for all $m$.
\end{itemize}
For each $n \in \{1,\dots, \log_2(M)-1\}$ do the following:
\begin{itemize}
	\item  Generate a list of $M2^{-n}-1$ matchings of $\{1,\dots, M2^{-n}\}$ that covers all pairs. For each of these matchings $\{\{a_1,b_1\},\{a_2,b_2\},\dots \}$, pair up $B^n_{a_i}$ and $B^n_{b_i}$. 
	\begin{itemize}
		\item  
		Split $B^n_a$ and $B^n_b$ into half again and for $c,d \in \{0,1\}$ do the following:
		Generate a list of $2^{n-1}$ matchings that gets all pairs with one element of $B^{n-1}_{2a+c}$ and one element of $B^{n-1}_{2b+d}$. Extend each of these matchings to a matching on $B^n_a \cup B^n_b$ in $2^{n-1}-1$ ways so that each pair in $B^{n-1}_{2a+1-c}$ and each pair in $B^{n-1}_{2b+1-d}$ is covered. 
		
		This gives a total of $\le4\times 4^{n-1}$ matchings that cover all quadruples of the form $(w,x,y,z)$ with $w \in B^n_a$, $x,y,z \in B^n_b$ and at least one element in both $B^{n-1}_{2b}$ and $B^{n-1}_{2b+1}$ (or similarly for $a\leftrightarrow b$).
		
	\end{itemize}
\end{itemize}
The total number of matchings is therefore
\[\sum_{n=1}^{\log_2 M} 4^{n-1} +\sum_{n=1}^{\log_2 M -1}M2^{-n}\times 4 \times 4^{n-1}
\le 
\frac{5M^2}{6},\]
as claimed.

\begin{figure}
	\centering
	\begin{tikzpicture}
		\tikzstyle{vertex}=[circle,fill=black,inner sep=0pt, minimum size=1.5mm]
		\pgfmathtruncatemacro{\M}{5}
		\foreach \j in {1,...,\M}{
			\begin{scope}[xshift=\j*3.2cm]
				\foreach \x in {1,2,...,\M}{
					\node[vertex] (\x) at (-360*\x/\M :1) {};
					\node at (-360*\x/\M :1.3) {\x};
					\pgfmathsetmacro{\ji}{\j-\x}
					\draw[thick] (\x) -- (-360*\ji/\M :1);
				}
			\end{scope}
		}
	\end{tikzpicture}
	
	\vspace{8mm}
	\begin{tikzpicture}[scale=0.8]
		\tikzstyle{vertex}=[circle,fill=black,inner sep=0pt, minimum size=1.5mm]
		\pgfmathtruncatemacro{\M}{6}
		\foreach \j in {1,...,\M}{
			\begin{scope}[xshift=\j*3.3cm]
				\foreach \x in {1,2,...,\M}{
					\node[vertex] (\x) at (-360*\x/\M :1) {};
					\node at (-360*\x/\M :1.3) {\x};
					\pgfmathsetmacro{\ji}{\j-\x}
					\draw[thick] (\x) -- (-360*\ji/\M :1);
				}
			\end{scope}
		}
	\end{tikzpicture}
	\caption{$M$ matchings that contain all pairs for (top) $M=5$ and (bottom) $M=6$. Each matching is a set of parallel lines on the circle.}
	\label{fig:matchingoncircle}
\end{figure}
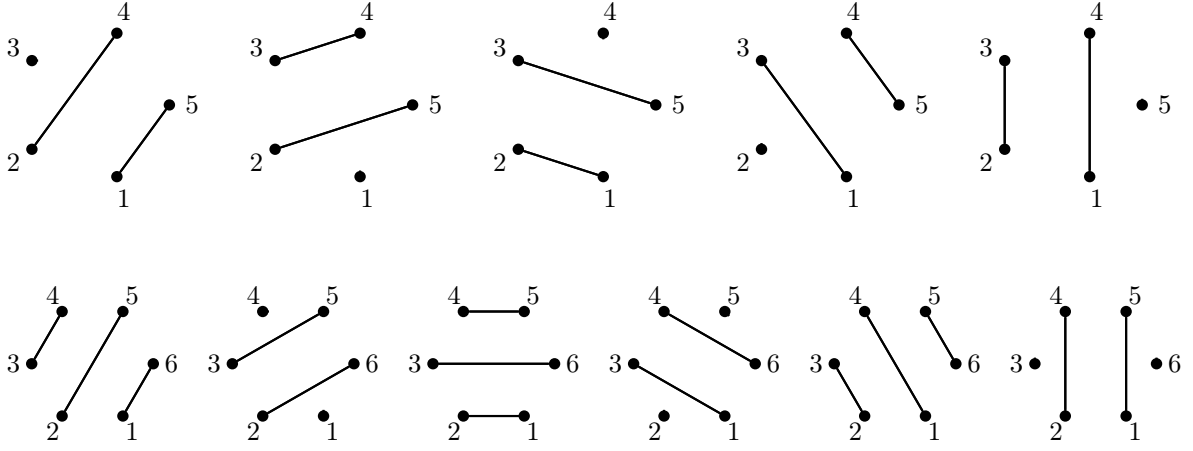

\textbf{(NC)}  We want to adapt the method of the previous section so that all matchings are non-crossing. We first describe how to get all pairs in $\{1,\dots, M\}$ using $M$ matchings. The $j$th matching pairs up $(i,j-i \mod M)$ for all $i$. Note that if $j-i \mod M = 0$, the $i-$th mode is not paired. These matchings can be visualized as parallel lines on circles as shown in \cref{fig:matchingoncircle}. 

We can use this method in all the places in the previous algorithm where we need to generate all pairs within a subset $B^{n-1}_{2m}$. This takes $2^{n-1}$ non-crossing matchings, compared to $2^{n-1}-1$ matchings previously, making minimal difference to the total.
We can also use this method when pairing up blocks $B^n_a$ and $B^n_b$.

However, we do not have an alternative non-crossing strategy to get all pairs with one element in $B^{n-1}_{2a+c}$ and one element in $B^{n-1}_{2b+d}$, which we were able to do with $2^{n-1}-1$ matchings. Instead, we can just use our non-crossing strategy to get all pairs in $B^{n-1}_{2a+c} \cup B^{n-1}_{2b+d}$ using $2^{n}$ matchings.

If we do this, the total number of non-crossing matchings is
\[\sum_{n=1}^{\log_2 M} 4^{n-1} +\sum_{n=1}^{\log_2 M -1}M2^{-n}\times 4 \times 2^{n-1}\times 2^n
\le 
\frac{M^2}{3}+M^2
=\frac{7M^2}{6},\]
giving an upper bound of $7M^2/3$ non-crossing measurements.

\subsubsection{Quartic terms that act on three modes}
The strategy discussed in the previous section allows all quartic fermionic terms to be measured. However, we can get a more efficient protocol if we only want to measure terms that act non-trivially on only 3 modes.  Physically, this corresponds to correlated hopping of the form $c_i^\dagger c_j n_k$ that may appear in the Hamiltonian. After the JW transformation, these are of the form 
\[X_i \left(\prod_{i<l<k}Z_l \right) Y_j Z_k\quad \text{ or } \quad Y_i \left(\prod_{i<l<k}Z_l \right) X_j Z_k, \]
where $i < k$ and $l \notin \{i,k\}$.

There are $2\binom{M}{2}(M-2)=M^3-O(M)$ terms of this form, and we can measure all of them in $2M\log(M)$ non-crossing measurements as follows.

Assume for simplicity that $M$ is a power of 2, $M=2^m$. We can construct $M$ non-crossing matchings such that all pairs occur in at least one matching, as shown in \cref{fig:matchingoncircle} and discussed in the upper bound for non-crossing strategies to measure all quartic terms in \cref{sec:upper_bounds}. For each of these matchings, we choose $2\log_2(M/2)$ measurement settings, where for each measurement setting we measure each pair in the matching in either the $\{XY,YX,ZZ\}$ basis or the $\{ZI,IZ\}$ basis. We want to do this such that for any two pairs in a matching, say $\{a,b\}$ and $\{c,d\}$, there is a measurement setting such that $\{a,b\}$ is measured in the $\{XY,YX,ZZ\}$ basis and $\{c,d\}$ is measured in the $\{ZI,IZ\}$ basis (and vice versa).
This can be done with a binary partitioning scheme. Explicitly, label each pair in the matching with a binary string $x \in \{0,1\}^{m-1}$.
For $j\in \{1,\dots,m-1\}$, we have two measurement settings: one where the pair with label $x$ is measured in the $\{XY,YX,ZZ\}$ basis if $x_j=0$ and in the $\{ZI,IZ\}$ basis if $x_j=1$; and another where we do the opposite -- i.e. we measure the pair labeled by $x$ in the $\{XY,YX,ZZ\}$ basis if $x_j=1$ and in the $\{ZI,IZ\}$ basis if $x_j=0$.
If we have two pairs with labels $x$ and $y$, then if $x$ and $y$ are distinct, they must differ in at least one bit, and so there is a measurement setting where they are measured in different bases.

\subsection{Summary}

In this section, we have discussed all the algorithms needed to efficiently simulate a material Hamiltonian in a quantum computer. Based on these ideas, we have built a compiler that is able to perform the different decompositions of terms into layers of quantum gates, thus allowing us to explore the circuit depth associated with different materials. In the next section we explain the structure and design of this compiler.


%% file: state_prep.tex
\subsection{State preparation}\label{sec:state_prep}

\subsubsection{Fock states}
The preparation of Fock states under the JW transform is straightforward, with each Fock state corresponding to a computational basis state, i.e., $$ \ket{b_0, b_1,..., b_m} \rightarrow \otimes_i \ket{ b_i } \;\; b_i \in \{0,1\}.$$ However in the case of the hybrid compact encoding, the Fock state is encoded in a stabilizer code space, which necessitates a non-trivial state preparation procedure.

In the case of the 2D hybrid encoding, each stabilizer can be decomposed into a product of a classical parity check (product of $Z$ operators) on the strings of data qubits $S_D =ZZZ...ZZ$, coupled to a four qubit operator $S_F= XYXY$ acting on the four face qubits (red ancillary qubits in \cref{fig:hybrid}). The set of stabilizers on the face qubits are identical to the stabilizers of the surface code, up to some local Pauli basis transformations \cite{bonilla2021xzzx, Dennis2002}. Thus if we wish to prepare a state where $S_D=1$ for all data qubit stabilizers, such as the vacuum Fock state $\ket{0}$, then the algorithmic overhead of this state preparation is at most that of preparing a surface code state.

A given surface code state can be prepared either via a unitary circuit, or by measuring and correcting stabilizers. In the case of unitary preparation, a circuit of depth of $2\mathcal{L}$ is required, where $\mathcal{L}$ is the max side length of the surface code \cite{Higgott2021}. In the case of the hybrid encoding $\mathcal{L}$ depends only on the number of sites $\{N_1, N_2, N_3\}$, and not on the mode number. More specifically, $\mathcal{L} = \textrm{max}(N_1, N_2)/2$ -- assuming $N_3$ is collapsed into a single site, as described in \cref{sec:Wannier_basis_mapping}. Similar considerations hold in the 3D encoding, however the underlying code generated by the support of the face qubits is not as well studied. We defer comment on unitary state prep in this fashion to a later date when we can present a more comprehensive understanding of this underlying code.

If one has access to intermediate measurements and classically conditioned circuit operations, then it may be more convenient to use a measure-and-correct scheme of preparation. In the case of preparing a surface code state the measure-and-correct scheme works by measuring stabilizers and performing depth-1 Pauli corrections conditioned on the outcomes of the measurements. These corrections appear as Pauli strings between pairs of syndromes \cite{Dennis2002}. However, in the case of preparing a Fock state, one does not need to prepare a surface code state, and may instead change the convention of the logical fermionic operations depending on the syndrome measurement. This gauge fixing saves having to actively perform the correction. 
An additional feature of encoding a fermionic Fock state is that the surface code stabilizers $S_F$ can be measured without introducing any additional ancillary qubits. Instead, one may employ one of the data qubits and measure in the $Z$ basis. After correcting or gauging out the syndromes, the Fock state prepared will depend on the syndrome produced, with a fermion occupying each data qubit which yielded $-1$ when measured in the $Z$ basis. Finally we note that in the case of preparing a Fock state, the data qubits need not be measured in the stabilizer, since any stabilizer acts on these qubits with a $Z$ operator. Thus the cost of coherently measuring the stabilizer is related only to its support on the ancillary face qubits. The reasoning described here applies in both the 2D and 3D encoding. In the 2D case the stabilizer on the face qubits is weight 4, while in the 3D case the stabilizer is weight 8. 


Given an encoding of a vacuum Fock state $\ket{0}$, any other Fock state can be reached by applying a layer of single qubit Pauli operators (and vice-versa), in an analogous fashion to how any computational basis state can be reached from any other by a layer of single qubit $X$ operators. To see this, we note that a product of Majoranas of the first kind ($\gamma_i$) applied to a vacuum state yields a Fock state (up to appropriate sign from normal ordering) 
$$\prod_i \gamma_i\ket{0} = \prod_i w_i^\dagger \ket{0}, $$
and any Fock state can be expressed this way.
In the hybrid compact encoding, any product of Majoranas corresponds to a product of Pauli operators, which can be applied as a single layer of single qubit unitaries, each unitary being a single qubit Pauli operator. 

The only caveat to the above discussion is that in some cases the hybrid compact encoding only represents even or odd parity Fock states. For example in the 2D encoding this happens when the number of non-trivial stabilizers is one more than the number of ancillary qubits, ie when the number of face qubits is less than half the number of faces. In this case the parity operator $\prod_i V_i$ is in the set of stabilizers and odd products of Majoranas do not admit a representation. However one may freely fix a convention for the parity of the encoding by changing sign conventions for specific edge operators. So one may prepare any Fock state by fixing the correct edge operator sign convention, thus fixing the parity sector, and then proceeding as described above.

Thus, using the methods outlined above, the minimal cost of preparing a Fock state is the circuit depth of coherently measuring the syndromes of the face qubit support of the stabilizers, followed by a depth-1 single qubit unitary operation -- which  executes any requisite corrections and performs the transformation from the vacuum Fock state to any other Fock state. With access to arbitrary two qubit operations a stabilizer of weight $w$ can be measured coherently in depth $2 \lceil \log_2(w) \rceil-1$. The generating stabilizers in the 2D case can be measured in a sequence of $4$ simultaneously measurable layers, this can be seen by noting that each face qubit has support on 4 generators. In the 3D case it is more challenging to determine the ideal decomposition since not every cycle of edges around a face need be a generating stabilizer, however the number of distinct face cycles that act on an individual face qubit is $10$, so we may multiply the circuit depth by this number to get an estimate on the number of layers. Counting only two-qubit gates, the depth of Fock state preparation is thus $12$ in the 2D case and estimated $50$ in the 3D case.

\subsubsection{Fermionic Gaussian states}

Another commonly prepared class of input fermionic states are the fermionic Gaussian states, which include Slater determinants. These states appear as the ground states of non-interacting fermionic Hamiltonians (those containing only quadratic terms) and also as ansatzes for ground states of interacting fermionic Hamiltonians in classical methods. As such they can often serve as good input states for quantum algorithms such as \gls{vqe}.   




The best known methods for preparing Gaussian states are designed for JW representations of fermionic systems and involve sequences of Givens rotations $\mathcal{G}_{ij}(\theta, \phi)$ and Boguliubov transformations $\mathcal{B}$ \cite{Jiang2018}:
\begin{align}
\mathcal{B}  := \gamma_F \prod_{i\neq L} V_i,\\
\mathcal{B}c_L \mathcal{B}^\dagger = c_L^\dagger,\\
\mathcal{B}c_i \mathcal{B}^\dagger = c_i,  \;\;\text{for } i \neq L,
\end{align}
where here $L$ is the index of the last mode in the choice of JW ordering. As discussed earlier, there can be cases where the hybrid compact encoding does not admit a representation of single Majoranas, in which case Bogoliubov transformations of this kind are not possible to implement, however one may always choose a parity sector and then restrict to the class of transformations that preserve parity. 

For the JW transform, preparing Gaussian states requires circuit depth at most $M-1$ using an algorithm of Jiang et al.\ \cite{Jiang2018} -- where the number of modes $M =N_{\mathrm{cells}}N_{\mathrm{modes/cell}}$. In the case of the 2D and 3D hybrid compact encodings the scaling of the depth of the circuit for preparing a Gaussian state can not be much worse than that of the JW transform, because one can always overlay a JW ordering on the full hybrid compact encoding and perform the Givens rotations in accordance with the best known methods for the JW transform. However, in this case the hybrid encoding introduces some overhead due to the higher Pauli weight of some of the edge operators -- namely those with support on face qubits -- and reduced opportunities for parallelization -- namely when attempting to perform Givens rotations on two pairs of modes whose edge operators have support on a common qubit. The algorithm of Jiang et al.\ acts on (at most) all consecutive modes at each step in an even-odd pattern. If we act on an even number of modes in total (as will always be the case if we consider spin) then at every other step, we only have 2-qubit gates. By \cref{sec:gate_complexity}, the depth of the step involving 3-qubit gates is 4, and involving 4-qubit gates is 6. To account for overlapping action on face qubits we have to stagger the Givens rotations on those edges. Therefore, the overall depth is at most $2*4\lceil (M-1)/2 \rceil + \lfloor (M-1)/2 \rfloor \approx 4.5M$ for the 2D hybrid encoding and  $2*6\lceil (M-1)/2 \rceil + \lfloor (M-1)/2 \rfloor \approx 6.5M$ for the 3D hybrid encoding.

%% file: Compiler_Description.tex
\section{Circuit compiler design}
\label{sec:compiler}

Here we outline the structure of the compiling algorithm which we use to compute the depth of the circuit resulting from applying either \gls{tds} or \gls{vqe} to a given material. The compiler computes the circuit depth of a Trotter step in \gls{tds}, or an ansatz layer in \gls{vqe}, both of which have similar structure.  The methods in the compiler can be separated in two main steps.

{\bf Precompilation step:} The precompiling step prepares the requisite data for the circuit compiler. It takes as input a fermionic Hamiltonian, a specification of which modes are associated with which sites, and a specification of the spatial layout of the sites (e.g., Hamiltonian constructed in \cref{sec:Physics}). Depending on user specification, the compiler builds the appropriately sized 3D or 2D hybrid encoding, and assigns modes in the encoding so that all modes on the same site are grouped together on the same \gls{jw} chain in the hybrid encoding, or if one is representing a 3D system on a 2D encoding, it collapses one axis down into a single site before assigning modes. See discussion in \cref{sec:qubit_rep}.

{\bf Compiling step:} In the compiling step the terms in the Hamiltonian are organized into a sequence of groupings via a series of decomposition subroutines. Starting with a single grouping, every decomposition subroutine decomposes each grouping into smaller groupings. Next, the terms are translated into Pauli operators and undergo a final decomposition. Finally the circuit cost for each grouping is totalled.

The decomposition subroutines are
\begin{itemize}
\item {\bf Common sites}: Groups together terms which act on exactly the same set of sites.

\item {\bf Mutually commuting terms}: Groups together mutually commuting terms, optimizing for the smallest number of groups using a pre-supplied graph colouring algorithm.

\item {\bf Fswap network}: Finds an optimized sequence of fermionic swaps on the modes of the encoding. Groups together any terms with sufficiently low-weight Pauli representation at every step of the fswap sequence. Optionally all modes are put back in their original location at the end of each sequence of fswaps. See subsection \ref{sec:swap_network_details} for more details.

\item {\bf Disjoint qubits}: Groups together terms which act on different sets of qubits, optimizing for the smallest number of groups using a pre-supplied graph colouring algorithm.
\end{itemize}

The choice of ordering and inclusion of particular decomposition routines will depend on a number of considerations:
\begin{itemize}
\item The decomposition into common sites is primarily useful when compiling the unit cell of a translationally invariant system (see subsection \ref{sec:compile_unit_cell}), and may be excluded otherwise.

\item The inclusion of a decomposition into mutually commuting terms is important for evaluating the Trotter error of a particular \gls{tds} circuit and may be excluded when considering \gls{vqe}. Depending on the importance of minimizing Trotter error, it may appear either before or after the fswap network routine.

\item Generally it is useful to include the fswap network routine, but in some cases the density of long range terms may be sufficiently small that the fswap network does not improve circuit depths, so it may be worth checking the compilation with and without the fswap network. 
\end{itemize}

\subsection{Fswap network implementation details} \label{sec:swap_network_details}
With respect to circuit depth, when implementing an interaction between two modes, it is preferable that they are adjacent within the fermionic encoding, as outlined in \cref{sec:time_evolution_by_terms}. 
By swapping modes through the graph of the encoding, we can ensure that modes which share an interaction will become adjacent at some point, in almost all cases.
In the case where the modes are arranged locally on a graph as opposed to a chain, this is not guaranteed, but can be obtained in practice, as we will describe in \cref{sec:swap_net_distance_min}.
Fswap operations can be performed efficiently on modes which are adjacent, i.e., which share an edge in the encoding (see \cref{sec:time_evolution_by_terms}), and fswaps can be performed in parallel provided no mode is involved in more than one fswap.
We therefore design an fswap network, i.e. a series of fswap layers interspersed with layers of interactions which are facilitated by the intermediate mode configurations. 
\par 

In particular, we seek the fswap network that enables implementation of all terms using the smallest number of fswap layers. After each fswap layer we implement all allowed terms (those on adjacent modes) before proceeding to the next fswap layer. This does not necessarily give the lowest quantum circit depth overall, but is a reasonable heuristic.
The compiler is equipped with two routines for determining the fswap layers required to facilitate interactions, which we now introduce. 

\subsubsection{Chain fswap network}
\label{sec:swap_net_chain}

The chain fswap network, as described in \cite{kivlichan18}, is applicable in the case where all relevant modes are arranged linearly, i.e., in a JW string of length $m$ modes (equivalently, $m$ data qubits). 
Fswap layers iteratively perform fswaps on even-indexed edges, followed by odd-indexed edges.
This mechanism ensures that every mode-pair is achieved within $m$ fswap layers.

\begin{figure}
    \centering
    \includegraphics[scale=0.6]{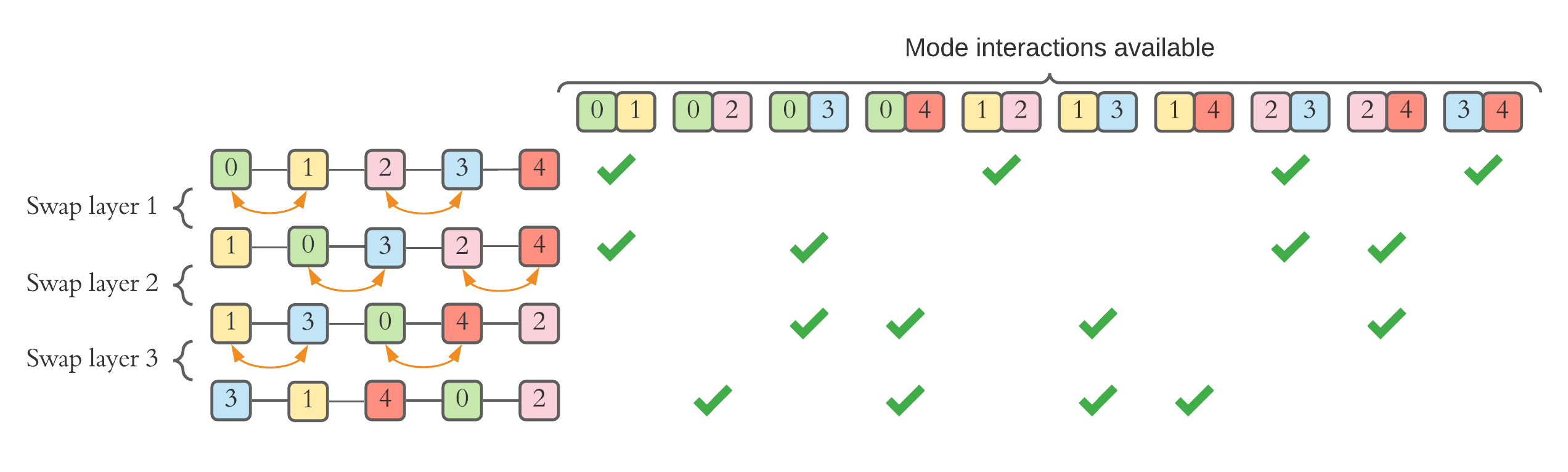}
    \caption{
        Fswap network example for a five-mode linear system, where we require all-to-all interactions, i.e., that each mode should be adjacent to all other modes at least once. 
        The chain on the left represent the configuration of the system, and the table on the right indicates whether a given interaction (between a pair of modes) is available at that configuration. 
        Green ticks indicate that the pair is available on the present configuration; by the end of the third layer, all required interactions have been implementable at least once.
        Fswap layers on the mode configuration (left) are ordered: the first layer performs fswaps on even-indexed edges, i.e., the $0^{\rm{th}}$ and $2^{\rm{nd}}$ of the chain; 
        the second layer performs fswaps on odd-indexed ($1^{\rm{st}}, 3^{\rm{rd}}$ edges), in a repeating pattern until either (i) all required interactions have been implemented or (ii) the chain has completely reversed after $m$ fswap layers.
    }
    \label{fig:swap_net_demo}
\end{figure}

For example, the fswap network in \cref{fig:swap_net_demo} aims to implement a set of all-to-all interactions among five modes, where each mode interacts with every other mode in the system. 
This can be described as the set of layers in \cref{tab:swap_net_demo_layers}, which must then be translated into circuit gates.

\begin{table}[h]
    \centering
    \begin{tabular}{lr}
        Layer type & Modes to interact/fswap \\
        \midrule
        Interaction &  $(0,1), (1,2), (2,3), (3,4)$ \\
        Fswap & $(0,1), (2,3)$ \\
        Interaction & $(0,3), (2,4)$ \\
        Fswap & $(0,3), (2,4)$ \\
        Interaction &  $(0,4), (1,3)$ \\
        Fswap & $(1,3), (0,4)$ \\
        Interaction & $(0,2), (1,4)$
    \end{tabular}
    \caption{Circuit layers corresponding to the fswap network in \cref{fig:swap_net_demo}.}
    \label{tab:swap_net_demo_layers}
\end{table}

\subsubsection{Distance minimising fswap network} 
\label{sec:swap_net_distance_min}
For a more general \emph{mode graph} where, for instance, the modes are not arranged linearly, we do not have such a straightforward strategy.
Moreover, the chain fswap network does not guarantee that quartic terms will all be facilitated: quartic terms consist of four Majorana indices, and therefore require that two mode-pairs are adjacent simultaneously, even if the two pairs are distant from each other on the encoding graph. 
The chain fswap network ensures that every pair will be adjacent, but not that every set of two pairs will be simultaneously adjacent, as required. 
Instead, given a set of required interactions, we devise an fswap network which seeks to find the optimal set of fswaps at each fswap layer, in order to facilitate as many interactions as possible on the subsequent interaction layer. 
That is, the distance of a mode graph $G$ with respect to the set of interaction terms required, $\mathcal{T}$, can be evaluated as
\begin{equation}
    \label{eqn:swap_net_total_distance}
    D(G) = \left( \sum_{t \in \mathcal{T}}  d(t | G)^p \right)^{1/p}, 
\end{equation}
where $d(t | G)$ is the cost of a single term $t$ evaluated with respect to the graph $G$, and $p>0$ is a hyperparameter described in \cref{eqn:swap_net_full_cost} of  \cref{sec:time_evolution_by_terms}.
The cost of each term is evaluated as the path length between the modes involved in the term, 
\begin{equation}
    \label{eqn:swap_net_internal_distance}
    d(t | G) = G\ttt{.path\_length}(t).
\end{equation} 
Then, a proposed fswap $s$ will result in a modified graph $G(s)$, yielding \begin{equation}
    d(t | G, s) = G(s)\ttt{.path\_length}(t). 
\end{equation}

\begin{figure}
    \centering
    \includegraphics[scale=0.6]{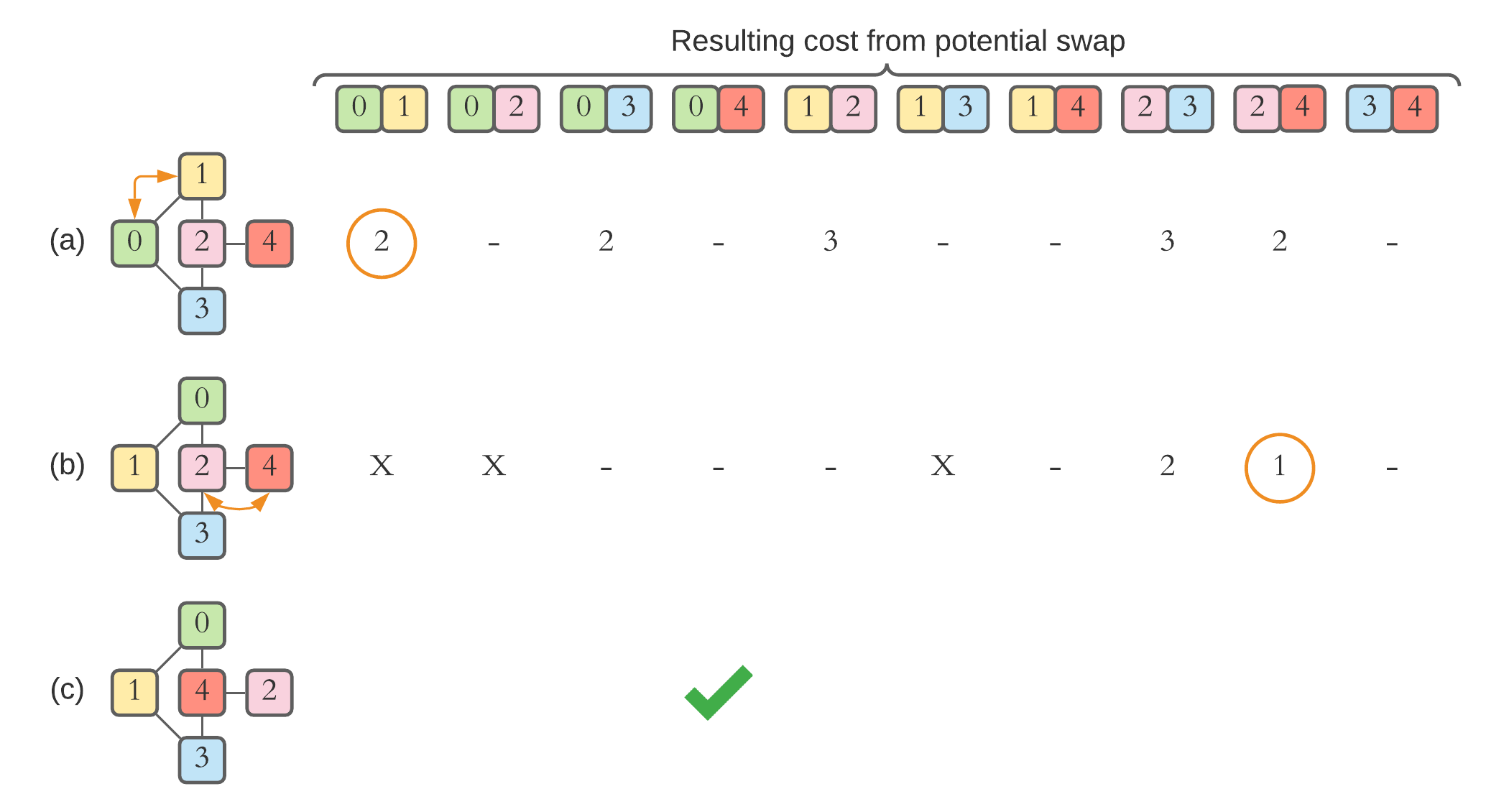}
    \caption{
    Distance minimising fswap network: determining the set of fswaps to apply in order to facilitate an interaction between modes $(0,4)$ on a small mode graph. 
    Mode graphs (left) represent modes as squares, while lines between modes indicate an edge, i.e., a connection in the fermionic encoding.
    In the initial configuration \textbf{(a)}, the shortest distance between modes $(0,4)$ is 3, e.g., the path via edges $\{ (0,1), (1,2), (2,4)\}$. 
    Iteratively, 
    (i) available fswaps $s \in \mathcal{E}$ are evaluated, i.e., $d(t | G,s)$ is computed $ \forall s \in \mathcal{E}$, reported as a path length in the table of the graphic;
    (ii) one of the the fswaps which yield the lowest cost within the round (circled, orange) is added to the fswap layer $\mathcal{S}$.
    Dashes indicate that such an edge is not present in the graph.
    \textbf{(b)}, The available edges are again evaluated and the best option is identified as the swap between modes (2,4); \ttt{X}s indicate that one of the modes in the proposed fswap has already been used and is therefore unavailable.
    \textbf{(c)}, $G(\mathcal{S})$, the mode graph after the fswap layer has been applied, showing the interaction $t=(0,4)$ is available (green tick).
    }
    \label{fig:swap_net_demo_graph}
\end{figure}

The edge list of $G$ -- labelled $\mathcal{E}$ -- defines the set of permitted fswaps for $G$; we can evaluate the cost of each available fswap independently, giving $\large\{ D(G(s)) \large\}_{s \in \mathcal{E} }$.
The edge which results in the lowest cost, $s_1$, is added to the present fswap layer, $\mathcal{S} = \{ s_1\}$.
However, we can perform numerous fswaps in the same layer, provided no mode is involved in more than one swap operation. 
We therefore evaluate the effect of swapping the remaining available edges, i.e. compute the set 
$\large\{ D(G(s)) \large\}_{s \in \mathcal{E} \setminus s_1 }$, and add the best fswap from this set to $\mathcal{S}$. 
This process is repeated until there are no remaining available fswaps, or it is no longer advantageous to include more fswaps, i.e., the cost of the proposed graph would increase from any available fswap.
\cref{fig:swap_net_demo_graph} shows the calculation of $d(t | G,s)$ for each $s \in \mathcal{E}$ for a small graph which requires a single interaction $\mathcal{T} = \{ (0,4)\}$.
In this example the fswap network is seeking an interaction between modes $(0,4)$, and finds that a single fswap layer $\mathcal{S} = \{(0,1), (2,4) \} $ facilitates the sole required interaction. 
\par 

This  distance minimising fswap network mechanism can be varied by defining an alternative total distance function, \cref{eqn:swap_net_total_distance}, or internal distance function, \cref{eqn:swap_net_internal_distance}. 
Furthermore, in realistic use cases (see, e.g., \cref{sec:model_ham_eg}), the distance function will be based on Majorana indices, of which two reside on each mode. 
Such terms are implementable if both Majoranas are on the same mode, or their corresponding modes are adjacent. 
Such considerations can be built into the design of $d(t | G)$.
In particular, the swap network can be designed to handle quartic interaction terms by specifying an internal distance function which is minimised when the graph permits the implementation of the quartic term. 

\subsubsection{Composite fswap network}
\label{sec:fswap_network}
The method described in \cref{sec:swap_net_distance_min} ought to produce a shorter fswap network overall, and is applicable for all mode graphs including chains. However, it is much more expensive to compute than the chain swap network, owing to the requirement to evaluate the cost of all remaining $t \in \mathcal{T}$ for each $s \in \mathcal{E}$, on potentially large graphs $G$. 
It is therefore preferable to rely on the chain fswap network of \cref{sec:swap_net_chain} when possible, which guarantees that all quadratic (and some quartic) interactions will be implementable within $m$ fswap layers (for a chain of length $m$ modes). 
The cases in which the chain method do not suffice are: 
\begin{easylist}[itemize]
& the mode graph of the system is not linear;
& some quartic interactions are not made implementable, i.e., when multiple edges are required to implement a single term, the chain method does not guarantee they will occur concurrently.
\end{easylist}
\par 

In practice then, we compose an fswap network from one or both of the above methods, depending on the required interactions and graph structure. 
Also note that the fswap networks we wish to construct will likely exist in a larger mode graph than they require. 
For instance, the entire system under study contains $M$ modes across several sites, but we are interested in simulating only a subset of $m < M$ modes, such as the interactions contained on single physical site.
In these cases, it is sensible first to reduce to the minimal graph which supports the $m$ modes we require to implement the interactions in $\mathcal{T}$.
This can be done by finding the  Steiner tree of the mode graph containing all the modes involved in any $t \in \mathcal{T}$: this can include some modes which are \emph{not} directly involved in any interaction, but which reside in between modes of interest.
We can compute the fswap network only upon this subgraph, but in practice the swaps will be applied in the space of the full graph. 
\par 

We employ the following strategy for a given mode graph $G$ and interactions $\mathcal{T}$. 

\begin{easylist}[enumerate]
& Isolate the modes $\mathcal{M}$ which are involved in \emph{any} interaction in $\mathcal{T}$:
&& construct a Steiner tree consisting only of $\mathcal{M}$, i.e., a subgraph of $G$;
&& if the Steiner tree has fewer modes than $G$, it will require fewer fswaps to traverse, so replace $G$ with the Steiner tree. 
& If $G$ is semi-Eulerian, i.e., there is a path containing each edge once (equivalently, a chain encompassing all modes):
&& form a new graph from the Euler path edges of $G$, and replace $G$; 
&& run the chain fswap network on $G$ for $\mathcal{T}$, resulting in the fswap network $S_1$;
&& throughout $S_1$, the terms $\mathcal{T}_1 \subseteq \mathcal{T}$ were implementable
&&& if the initial $\mathcal{T}$ contains only quadratic terms, $\mathcal{T}_1 = \mathcal{T}$, and the fswap network composition can terminate;
&&& in general, some terms remain, so replace $\mathcal{T}$ with $\mathcal{T} \setminus \mathcal{T}_1$.
& Again, attempt to reduce $G$ to a Steiner tree based on the required interactions $\mathcal{T}$.
& Pass $G, \mathcal{T}$ to the distance-minimising fswap network to produce $S_2$.
& Combine the generated fswap networks into a single fswap network, $S \gets \{ S_1, S_2 \}$.
& Restore the initial mode graph $G$
&& After all fswap and interaction layers of $S$ are implemented, reverse all fswaps in $S$ so that the final configuration matches the initial configuration. This is necessary to facilitate the circuit tiling described next in \cref{sec:compile_unit_cell}.
\end{easylist}


\subsection{Compiling unit cells of translationally invariant Hamiltonians }\label{sec:compile_unit_cell}

Although the compiler is capable of handling a fully populated multi-site material Hamiltonian, it is usually more efficient to perform the compilation on a unit cell and its neighbours. This introduces a few subtleties which are worth explaining. 

\begin{figure}%
\centering
\begin{minipage}{1.1in}%
\includegraphics[scale=0.25]{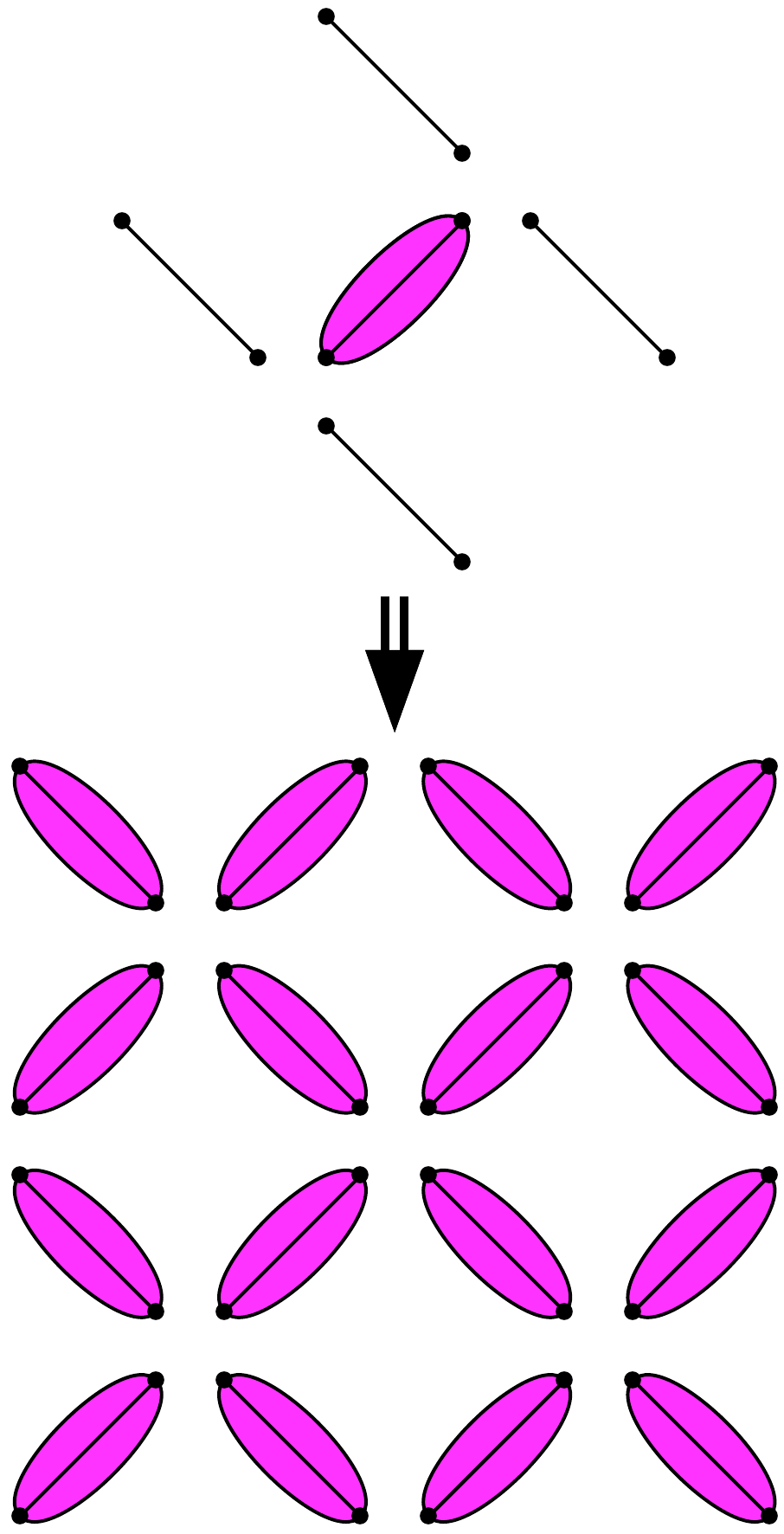}
\end{minipage}%
\begin{minipage}{1.1in}%
\includegraphics[scale=0.25]{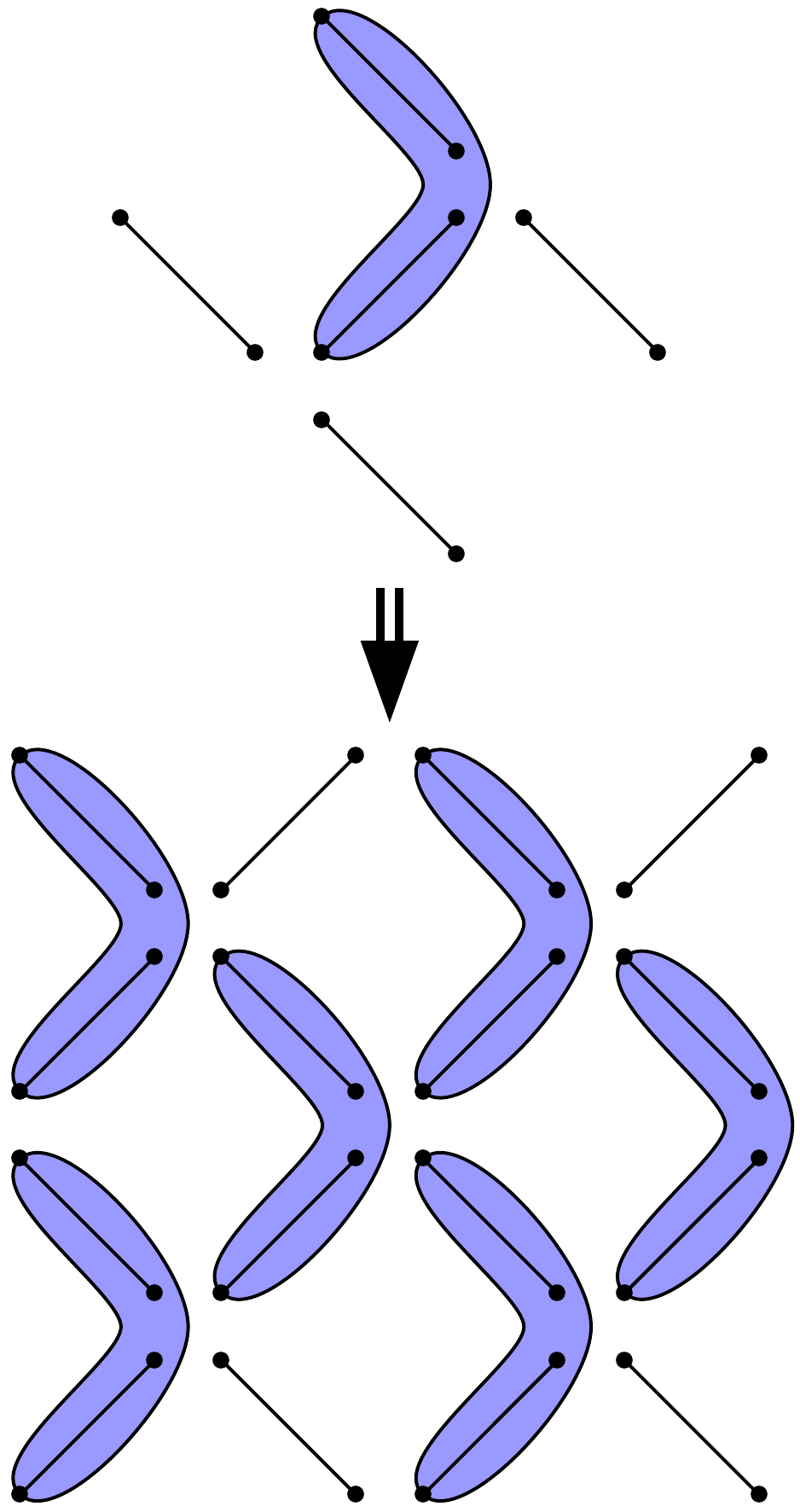}
\end{minipage}%
\begin{minipage}{1.1in}%
\includegraphics[scale=0.25]{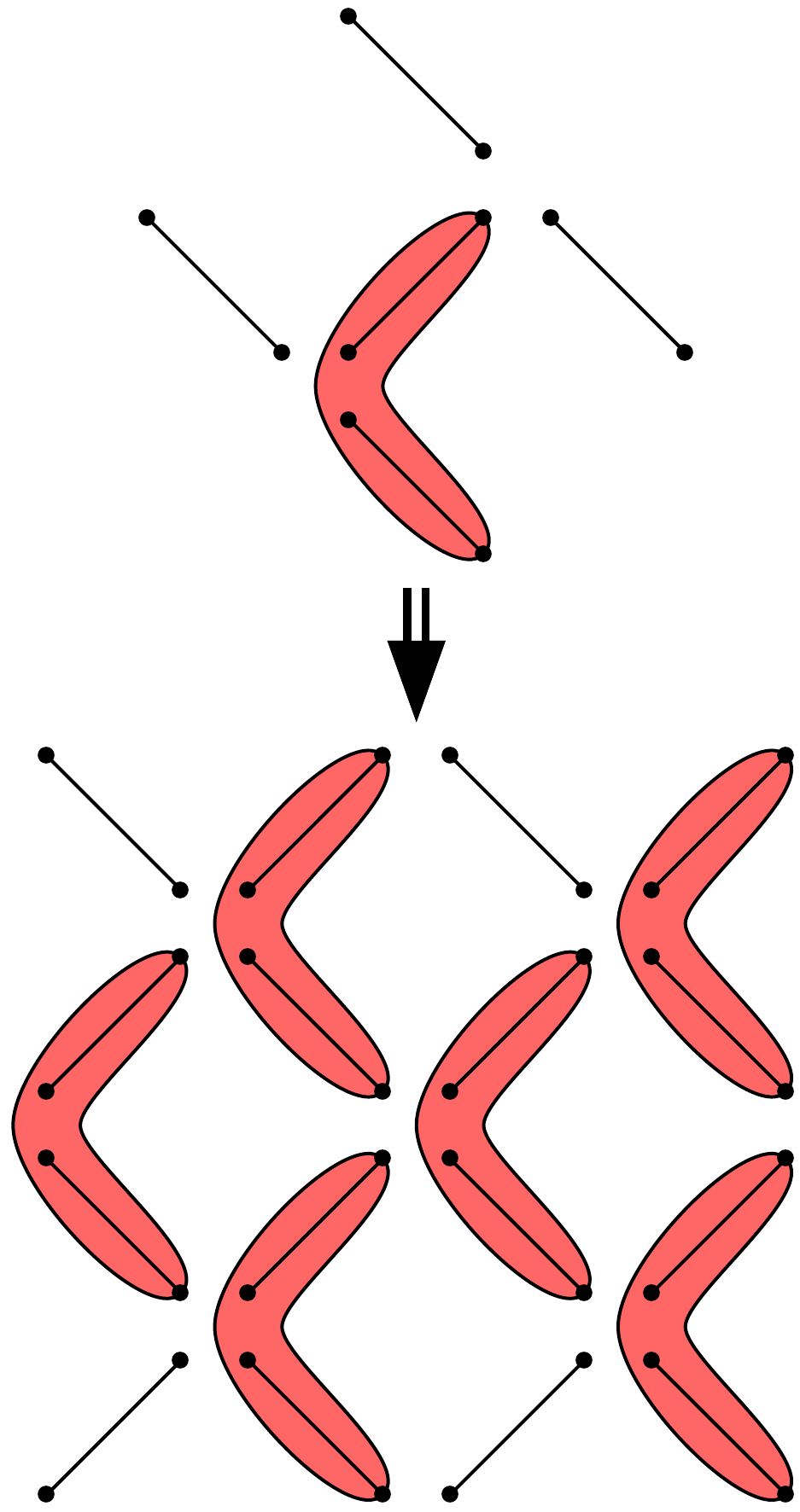}
\end{minipage}%
\begin{minipage}{1.1in}%
\includegraphics[scale=0.25]{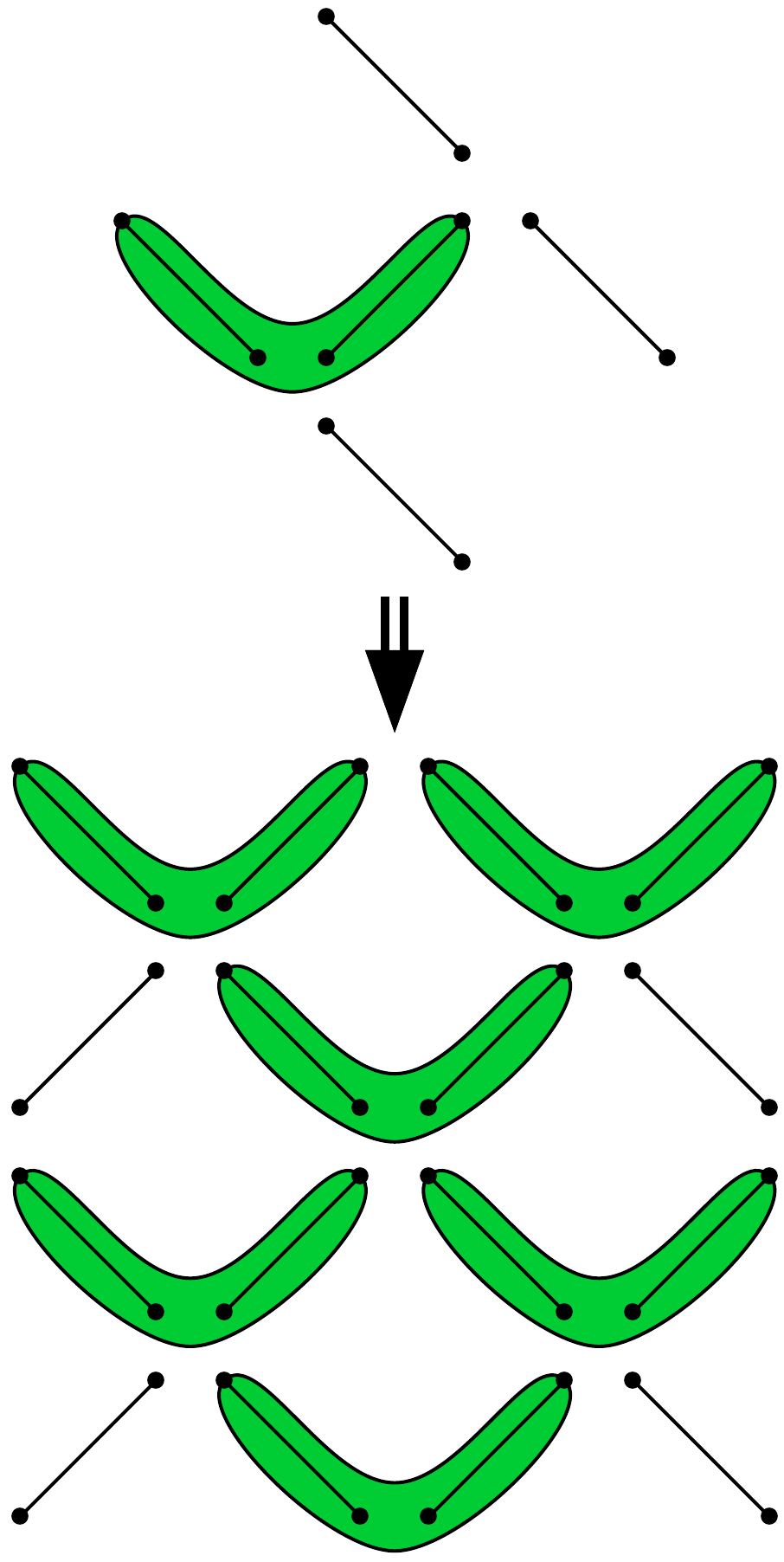}
\end{minipage}%
\begin{minipage}{1.1in}%
\includegraphics[scale=0.25]{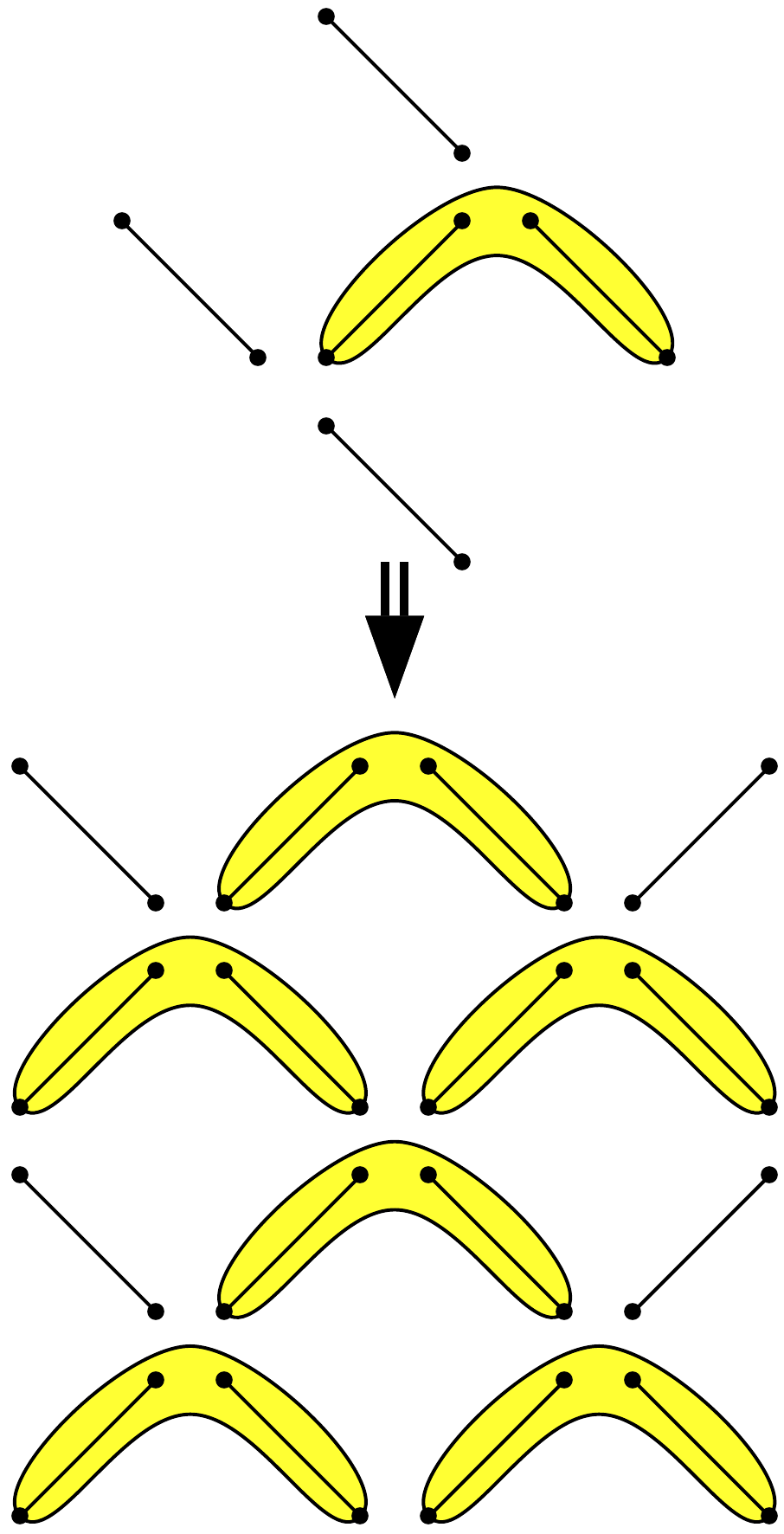}
\end{minipage}%
\caption{An illustration of how the common site decomposition on a unit cell translates to the full system for nearest neighbour interactions on the 2D hybrid encoding.}%
\label{fig:2dtiling}%
\end{figure}

\begin{figure}
\centering
\includegraphics[scale=0.5]{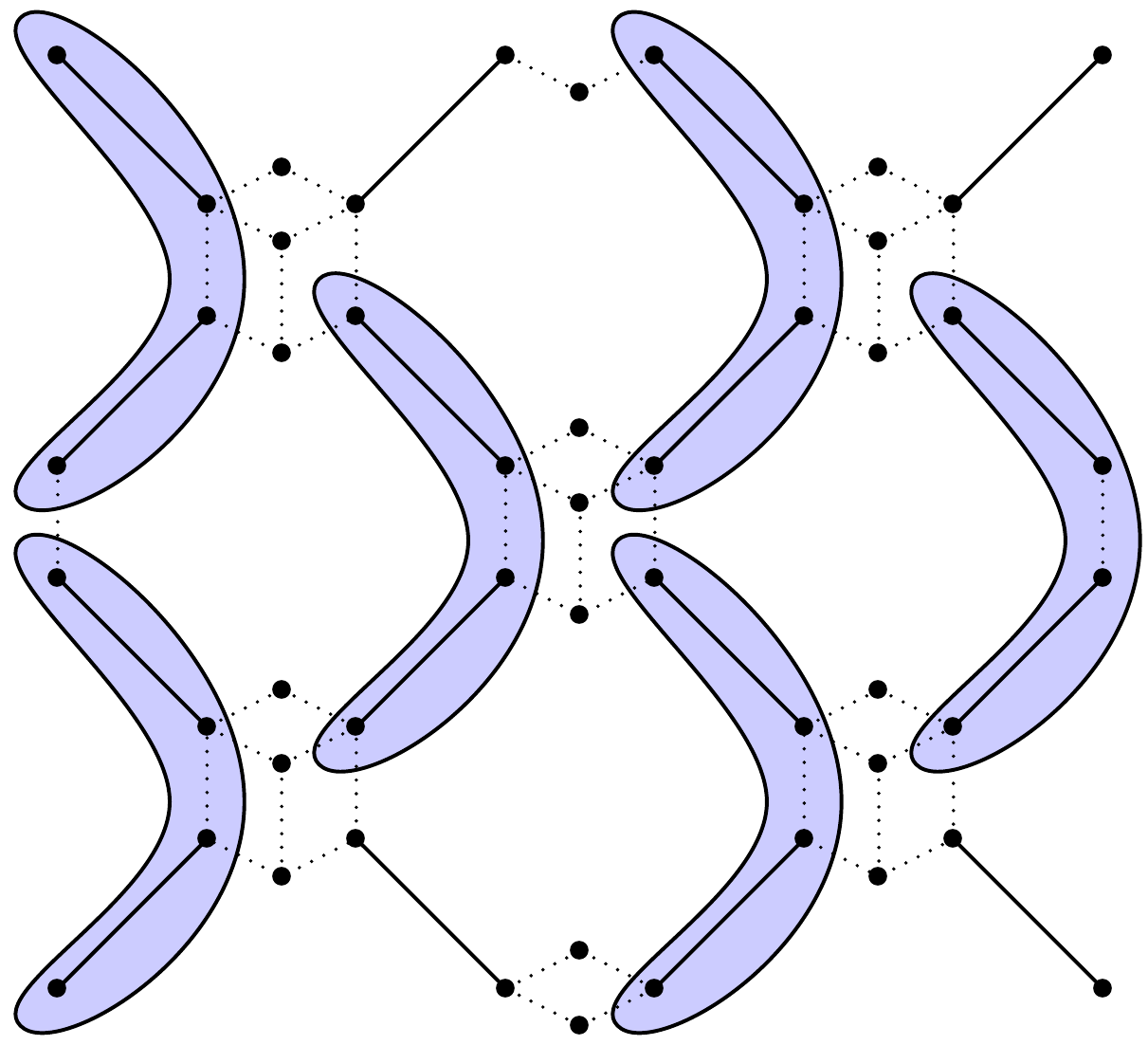}
\caption{An illustration of how one of the possible common site decomposition on a unit cell translates to the full system for nearest neighbour interactions on the 3D hybrid encoding. Here the tiling is done in the orthographic plane.}%
\label{fig:3dtiling}%
\end{figure}

The premise behind performing the compilation on a unit cell is that a judicious choice of groupings of terms in $\mathcal{T}$ onto common sites will yield a circuit depth that reflects the circuit depth of the fully populated system. This holds when each grouping of terms onto common sites can be uniformly tiled onto the full system in a parallelizable fashion, so that the circuit depth of each grouping is the circuit depth of the full tiling. This is illustrated in  \cref{fig:2dtiling} and \cref{fig:3dtiling} for nearest neighbour terms.

Because the fermionic encoding does not have exactly the same translational symmetry as the material, some care needs to be taken in how the groupings are ultimately tiled, and how the modes are ordered in the encoding. In particular, the choice of tiling needs to ensure that two parallelized groupings are not using the same ancillary face qubits at the same time. Furthermore, the modes in the encoding of the unit cell need to be ordered in such a way that their relative positions within a grouping are consistent throughout the tiling of that grouping. 

In the case of next nearest neighbour terms, one can always consider a larger cell which includes more sites, however again care should be taken in the precompiling stage to remove any terms that would be double counted under tiling, to ensure that mode ordering is consistent, and to ensure the terms are grouped in a way that can be uniformly tiled. Currently the compiler is not equipped to handle this case, and so instead any terms that are beyond nearest neighbour are separated out and used to populate a complete lattice, described in \cref{sec:compiling_nnn}. 
The remaining compiler routines are then applied to this remaining collection of terms. Unfortunately this yields sub-optimal circuit depths and can be quite computationally taxing.  Further improvements on the logic regarding tiling are currently in development and we expect circuit depths on terms beyond nearest neighbour to significantly improve after these changes.

\subsubsection{Compiling beyond-nearest-neighbour terms}
\label{sec:compiling_nnn}
As detailed in \cref{sec:compile_unit_cell}, terms beyond next nearest neighbour, denoted NNN+,  must be included explicitly in the set of terms used to cost the overall circuit corresponding to the lattice $\mathcal{L}$.
That is, the subset of terms $\mathcal{T}_{\mathrm{NNN+}} \in \mathcal{T}$ must be repeated throughout $\mathcal{L}$ wherever they are compatible, i.e. wherever they do not exceed the boundaries of $\mathcal{L}$.
Starting from a term $t\in \mathcal{T}_{\mathrm{NNN+}}$, which by construction involves the central site $s_0 \in \mathcal{L}$, we attempt to tile $t$ upon $\mathcal{L}$ by the following steps:
\begin{easylist}[itemize]
    & Determine the structure of $t$, i.e., the sites and orbitals involved in the interaction, and their relative positions.
    & For each site $s \in \mathcal{L}$, construct a new term $t^{\prime}$ retaining the same structure of $t$ by translating all the sites of $t$ by the vector $\bm{v}= s - s_0$.
    & if $t^{\prime}$ extends beyond the lattice, i.e., it involves a site $s'\notin \mathcal{L}$, it is discarded. In other words, the Hamiltonian is truncated to terms completely contained in $\mathcal{L}$. For instance let us consider the following case:
    && $\mathcal{L}$ is a $3 \times 3 \times 3$ lattice, i.e., it consists of the points $(n_1,n_2,n_3)$ of a Cartesian grid with $n_i \in \set{0,1,2}$.
    && $t$ involves the central site $s_0 = (1,1,1)$, and the central-north site at $s_N = (1,1,2)$, such that $t$ contains the relative vector of $t$ is $\bm{v}_N = s_N - s_0 = (0,0,1)$.
    && Then, tiling $t$ to a new site, say $s^{\prime} = (2,2,2)$, would involve the site $s^{\prime} + \bm{v} = (2,2,3)$, but $s^{\prime} \notin \mathcal{L}$.
    && Therefore the new term is not included in $\mathcal{T}_{\mathrm{NNN+}}$.
    & If $t^{\prime}$ is valid within $\mathcal{L}$, then add it to $\mathcal{T}_{\mathrm{NNN+}}$.
    
\end{easylist}

Following this tiling procedure, $\mathcal{T}_{\mathrm{NNN+}}$ now has many more terms and dominates the onsite/nearest-neighbour terms.
In dealing with $\mathcal{T}_{\mathrm{NNN+}}$, we do not group terms in advance of costing, since the advantage of this step is to arrange subsets of terms which facilitate the parallelism outlined in \cref{sec:compile_unit_cell} which is not available for terms in $\mathcal{T}_{\mathrm{NNN+}}$.
Moreover, the interactions in this set are large-range by definition, i.e., they extend across at least three sites of $\mathcal{L}$, necessitating either high-cost circuit terms or many fswaps in order to achieve low-cost terms. 
Since we do not have groupings, here we opt to omit the swap network stage of the compiler since it would have to act on the entire space of $\mathcal{L}$ which is unlikely to outperform the direct cost of the terms, although it is an open question whether a swap routine could improve the depth here. 
For these reasons, the depth associated with $\mathcal{T}_{\mathrm{NNN+}}$ is expected to dominate the circuit for simulating the Hamiltonian, despite not dominating the total number of terms.
We are currently working to improve the compiler to handle NNN+ terms, by incorporating logic which is able to generically understand the non-trivial relationship between the translational symmetry of the physical system and the translational symmetry of the given encoding. This will allow for the grouping and tiling of beyond-nearest-neighbour terms in the same fashion as onsite and nearest-neighbour terms -- which is expected to yield significant reduction in circuit depths for the unitary execution of NNN+ terms.

\subsection{Circuit costing}
After applying the decomposition routines, each grouping is given a circuit cost according to the terms it contains. The depth cost of an individual grouping is given by the largest circuit depth of an individual term in the group, which may depend on the gate set available in hardware. Assuming access to arbitrary 2-qubit rotations yields a depth of $ 2 \ceil{\log_2(w)}-1$, where $w$ is the weight of the Pauli term (see \cref{sec:gate_complexity}). Summing this cost over all groups gives the total depth. In order to account for the fact that 2-qubit gates are given cost 1, we modify the disjoint qubit decomposition routine to allow  any terms acting on exactly the same two qubits to be kept together in the same grouping. This way the coster treats them as though they are performed in parallel. 

The fermionic swap network will introduce a number of fswap gates. These can be handled in exactly the same fashion as the time evolution operations coming from the Hamiltonian. See \cref{sec:gate_complexity} for details.


%% file: Results.tex
\section{Results}
\label{sec:results}

In this section we present several results about the circuit costs of different materials using the methods discussed in this paper.
To familiarise the reader with our approach, in \cref{sec:model_ham_eg} we walk through a system simple enough that all the components can be understood easily. This system consists of a 2D lattice of unit cells with $N_{\mathrm{orbitals/cell}} = 2$ orbitals per unit cell. We include this discussion as an aid to grasp the algorithmic concepts developed in this work. This simple system allows one to detach the analysis of the circuit complexity from the physics, but it does not answer satisfactorily how to generate a Hamiltonian instance related to a particular material. In \cref{sec:full_an_SRVO3} we describe in full detail the whole construction developed in this work, from a \gls{dft} analysis all the way down to the circuit decomposition, for the strongly correlated material strontium vanadate (SrVO$_3$).
The results for different materials and assumptions are shown in \cref{sec:res_materials}.

\subsection{Circuit analysis of a simple example}
\label{sec:model_ham_eg}

Here we discuss a simple model of immobile impurity levels coupled to mobile electrons (the bath) in 2D and show the results of applying our circuit compiler to it. The unit cell contains one impurity level and one bath mode, and the bath modes can hop to neighboring sites (see \cref{fig:system_ex}). We first concentrate on a single unit cell, where we discuss in detail the procedure of mapping a fermionic model into a Pauli Hamiltonian, and the use of the \gls{jw} string and fswap networks. In \cref{sec:many-uc_ex} we discuss tiling this unit cell across the lattice, and the advantages of the hybrid encoding. Each disconnected unit cell is easily diagonalisable on a classical computer.

\subsubsection{One unit cell}

\begin{figure}[t]
    \centering  
    \includegraphics[width=0.8\linewidth]{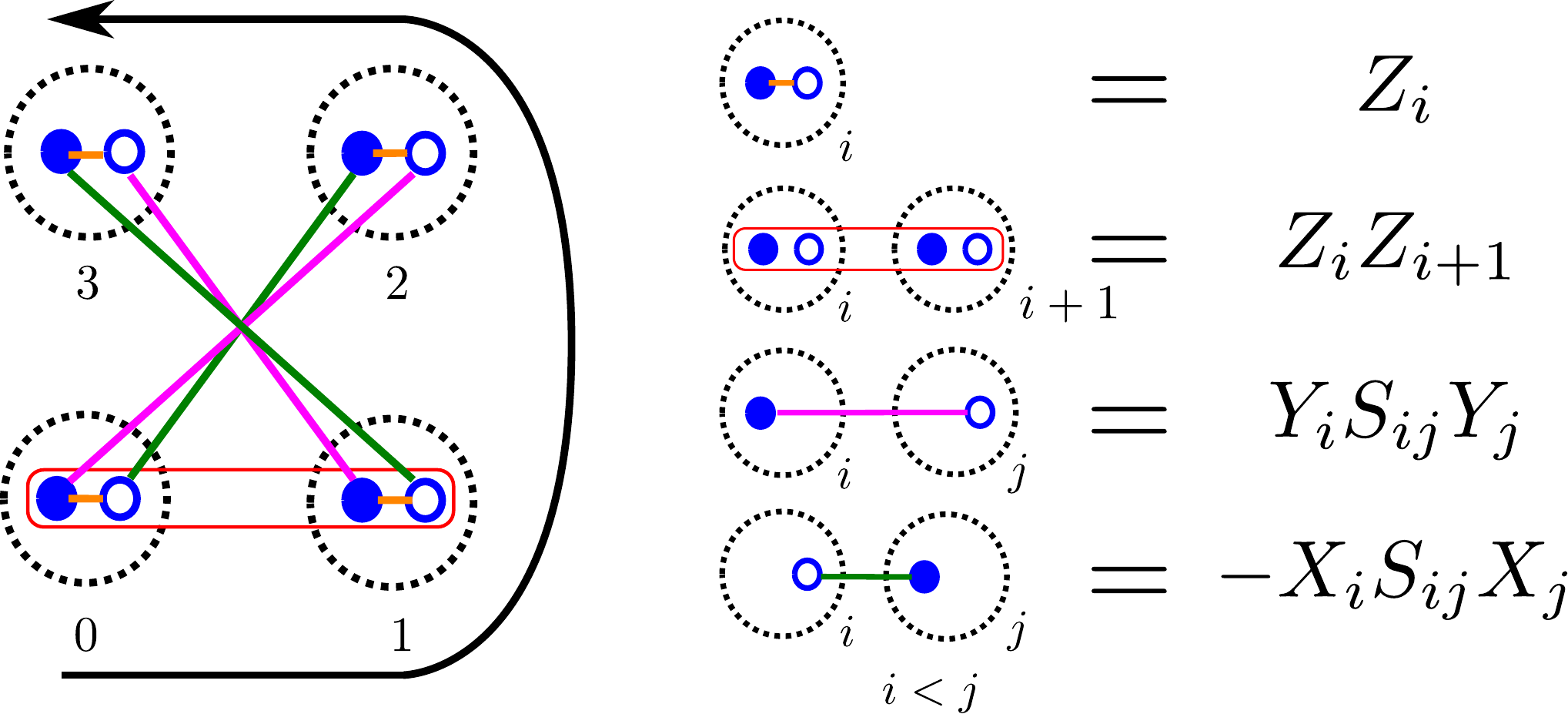}
    \caption{(Left) Connectivity graph of a single unit cell. Each mode is represented by dashed circles. Filled (open) blue circles inside represent the even (odd) Majoranas at that site. Quadratic interactions are depicted as lines between Majorana modes. The quartic interaction is represented by the red rectangle containing four modes. The black arrow denotes the \gls{jw} ordering. (Right) Four different types of interactions in the Hamiltonian and their map into Pauli monomials. S$_{ij}$ is a string of Pauli Z between $i$ and $j$ (see main text for more details). }
    \label{fig:system_ex}
\end{figure}

A system with a single unit cell consists of two electrons, one being the impurity electron of spin $\sigma$ (created and annihilated by $d^\dagger_\sigma$ and $d_\sigma$, respectively) and the other being the bath electron of spin $\sigma$ (created and annihilated by $c^\dagger_\sigma$ and $c_\sigma$, respectively). This corresponds to the simple Hamiltonian in complex fermion form
\begin{equation}\label{eq:ham_example}
    H_{\rm cell}=\sum_{\sigma=\uparrow,\downarrow}( \epsilon_\sigma c^\dagger_\sigma c_\sigma + \Delta(c^\dagger_\sigma d_\sigma+d^\dagger_\sigma c_\sigma)+\epsilon^d_\sigma d^\dagger_\sigma d_\sigma) + U d^\dagger_\uparrow d^\dagger_\downarrow d_\downarrow d_\uparrow,
\end{equation}
where  $\epsilon_\sigma$, $\epsilon_\sigma^d$, $\Delta$, and $U$ are real parameters. This system can be represented by 4 modes. Relabelling $(d_\uparrow,d_\downarrow,c_\uparrow,c_\downarrow)=(a_0,a_1,a_2,a_3)$ and introducing the Majorana fermion representation $a_j=(\gamma_{2j}+i\gamma_{2j+1})/2$ the Hamiltonian in \cref{eq:ham_example} (up to an overall constant) becomes
\begin{equation}\label{eq:ham_example_maj}
    H_{\rm cell}=\sum_{j=0}^3 i\frac{\epsilon_j}{2} \gamma_{2j}\gamma_{2j+1} + \sum_{j=0,1}i\frac{\Delta}{2}(\gamma_{2j}\gamma_{2j+5}-\gamma_{2j+1}\gamma_{2j+4}) -\frac{U}{4}\gamma_0\gamma_1\gamma_2\gamma_3,
\end{equation}
with parameters $(\epsilon_0,\epsilon_1,\epsilon_2,\epsilon_3)=(\epsilon_\uparrow^d+\frac{U}{2},\epsilon_\downarrow^d+\frac{U}{2},\epsilon_\uparrow,\epsilon_\downarrow)$. The particular connectivity structure in this example maps straightforwardly to the fermionic encoding given by a single \gls{jw} line. As we will see in more complex situations, this encoding can be hybridized with the compact encoding to generate a more efficient mapping, i.e., one that reduces the operator weight in term of the Pauli operators. In what follows, we will use the \gls{jw} string depicted in \cref{fig:system_ex}\footnote{The reader may note that for the particular connectivity graph of this problem, a better \gls{jw} ordering would be produced by relabelling $2\rightarrow 0$ and $0 \rightarrow 2$, as there all the interactions would be between nearest neighbors. We intentionally choose the ordering shown above, as it motivates the introduction of larger weight operators, which are needed in generic situations where the connectivity graph does not consist of a simple line.}.

 In this system we have two type of fundamental interactions: $\gamma_{2m}\gamma_{2l+1}$ or $\gamma_{2m+1}\gamma_{2l}$ (with $m\leq l$). The quartic interaction is a product of these fundamental ones. Interaction terms like $\gamma_{2i}\gamma_{2j}$ or $\gamma_{2i-1}\gamma_{2j+1}$ are not present because the original Hamiltonian is Hermitian and invariant under complex conjugation $\mathcal{K}$ (i.e., $\mathcal{K}H\mathcal{K}^{-1}=H$, with $\mathcal{K}i\mathcal{K}^-1=-i$, where $i$ is the imaginary number). This antiunitary symmetry acts like usual time-reversal symmetry on spinless fermions. We find 
\begin{equation}\label{eq:maj_trans}
    \mathcal{K}a_j\mathcal{K}^{-1}=a_j\rightarrow \mathcal{K}\gamma_{2j}\mathcal{K}^{-1}=\gamma_{2j}, \quad \mathcal{K}\gamma_{2j+1}\mathcal{K}^{-1}=-\gamma_{2j+1}.
\end{equation}
Hermitian quadratic operators have the form $i\gamma_a\gamma_b$. For these terms to be invariant under $\mathcal {K}$, they have to contain an odd number of Majoranas with odd index, according to \cref{eq:maj_trans}. This shows the role of symmetry in restricting the type of operators present in the Hamiltonian.
Interactions are uniquely specified by the structure of the  Majorana monomial, and can be mapped directly to the Pauli algebra under the \gls{jw} encoding. There are three possibilities:
$i\gamma_{2m}\gamma_{2m+1}$ is mapped into $Z_m$, $-i\gamma_{2m}\gamma_{2l+1}$ to $Y_{m}S_{ml} Y_{l}$, and $-i\gamma_{2m-1}\gamma_{2l}$ to $-X_mS_{ml}X_l$ (with $m\leq l$). Here, $S_{ij}\equiv \prod_{i<k<j}Z_k$ is a string of Pauli $Z$s between the mode $i+1$ and the mode $j-1$,  
For further details see \cref{sec:qubit_rep}.

The Hamiltonian $H_{\mathrm{cell}}$ in the Pauli representation is
\begin{equation}\label{eq:ham_paul_ex}
    H_{\rm cell}=\sum_{j=0}^3 \frac{\epsilon_j}{2} Z_j - \sum_{j=0,1}\frac{\Delta}{2}(Y_jS_{j,j+2}Y_{j+2}+X_jS_{j,j+2}X_{j+2}) +\frac{U}{4}Z_0Z_1. 
\end{equation}

For quantum algorithms like \gls{vqe} or \gls{tds}, we need to create a unitary of the form $U=e^{i\theta H_{\rm cell}}$, where $\theta$ is some parameter. In a quantum computer, the accessible operations form a fixed subset of gates, from which any possible unitary on the whole system can be approximated. Usually these consists of 1-qubit and 2-qubit gates. Using the Suzuki-Trotter formula, we can approximate the unitary $U$ by a series of simpler unitaries that can be implemented in the quantum computer. This decomposition has to be done in a way that minimizes the depth of the circuit constructed.

We wish to know the circuit depth of the unitary generated by the Pauli string for a given interaction. 
Our circuit depth will reflect the number of layers of 2-qubit gates required to run a given algorithm for the specified Hamiltonian. If only a single qubit is involved, i.e., the weight is $w=1$, we only require 1-qubit gates, which we assume access to at no cost.
In general, for Pauli strings with weight $w$, we saw in \cref{sec:gate_complexity} that such a circuit can be performed with \emph{cost}
\begin{equation}\label{eqn:monomial_cost}
    c(w) = 2 \ceil{\log_2 w} -1
\end{equation}
sublayers of 2-qubit gates.
However, we cannot concurrently perform operations which involve the same qubits: that is, we must separate the interactions into lists, each of which contains terms with disjoint support. 
We therefore separate the terms in \cref{eq:ham_paul_ex} into a number of \emph{layers}, where each layer can consist of several terms that can be implemented simultaneously. 
The cost of any layer is the depth, $d$, of the most expensive term in that layer. 
The overall depth is therefore 
\begin{equation}
    d = \sum_{l\in {\rm layers}} \max_w \{ c(w) \}_l.
\end{equation}

\begin{figure}[t]
    \centering  
    \includegraphics[width=\linewidth]{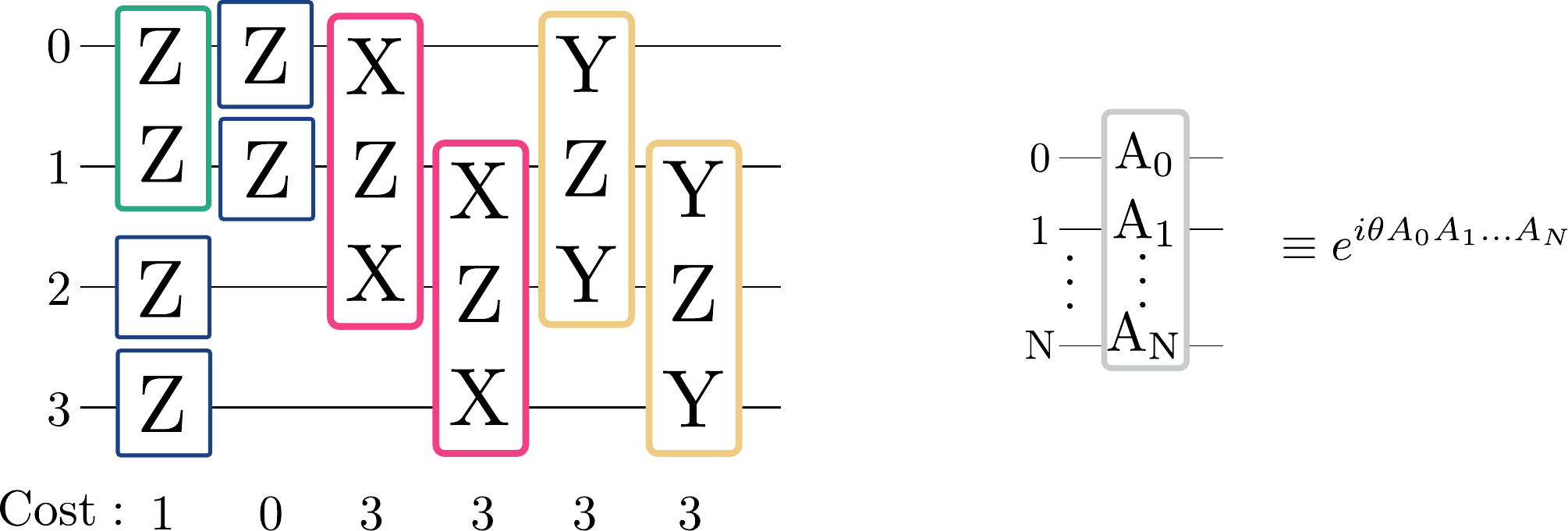}
    \caption{(Left) Circuit decomposition to implement all the terms of the Hamiltonian in  \cref{eq:ham_paul_ex} once. The cost of each layer of gates is shown below.(Right) Circuit notation used. Each box in the left represents a unitary. The parameters $\theta$ used may vary depending on the particular algorithm.}
    \label{fig:circuit_ex}
\end{figure}

We break down the set of interactions in \cref{eq:ham_paul_ex} into layers and show the resulting circuit in \cref{fig:circuit_ex}. The overall circuit depth for implementing all the terms of this Hamiltonian once is therefore $d=13$ rounds of 2-qubit gates. 

\newpage

\subsubsection{Fermionic swap networks}\label{sec:model_ham_swap_net}

Many terms in the circuit represented in \cref{fig:circuit_ex} have cost 3, since their Pauli string crosses a third qubit, whereas monomials involving only adjacent qubits have depth 1. 
It may therefore be advantageous to rearrange the modes before implementing interactions: \emph{fermionic swap} (fswap) operations can be used to move modes through the graph.
Although they also incur a cost (as described in \cref{sec:swap_network_details}), here we assume they have uniform cost $d=1$ . However, if they successfully enable low-weight operations for all other interactions, they can prove beneficial overall. 

\begin{figure}[t]
    \centering
    \includegraphics[width=\linewidth]{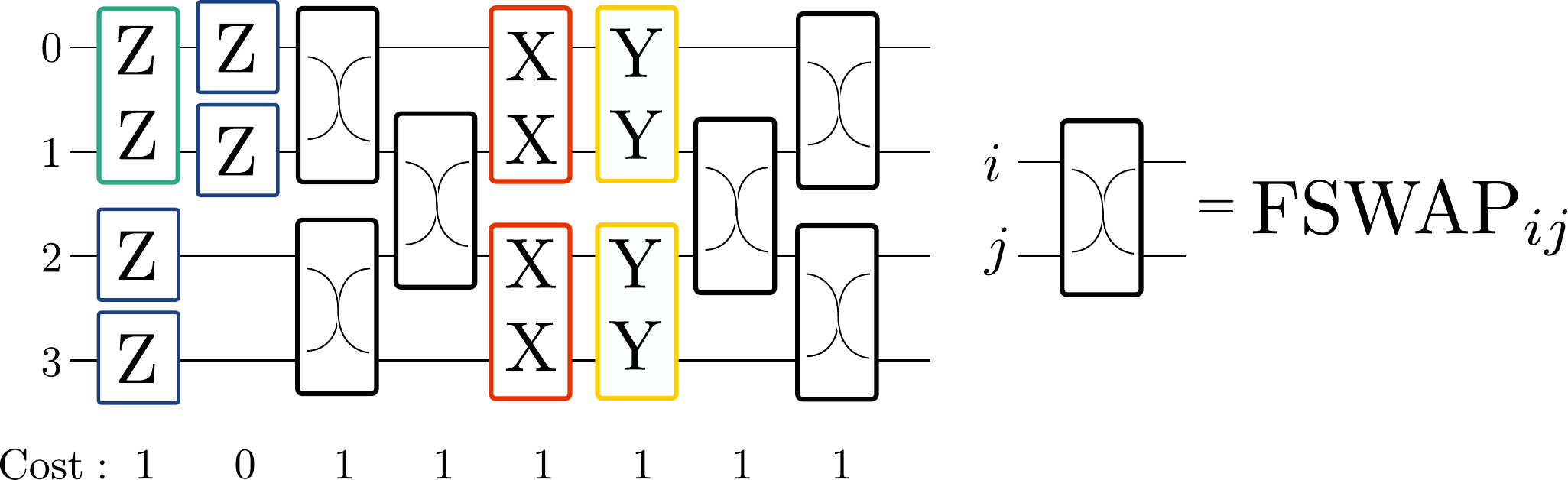}
    \caption{(Left) Circuit implementing the unitaries generated by all the terms in the Hamiltonian. In this approach we use fermionic swap networks to decompose high weight operators into
    two qubit gates. If the cost of doing  fermionic swap gates is small, this strategy can  be advantageous. Further gains can be achieved by merging two-qubit gates -- e.g., the gates generated by $XX$ and $YY$ at the centre of the circuit -- which is preferred in a cost model where {\bf any} 2-qubit gate has the same cost $d=1$.
    (Right) Diagram for the fermionic- swap gate.}
    \label{fig:depth_bar_with_swaps}
\end{figure}

The initial mode configuration can allow implementation of the terms 
$\{ Z_0Z_1, Z_i\},$ 
but as above, we must split these into sublayers whose constituents have disjoint support, i.e., we create two layers 
$ \{ Z_0Z_1, Z_2, Z_3\}; \{Z_0, Z_1\}. $ 
Thereafter, we run two fswap layers. Denoting the fswap between modes $a$ and $b$ as FSWAP$_{ab}$, we apply the layers  $\{\rm{FSWAP}_{01},\rm{FSWAP}_{23} \};\{\rm{FSWAP}_{12}\}$, which maps the original \gls{jw} ordering $(0,1,2,3)$ into the ordering $(1,3,0,2)$. In this new ordering the operators $\gamma_{0}\gamma_{5}$, $\gamma_2\gamma_7$,
$\gamma_1\gamma_4$, and $\gamma_3\gamma_6$ all represent interactions between Majoranas in nearest neighbor modes and as such, they do not have strings of $Z$ operators attached. This allows (after splitting into disjoint-support sublayers) the interactions 
$$ \{ Y_0Y_2, Y_1Y_3\}; \{ X_0X_2, X_1X_3 \}.$$
We also must reverse the fswaps in order to recover the starting mode configuration, so after the final interaction layer, there are two further fswap layers. This step is useful if one wishes to apply the same exact circuit many times, but can be relaxed if we may start from a scrambled configuration in the next step. However, care should be taken when applying the compiler to ensure that translational invariance is preserved, and if modes are scrambled at the beginning of a given layer then a new compilation must be performed for that layer. Therefore, generally the simplest strategy is to return modes to their original configuration at the end of a given Trotter or VQE layer.
The swapped circuit and its associated costs are depicted in \cref{fig:depth_bar_with_swaps}. 
We include the cost of the fswap layers since it contributes to the overall circuit depth. 
The total depth using this method is then $d=7$, verifying that the generation of an fswap network is justified here, since it leads to a decrease in circuit depth compared with the prior method which gave $d=13$ in \cref{fig:circuit_ex}. 

\subsubsection{Many unit cells}
\label{sec:many-uc_ex}

\begin{figure}[t]
    \centering
    \includegraphics[width=\linewidth]{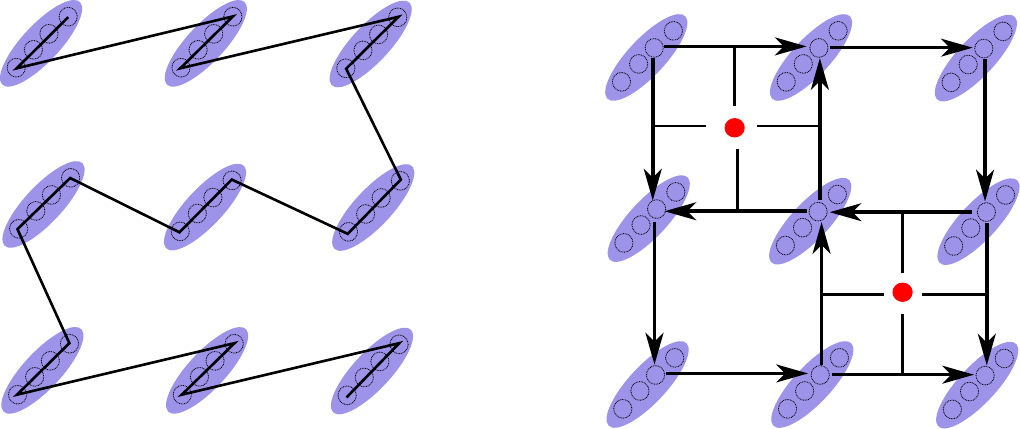}
    \caption{(Left) \gls{jw} ordering across a 2D lattice. Neighboring unit cells (ellipses) can end very far away in the \gls{jw} chain, leading to a large operator weight for physical processes that happen locally. (Right) Using extra ancilla qubits (red circles), the hybrid encoding can generate low-weight Pauli operators for the intracell processes. For the intercell processes, \gls{jw} ordering and fswap networks can implement all the required terms.}
    \label{fig:ex_lattice}
\end{figure}

We are now in a position to discuss the generalized model of impurity levels coupled to mobile electrons.
Consider the Hamiltonian
\begin{equation}
    H=\sum_{k\in {\rm cells}} H^{(k)}_{\rm cell}+t\sum_{\langle i,j\rangle}( c_{i,\sigma}^\dagger c_{j,\sigma}+ c_{j,\sigma}^\dagger c_{i,\sigma}), 
\end{equation}
where $H^{(k)}_{\rm cell}$ is the Hamiltonian of the $k^{\rm th}$ cell, and the second term represents the intercell coupling, produced by the hopping of bath modes between nearest neighbor cells in 2D. Following the procedure outlined above has some drawbacks in this larger system. Although the fermions hop locally in the 2D system, after including a \gls{jw} ordering that maps the system into a line, some local interactions become very non-local (see \cref{fig:ex_lattice}). This ultimately increases the circuit depth by a factor that depends on the whole size of the system.

To keep the operator weight of Pauli strings independent of the system size, we exploit the hybrid encoding introduced in \cref{sec:qubit_rep}.
In this encoding, we add extra ancilla qubits that allow to represent the fermion algebra with low-weight Pauli operators. The price to pay is that more qubits are needed. In materials the interactions (understood generically as electron-electron interactions or hopping) between the modes in the unit cell are expected to be more dense than the intercell interactions, meaning that inside the cell the fswap protocol could be very beneficial to bring modes together (as discussed in \cref{sec:swap_network_details}). The hybrid encoding can minimise the use of the extra qubits by using them to implement the interactions between different cells, which are expected to be sparse.

\subsection{Full-stack analysis: Strontium vanadate}
\label{sec:full_an_SRVO3}

\begin{figure}[ht]
    \centering
    \includegraphics[scale=1]{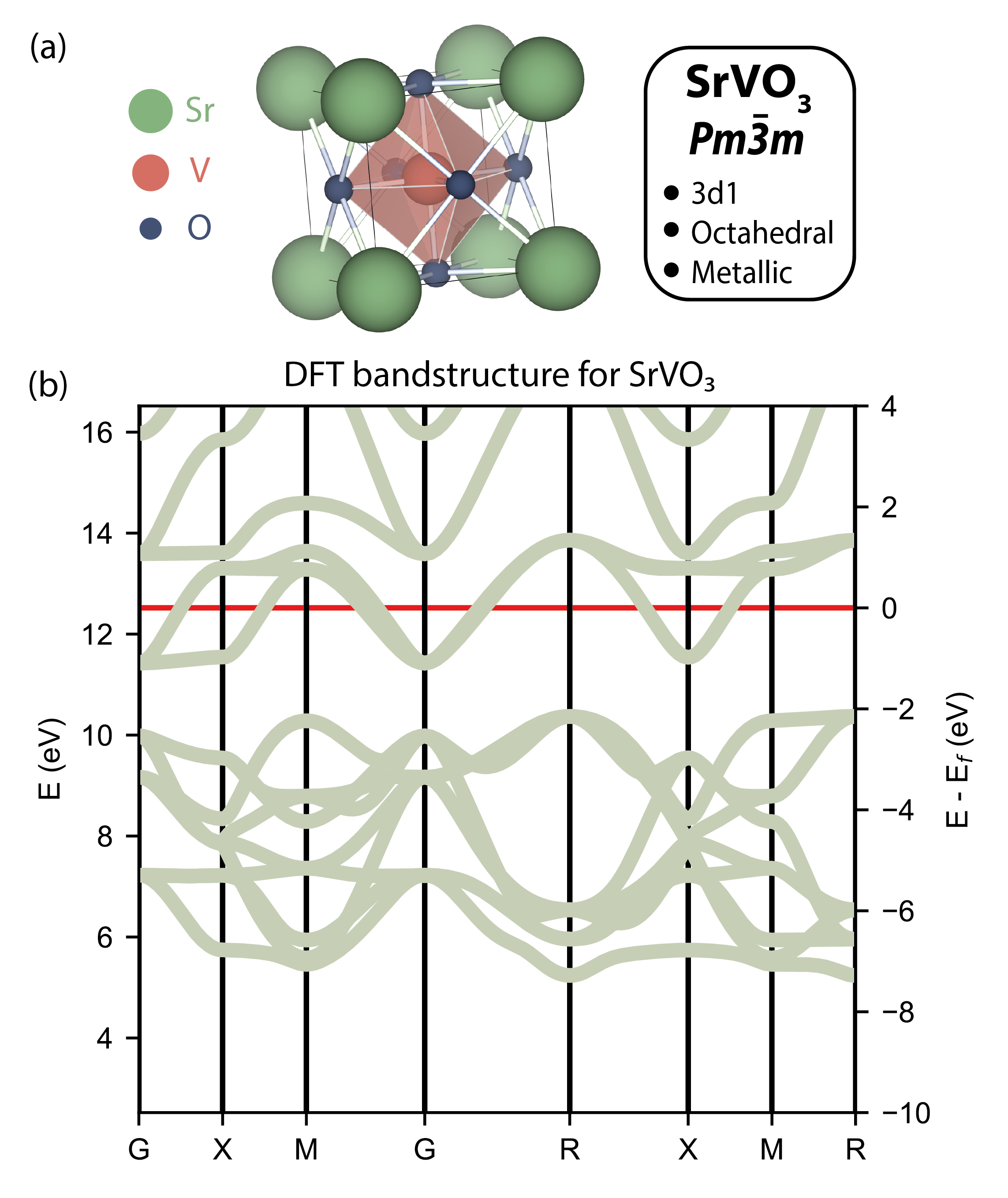}
    \caption{
    (a) The structural unit cell of SrVO$_3$ with spacegroup \emph{Pm$\overline{3}$m}, 3d1 nominal electronic valence configuration, and octahedral coordination environment. In this configuration, the compound is a metal. (b) The ground state electronic bandstructure as predicted using \gls{dft} along the high symmetry path in the Brillouin zone. The red line indicates the position of the Fermi level $E_f$.
    }
    \label{fig:svo_structure}
\end{figure}

In this section we present a full-stack analysis of the transition metal
perovskite oxide SrVO$_3$. Its unit cell is shown in \cref{fig:svo_structure}(a). 
Materials in the perovskite oxide family with chemical formula ABO$_3$ form basic components 
of the Earth's mantle, are central to many technological applications, and pioneer current 
research efforts to design bespoke materials with multifunctional properties. They can exhibit a wide range of physical properties, spanning insulating, semiconducting, and metallic characteristics as well as a superconducting, correlated, multiferroic, and ferroelectric phases, and highly controllable transitions between them. 
For example, synthetic perovskite oxides are used in batteries \cite{suntivich2011design}, qubits \cite{liu2021coherent,},  
high-temperature superconductors \cite{bednorz1986}, solar cells \cite{park2015perovskite},
semiconductors \cite{perovskite_semiconductor}, spin switches \cite{lupo2021slater}, multiferroics \cite{spaldin_ferroic} and ferroelectrics \cite{ye2018metal}. In particular, 
SrVO$_3$ has been used as components in both the anode \cite{svo_anode} and cathode \cite{svo_cathode} of Li-ion batteries. 
Their widespread usage is a result of the remarkable stability of the ABO$_3$ chemical structure \cite{Filip5397}, 
and allows for highly tunable functional properties as a result of being able to control and couple the many electronic degrees of freedom, such as orbital, charge, and lattice in a large phase space of chemical species. 
We start by presenting the \gls{dft} results and their subsequent mapping to \glspl{mlwf}, followed by the computation of the single-body and two-body matrix elements that parametrise the complex fermionic Hamiltonian of a periodic system. We conclude with a discussion on the calculation of
the quantum complexity for this Hamiltonian by calculating its circuit depth.

We first describe the results as obtained from \gls{dft} and the subsequent
Wannierisation procedure. All \gls{dft} calculations were performed using the plane-wave code Quantum Espresso \cite{QE_2009, Giannozzi_2017}, version 6.8, together with the GGA-PBE
exchange correlation functional \cite{GGA}. Atomic cores were treated using the ONCVPSP pseudpotential library \cite{ONCVPSP} with
valence configuration
Sr(4\emph{s}4\emph{p}4\emph{d}5\emph{s}5\emph{s}5\emph{p}),
V(3\emph{s}3\emph{p}3\emph{d}4\emph{s}), and O(2\emph{s}2\emph{p}). The
plane-wave basis representation is used for the wavefunctions, with a cutoff of
$E_{\mathrm{cut}} = 400$ eV (see \cref{{cut_off}}). We use a $4 \times 4 \times 4 $ $\Gamma$-centered k-point mesh in the
Brillouin zone for $k$-point sampling. Structural degrees of freedom are relaxed
until all forces are smaller than 1 mRyd/a.u. (see \cref{eq:forces}). Subsequently, the generation of
\glspl{mlwf} is performed with Wannier90 \cite{w90}. 

\cref{fig:svo_structure}(b) presents the electronic bandstructure of SrVO$_3$
along the high symmetry Brillouin zone path. The Fermi level is indicated by the
horizontal red line, which intersects a triply degenerate band with bandwidth
$\sim 2$ eV, and is thus representative of a metallic system. We also note the
set of 9 non-degenerate bands in the range $[-8,-2]$ eV (where $E_f$ is zeroed).
 
\begin{figure}[ht]
    \centering
    \includegraphics[width=\linewidth]{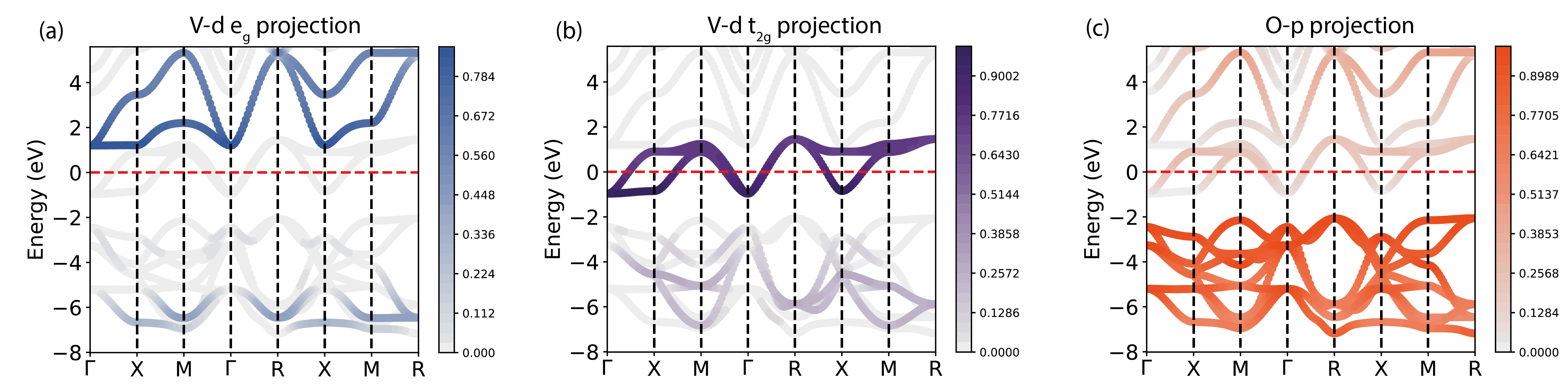}
    \caption{Fatbands projections on the \gls{dft} electronic band structure for SrVO$_3$ using projectors corresponding to (a) V-d e$_g$ orbitals,  (b) V-d t$_{2g}$ orbitals, and (c) O-p orbitals.}
    \label{fig:svo_fatbands}
\end{figure}

 To identify the orbital character of the electronic structure in
\cref{fig:svo_structure} we perform a fatbands projection analysis according to
\cref{eq:fatband} using d-orbital projections for the vanadium ions and
p-orbital projections for the oxygen ions. These results are presented in
\cref{fig:svo_fatbands}, where we see that the states of the triply degenerate
band at the Fermi level belong to the d-states of vanadium. Furthermore, we see
from \cref{fig:svo_fatbands}(a) and (b) that the full d-manifold of states
splits into the e$_g$ and t$_{2g}$ crystal field subgroups of the overall full d (rotation)
group, where the t$_{2g}$ states are responsible for the conduction electrons.
Moreover, the states below the Fermi level in the range $[-8,-2]$ eV are
primarily O-p type. Knowing the orbital character of the electronic states is a
crucial step in determining the necessary initial projectors used for generating
\glspl{mlwf}, which affects the numerical stability of the Wannierisation procedure.
Additionally, it is an essential component in determining the physical
interpretation of these electronic states, required for comparing against
experimental results.

\begin{figure}[ht]
    \centering
    \includegraphics[scale=0.7]{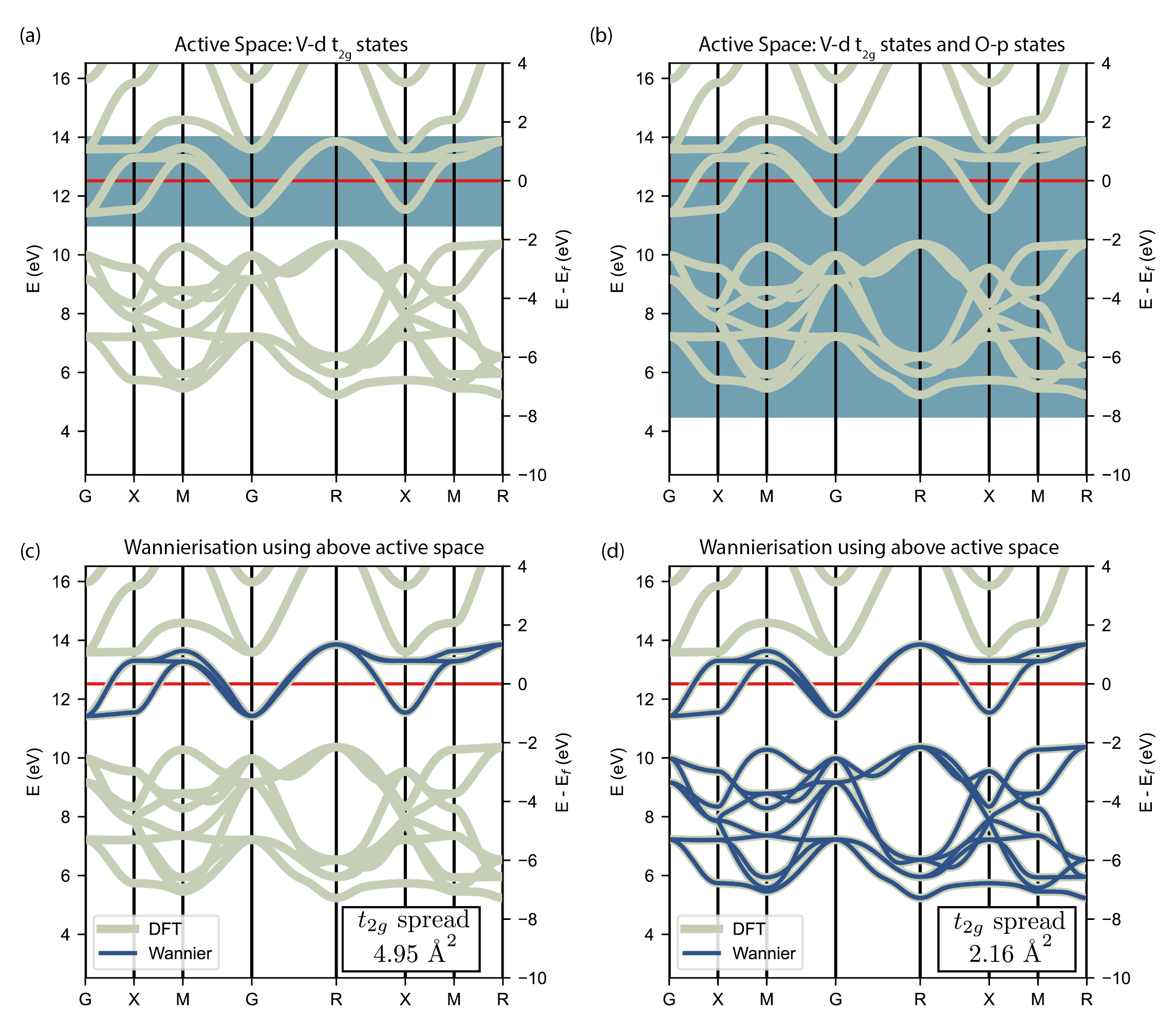}
    \caption{Two choices of active space for the Wannierisation protocol consisting of (a) 3 V-d t$_{2g}$ states and (b) 3 V-d t$_{2g}$ states, as well as an additional 9 O-p states. The bands obtained via the Wannierization procedure for the two choices of active space are shown in panels (c) and (d). In the bottom-right corners we highlight the spread -- as defined in \cite{Marzari12} -- of the \glspl{mlwf} corresponding to the t$_{2g}$ levels in the two cases, which is (c) 4.95 \AA  and (d) 2.16 \AA, respectively.
    }
    \label{fig:svo_wannier}
\end{figure}

Knowing the orbital character of the electronic bands near the Fermi level
allows us to reliably choose regions of the bandstructure over which an active
space can be chosen, and which can be used to parametrise fermionic Hamiltonians. The
upper panel of \cref{fig:svo_wannier} presents two possible choices of active
space for SrVO$_3$. In \cref{fig:svo_wannier}(a) we illustrate the active space
containing the 3 V-t$_{2g}$ orbitals, while in \cref{fig:svo_wannier}(b) we show
the active space that contains the 3 V-t$_{2g}$ and 9 O-p orbitals. Accounting
for spin degeneracy, the t$_{2g}$ active space contains 6 modes, while the
t$_{2g}$+p active space contains 24 modes. Choosing a larger window over which
the Wannier functions can be generated allows for their maximal localisation,
but at the price of including additional modes, while wannierising over smaller
active spaces results in a larger total spread of the resultant Wannier functions.
In \cref{fig:svo_wannier}(c)-(d) we show the corresponding \glspl{mlwf}, highlighting a
larger total spread for the t$_{2g}$ active space compared to the t$_{2g}$+p
active space. 
 
\begin{figure}[ht]
    \centering
    \includegraphics[scale=0.8]{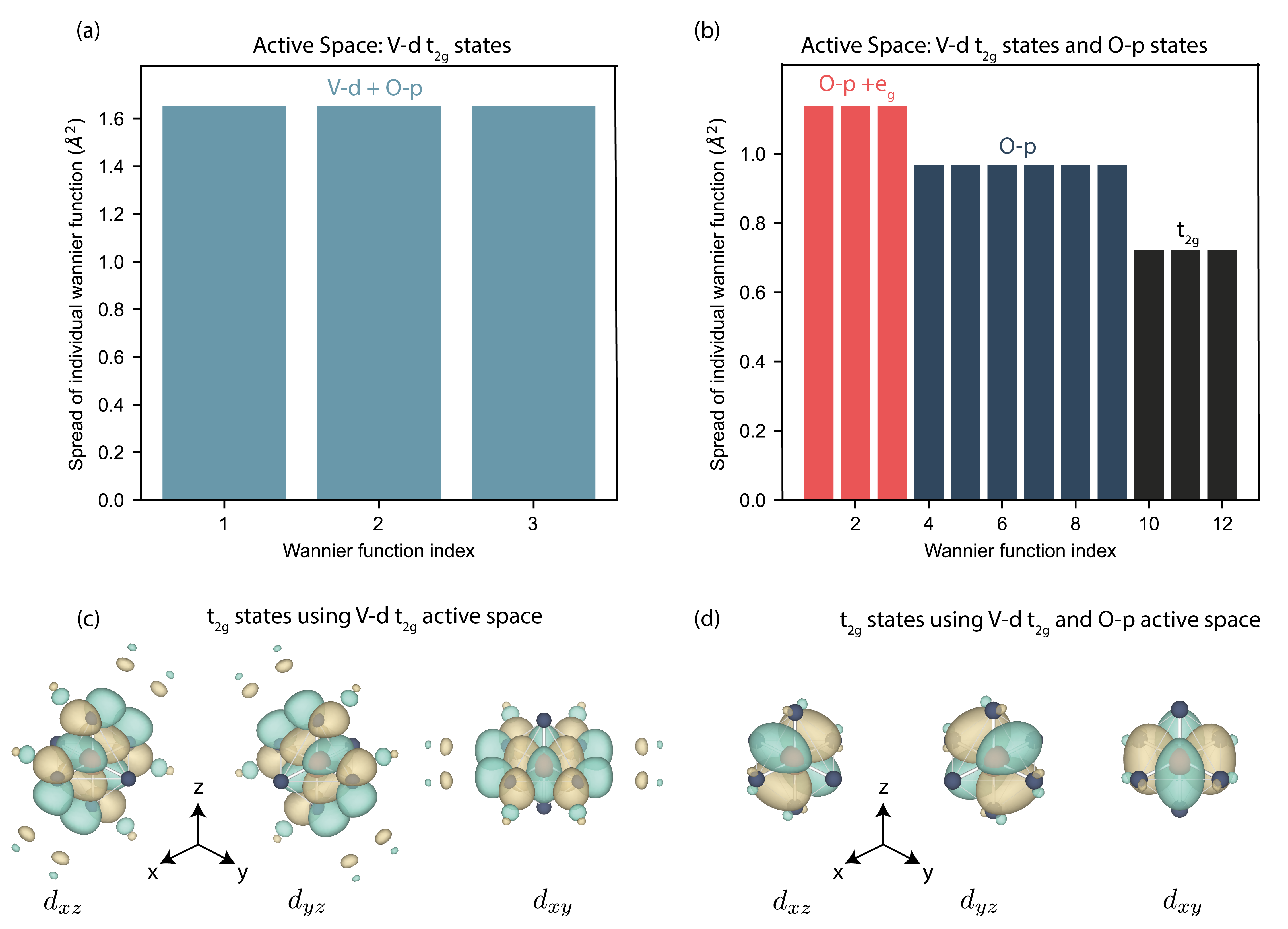}
    \caption{The spread of each Wannier function for the subspaces consisting of (a) 3 V-d t$_{2g}$ states and (b) 3 V-d t$_{2g}$ states with an additional 9 O-p states, as illustrated in \cref{fig:svo_wannier}. The isosurface plot of the corresponding \glspl{mlwf} for the (c) small and (d) large active spaces. Here, the isosurface value has been set to 0.325.} 
    \label{fig:svo_spreads}
\end{figure}

Furthermore, \cref{fig:svo_spreads}(a)-(b) illustrates the individual spreads of
the generated Wannier functions. By wannierising over the large energy window,
we see that the smallest spreads belong to the V-t$_{2g}$ orbitals, and
increases for the O-p states. \cref{fig:svo_spreads}(c)-(d) presents the
corresponding t$_{2g}$ Wannier functions on a real space grid, highlighting the
clear differences in their locality depending on the size of the active space
considered. Using the t$_{2g}$ active space only results in Wannier functions
which extend into neighbour and nearest-neighbour cells, thus resulting in
significant overlap with the Wannier functions in those cells. Whereas the
larger active space results in localised Wannier functions that do not extend into neighbouring cells.

\subsubsection{Computation of hopping matrix and Coulomb tensor coefficients }
\label{sec:HM_and_CT_coefficients}

Once selected the active space and built the corresponding \glspl{mlwf}, we can proceed with the calculation of the hopping matrix and Coulomb tensor coefficients as defined in \cref{eq:Wannier_T_V_core}. As explained in \cref{sec:DFT}, in the usual \gls{dft} procedure the ionic potential $\tilde{U}(\bm{r})$ for the electrons in the active space is replaced by the effective Kohn-Sham potential $\tilde{U}_{\mathrm{eff}}(\bm{r})$. The latter contains the entangled contributions arising from the original ionic potential, the screening effects of core electrons, and part of the electron-electron interaction. Therefore, it is convenient to rewrite the hopping matrix and Coulomb tensor for the electrons in the active space as
\begin{subequations}
\label{eq:Wannier_T_V_active_space}
    \begin{align}
    T(\bm{R})_{mn} &= \int d\bm{r} \mathcal{W}^{\bm{R}}_{m,\sigma}(\bm{r}) \left[-\frac{\hbar^2\nabla^2}{2m}+
    \tilde{U}_{\mathrm{eff}}(\bm{r})\right]\mathcal{W}^{\bm{0}}_{n,\sigma}(\bm{r}), \\
    \tilde{V}_{slmn}^{(\bm 0,\bm R_2,\bm R_3,\bm R_4)}&=\frac{1}{2}
    \int d\bm{r} \int d\bm{r}'\mathcal{W}_{s, \sigma}^{\bm{0}}(\bm{r})\mathcal{W}_{l, \sigma'}^{\bm{R}_2}(\bm{r}')W(|\bm{r}-\bm{r}'|)\mathcal{W}_{m, \sigma'}^{\bm{R}_3}(\bm{r}')\mathcal{W}_{n, \sigma}^{\bm{R}_4}(\bm{r}),\label{eq:Wannier_CT_core_W}
    \end{align}
\end{subequations}
where $W(|\bm{r}-\bm{r}'|)$ represents the part of the electron-electron Coulomb potential which is not captured by the \gls{dft} procedure. Many sophisticated approaches have been designed to mitigate the double-counting emerging from this decomposition and to obtain accurate values of the screened Coulomb potential \cite{Imada10, Anisimov1997, Aryasetiawan1998, dft_dmft_rev} but they are beyond the scope of this work. Instead, one simpler approach is to consider a Thomas-Fermi screened interaction potential of the form $W(|\bm{r}-\bm{r}'|) = q_e/(4\pi\epsilon_0) e^{-\mu_{\mathrm{TF}}|\bm{r}-\bm{r}'|}/|\bm{r}-\bm{r}'|$, with $\mu_{\mathrm{TF}}$ a material dependent inverse screening length \cite{Ribic2014}. For the sake of simplicity, in what follows we will focus on the unscreened case only, corresponding to setting $\mu_{\mathrm{TF}}= 0$ (i.e., we assume $V(|\bm{r}-\bm{r}'|) = W(|\bm{r}-\bm{r}'|)$). This will result in more, stronger, and longer-range interactions and, therefore, the results we will show below represent an upper bound for the quantum circuit complexities for the simulation of real materials. The effects of screening will be addressed in future work.

All the steps to calculate the hopping matrix and Coulomb tensor coefficients are described in detail in \cref{app:Hamiltonian_coeff_pipeline}. In what follows, we will give a brief description of the most important steps and analyse the outputs for the case of SrVO$_3$ with 3 V-t$_{2g}$ states in its active space introduced in the previous section. Importantly, we will work within the following assumptions:

\begin{enumerate}
    \item We will consider non-magnetic materials only, so that the two spin sectors are degenerate;
    \item Hopping matrix and Coulomb tensor coefficients involving unit cells which are nearest neighbors of order larger than $n_0$ and $n_{\mathrm{int}}$, respectively, are negligible;
    \item Hopping matrix and Coulomb tensor coefficients whose absolute value is smaller than a pre-determined threshold are negligible.  
\end{enumerate}

\begin{figure}[ht]
    \centering
    \includegraphics[width=0.6\textwidth]{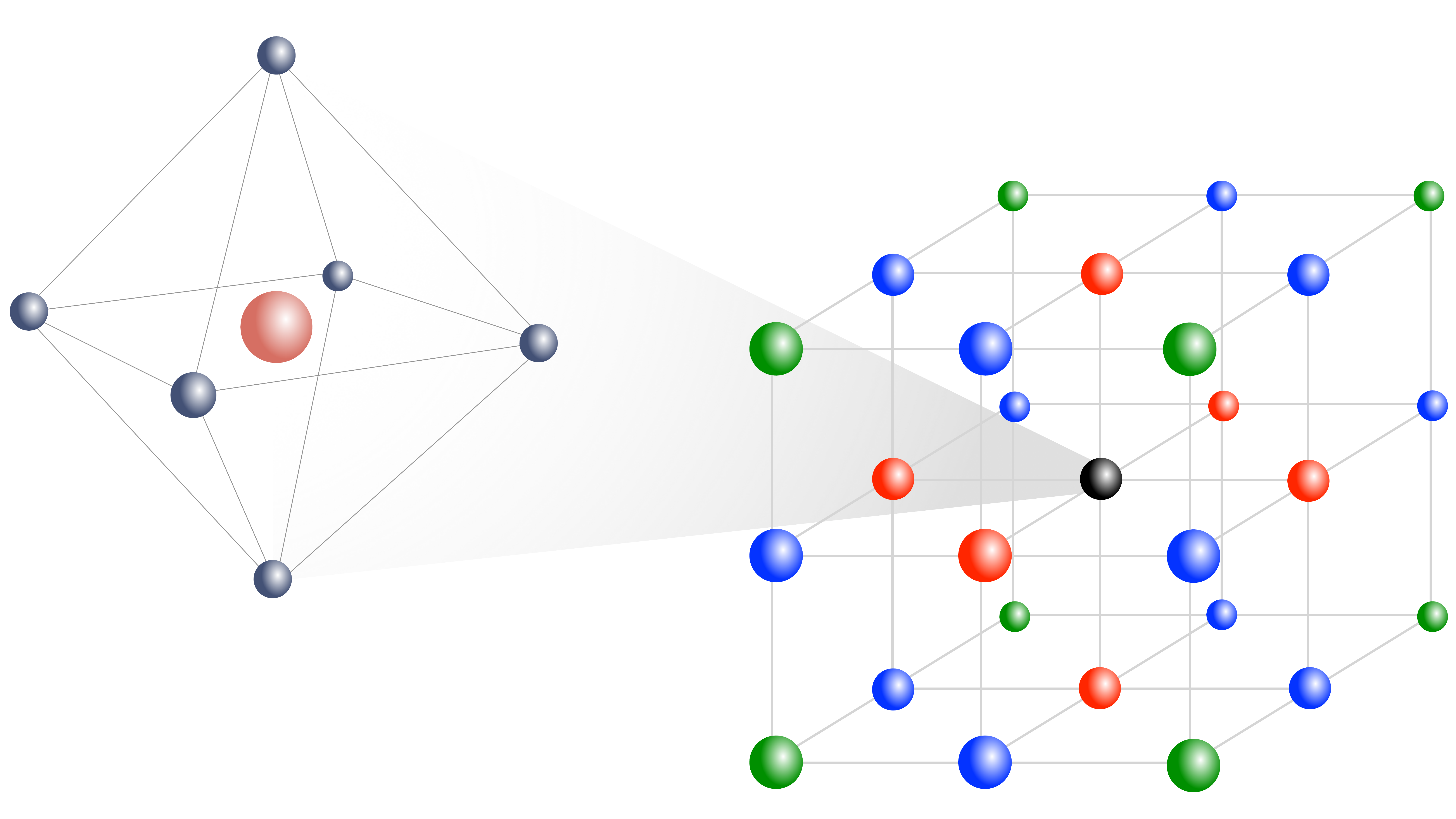} 
    \caption{
    Unit cell and motif for SrVO$_3$. In the unit cell, on the left, only the atoms whose atomic orbitals contribute significantly to the bands in the selected active space are shown, namely the central vanadium atom (big orange sphere) surrounded by six oxygen atoms (small blue spheres) (see \cref{fig:svo_structure}). The full SrVO$_3$ lattice, a portion of which is shown on the right hand side, can be obtained by translating the unit cell by multiple integers of the lattice vectors. The motif of order $n=1$ is formed by the central cell (black) and the 6 nearest neighbors (red), the motif of order $n=2$ includes also the 12 next-nearest neighbors (blue), and the motif of order $n=3$ is obtained by adding the $8$ next-to-next-nearest neighbours (green).
    }
    \label{fig:motif_srvo}
\end{figure}

The second assumptions takes advantage of the real-space localisation of the \glspl{mlwf} and implies that the hopping matrix and the Coulomb tensor coefficients are calculated on motifs of order $n_0$ and $n_{\mathrm{int}}$, respectively. Recalling the definitions of \cref{sec:motif}, they are denoted by $\mathcal{N}^{n_0}_O$ and $\mathcal{N}^{n_{\mathrm{int}}}_O$. In \cref{fig:motif_srvo} we show the motifs of order 1,2, and 3 for SrVO$_3$. Each site of the motif corresponds to a unit cell of the material. We denote with $N_{\mathrm{cells/motif}}$ the number of unit cells per motif and with and $N_{\mathrm{orbitals/cell}}$ the number of orbitals per cell (i.e., the number of bands contained in the chosen active space). Note that $N_{\mathrm{cells/motif}} = \mathrm{max}( \mathrm{dim}(\mathcal{N}^{n_0}_{\bm{0}}), \mathrm{dim}(\mathcal{N}^{n_\mathrm{int}}_{\bm{0}}))$. The total number of complex fermion modes per motif is $M = 2N_{\mathrm{cells/motif}}N_{\mathrm{orbitals/cell}}$.

\paragraph{Hopping matrix.} The hopping matrix $T(\bm{R})$ can be directly obtained from the output of the Wannierisation procedure described in the previous section as performed by Wannier90. For instance, the hopping of electrons from the central unit cell at $\bm{R} = \bm{0}$ to the cells at $\bm{R}_A = \bm{R}_1$ and $\bm{R}_B = \bm{R}_1 + \bm{R}_2$ (with, in general, $\bm{R} = n_1 \bm{R}_1 + n_2 \bm{R}_2 + n_3 \bm{R}_3 $; see \cref{eq:R_lattice}) is described by the two matrices
\begin{equation}
    T_{(100)}[\mathrm{eV}]=
    \begin{pmatrix}
        -0.26 & 0     & 0\\
        0     & -0.26 & 0\\
        0     & 0     & -0.26\\
    \end{pmatrix},
    \quad
    T_{(110)}[\mathrm{eV}]=
    \begin{pmatrix}
        0.006 & 0.009 & 0\\
        0.009 & 0.006 & 0\\
        0     & 0     & -0.082\\
    \end{pmatrix},
\end{equation}
respectively. 
Here, we introduced the notation $T_{(n_1n_2n_3)}:= T(\bm{R}) $. Noting that $|\bm{R}_A| < |\bm{R}_B|$, the equation above also shows that, in general, the value of the hopping matrix coefficients decrease as a function of the distance between the cells and therefore a nearest neighbor approximation is well justified. 

\begin{figure}[ht]
 \centering
 \includegraphics[width=0.48\textwidth]{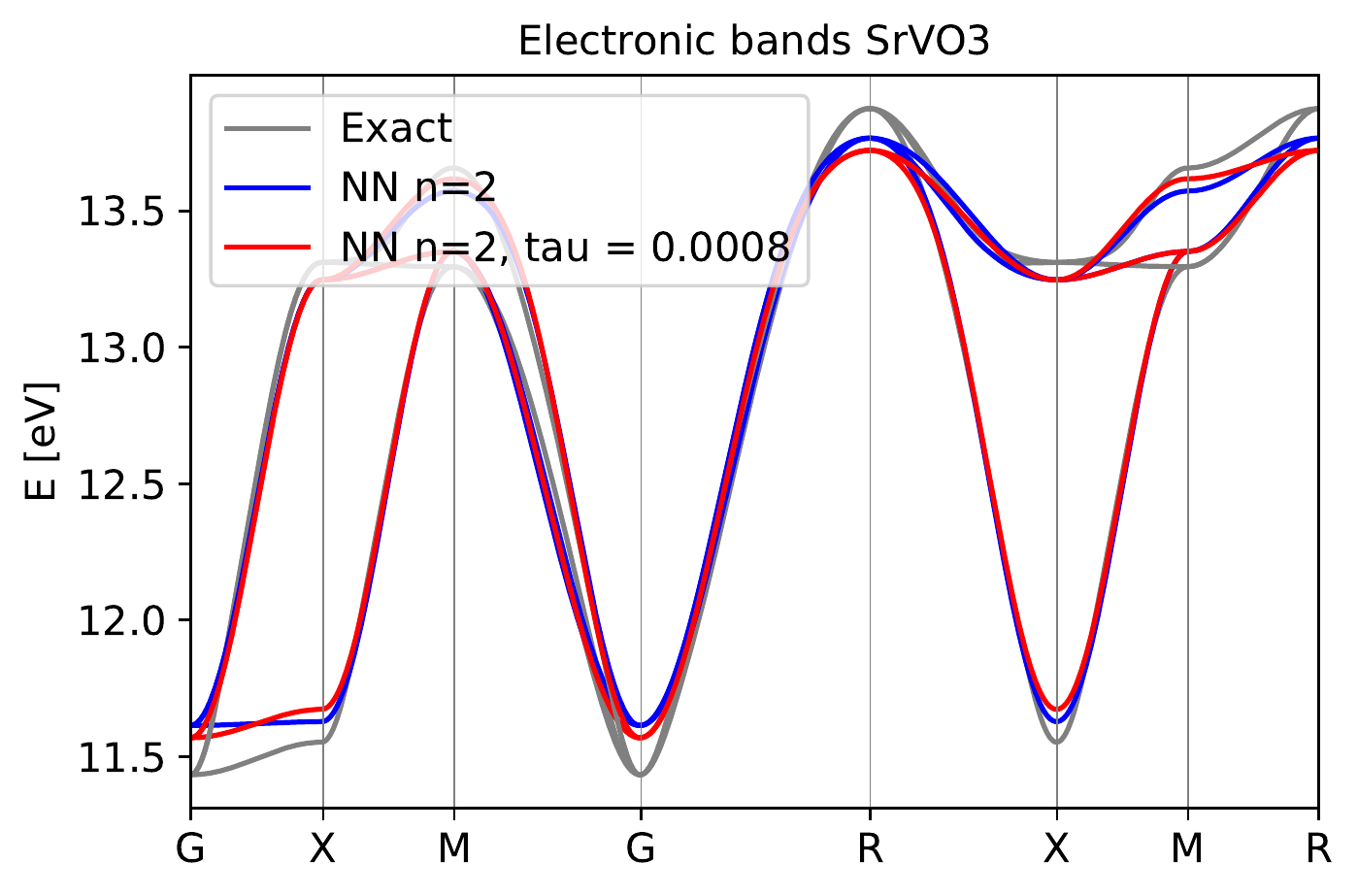} 
 \quad
 \includegraphics[width=0.48\textwidth]{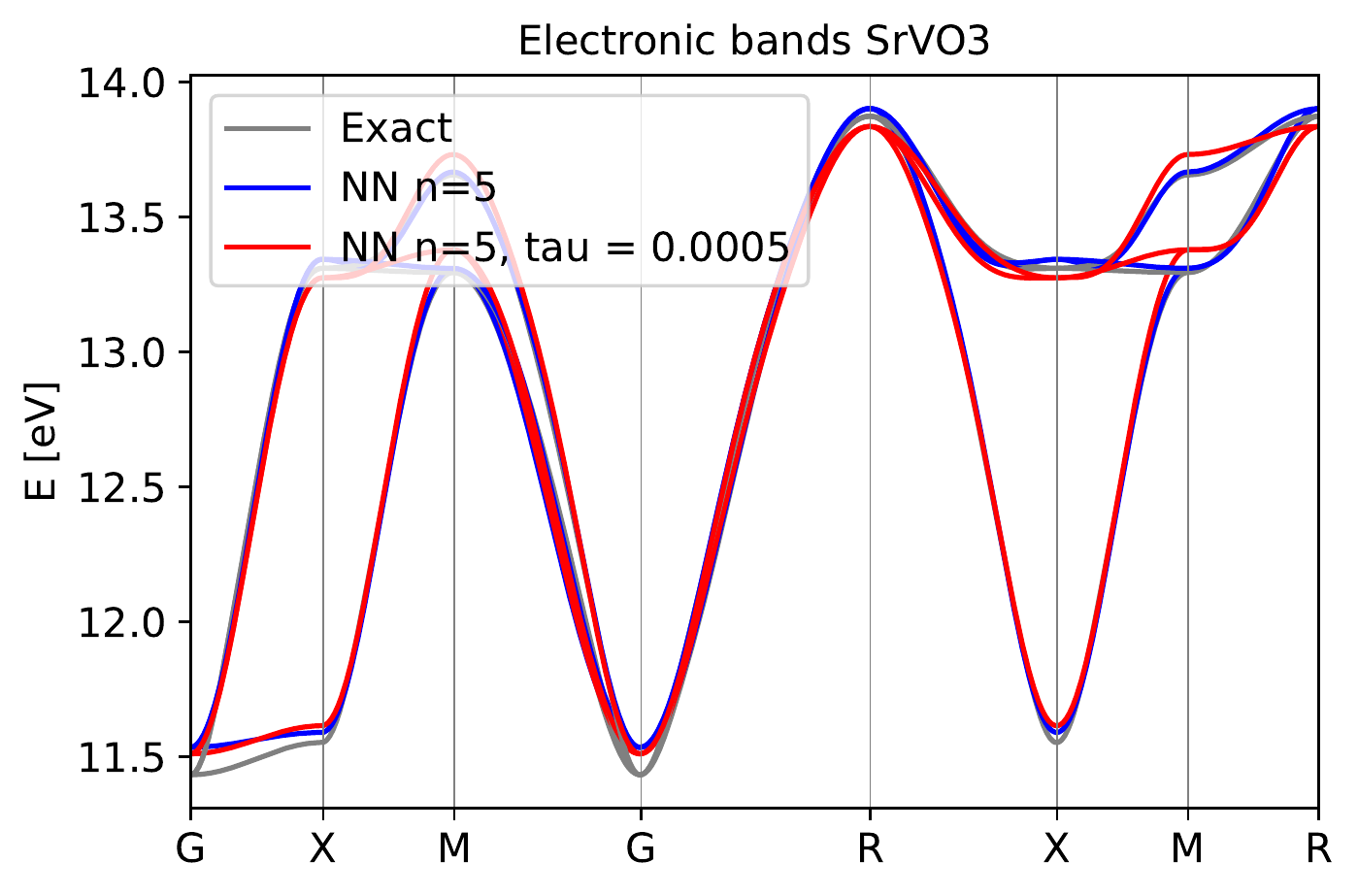} 
 \caption{Comparison between the exact electronic band structure of SrVO$_3$, $\varepsilon_i(\bm{k})$ (grey lines), the one obtained by truncating the hopping matrix $T(\bm{R})_{mn}$ to a motif of order $n_0$, $\varepsilon^{(n_0)}_i(\bm{k})$ (blue lines), and the one obtained from the filtered hopping matrix $\tilde{T}(\bm{R})_{mn}$, $\overline{\varepsilon}^{(n_0)}_i(\bm{k})$ (red lines), for (left panel) $n_0 = 2$ and (right panel) $n_0 = 5$. Here, $\tau_{0}$ is determined in a such a way that $t_0$ coincides with the largest absolute value of $T(\bm{R})_{mn}, \forall\bm{R}\notin\mathcal{N}^{n_0}_{\bm{0}}$.}
 \label{fig:srvo_tb_order}
\end{figure}

In the next step we determine the order $n_0$ of the motif to be used in the truncation of the hopping matrix according to the second assumption discussed above. We do that by setting, for a given $n_0$, $T(\bm{R})_{mn} = 0, \forall \bm{R}\notin\mathcal{N}^{n_0}_{\bm{0}}$ and comparing the band structure obtained from the truncated hopping matrix with the original one. Here, we recall that $\mathcal{N}^{n_0}_{\bm{0}}$ denotes the set of lattice vectors corresponding to those unit cells which are nearest neighbors of order $\leq n_0$ with respect to the central one. The band structure obtained from truncated hopping matrix is given by the eigenvalues of the matrices
\begin{equation}
    h(\bm{k})_{mn} = \sum_{\bm{R}\in \mathcal{N}^{n_0}_{\bm{0}}}e^{i\bm{k}\cdot\bm{R}}T(\bm{R})_{mn}
\end{equation}
for values of $\bm{k}$ along the high symmetry path in the Brillouin zone. For SrVO$_3$, the bands obtained from this approximation are shown in blue in \cref{fig:srvo_tb_order} for (a) $n_0=2$ and (b) $n_0 = 5$. To obtain a consistent approximation, we then
introduce the following filtered hopping matrix
\begin{equation}
    \overline{T}(\bm{R})_{mn}=\begin{cases}
    T(\bm{R})_{mn} & \text{if } |T(\bm{R})_{mn}| \geq t_0\\
    0 & \text{otherwise}
    \end{cases},
\end{equation}
where we set to zero all the coefficients $T(\bm{R})_{mn}$ smaller than a threshold $t_0 = \tau_0 \times \mathrm{max}|T(\bm{R})_{mn}|$, $\forall \bm{R}\notin\mathcal{N}^{n_0}_{\bm{0}}$. The latter is obtained from the largest of the absolute values of the hopping coefficients involving sites with nearest neighbour order $>n_0$ with respect to the central cell. For SrVO$_3$ this further step results in the red bands in \cref{fig:srvo_tb_order} which represent a good approximation of the truncated bands for both values of $n_0$. In this case, as shown in \cref{tab:srvo_HM_numbers}, the two approximations combined allow us to reduce the number of non-zero hopping matrix coefficients $T(\bm{R})_{mn}$ from $81$ to $33$ for $ n_0=2 $ and from $296$ to $93$ for $n_0=5$.

\begin{table}
    \centering
    \begin{tabular}{cccc}
    \toprule
    $n_{0} $ & All & Non-zero & $|T(\bm{R})_{mn}|>t_{0}$ \\
    \midrule
    1 & 63 & 21 & 15 \\ 
    2 & 171 & 81 & 33 \\ 
    5 & 513 & 296 & 93 \\
    \bottomrule
    \end{tabular}
    \caption{Number of total (second column), non-zero (third column), and filtered hopping matrix coefficients (fourth column) for SrVO$_3$ for different values of the motif order, $n_0 = 1, 2, 5$, respectively.}
    \label{tab:srvo_HM_numbers}
\end{table}

In order to determine $n_0$ in systematic way, we look for the minimum value of $n_0$ such that the distance between the exact bands $\varepsilon_{i}(\bm{k})$ and the ones obtained from the filtered truncated hopping matrix $\overline{T}(\bm{R})_{mn}$, $\overline{\varepsilon}^{(n_0)}(\bm{k})$, is smaller than a pre-determined tolerance. In our case, we choose to measure the distance between the bands according to~\cite{Garrity2021}

\begin{equation}
    \mathcal{D}(n_0) = \mathrm{max}_{i,\bm{k}}\left|\varepsilon_{i}(\bm{k}) - \overline{\varepsilon}^{(n_0)}_i(\bm{k})\right|,
\end{equation}
where $i$ is the band index and $\bm{k}$ is sampled from a regular grid on the Brillouin zone. In what follows, for each material we will use the smallest value of $n_0$ such that $\mathcal{D}(n_0) \leq 0.5\ \mathrm{eV}$.

\paragraph{Coulomb tensor.} To obtain the Coulomb tensor coefficients $\tilde{V}^{(\bm{0},\bm{R}_2,\bm{R}_3,\bm{R}_4)}_{slmn}$ one has to compute numerically the $6-$dimensional integral in \cref{eq:Wannier_CT_core_W} using the Wannier functions generated by Wannier90 for each value of the orbital and site indices. For a standard material this would result in an extremely large number of integrals and, consequently, long classical computational time. 
However, thanks to the localisation of the \glspl{mlwf}, only a few of these integrals are important in determining the properties of a material. Therefore, as stated in the second working assumption above, we only consider coefficients involving sites which are reciprocally nearest neighbors of order $\leq n_{\mathrm{int}}$ and set to zero all the others. Using the notation introduced above, this means that only the coefficients $\tilde{V}^{(\bm{R}_1,\bm{R}_2,\bm{R}_3,\bm{R}_4)}_{slmn}$ with $\bm{R}_1,\bm{R}_2,\bm{R}_3,\bm{R}_4 \in \mathcal{N}^{n_{\mathrm{int}}}_{\set{\bm{0},\bm{R}_1,\bm{R}_2,\bm{R}_3,\bm{R}_4}}$ are non-zero. As described in details in \cref{app:sec:CT_pipeline_CS}, the symmetry properties of the Coulomb tensor given in \cref{eq:CT_equivalent_configs} make it possible to identify a minimal set of unique site configurations from which all of the other can be reconstructed at a later stage. For instance, since $\tilde{V}^{(\bm{0}, \bm{0}, \bm{R}, \bm{0})}_{slmn} = \tilde{V}^{(\bm{0}, \bm{0}, \bm{0}, \bm{R})}_{lsnm}$ it is sufficient to compute only coefficients with site structure $(\bm{0}, \bm{0}, \bm{0}, \bm{R})$. For each unique site configuration, exploiting again \cref{eq:CT_equivalent_configs}, we then determine all the non-equivalent orbital configurations required to obtain the full Coulomb tensor. As an example, for a coefficient with site structure $(\bm{0}, \bm{0}, \bm{R}, \bm{R})$ the orbital configurations $(s,l,m,n)$ and $(l,s,n,m)$ are equivalent and, therefore, $\tilde{V}^{(\bm{0}, \bm{0}, \bm{R}, \bm{R})}_{slmn} = \tilde{V}^{(\bm{0}, \bm{0}, \bm{R}, \bm{R})}_{lsnm}$. Hence, for a given site configuration, we need to compute the Coulomb tensor coefficients only for a restricted set of orbital configurations. This procedure, which is based on fundamental symmetry properties of the Coulomb tensor, allows us to significantly reduce the number of integrals to be computed. For SrVO$_3$, the latter goes from $5913$ to $831$ for $n_{\mathrm{int}}=1$ and from $101169$ to $3180$ for $n_{\mathrm{int}}=2$. See the ``Unique'' column in \cref{tab:srvo_CT_numbers}.

\begin{figure}[ht]
 \centering
 \includegraphics[width=0.48\textwidth]{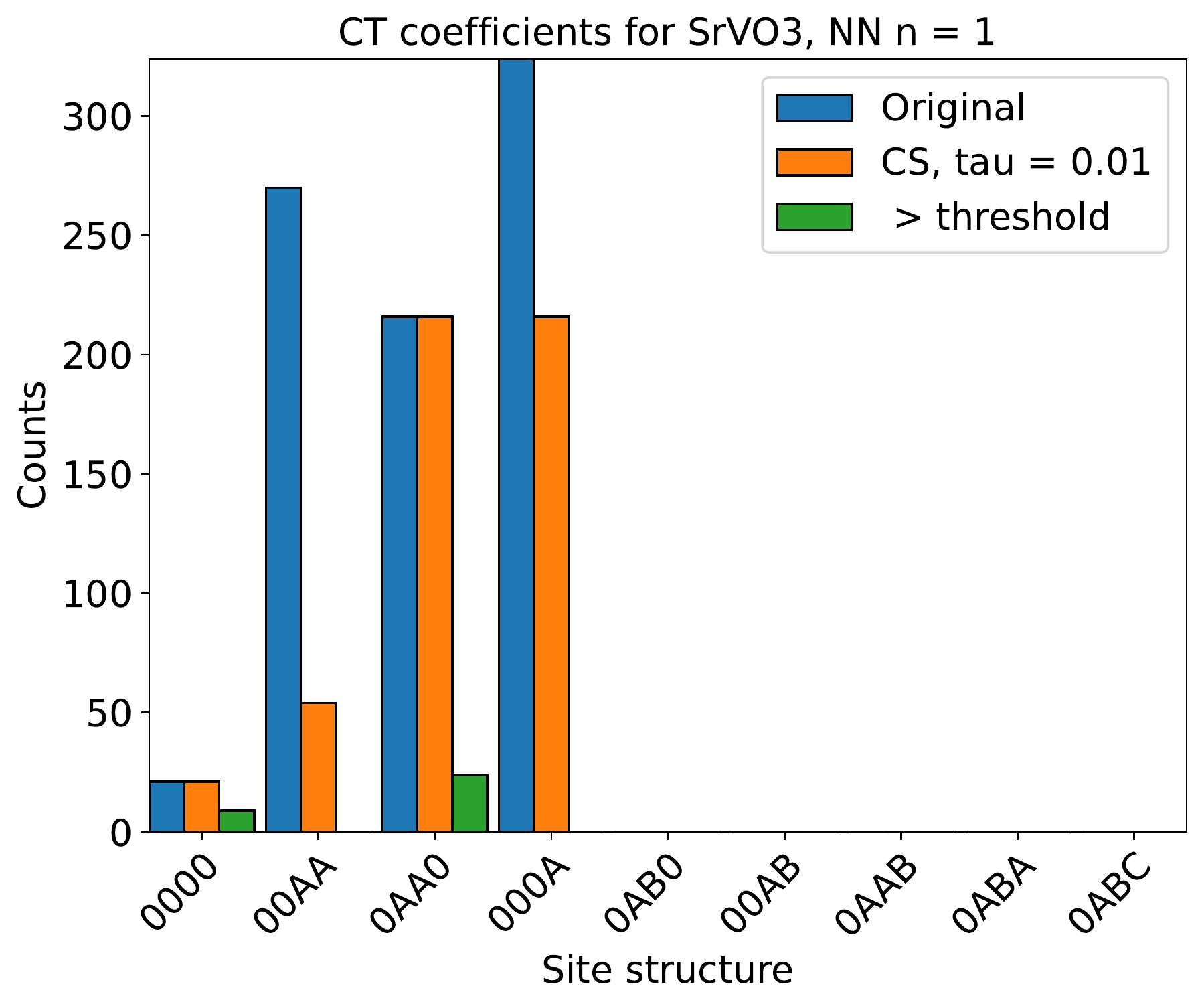} 
 \quad
 \includegraphics[width=0.48\textwidth]{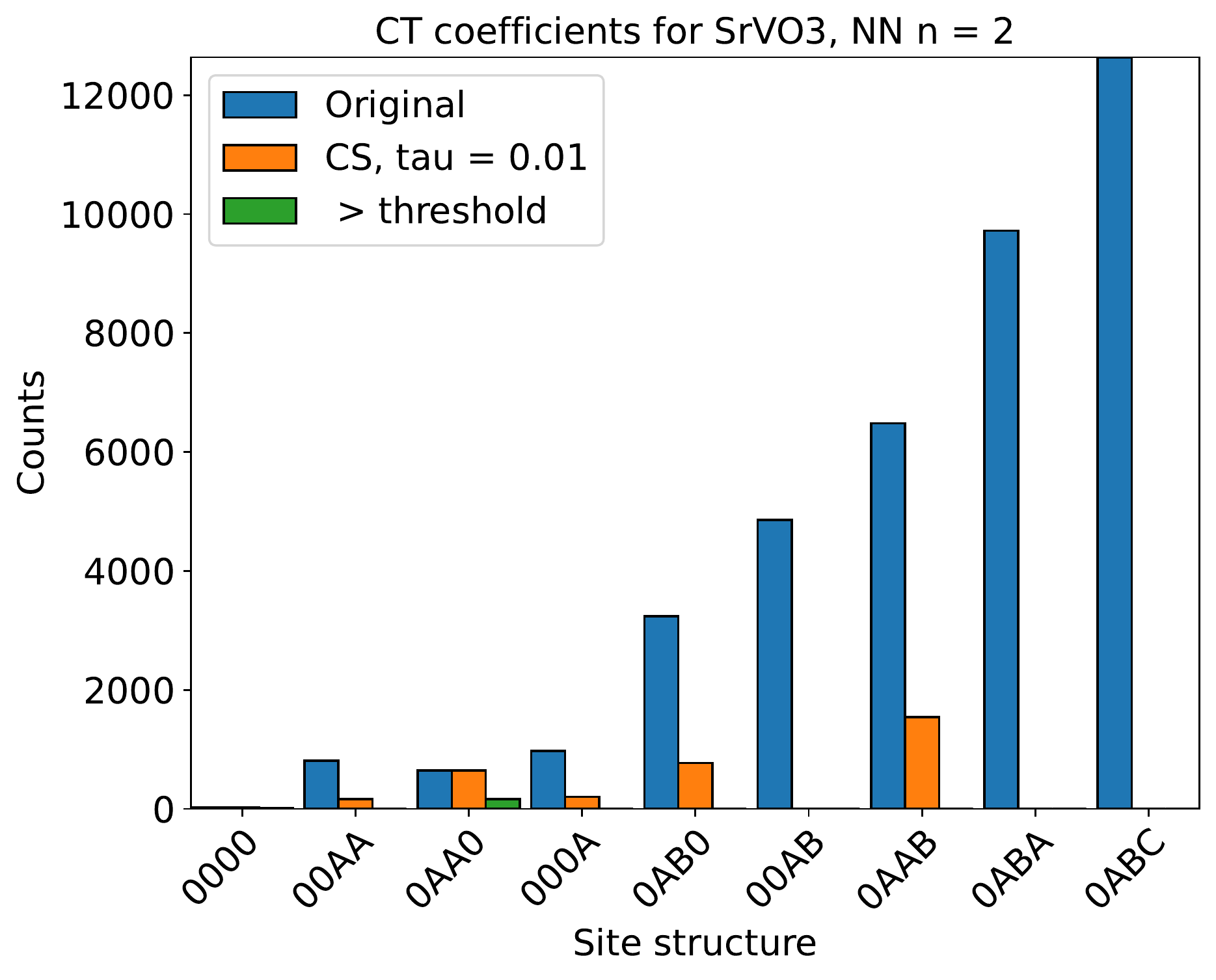} 
 \caption{
 Number of unique Coulomb tensor coefficients with site structure $ABCD$ (i.e., $\tilde{V}^{(\bm{R}_A,\bm{R}_B, \bm{R}_C, \bm{R}_D)}_{slmn}$ with $s,l,m,n$ being unique orbital configurations of the site structure $(\bm{R}_A,\bm{R}_B, \bm{R}_C, \bm{R}_D)$) to be computed without (blue) and with (orange) the Cauchy-Schwarz inequality for (left panel) $n_{\mathrm{int}} = 1$ and (right panel) $n_{\mathrm{int}} = 2$. In both panels, the green bars denote the number of coefficients per site structure whose actual absolute value is greater than the threshold, with  $\tau_{\mathrm{int}} = 0.01$.
 }
 \label{fig:srvo_CS_CT_coefficients}
\end{figure}

As stated above, we also assume that the physical properties of the system are determined by those Coulomb tensor coefficients whose absolute value is larger than a given threshold. We fix the latter as $t_{\mathrm{int}} = \tau_{\mathrm{int}} \times \mathrm{max}|\tilde{V}^{(\bm{0}, \bm{R}_2, \bm{R}_3, \bm{R}_4)}_{slmn}|$, with $\bm{R}_2, \bm{R}_3, \bm{R}_4 \in \mathcal{N}^{n_{\mathrm{int}}}_{\set{\bm{0}, \bm{R}_2, \bm{R}_3, \bm{R}_4}}$ and $s,l,m,n \in \set{1,...,M}$ and define the filtered Coulomb tensor as
\begin{equation}
\label{eq:Coulomb_tensor_filtering}
    \overline{V}^{(\bm{0}, \bm{R}_2, \bm{R}_3, \bm{R}_4)}_{slmn} = \begin{cases}
    \tilde{V}^{(\bm{0}, \bm{R}_2, \bm{R}_3, \bm{R}_4)}_{slmn} & \text{if } |\tilde{V}^{(\bm{0}, \bm{R}_2, \bm{R}_3, \bm{R}_4)}_{slmn}| \geq t_{\mathrm{int}}\\
    0 & \text{otherwise}
    \end{cases}.
\end{equation}
This approximation can be conveniently combined with the Cauchy-Schwarz inequality introduced in 
\cref{sec:CS_ineq} to further reduce the number of integrals which one needs to compute. Indeed, \cref{eq:CT_Cauchy-Schwarz_1} implies that 
\begin{equation}
    \label{eq:CS_pipeline}
    |\overline{V}^{(\bm{R}_1, \bm{R}_2, \bm{R}_3, \bm{R}_4)}_{slmn}|^2\leq \overline{V}^{(\bm{R}_1, \bm{R}_1, \bm{R}_4, \bm{R}_4)}_{ssnn} \overline{V}^{(\bm{R}_3, \bm{R}_3, \bm{R}_2, \bm{R}_2)}_{mmll}.
\end{equation}
Hence, thanks to the discrete translational invariance, if all the coefficients with site structure $(\bm{0},\bm{0},\bm{R},\bm{R})$ (with $\bm{R}\in\mathcal{N}^{n_0}_{\bm{0}}
$) are known one can a priori exclude from the computation all those coefficients with site structure $(\bm{R}_1, \bm{R}_2, \bm{R}_3, \bm{R}_4)$ and orbital configuration $(s,l,m,n)$ such that
\begin{equation*}
    \overline{V}^{(\bm{R}_1, \bm{R}_1, \bm{R}_4, \bm{R}_4)}_{ssnn} \overline{V}^{(\bm{R}_3, \bm{R}_3, \bm{R}_2, \bm{R}_2)}_{mmll} < t_{\mathrm{int}},
\end{equation*}
since \cref{eq:CS_pipeline} guarantees that they are going to be smaller than the threshold. The advantages deriving by using this procedure are shown in \cref{fig:srvo_CS_CT_coefficients} for SrVO$_3$ with $\tau_{\mathrm{int}} = 0.01$, $n_{\mathrm{int}} = 1$ and $n_{\mathrm{int}} = 2$. In this case, the Cauchy-Schwarz inequality allows us to further reduce the number of integrals to be computed from a total of $831$ to $444$ ($60$ of which will turn to have absolute value greater than the threshold $t_{\mathrm{int}}$ after evaluation) for $n_{\mathrm{int}} = 1$ and  from $39387$ to $3180$ ($168$ of which have absolute value greater than the threshold $t_{\mathrm{int}}$) for $n_{\mathrm{int}} = 2$. See the ``Unique+Cauchy-Schwarz'' and the last columns in \cref{tab:srvo_CT_numbers}.

Finally, to obtain a consistent approximation for the Coulomb tensor, we perform an additional filtering step: We focus on the coefficients of the tensor with $n_{\mathrm{int}} = 2$ not included in the tensor with $n_{\mathrm{int}} = 1$ (i.e., coefficients $\overline{V}^{(\bm{R}_1, \bm{R}_2, \bm{R}_3, \bm{R}_4)}_{slmn}$ such that at least one of $\bm{R}_1, \bm{R}_2, \bm{R}_3, \bm{R}_4$ is not in $ \mathcal{N}^{n_{\mathrm{int}}}_{\set{\bm{0}, \bm{R}_1, \bm{R}_2, \bm{R}_3, \bm{R}_4}} $) and we look for the one with the largest absolute value; then, we use the latter as a new threshold $t_{\mathrm{int}}'$ to filter once more the Coulomb tensor with $n_{\mathrm{int}} = 1$ as we did in \cref{eq:Coulomb_tensor_filtering}. 

\begin{table}
\centering
\begin{tabular}{ccccc}
\toprule
$n_{\mathrm{int}}$ & All  & Unique & Unique+Cauchy-Schwarz & Unique  $ |\tilde{V}^{(\bm{R}_1,\bm{R}_2, \bm{R}_3, \bm{R}_4)}_{slmn}| > t_{\mathrm{int}}$ \\
\midrule
1 & 5913 & 831 & 444 & 60\\ 
2 & 101169 & 39387 & 3180 & 168 \\
\bottomrule
\end{tabular}
\caption{
Number of the Coulomb tensor coefficients to be computed for SrVO$_3$ using the various procedures described in the main text for $n_{\mathrm{int}} = 1 $ and $n_{\mathrm{int}} = 2 $: total number of coefficients (second column), number of unique coefficients (third column), number of unique coefficients after employing the Cauchy-Schwarz inequality (fourth column), and number of unique coefficients above the threshold (fifth column).
}
\label{tab:srvo_CT_numbers}
\end{table}

\paragraph{Single-index Hamiltonian} From the hopping matrix and the Coulomb tensor evaluated as described above it is straightforward to construct the motif Hamiltonian in the Wannier basis (see \cref{Ham_wannier}). To proceed, it is convenient to make the (trivial) spin structure of the coefficients explicit and to build a completely anti-symmetric Hamiltonian. This can be done by introducing the following spinful Hopping matrix and Coulomb tensor coefficients,
\begin{equation}
    \overline{T}_{(\bm{R}_1,m,\sigma_1),(\bm{R}_2, n, \sigma_2)} = 
    \begin{cases} 
    \overline{T}(\bm{R}_1-\bm{R}_2)_{mn} & \text{if } \sigma_1 = \sigma_2\\
    0 & \text{otherwise}
    \end{cases}
\end{equation}
and 
\begin{equation}
    \overline{V}_{(\bm{R}_1,s,\sigma_1),(\bm{R}_2,l,\sigma_2),(\bm{R}_3,m,\sigma_3),(\bm{R}_4,n,\sigma_4)} = 
    \begin{cases}
    \overline{V}^{s,(\bm{R}_1,\bm{R}_2,\bm{R}_3,\bm{R}_4)}_{slmn} & \text{if } \sigma_1 = \sigma_2 = \sigma_3 = \sigma_4\\
    \frac{1}{2}\overline{V}^{(\bm{R}_1,\bm{R}_2,\bm{R}_3,\bm{R}_4)}_{slmn} & \text{if } \sigma_1 = \sigma_3 \text{ and } \sigma_2 = \sigma_4\ (\sigma_1 \neq \sigma_2)\\
    -\frac{1}{2}\overline{V}^{(\bm{R}_2,\bm{R}_1,\bm{R}_3,\bm{R}_4)}_{lsmn} & \text{if } \sigma_1 = \sigma_4 \text{ and } \sigma_2 = \sigma_3\ (\sigma_1 \neq \sigma_2) \\
    0 & \text{otherwise}
    \end{cases},
\end{equation}
respectively. Here, $ \overline{V}^{s,(\bm{R}_1,\bm{R}_2,\bm{R}_3,\bm{R}_4)}_{slmn} = \big[\overline{V}^{(\bm{R}_1,\bm{R}_2,\bm{R}_3,\bm{R}_4)}_{slmn}-\overline{V}^{(\bm{R}_2,\bm{R}_1,\bm{R}_3,\bm{R}_4)}_{lsmn}\big]/2$. Each site-mode-spin triplet $(\bm{R}_i, i, \sigma_i)$ is then mapped to a single-index $\alpha_i\in\{1,...,2M\}$, with $M$ the number of complex fermion modes per motif, so that the motif Hamiltonian can be written as
\begin{equation}
    H = \sum_{\alpha,\beta} \overline{T}_{\alpha\beta} w^\dagger_\alpha w_\beta + \sum_{\alpha,\beta,\gamma,\delta} \overline{V}_{\alpha\beta\gamma\delta} w^\dagger_\alpha w^\dagger_\beta w_\gamma w_\delta.
\end{equation}

Finally, as discussed in \cref{sec:Wannier_basis_mapping}, we introduce the Majorana basis operators as $w_\alpha = ( \gamma_{\alpha}+i\bar{\gamma}_{\alpha})/2$ and $w^\dagger_\alpha = ( \gamma_{\alpha}-i\bar{\gamma}_{\alpha})/2$, in terms of which the full motif  Hamiltonian reads
\begin{equation}
    \label{eq:H_Majorana}
    H_M = \sum_{k \in \{0,1\}^{2M}} \alpha_k \prod_j \gamma_j^{k_{2j}} \bar{\gamma}_j^{k_{2j+1}} \;,\; |k| \in \{2,4\}.
\end{equation}
This can then be used as an input for the \gls{vqe} and \gls{tds} algorithm described in \cref{sec:vqe_alg} and \cref{sec:compiler}. As discussed in \cref{sec:compile_unit_cell}, \cref{eq:H_Majorana} can be used to tile a system of any size, without increasing the depth of the layer of quantum gates that implements these interactions.

\subsection{Circuit analysis}\label{sec:circuit_analysis}
In this section we apply the circuit compilation techniques described throughout this manuscript to a series of materials, to assess the feasibility of simulating those materials on quantum hardware. 
We begin in \cref{sec:circ_srv} by analysing \gls{srv} in detail, since we have described it completely in \cref{sec:full_an_SRVO3}, followed by four other materials in \cref{sec:res_materials}. 
Finally, we analyse the number of measurement rounds that would be necessary to approximate the observables of interest for each material, under the set of measurement strategies described in \cref{sec:measurements}.

\subsubsection{Strontium vanadate}\label{sec:circ_srv}
Having the motif Hamiltonian of SrVO$_3$ in Majorana form as above, we map it to Pauli operators following the hybrid encoding introduced in \cref{sec:qubit_rep}. To do this, we first recall that the hybrid encoding acts on the sites of a Cartesian grid. Thus, we have to consider the Cartesian motif introduced in \cref{sec:motif}. 
 
\begin{figure}
    \centering
    \includegraphics[scale=1]{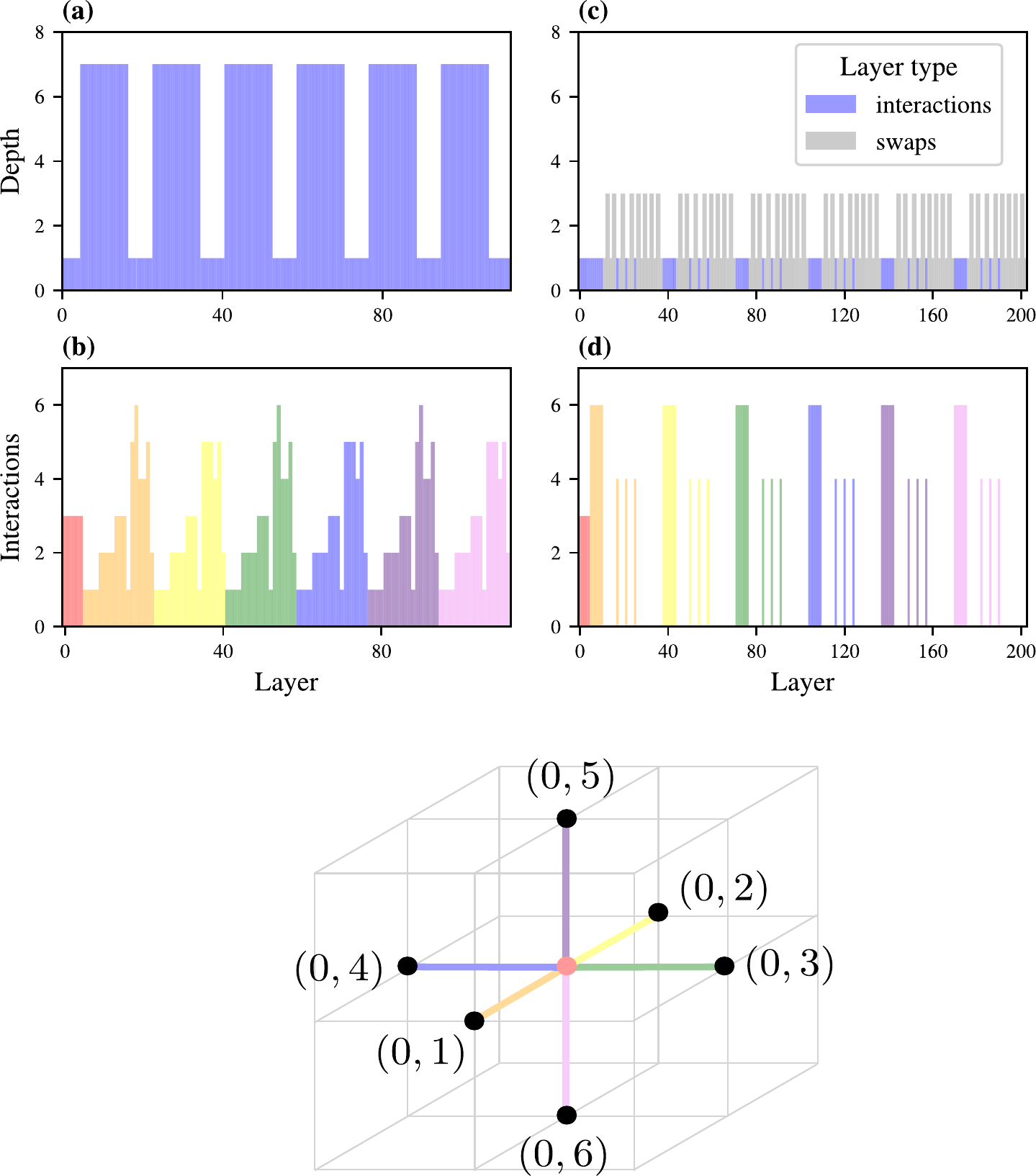}
    \caption{
        Circuit breakdown for \gls{srv} for onsite and nearest neighbour terms. Top:
        \textbf{(a)} Circuit depth cost per layer in an approach without fswaps. 
        \textbf{(b)} The number of interactions implemented in each layer. 
        \textbf{(c)} Circuit depth cost per layer using fswaps.
        \textbf{(d)} Breakdown of the number of interactions per layer using fswaps. Here, layers with zero interactions implement fswaps.
        The colour code in \textbf{(b)} and \textbf{(d)} is the same and denotes the pair of sites $(0,x)$ involved in an interactions, as shown in the bottom diagram, with $x$ increasing to the right.
        Bottom: Mapping of the \gls{srv} motif of order $n=1$ (consisting of the cells highlighted by the pink and black dots) to the $3\times3\times3$ Cartesian motif (compare with \cref{fig:motif_srvo}). Each coloured line denote the real-space interaction $(0,x)$ between the central cell (labeled by $0$ in the Cartesian motif) and the $x-$th neighbouring cell. 
        The compiler groups terms by their sites in advance, so the final circuit layering will have sections dedicated to each site grouping. 
    }
    \label{fig:srvo3_analysis}
\end{figure}

The latter contains $N_C$ site, of which $N_D = N_C -  N_{\mathrm{cells/motif}} $ are not contained in the material real-space motif. This results in $M_D = N_D N_{\mathrm{modes/cell}} $ additional complex fermion modes, with $N_{\mathrm{modes/cell}} = 2N_{\mathrm{orbitals/cell}}$ for a total of $M_C = M + M_D$ modes. The latter do not enter the motif Hamiltonian and, therefore, there are no interactions involving such modes. However, they must be kept in order to apply the tiling algorithm described in \cref{sec:compile_unit_cell}, introducing a qubit overhead of $2M_D$. In particular, the mapping of the order $n=1$ motif of SrVO$_3$ (consisting of the black and red sites in \cref{fig:motif_srvo}) to the corresponding $3\times3\times3$ Cartesian motif is shown in the bottom panel of \cref{fig:srvo3_analysis}. 

Each site of the Cartesian motif is labeled by a single site-index $x\in \set{0,...,N_C-1}$, with $x=0$ denoting the central cell, $x\in \set{1,...,6}$ labeling the 6 nearest neighbours and $x\in \set{7,...,26}$ corresponding to the $N_D = 27-7 = 20$ additional sites forming the Cartesian motif. Since SrVO$_3$ has a cubic lattice, the nearest-neighbour order of the sites is the same in the real-space and Cartesian motif, i.e., nearest neighbours of order $n$ in real space are mapped onto nearest neighbours of order $n$ in the Cartesian grid.

The $2M_C$ Majorana modes of the Cartesian motif are then encoded into qubits via the hybrid encoding discussed in \cref{sec:qubit_rep}. The total number of qubits required to simulate the Cartesian motif is $N_Q = 2M_C + N_F$, with $N_F$ being the number of auxiliary face qubits.
After that the circuit is compiled with respect to the algorithm of interest, e.g., \gls{vqe}. 
As outlined in \cref{sec:compiler}, a series of compilation routines are applied to the set of interactions to derive the final circuit, including the fswap network described in \cref{sec:swap_network_details}. As explained there, it is important to separate the interactions based on the sites of the real-space motif upon which they act. This can be seen in the diagram at the bottom of \cref{fig:srvo3_analysis}, where we show the location on the Cartesian motif of the real-space interaction terms involving the central cell located at $0$ and the nearest neighbour one at $x$ (denoted by the pair $(0,x)$), for $x\in \set{1,...,6}$.

In \cref{fig:srvo3_analysis}\textbf{(a)} we show the circuit depth breakdown (for onsite and nearest-neighbour terms for simplicity) by layer without including fswaps, similar to the analysis in \cref{sec:model_ham_eg}. 
In \cref{fig:srvo3_analysis}\textbf{(b)} we can see how the circuit layers are distributed among terms acting on various sites. 
In this material, all interactions involve at most two sites: the central site and one neighbour.
In contrast, including fswaps, we see in \cref{fig:srvo3_analysis}\textbf{(c)} that the depth of each layer reduces, but we have more layers.
In this case, the majority of circuit depth is spent on implementing fswap operations, also listed in \cref{tab:srvo3_data}. 
In general, it is difficult to know a priori whether the fswap network will be beneficial to the depth. In our case, we run the circuit compiler both with and without the fswap network to determine the shortest circuit which should simulate the same process. 

In the same way we can employ the circuit compiler to evaluate different algorithms; in particular, we consider \gls{vqe} and \gls{tds}, and compile both with and without the aid of an fswap network. This is reported in \cref{tab:srvo3_data}. 
We see that the swap network yields lower circuit depths in each case, as elucidated in \cref{fig:srvo3_analysis}. Overall the shortest available circuit corresponds to \gls{vqe} including an fswap network: for the remainder of this section we will consider \gls{vqe} in particular, but it is straightforward to run the alternative compilations, which are listed in \cref{sec:apdx_materials}.

\begin{table}
    \centering
    \input{Figs_Results/materials_results/srvo3_only}
    \caption{
        Circuit depth analysis for \gls{srv}, compiling \gls{vqe} and \gls{tds}.
        Each algorithm is compiled with and without the use of fswap networks (asterisks denote the fswap network compilation is omitted).
        We report the breakdown of the circuit depth into the implementation of interactions and swaps.
        Interactions and depths are reported separately for (i) onsite and nearest-neighbour (NN) terms, which can be tiled generically across the encoding and therefore run in parallel, and (ii) next-nearest-neighbour and beyond (denoted NNN+) terms, which can not be tiled on the encoding and therefore must be costed explicitly. 
        Properties of \gls{srv} are given in \cref{tab:mat_props}, and the number of terms of each Hamiltonian type are listed in \cref{tab:n_terms_by_type}.
    }
    \label{tab:srvo3_data}
\end{table}

\subsubsection{Trotter error}\label{sec:trotter_err}


In the case of \gls{tds} we may also compute an estimate of the Trotter error. This is done using an upper bound estimate given by Eq.\ (26) in  \cite{clinton2021hamiltonian},
\begin{equation}\label{clinton}
\epsilon \leq C_1 \frac{T \delta^p}{(p+1)!} + C_2 \frac{T}{\delta} I(N)
\end{equation}
with
\begin{align}
C_1&:=n p B_p^2 \Lambda^{p-1}N [M H_p-B_p + B_p(N/\Lambda)]^{p-1}\\
C_2&:=n B_p^2 (M H_p \Lambda)^p N[(S_pM)^2-S_pM].
\end{align}
Here, $T$ is the target time; $p$ is the Trotter order (currently the compiler is only designed to handle $p=1$); $N$ is the largest number of terms in a given mutually commuting grouping; $M$ is the number of mutually commuting groupings; $\Lambda$ is a bound on the largest norm of any mutually commuting grouping; $\delta$ is the Trotter step size; $n$ is the maximum number of terms from a single grouping that do not commute with a given term in another grouping, and $B_p$, $H_p$, $S_p$, and $I$ are defined in the original work. This equation may be inverted to compute the required circuit depth of a \gls{tds} simulation for a given target time $T$ and Trotter error $\epsilon$.

It is important to emphasize here that the bounds on the Trotter error will depend on the order in which the terms in the Hamiltonian are executed. However the order of execution will depend on the choices taken by the compiler. And the quality of the Trotter error plays a role in the multi-layer circuit depth of executing time dynamics for some fixed time. As such one may wish to prioritize minimizing Trotter error over minimizing the  circuit depth of a single layer by modifying the execution of the compiler. One concrete way this can be done is by decomposing into layers of mutually commuting terms first before proceeding with executing any swap network protocols.

We include here a comparative analysis of two compiler strategies to examine their relative efficacy in reducing overall circuit depth of TDS. One strategy prioritizes the circuit depth of a single layer by decomposing into commuting layers after performing swap networks, and the other prioritizes Trotter error by decomposing into commuting layers before performing swap networks.  In this case we fix the Trotter order $p=1$. Given a single layer circuit produced by the compiler using a particular compiler strategy, and given a fixed Trotter error, we numerically invert equation (\ref{clinton}) to compute the requisite Trotter step size $\delta$. For order 1 Trotter formulae, the total circuit depth to evolve for time $T$ is given by $DT/\delta$ where $D$ is the depth of a single layer. 
Thus we can compare the quality of different compiler choices by considering the Trotter ratio $D/\delta$. 
This is shown in \cref{fig:trotter_ratio}. 
It is important to emphasize that these values do not indicate the actual circuit depth of a practical time dynamics simulation, since the choice of evolution time $T$ will depend on the details of the algorithm as well as the spectral norm of the Hamiltonian, which in our analysis has not been normalized. 

\begin{figure}%
\centering
\begin{minipage}{0.45 \linewidth}%
\includegraphics[scale=0.55]{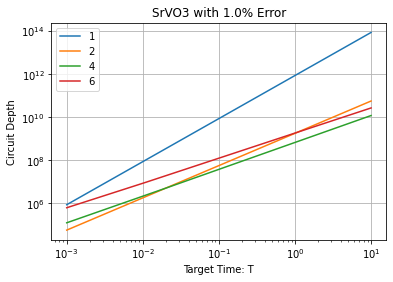}
\end{minipage}%
\hfill
\begin{minipage}{0.45 \linewidth}%
\includegraphics[scale=0.55]{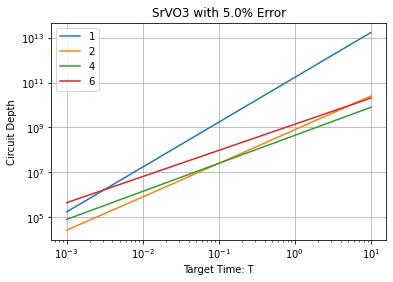}
\end{minipage}%
\vspace{1em} 
\begin{minipage}{0.45 \linewidth}%
\includegraphics[scale=0.55]{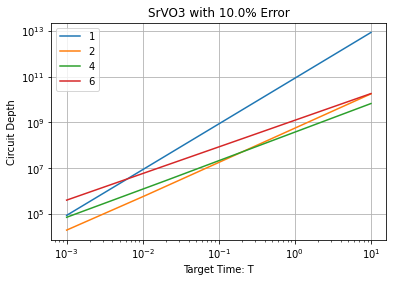}
\end{minipage}%
\caption{A comparison of the effect of Trotter order on the total depth of a time dynamics simulation of SrVO$_3$ for a range of target simulation times and a range of target Trotter errors, using the Trotter layer circuit compiled by prioritizing Trotter error. These figures are meant to illustrate relative performance of different Trotter orders, and do not necessarily correspond to the timescales which may be desired in experiments. }%
\label{fig:trotter_analysis}%
\end{figure}

\begin{figure}
    \centering
    \includegraphics{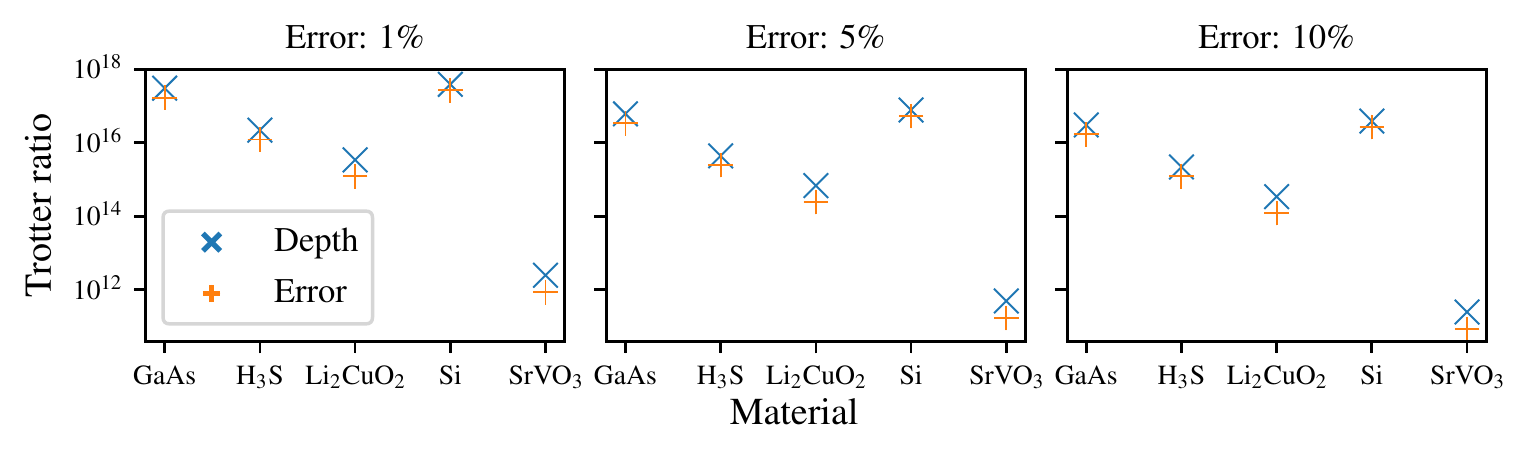}
    \caption{
        Trotter ratio for each material, showing the outcomes of the circuit compiler when alternative decomposition strategies are employed, namely which prioritise the circuit depth or Trotter error of a single layer. 
        Trotter ratio is defined in the text as $D/\delta$, where $D$ is the circuit depth computed by the compiler for a single layer and $\delta$ is the Trotter step size. These figures are meant to illustrate relative performance of different compiler routines, and not as indicators of expected circuit depth. Furthermore this analysis only holds for Trotter order $p=1$, which does not yield optimal Trotter error rates. For alternative analyses which consider $p>1$ see \cref{fig:trotter_analysis}.
    }
    \label{fig:trotter_ratio}
\end{figure}

Additionally we include here a comparison of how different Trotter orders $p$ influence total circuit depth for SrVO$_3$ -- see  \cref{fig:trotter_analysis}. This is calculated by upper bounding the circuit depth using the first order Trotter layers computed by the compiler. It is important to emphasize that these plots are meant to illustrate the effect of different Trotter orders, and are not necessarily prescriptive of a likely range of useful target times. This is because the particular target time may depend on the application, and additionally the Hamiltonians are neither normalized, nor is an appropriate characteristic time scale for the material known ahead of time.

\subsubsection{Further materials}
\label{sec:res_materials}
By focusing on \gls{vqe} and including an fswap network in each case, we can follow the same procedure outlined for \gls{srv} in \cref{sec:circ_srv} to construct circuits for the simulation of a number of materials, summarised in \cref{tab:materials_analysed}. We choose this as a representative sample of materials across a minimal but technologically relevant chemical space, spanning from light materials such as hydrogen and lithium to well known correlated ions like copper. 
Again we see that the circuit depth is dominated by the cost of implementing NNN+ terms, owing to the explicit inclusion of all terms in the lattice outlined in \cref{sec:compiling_nnn}, while onsite/NN terms are relatively cheap because they can be performed in parallel (\cref{sec:compile_unit_cell}).
In particular, the larger circuit depth observed for the superconductor H$_3$S relative to the other materials observed is due to the large bandwidth of states considered in the active space, which spans $\approx 40$ eV, roughly four times as large as of the other materials under consideration. This requires many more NNN+ terms to reproduce the Kohn-Sham electronic bandstructure within the considered tolerance.
These resource requirements are orders of magnitude lower than previous efforts, and can be improved further by (i) reducing the number of NNN+ terms through effective screening; (ii) invoking a translationally invariant fermionic encoding that integrates fully the connectivity structure of the Hamiltonian, beyond nearest neighbours such that the NNN+ terms are generically tilable in the same way as the onsite/NN terms.

\begin{table}
\centering
\input{Figs_Results/materials_results/vqe_results}
\caption{
    Circuit resources required for a representative set of materials. 
    We list the number of qubits and circuit depth required for a single layer of \gls{vqe}.
    The number of qubits reported is a function of both the number of modes of the Cartesian motif and of the number auxiliary face qubits introduced by the hybrid encoding.
    We report the breakdown of the circuit depth into the depth spent implementing interactions/swaps by whether the terms are onsite/NN or NNN+. 
    The circuit depths reported refer to a single layer of \gls{vqe}, without accounting for state preparation. 
    Properties of the listed materials are given in \cref{tab:mat_props}, and the number of terms of each Hamiltonian type are listed in \cref{tab:n_terms_by_type}.
}
\label{tab:materials_analysed}
\end{table}
\subsubsection{Measurement rounds}
\begin{figure}
    \centering
    \includegraphics{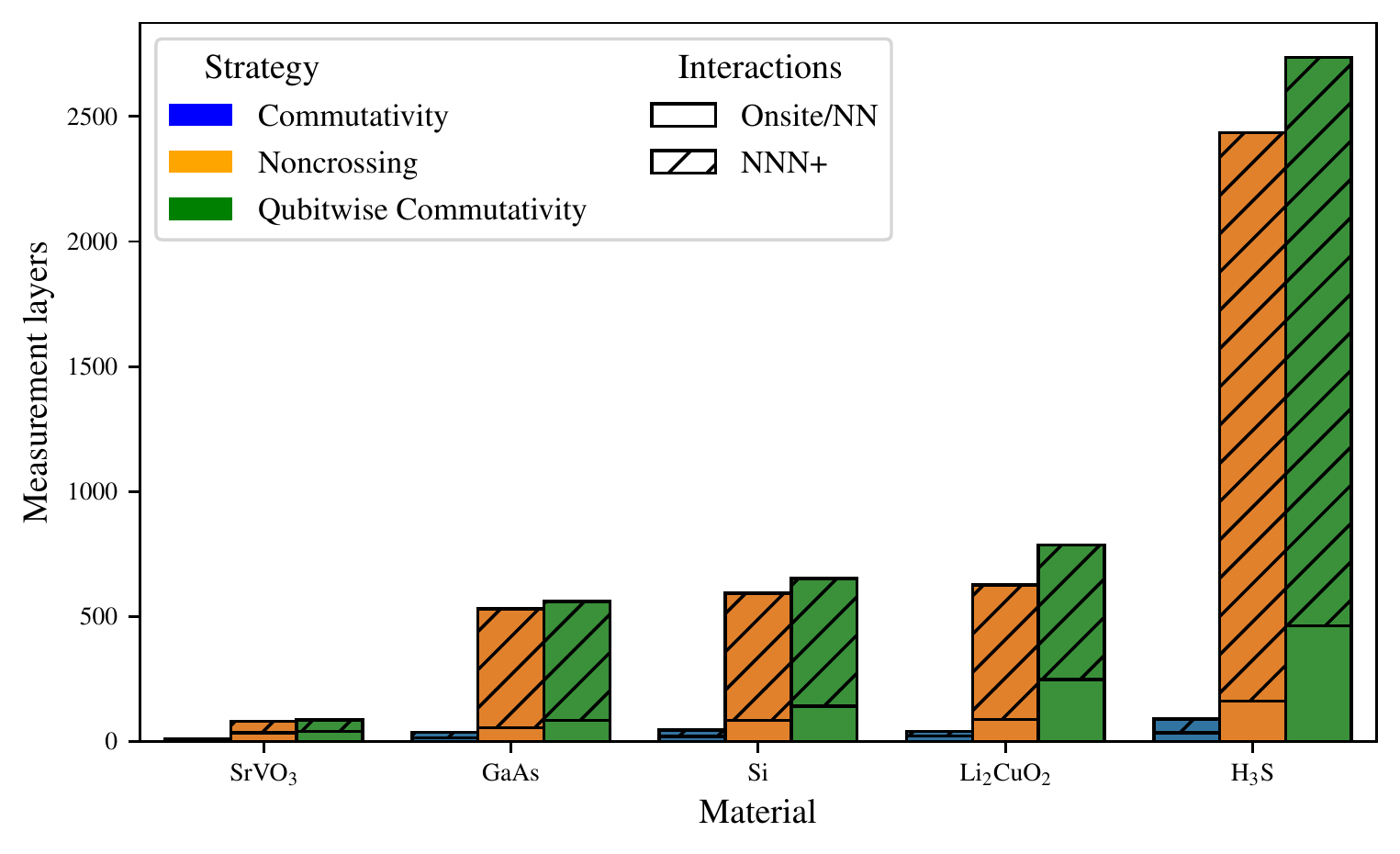}
    \caption{
    Measurement layers: the number of different measurements required in order to measure the terms included in the Hamiltonian. 
    Various measurement strategies are presented, split by the types of Hamiltonian terms involved in each measurement layer. The noncrossing case always measures NNN+ terms using a QWC strategy.
    Properties of the listed materials are given in \cref{tab:mat_props}.
    These data are listed explicitly in \cref{tab:msmt_layers}. 
    }
    \label{fig:msmt_layers}
\end{figure}

We analyse the number of measurement layers required for each of the studied materials using each of the measurement strategies described in \cref{sec:measurements}, again breaking into the subset required for onsite/NN terms and NNN+ terms in \cref{fig:msmt_layers} (and listed in \cref{tab:msmt_layers}).
There are three key messages: 
\begin{enumerate}[label=(\roman*)]
    \item The commuting terms measurement strategy is optimal, and far superior to alternative methods, as expected. However, it is difficult to prepare the appropriate bases in which to measure the sets of commuting terms in practice.
    \item The noncrossing-terms strategy outperforms qubitwise commutativity, especially when considering only onsite/NN terms. 
    \item When a noncrossing or QWC strategy is used, similarly to the circuit depth, the NNN+ terms demand many more measurement layers than the onsite/NN counterpart. Currently, even if a noncrossing strategy is used for onsite/NN terms, we are forced to use QWC for NNN+ terms. The measurement cost could be substantially reduced by developing an appropriate generalisation of the noncrossing concept to these terms.
\end{enumerate}

%% file: Figs_Results/materials_results/srvo3_only.tex
\begin{tabular}{llrrrrr}
\toprule
         &  & \multicolumn{2}{c}{\ \ Onsite/NN} & \multicolumn{2}{c}{\ \ \ \ \ \ \  NNN+} & Total \\
         &  & Interactions & Swaps & Interactions & Swaps &  \\
 & Algorithm &              &       &              &       &       \\
\midrule
\multirow{4}{*}{SrVO$_3$} & TDS &           72 &   336 &          700 &     0 &  1108 \\
         & TDS$^{*}$ &          552 &     0 &          700 &     0 &  1252 \\
         & VQE &           72 &   336 &          476 &     0 &   884 \\
         & VQE$^{*}$ &          555 &     0 &          476 &     0 &  1031 \\
\bottomrule
\end{tabular}

%% file: Figs_Results/materials_results/vqe_results.tex
\begin{tabular}{llrrrrr}
\toprule
         &  & \multicolumn{2}{c}{\ \ Onsite/NN} & \multicolumn{2}{c}{\ \ \ \ \ \ \  NNN+} &  Total \\
         &  & Interactions & Swaps & Interactions & Swaps &  \\
 & Qubits &              &       &              &       &        \\
\midrule
GaAs & 1120 &           98 &   564 &         7191 &     0 &   7853 \\
H$_3$S & 1870 &          126 &  1088 &        36126 &     0 &  37340 \\
Li$_2$CuO$_2$ & 1024 &          123 &  1544 &         6710 &     0 &   8377 \\
Si & 1120 &          112 &   592 &         7857 &     0 &   8561 \\
SrVO$_3$ & 180 &           72 &   336 &          476 &     0 &    884 \\
\bottomrule
\end{tabular}

%% file: Outlook.tex
\section{Outlook}
\label{sec:outlook}

Simulating materials is a promising application for quantum computers.
The progress reported here combines a number of complementary approaches across the full quantum materials simulation stack that, when combined together, dramatically reduce the quantum circuit depth requirements by orders of magnitude compared to naive baseline estimates. Crucially, the design process produces quantum circuit depths for Trotter and VQE layers which are independent of the material's size by taking advantage of the locality of material Hamiltonians.  

We expect that our proposed framework for materials simulation on quantum computers can be enhanced further by continuing to incorporate physically motivated structure into the choices of fermionic encodings, basis representations, and swap network protocols. Examples of such considerations which may be particularly fruitful include
\begin{itemize}
    \item a more sophisticated incorporation of point and spin symmetries into the choice of fermionic encoding;
    \item a more careful consideration of the choice of ordering of modes in the hybrid chains;
    \item a more aggressive analysis of how many terms in the Hamiltonian may be truncated --  in particular, the systematic incorporation of electron screening effects in many materials truncates the long-range effects of the Coulomb interaction, which can be computed using classical approaches such as RPA \cite{RPA} and constrained DFT \cite{cLDA}. It is our expectation that this will decrease the number of non-local quartic interactions needed to be considered in the quantum simulation;
    \item further localisation of the single-particle Wannier states, such as with the recently proposed selected columns of the density matrix maximal localisation technique \cite{scdm};
    \item a tighter feedback loop between the choice of fermionic basis and the resulting form of the qubit Hamiltonian;
    \item  properly accounting for the translational symmetries of NNN+ terms in the fermionic encoding;
    \item identifying early termination heuristics for the fswap networks.
\end{itemize}

 One consequence of this work is the identification of certain materials, from the small set we have considered, which are particularly well suited to quantum simulation due to the details of their physics -- such as SrVO$_3$. Beyond reducing circuit depths and improving error mitigation techniques, identifying the appropriate physical systems which are best suited for simulation on NISQ devices is essential. The development of the tools described in this work can be used to allow the application of data-driven and high-throughput techniques to understand the classes of materials most amenable to quantum simulation.

It is possible that the strategies employed in this work to tackle materials may also be applicable to other quantum systems such as chemicals -- particularly those with some kind of extended spatial structure.  We also mention that the computation of ground state properties for materials may be more efficiently realised on classes of model Hamiltonians, such as those with Slater-Kanamori interactions, within approximate embedding formalisms, which have enjoyed recent success using classical algorithms \cite{dft_dmft_rev}, but are ultimately constrained by the exponentially scaling Hilbert space with their system size. 

Finally, this work has made some assumptions about the structure of the quantum hardware, the most significant of which is the assumption of all-to-all hardware connectivity. This assumption, while valid for certain architectures (such as some ion traps), does not hold universally; it was made largely to reduce the complexity of performing the analysis, so as to more rapidly demonstrate the benefits of the strategies we are proposing. We plan to continue to iterate on the design of the compiler to incorporate specific hardware layouts, and introduce additional strategies to improve circuit depths in these cases. However it is important to recognise that the horizon of quantum hardware design is extremely broad, and it is likely that the development of hardware may well be done in an anticipation of specific algorithmic requirements. Developing tools such as the one given here can help better identify what these requirements may look like.

 Our results show that considering seriously the structure of the physical problem at hand and incorporating these considerations into the design of quantum algorithms can dramatically accelerate progress towards quantum advantage. 

\subsection*{Acknowledgements}

We would like to thank Tom Ellaby, Glenn Jones, Ludovic Briquet, Maud Einhorn, Christopher Savory, and David Scanlon, and also the rest of the Phasecraft team for many helpful and insightful discussions. This project has received funding from the European Research Council (ERC) under the European Union's Horizon 2020 research and innovation programme (grant agreement No.\ 817581), and was supported by InnovateUK grant 133990, ``Quantum computing for battery materials''. Google Cloud credits were provided by Google via the EPSRC Prosperity Partnership in Quantum Software for Modeling and Simulation (EP/S005021/1).


%% file: Appendix.tex
\section*{Appendix}

\section{Baseline for qubit requirements and gate depth of materials}
\label{app:previous_method}
Here, we briefly give a naive estimate of the resources (i.e., qubit number and circuit depths) required to simulate a material on a quantum computer using general existing methods and without taking advantage of the tailored approach we exploited in the main text.

We consider a material with periodic boundary conditions, discrete translational symmetry, and lattice volume $V$ (i.e., with a total of $V$ unit cells). In the spirit of the Born-Oppenheimer approximation \cite{Cederbaum_BO}, we assume a stationary atomic configuration and we are interested in simulating the electronic degrees of freedom of the system. For the sake of simplicity, we may choose to describe the material in the Bloch basis introduced in \cref{sec:Physics}. In analogy with the procedure described in \cref{sec:Kohn-Sham}, we assume that we are able to identify a set of active bands where the relevant physics takes place. In this case we are concerned with the Hamiltonian restricted to the $m$ modes indexed by those bands. These $m$ bands are determined by identifying the last occupied atomic orbital for each atom in the material and then including all occupied orbitals for each atom. Accounting for spin, the total number of active fermionic modes is $m=2Vb$, where $b$ is the number of bands. 

One may represent the fermionic system on the qubits of a quantum computer using the Jordan-Wigner (JW) transform in accordance with some pre-chosen linear ordering of the fermionic modes (see \cref{sec:qubit_rep}). This requires one qubit per mode. We denote by $H'$ the Hamiltonian expressed as a sum of Pauli operators on the qubit system corresponding to the JW transform of $H$. 

We now wish to estimate the circuit depth of a potential quantum algorithm for such a representation. As a benchmark we consider the circuit depth of the following unitary circuit, as used in the main text:
\begin{equation}
    U = \prod_j \mathrm{exp}(i\alpha_j H'_j),
\end{equation}
where we are free to choose the ordering of the product. Circuits of this type appear as subroutines in both VQE (under a Hamiltonian variational ansatz) and time dynamic simulation (TDS). In both cases, these subroutines are repeated multiple times, introducing an additional multiplicative overhead to the circuit depth, which we will not detail here.

A consequence of choosing the JW transform is that many of the terms in $H'$ operate on a large number of qubits. In particular, fermionic operations corresponding to interactions between fermionic modes far from each other in the aforementioned linear ordering precipitate costly circuit decompositions of individual $\mathrm{exp}(i\alpha_j H'_j)$ terms.

In this Appendix we consider two previously known methods of implementing the desired interactions. The first is simply to implement the terms in $H'$ in sequence, via the logarithmic-depth circuit for computing parities described in \cref{sec:gate_complexity}. Each term can be implemented in depth at most $\lceil \log_2 m \rceil - 1$ (given that we are allowed all-to-all interactions) leading to an overall depth of at most $T(\lceil \log_2 m \rceil - 1)$ for a Hamiltonian with $T$ terms. Note that it may be possible to reduce the complexity somewhat by implementing some of these terms in parallel, and using the fact that many of the terms act nontrivially on fewer than $m$ qubits; we do not explore these further here.

The second method was proposed in~\cite{microsoft_patent}, and allows all quartic interactions to be implemented in quantum circuit depth approximately $O(m^3)$. More precisely, assuming that each quartic term requires 2-qubit depth 3 (as used in \cref{sec:gate_complexity}), a lower bound on the overall 2-qubit gate depth is approximately
\begin{equation}
\label{app:eq:UB2}
    0.76 m^{3.06} + \frac{12T}{m},
\end{equation}
where we estimate the total cost by summing the cost of swap layers from Fig, 7E of \cite{microsoft_patent} (taking the fit up to 400 qubits), and make the optimistic assumption that the $T$ terms can be optimally parallelised in between the swap layers, such that we implement $m/4$ of them at each layer, each with 2-qubit gate depth 3.

It remains to compute the total number of terms $T$. We will assume for simplicity that there are only quartic terms, because in practice quadratic terms can likely be implemented at a lower-order cost during the course of the algorithm to implement the quartic terms.

In terms of the complex fermion creation and annihilation operators $f^\dagger_{\bm{k},b,\sigma}$ and $f_{\bm{k},b,\sigma}$, the quartic interactions in the Bloch basis are (see, e.g., second term in \cref{Ham_band_basis})
\begin{equation}
    H_{\mathrm{int}}=\sum_{\bm{k}_i,b_i}{V}^{\bm{k}_1 \bm{k}_2 \bm{k}_3 \bm{k}_4}_{b_1 b_2 b_3 b_4}\delta_{\bm{k}_1+\bm{k}_2,\bm{k}_3+\bm{k}_4}\sum_{\sigma,\sigma'}f^\dagger_{\bm{k}_1,b_1,\sigma}f^\dagger_{\bm{k}_2,b_2,\sigma'}f_{\bm{k}_3,b_3,\sigma'}f_{\bm{k}_4,b_4,\sigma},
\end{equation}
where $\bm{k}_i$  are Bloch momenta, $b_i$ are band indices and $\sigma$ is the spin. Here $\delta_{\bm{k}_1+\bm{k}_2,\bm{k}_3+\bm{k}_4}$ enforces explicitly the Bloch momentum conservation (up to lattice vectors). To ease the count, it is useful to write $H_{\mathrm{int}}$ in terms of singlet and triplet scattering terms (valid in presence of time reversal symmetry)
\begin{equation*}
    H_{\mathrm{int}}=\sum_{\alpha_i}\delta_{\bm{k}_1+\bm{k}_2,\bm{k}_3+\bm{k}_4}\left(V^{+}_{\alpha_1 \alpha_2 \alpha_3 \alpha_4}\psi^\dagger_S(\alpha_1,\alpha_2)\psi_S(\alpha_3,\alpha_4)+V^{-}_{\alpha_1 \alpha_2 \alpha_3 \alpha_4}\sum_{a=0,\uparrow,\downarrow} \psi^\dagger_a(\alpha_1,\alpha_2)\psi_a(\alpha_3,\alpha_4)\right),
\end{equation*}
where we have defined the super-index $\alpha_i =(\bm{k}_i,b_i)$ that can take $Vb$ different values and $V^\pm_{\alpha_1 \alpha_2 \alpha_3 \alpha_4} \equiv \frac{1}{2}({V}^{\bm{k}_1 \bm{k}_2 \bm{k}_3 \bm{k}_4}_{b_1 b_2 b_3 b_4} \pm {V}^{\bm{k}_1 \bm{k}_2 \bm{k}_4 \bm{k}_3}_{b_1 b_2 b_4 b_3})$. The singlet and triplet operators are
\begin{align}
    \psi_S(\alpha_1,\alpha_2)&=\frac{1}{\sqrt{2}}(f_{\bm{k}_1,b_1,\uparrow}f_{\bm{k}_2,b_2,\downarrow}-f_{{\bf k_1},b_1,\downarrow}f_{\bm{k}_2,b_2,\uparrow}),\\
    \psi_0(\alpha_1,\alpha_2)&=\frac{1}{\sqrt{2}}(f_{\bm{k}_1,b_1,\uparrow}f_{\bm{k}_2,b_2,\downarrow}+f_{\bm{k}_1,b_1,\downarrow}f_{\bm{k}_2,b_2,\uparrow}),\\ 
    \psi_\sigma(\alpha_1,\alpha_2)&=f_{\bm{k}_1,b_1,\sigma}f_{\bm{k}_2,b_2,\sigma}.
\end{align}
For the singlet operator above, we can choose $\binom{Vb}{2}$ values for the pair of indices $\alpha_1$ and $\alpha_2$ that give a different operator plus $Vb$ choices when $\alpha_1=\alpha_2$. On the other hand, for any of the triplet operators one can only choose $\binom{Vb}{2}$ values of the pair $(\alpha_1,\alpha_2)$ that generate a different operator. Note that $\psi_a(\alpha,\alpha)=0$.

Then, the overall bound is given by
\begin{equation}
    T = \frac{1}{V}\left[\left[\binom{Vb}{2}+ Vb\right]^2+3\binom{Vb}{2}^2\right]=
    V^3b^4-V^2b^3+Vb^2.
\end{equation}

Here, the reduction by a factor of $V$ is due to lattice momentum conservation. Additional symmetries of the Hamiltonian may introduce more savings in the number of terms (with the size of the savings being proportional to the size of the symmetry), but few are likely to be as large as the lattice translation symmetry.

{Using the fact that a Hamiltonian with $T$ terms can be implemented with an overall depth of at most $T(\ceil{\log_2m}-1)$ and \cref{app:eq:UB2}, we can derive two upper bounds on the quantum circuit depth. In terms of $V$ and $b$, they take the forms}
\begin{align}\label{eq:bounds}
    \mathrm{UB}_1 &= (V^3 b^4 - V^2 b^3 + Vb^2) \lceil \log_2(Vb)\rceil \\
    \mathrm{UB}_2 &= 6.34 (Vb)^{3.06} + 6(V^2 b^3 - V b^2 + b),
\end{align} 
respectively. 
We can also put a crude lower bound on the quantum circuit complexity of any method based on the JW transform. If we have $T$ quartic terms and $m$ qubits, we can implement at most $m/4$ terms at each step. Assuming again that each quartic term requires 2-qubit depth 3, we require 2-qubit gate depth of at least $12T/m$, or in terms of $V$ and $b$, at least
\begin{equation}
    \mathrm{LB} = 6(V^2 b^3 - V b^2 + b).
\end{equation}

In order to compare these naive estimates with the results reported in \cref{sec:results}, in \cref{table:estimates_intro} of the main text
we considered a system consisting of $V=3^3$ unit cells for SrVO$_3$ and $V=5^3$ for the rest of the materials and reported the better of the two upper bounds in each case.

The assumptions that we use for the number of bands are shown in Table \ref{tab:elements_bands}. Although other estimates for the number of bands are possible, note that for a large piece of material $V\gg b$ and dominates in (\ref{eq:bounds}).

\begin{table}
    \centering
\begin{tabular}{clc}\toprule
     Element & electron config. & orbitals considered \\\midrule
     Gallium & [Ar] 3d$^{10}$4s$^2$4p$^1$ & 3 (from p) \\
     Arsenic & [Ar] 3d$^{10}$4s$^2$4p$^3$ & 3 (from p) \\
     Hydrogen & 1s$^1$ & 1 (from s) \\
     Sulfur & [Ne] 3s$^2$3p$^4$ & 3 (from p) \\
     Lithium & [He] 2s$^1$ & 1 (from s) \\
     Copper & [Ar] 3d$^{10}$4s$^1$ & 6 (from d and s)\\
     Oxygen & [He] 2s$^2$2p$^4$ & 3 (from p) \\
     Silicon & [Ne] 3s$^2$3p$^2$ & 3 (from p) \\
     Strontium & [Kr] 5s$^2$ & 1 (from s) \\
     Vanadium & [Ar] 3d$^3$4s$^2$ & 6 (from d and s)
\end{tabular}\caption{Electron orbitals of different elements. To obtain the number of bands for the different materials presented in the main text, we use the orbitals shown here and multiply by the number of times an element appears in the the chemical formula of the material. We use GaAs (4Ga+2As); H$_3$S (3H+S); Li$_2$CuO$_2$ (2Li+Cu+2O); SrVO$_3$ (Sr+V+3O); Si (Si). }\label{tab:elements_bands}
\end{table}

\section{Exponentially localized Wannier functions}
\label{app:MLWFs}

Here we discuss the two conditions for the existence of maximally localized Wannier functions (MLWFs). Recalling the definition of the Wannier functions introduced in \cref{sec:Wannier_functions}, we have
\begin{equation}
 \mathcal{W}^{\bm{0}}_{s, \sigma}(\bm{r})=\sum_{\bm{k},n}u_{\bm{k},n, \sigma}(\bm{r})U_{ns}(\bm{k})e^{-i\bm{k}\cdot\bm{r}} \equiv 
\sum_{\bm{k}}v_{\bm{k},{s}, \sigma}(\bm{r})e^{-i\bm{k}\cdot\bm{r}},
\end{equation}
where $v_{\bm{k},{s}, \sigma}(\bm r)$ are quasi-Bloch functions. We can use the analyticity of the quasi-Bloch functions $v_{\bm{k},{s},\sigma}(\bm r)$ as a function of the crystal momentum $\bm{k}$ to show that the Wannier functions are localized, using the following result \cite{Cloizeaux1,Cloizeaux2,Nenciu1983}

\textbf{Theorem 1 (Cloizeaux):} Let $f(\bm k)$ be a function of the $n-$dimensional complex vector $\bm k=\bm k'+i\bm k''$ defined in the $n-$torus with periods $\bm b_j$ $(j= 1,\dots, n)$, i.e., $f(\bm k +\bm b_j)=f(\bm k)$. 
If $f(\bm k)$ is an analytic function of $\bm k$ in a strip defined by $|\bm k'' |< A$, with $A$ a positive real number, then:
\begin{enumerate}
 \item $f(\bm k)$ can be expanded in a convergent Fourier series in this domain as
 \begin{equation}
 f(\bm k) =\sum_{\bm R}e^{i\bm k \cdot \bm R}g(\bm R), 
\end{equation}
where $\bm R= \sum_j n_j \bm r_j$ is a reciprocal lattice vector to $\bm k$, with $ \bm r_j $ satisfying
$\bm b_j \cdot \bm r_l = 2\pi \delta_{jl}$, and 
\item the Fourier coefficients $g(\bm R)$ satisfy $\lim_{|\bm R|\rightarrow\infty} e^{b|\bm R|} g(\bm R)=0$ for any $0<b < A$.
\end{enumerate}
Conversely, if the Fourier coefficients have this asymptotic behavior, the series converges and is analytic in the region $|\bm k''|<A$.

\begin{figure}[ht]
 \includegraphics[width=\linewidth]{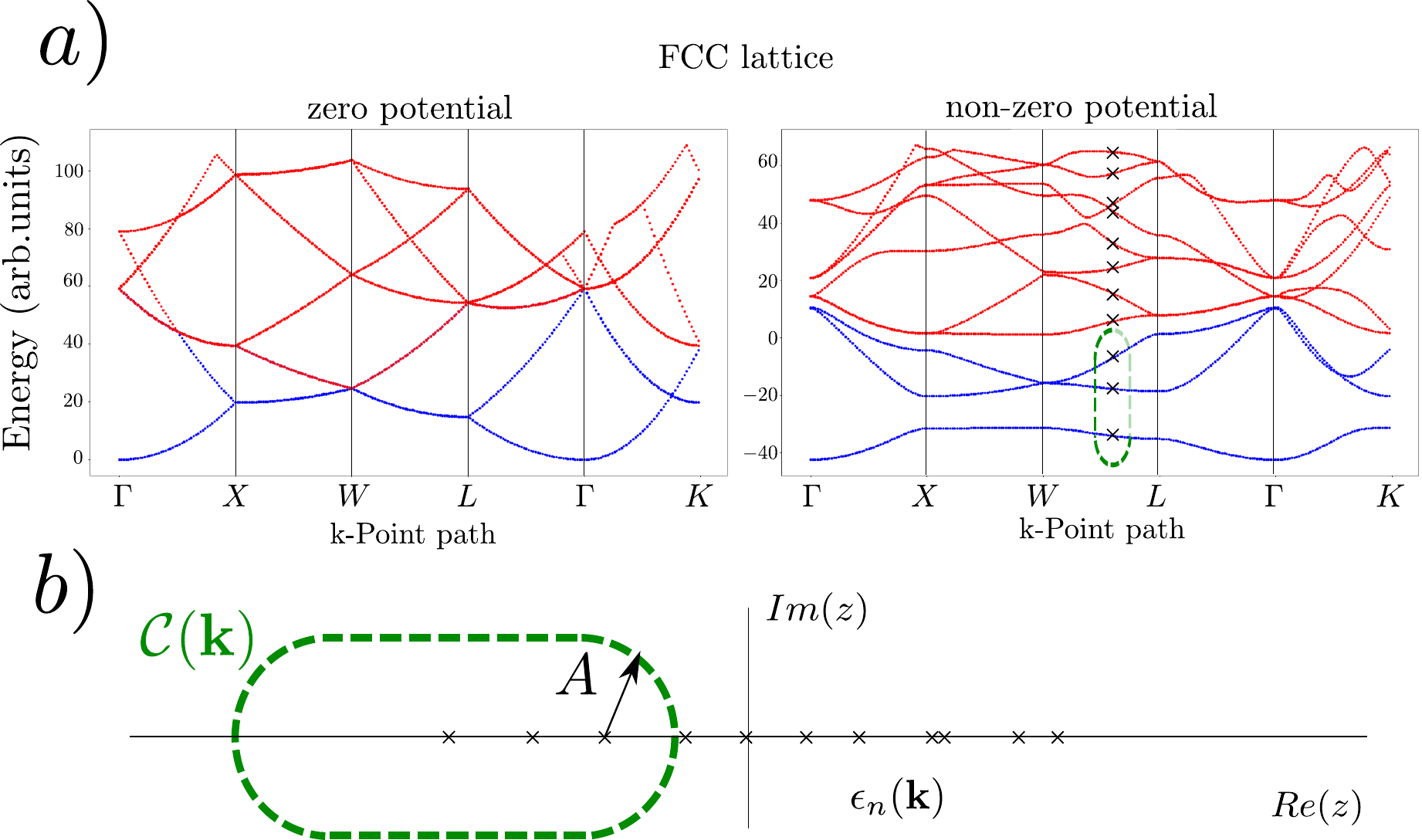}
 \caption{ \textbf{(a)} left. Band structure for face centered cubic (FCC) lattice, in the empty lattice approximation ($U_{\bm G}=0$) across a high symmetry path. The three bottom bands are highly degenerate for many values of the crystal momentum. \textbf{(a)} right. Bands structure for FCC lattice, with $U_{\bm G}=100\sum_a \frac{e^{i\bm r_a\cdot \bm G}}{\bm G}$, where the sum runs over the atoms in the unit cell. We observe that the bands split at high symmetry points, and a gap (for each $\bm k$) develops between the three lower bands and the upper energy bands. Considering each band as the real cross-section of a complex eigenenergy parameterized by $\bm k = \bm k' +i \bm k''$, we can define a projector onto the lower three bands by using the Riesz projector defined in \cref{Riesz} with $\mathcal{C}(\bm{k})$ given by the dashed green contour.
 \textbf{(b)} The integration contour $\mathcal{C}(\bm{k})$ for the Riesz projector. The width of the strip in the imaginary direction where the projector is analytic is determined by the gap $A$ between the active bands and the rest.}
 \label{fig:projector}
\end{figure}

Clearly, if we can show that a quasi-Bloch function $v_{\bm k, s, \sigma}(\bm r)$ is indeed an analytic function of the crystal momentum, then the corresponding Wannier function $\mathcal{W}_{s,\sigma}(\bm r)$ will be exponentially localized as a function of $\bm R$, due to Theorem 1 above.
The quasi-Bloch functions $v_{\bm k, s, \sigma}(\bm r)$ are associated with the single-particle energies, which are analytic except for points where the bands are degenerate. In that case, the energy surface in the complex plane has a branch-cut. We can define the projector onto the considered bands as
\begin{equation}\label{Riesz}
 P(\bm k)=\frac{1}{2\pi i}\int_{\mathcal{C}(\bm k)}\frac{dz}{z- H_0(\bm k)},
\end{equation}
where the contour $\mathcal{C}(\bm k)$ encloses the bands in the active space (see \cref{fig:projector}). Here $H_0(\bm k)$ is the non-interacting part of the Hamiltonian.

The gap $A$ between the last band considered and the first band outside the active space determines the decay of the exponential localization of the Wannier functions. It has been shown \cite{Brouder2007} that if the sum of Chern numbers on the bands belonging to the active space is zero, then the quasi-Bloch functions exist. The Chern number of a group of bands is defined as
\begin{equation}
 C_\ell = \frac{i}{2\pi}\sum_{i<j}\int_{BZ} B^{ij}(\bm k)dk_i\wedge dk_j, 
\end{equation}
where the integration is performed in the Brillouin zone. The Berry connection $B^{ij}(\bm k)$ characterizes the change of frame between different points in the Brillouin zone and is given by
\begin{equation}
 B^{ij}(\bm k)={\rm Tr}\left(P(\bm k)\left[\frac{\partial P(\bm k)}{\partial k_i},\frac{\partial P(\bm k)}{\partial k_j}\right]\right).
\end{equation}

Therefore, the two following conditions are sufficient to ensure the existence of localized Wannier functions:

\begin{enumerate}
 \item Vanishing Chern numbers. This condition is automatically satisfied in a system with time-reversal symmetry. Note that, however, systems without time-reversal symmetry can still have vanishing Chern number.
 \item Existence of a large gap between the bands in the active space. Note that, in this case, the system does not necessarily represent an insulator, as the Fermi energy can lie within the active space, which is separated from the rest of the bands.
\end{enumerate}

\section{Optimality of the Jordan-Wigner transform and fswap networks}
\label{app:jw}

Here we show that, perhaps surprisingly, if our goal is to implement time-evolution according to an arbitrary set of quadratic fermionic interactions (in an order of our choice), then an algorithm combining the Jordan-Wigner (JW) transform and fermionic swap (fswap) networks is close to optimal (in a particular sense). This algorithm, which was described in~\cite{kivlichan18}, was proven optimal among fswap network algorithms within the JW transform in~\cite{hagge2021}; here, by contrast, we prove it is close to optimal among all algorithms (of a somewhat restricted form) within any encoding.

To achieve this, we consider the following setup:

\begin{enumerate}
    \item We are given a fermionic Hamiltonian on $m$ modes of the form
    \begin{equation}
        H = \sum_{j,k} H_{jk} = \sum_{j,k} h_{jk} (c^\dag_j c_k + c^\dag_k c_j).
    \end{equation}
    \item Our goal is to implement the unitary transformation
    \begin{equation}
        U = \prod_{j,k} e^{i \theta_{jk} H_{jk}},
    \end{equation}
    with $\theta_{ij}$ a set of parameters, where we can take the product in an arbitrary order of our choice. This is sufficient to implement the Hamiltonian variational ansatz in VQE, for example.
    \item Our goal is to minimise the 2-qubit gate depth required to implement this transformation. We assume that an arbitrary 2-qubit gate has unit cost.
    \item We assume that each mode is associated with one or more qubits, such that at least one of these is a ``data'' qubit; and that to implement each interaction on 2 modes requires at least one 2-qubit gate across all the associated data qubits.
\end{enumerate}

To implement this using the JW transformation, we order the fermionic modes arbitrarily and associate each one with a qubit. We then use the fswap network proposed by Kivlichan et al.~\cite{kivlichan18}, which alternates between swapping pairs of adjacent modes starting at even and odd positions. We can combine time-evolution according to the corresponding terms in the Hamiltonian with these fswap operations at no additional cost, as we are assuming that all 2-qubit operations are equivalent. After $m$ layers of fswap operations, all pairs of modes have interacted and the order is returned to the original order, albeit reversed (which can be dealt with by acting on the modes in reverse order in the next step of the algorithm, if desired). Therefore, the algorithm has 2-qubit gate depth $m$.

To obtain a corresponding lower bound, we observe that in the worst case, there are $\binom{m}{2}$ interactions to be implemented in total. At each step, at most $\lfloor m/2 \rfloor$ interactions can be implemented, because each interaction involves at least one 2-qubit gate across the data qubits. If there are more than $m$ data qubits, this bound continues to hold, as we can select one qubit per mode to produce the set on which we need to implement the interactions. Therefore, in 2-qubit gate depth $k$ we can implement at most $k \lfloor m/2\rfloor$ interactions. Implementing all the interactions then requires $k \ge m-1$ (if $m$ is even) or $k \ge m$ (if $m$ is odd).

\section{Hamiltonian coefficients pipeline}
\label{app:Hamiltonian_coeff_pipeline}
In this Appendix we describe in details the steps required to generate all the hopping matrix (HM) and Coulomb tensor (CT) coefficients for an arbitrary material in the Wannier basis. 

\subsection{Full Hamiltonian of an electronic system on a lattice}
\label{app:sec:full_H}

AWe consider a material consisting of $N_{\mathrm{cells}}$ sites arranged in a translationally invariant (Bravais) lattice $\mathcal{G}$. Working within the Born-Oppenheimer approximation \cite{Cederbaum_BO}, we assume that its electronic problem has been restricted to an active space spanned by $N_{\mathrm{orbitals/cell}}$ Wannier functions (WFs), $\mathcal{W}^{P}_{i,\sigma}( \bm{r})$ per cell, with $P\in\mathcal{G}$, $i\in\set{1,...,N_{\mathrm{orbitals/cell}}}$ labeling the various orbitals, and $\sigma\in\set{\uparrow, \downarrow}$ the spin quantum number. By choosing a coordinate system with origin $O$, each point $P\in\mathcal{G}$ can be generated by a discrete translation by a vector $\bm{R}_P = n_1^P \bm{R}_1 + n_2^P \bm{R}_2 + n_3^P \bm{R}_3 \in \mathcal{B} $, with $\bm{R}_a$, $a=1,2,3$, the primitive vector of $\mathcal{G}$ and $\mathcal{B} = \set{\bm{R} = n_1 \bm{R}_1 + n_2 \bm{R}_2 + n_3 \bm{R}_3 | n_a \in \mathbb{Z}, a=1,2,3}$ the set of all the possible discrete translations on $\mathcal{G}$. Hence, $\bm{R}_O = \bm{0}$ and $\mathcal{W}^{P}_{i,\sigma}(\bm{r})=\mathcal{W}_{i,\sigma}(\bm{r}-\bm{R}_P)$.

The full Hamiltonian for the most general electronic system on the lattice $\mathcal{G}$ is 
\begin{equation}
    \label{app:eq:full_H}
    H = H_0 + H_{\mathrm{int}},
\end{equation}
with quadratic and quartic contributions given by
\begin{subequations}
\label{app:eq:H0_Hint}
    \begin{align}
        H_0 &= \sum_{\sigma_1,\sigma_2}\sum_{A,B}\sum_{i,j} t_{(A,i,\sigma_1),(B,j,\sigma_2)} w^\dagger_{A, i, \sigma_1} w_{B, j, \sigma_2},\\
        H_\mathrm{int} &= \sum_{\sigma_1,\sigma_2,\sigma_3,\sigma_4}\sum_{A,B,C,D}\sum_{i,j,k,l} V_{(A,i,\sigma_1),(B,j,\sigma_2),(C,k,\sigma_3),(D,l,\sigma_4)} w^\dagger_{A, i, \sigma_1} w^\dagger_{B, j, \sigma_2} w_{C, k, \sigma_3} w_{D, l, \sigma_4},
    \end{align}
\end{subequations}
with $\sigma_1, \sigma_2, \sigma_3, \sigma_4\in\{\downarrow, \uparrow\}$, $A,B,C,D\in\mathcal{G}$, and $i,j,k,l \in \{1,...,N_{\mathrm{orbital/cell}}\}$.  Here, $w^{(\dagger)}_{A, i, \sigma}$ destroys (creates) an electron at site $A$ with mode $i$ and spin $\sigma$. Finally, $t$ and $V$ denote the spinful HM and the CT, respectively.

In what follows we will make the following assumptions:

\begin{itemize}
	\item \textbf{Non-magnetic material (NMM) approximation}: we will consider materials without spin-orbit coupling and in the absence of magnetic fields. In this case, the spin up and spin down sectors are degenerate. From here on, we will therefore omit the spin index in the WFs, i.e., $\mathcal{W}^{P}_{i,\sigma}(\bm{r}) := \mathcal{W}^{P}_{i}(\bm{r})$. Note that this assumption can be relaxed by generalizing \cref{app:eq:spinful_HMCs} and \cref{app:eq:spinful_CTCs} below. 
	
	\item \textbf{Real Wannier functions}: we will assume that the WFs $\mathcal{W}^{P}_{i,\sigma}(\bm{r})$ are real. This is always  true if maximally localized Wannier functions can be built (see \cref{app:MLWFs}) \cite{Brouder2007}.
	
	\item \textbf{Nearest-neighbour (NN) approximation:} in calculating the HM and CT coefficients, we will consider the following NN approximation of order $n$. Thanks to the localization properties of the MLWFs, this is a reasonable assumption for all the materials we considered in the main text. By looking at a lattice site $P\in\mathcal{G}$ we group the other sites of the lattice in ascending order of Euclidean distances from $P$, namely $0=d_0<d_1<d_2<...<d_n$. All the sites of the lattice having the same distance $d_n$ from $P$ represent the nearest neighbours of order $n$ of $P$. Then, in the NN approximation of order $n$ we compute only those  HM (CT) coefficients involving lattice sites $A$ and $B$ ($A$, $B$, $C$, $D$) which are NN of order $\leq n$ with respect to each other, i.e., such that $|P_1 - P_2|\leq d_n$,  $\forall P_1,P_2\in\{A,B\}$ ($\forall P_1,P_2\in\{A,B,C,D\}$) and set $t^{AB}_{ij}=0$  ($V^{ABCD}_{ijkl}=0$) otherwise. Here, $|...|$ denotes the Euclidean distance between two sites. In particular, we introduce the general notation $\mathcal{N}^n_{\mathcal{G}'} = \set{P\in \mathcal{G}| |P-P'| \leq d_n,\ \forall P' \in \mathcal{G}'}$, with $\mathcal{G}'\subseteq \mathcal{G}$, and $N^n_{{\mathcal{G}'}} = \mathrm{dim}(\mathcal{N}^n_{\mathcal{G}'})$, to denote the set of sites of $\mathcal{G}$ which are NN of order $\leq n$ with respect to all the sites of $\mathcal{G'}$ and their total number, respectively.
\end{itemize}

Within these approximations, we have that the coefficients of the spinful HM  in \cref{app:eq:H0_Hint} are defined as
\begin{equation}
	\label{app:eq:spinful_HMCs}
	t_{(A,i,\sigma_1),(B,j,\sigma_2)}=
	\begin{cases}
		T^{AB}_{ij} & \text{if } \sigma_1 = \sigma_2\\
		0 & \text{otherwise}
	\end{cases},
\end{equation}
where the \emph{bare}\footnote{In what follows we will omit this specification and we will implicitly assume that all the HM and CT coefficients but the \emph{spinful} ones are \emph{bare} coefficients.} HM coefficients can be computed as
\begin{equation}
	T^{AB}_{ij}=\int d\bm{r} \mathcal{W}_i(\bm{r}-\bm{R}_A) h_{\mathrm{sp}}\mathcal{W}_j(\bm{r}-\bm{R}_B).
\end{equation}
Here, $h_{\mathrm{sp}} = -\hbar^2\nabla^2/(2m) + \tilde{U}_{\mathrm{eff}}(\bm{r})$ is the single particle Hamiltonian, with $\tilde{U}_{\mathrm{eff}}(\bm{r})$ the effective Kohn-Sham potential (implicitly) derived within the Density Functional Theory (DFT), as explained in  \cref{sec:DFT}). Thanks to the discrete translational invariance of the lattice, all the HM coefficients can be obtained from the \emph{core} ones, which are defined by setting $A=O$ in the equation above, 
\begin{equation}
    \label{app:eq:core_HM}
	T^{OB}_{ij}=\int d\bm{r} \mathcal{W}_i(\bm{r}) h_{\mathrm{sp}}\mathcal{W}_j(\bm{r}-\bm{R}_B).
\end{equation}

The coefficients of the \emph{spinful} CT in \cref{app:eq:H0_Hint}, $ V_{(A,i,\sigma_1),(B,j,\sigma_2),(C,k,\sigma_3),(D,l,\sigma_4)}$, are defined in terms of the \emph{bare} CT coefficients, $V^{ABCD}_{ijkl}$, as
\begin{equation}
\label{app:eq:spinful_CTCs}
    V_{(A,i,\sigma_1),(B,j,\sigma_2),(C,k,\sigma_3),(D,l,\sigma_4)} = 
    \begin{cases}
        V^{s,ABCD}_{ijkl} & \text{if } \sigma_1 = \sigma_2 = \sigma_3 = \sigma_4 \\
        \frac{1}{2}V^{ABCD}_{ijkl} & \text{if } \sigma_1 = \sigma_3 \text{ and } \sigma_2 = \sigma_4\ (\sigma_1 \neq \sigma_2)\\
        -\frac{1}{2}V^{BACD}_{jikl} & \text{if } \sigma_1 = \sigma_4 \text{ and } \sigma_2 = \sigma_3\ (\sigma_1 \neq \sigma_2)\\
        0 & \text{otherwise}
    \end{cases},
\end{equation}
with $ V^{s,ABCD}_{ijkl} = (V^{ABCD}_{ijkl} - V^{BACD}_{jikl})/2$. In turn, the CT coefficients can be obtained as
\begin{equation}
    V^{ABCD}_{ijkl} = \frac{1}{2}\int d\bm{r}\int d\bm{r}' \mathcal{W}_i(\bm{r}-\bm{R}_A) \mathcal{W}_j(\bm{r}'-\bm{R}_B)V(|\bm{r}-\bm{r}'|)\mathcal{W}_k(\bm{r}'-\bm{R}_C)\mathcal{W}_l(\bm{r}-\bm{R}_D),
    \label{app:eq:bare_CTCs}
\end{equation}
with $V(|\bm{r}-\bm{r}'|) $ the (screened) Coulomb potential.
Note that, by exploiting the discrete translational invariance of the lattice, we can set $A = O$ and get the \emph{core} CT coefficients
\begin{equation}
    C^{OBCD}_{ijkl} \equiv V^{OBCD}_{ijkl} = \frac{1}{2}\int d\bm{r}\int d\bm{r}' \mathcal{W}_i(\bm{r}) \mathcal{W}_j(\bm{r}'-\bm{R}_B)V(|\bm{r}-\bm{r}'|)\mathcal{W}_k(\bm{r}'-\bm{R}_C)\mathcal{W}_l(\bm{r}-\bm{R}_D).
    \label{app:eq:core_bare_CTCs}
\end{equation}
These coefficients can be computed via Monte Carlo (MC) integration and the full CT can then be retrieved from the core coefficients thanks to the lattice translational invariance (see below). 

\paragraph{Cauchy-Schwarz inequality}

To reduce the number of CT coefficients to be computed, we will exploit a Cauchy-Schwarz (CS) inequality between the coefficients themselves. This is obtained by re-writing \cref{app:eq:bare_CTCs} as an inner product,
\begin{align}
\label{app:eq:CT_innerproduct}
    V^{ABCD}_{ijkl} = \langle \rho^{AD}_{il}, \rho^{CB}_{kj}\rangle \equiv \frac{1}{2}\int d\bm{r} \int d\bm{r}' V(|\bm{r}-\bm{r}'|) \rho^{AD}_{il}(\bm{r}) \rho^{CB}_{kj}(\bm{r}'),
\end{align}
with $\rho^{AB}_{ij}(\bm{r}) \equiv \mathcal{W}_i(\bm{r}-\bm{R}_A) \mathcal{W}_j(\bm{r}-\bm{R}_B) $. Since $\tilde{V}(|\bm{r}-\bm{r}'|) $ is always a positive definite kernel, it can be shown that the inner product we have just introduced is well-defined and that the following CS inequality holds
\begin{equation}
\label{app:eq:CS_ineq}
    |V^{ABCD}_{ijkl}|^2 \leq V^{AADD}_{iill}V^{CCBB}_{kkjj}.
\end{equation}
In what follows we will show that, assuming that all the coefficients smaller than a given threshold are negligible, this inequality can be exploited to determine a priori which coefficients of the CT need to be computed via MC integration. 

\paragraph{Symmetry properties of the Coulomb tensor.} For the sake of simplicity we now introduce the composite indices $\lambda_i$ grouping together the site and orbital indices (e.g., $\lambda_1 = (A,i)$). From \cref{app:eq:bare_CTCs}, by exploiting the hermiticty of the CT, i.e., $V_{\lambda_1\lambda_2\lambda_3\lambda_4} = V_{\lambda_4\lambda_3\lambda_2\lambda_1} $, the swap symmetry $ V_{\lambda_1\lambda_2\lambda_3\lambda_4} = V_{\lambda_2\lambda_1\lambda_4\lambda_3} $, and the reality of the Wannier functions, one can derive the following identities (see also \cref{sec:fermion_integral_symmetries})
\begin{equation}
\label{app:eq:CT_identities}
    V_{\lambda_1\lambda_2\lambda_3\lambda_4} = V_{\lambda_4\lambda_3\lambda_2\lambda_1} = V_{\lambda_2\lambda_1\lambda_4\lambda_3} = V_{\lambda_4\lambda_2\lambda_3\lambda_1} = V_{\lambda_1\lambda_3\lambda_2\lambda_4} =
    V_{\lambda_3\lambda_4\lambda_1\lambda_2} = V_{\lambda_3\lambda_1\lambda_4\lambda_2} = V_{\lambda_2\lambda_4\lambda_1\lambda_3},
\end{equation}
which can be exploited to further reduce the number of CT coefficients we have to compute. 

\subsection{Motif Hamiltonian} 
\label{app:sec:motif_H}

As discussed in \cref{sec:HM_and_CT_coefficients} of the main text, the real-space localisation of the WFs allows us to focus on the Hamiltonian of small portion of the material consisting of a central cell and its NNs up to a given order $n$, which we refer to as a \emph{motif} of order $n$. The whole system (and its full Hamiltonian) can then be obtained via the tiling procedure described in \cref{sec:compile_unit_cell}. Note that, thanks to the discrete translational invariance of the lattice, we need to include in the motif Hamiltonian only those HM and CT coefficients involving the central unit cell at least once. That is, onsite interactions, and inter- and intra-cell hoppings involving exclusively cells different from the central ones are excluded. The motif Hamiltonian is hence given by $H^m = H^m_0 + H^m_{\mathrm{int}}$, with 

\begin{subequations}
	\label{app:eq:H0_Hint_motif}
	\begin{equation}
		\label{app:eq:motif_H0}
		H^m_0 = \sum_{\sigma_1,\sigma_2} \sum_{B} \sum_{i,j} \left[t^{m}_{(O,i,\sigma_1),(B,j,\sigma_2)} w^\dagger_{O,i,\sigma_1} w_{B,j,\sigma_2} + t^{m}_{(B,j,\sigma_2), (O,i,\sigma_1)} w^\dagger_{B,j,\sigma_2} w_{O,i,\sigma_1}\right],
	\end{equation}
	\begin{align}
		\label{app:eq:Hint_motif}
		H^m_\mathrm{int} = \sum_{\sigma_1,\sigma_2,\sigma_3,\sigma_4}\sum_{B,C,D}\sum_{i,j,k,l}&\left[ V^{m}_{(O,i,\sigma_1),(B,j,\sigma_2),(C,k,\sigma_3),(D,l,\sigma_4)} w^\dagger_{O, i, \sigma_1} w^\dagger_{B, j, \sigma_2} w_{C, k, \sigma_3} w_{D, l, \sigma_4}\right.\nonumber\\
		& + V^{m}_{(B,j,\sigma_2),(O,i,\sigma_1),(C,k,\sigma_3),(D,l,\sigma_4)} w^\dagger_{B, j, \sigma_2} w^\dagger_{O, i, \sigma_1} w_{C, k, \sigma_3} w_{D, l, \sigma_4}\nonumber\\
		& + V^{m}_{(B,j,\sigma_2),(C,k,\sigma_3),(O,i,\sigma_1),(D,l,\sigma_4)} w^\dagger_{B, j, \sigma_2} w^\dagger_{C, k, \sigma_3} w_{O, i, \sigma_1} w_{D, l, \sigma_4}\nonumber\\
		& + \left. V^{m}_{(B,j,\sigma_2),(C,k,\sigma_3),(D,l,\sigma_4),(O,i,\sigma_1)} w^\dagger_{B, j, \sigma_2} w^\dagger_{C, k, \sigma_3} w_{D, l, \sigma_4} w_{O, i, \sigma_1}\right],
	\end{align}
\end{subequations}
with $B,C,D \in \mathcal{N}^n_{\set{O}}$. Here, $t^{m}_{(A,i,\sigma_1),(B,j,\sigma_2)}$ and $V^{m}_{(A,i,\sigma_1),(B,j,\sigma_2),(C,k,\sigma_3),(D,l,\sigma_4)}$ are defined as in \cref{app:eq:spinful_HMCs} and \cref{app:eq:spinful_CTCs}, respectively, with $A,B,C,D \in \mathcal{N}^n_{\set{O,A,B,C,D}}$. 

In what follows, we will illustrate the various stages of the calculation of the HM and CT of the motif Hamiltonian within the real Wannier functions, NMM, and order $n$ NN approximations described in \cref{app:sec:full_H}.

\subsubsection{Hopping matrix coefficients}
\label{app:sec:HM_pipeline}

\paragraph{Stage H1: core HM coefficients.} In the first stage, the \emph{core} HM coefficients $T^{OB}_{ij} $, with $B\in\mathcal{G'}$, introduced in \cref{app:eq:core_HM} are read from the output of DFT/Wannier90 calculations. Here, $\mathcal{G'}$ is determined by size of the lattice considered in the  DFT simulation. 

\paragraph{Stage H2: filtered core HM coefficients.} In the second stage, the core HM coefficients $T^{OB}_{ij} $ are restricted to NNs of order $n_0$, i.e., for lattice sites $B\in\mathcal{N}^{n_0}_{\set{O}}$. To determine $n_0$, we compare the exact bands $\varepsilon_{i}(\bm{k})$ with the ones obtained from the \emph{filtered} order $n$ NN HM, $\overline{\varepsilon}^{(n_0)}_{i}(\bm{k})$. The latter are given by the eigenvalues of the matrices 
\begin{equation}
	h(\bm{k})_{ij} = \sum_{\bm{B}\in \mathcal{N}^{n_0}_{\bm{0}}}e^{i\bm{k}\cdot\bm{R}}\overline{T}^{OB}_{ij},
\end{equation}
with filtered HM coefficients
\begin{equation}
	\overline{T}^{OB}_{ij} = 
	\begin{cases}
		T^{OB}_{ij} & \text{if } |T^{OB}_{ij}| \geq t_0\\
		0 &\text{otherwise},
	\end{cases}
\end{equation}
and threshold $t_0 = \mathrm{max}|T^{OB}_{mn}|$, $\forall m,n$ and $\forall B\notin\mathcal{N}^{n_0}_{\set{O}}$ (i.e., $t_0$ is given by the largest absolute value of the HM coefficients which are NN of order $>n_0$ with respect to $O$). In what follows, an overlying bar will always be used to denote filtered quantities. To make the comparison quantitative, we measure the distance between bands as~\cite{Garrity2021}
\begin{equation}
	\mathcal{D}(n_0) = \mathrm{max}_{i, \bm{k}} \left| \varepsilon_{i}(\bm{k}) -\overline{\varepsilon}^{(n_0)}_{i}(\bm{k}) \right|
\end{equation}
and we determine $n_0$ as the minimum integer such that $\mathcal{D}(n_0) \leq \eta_\varepsilon$, with $\eta_\varepsilon$ a pre-determined threshold (for instance, in the main text we set  $\eta_\varepsilon = 0.5$ eV), for values of $\bm{k}$ sampled from a regular grid in the Brillouin zone.

\paragraph{Stage H3: motif HM coefficients.} In the third stage, the filtered \emph{motif} HM coefficients $\overline{T}^{OB}_{ij}$ and $\overline{T}^{BO}_{ij}$ with $B \in \mathcal{N}^{n_0}_{\set{O}}$ [as defined in \cref{app:eq:motif_H0}] are retrieved via
\begin{equation}
    \overline{T}^{BO}_{ij} = \overline{T}^{OB}_{ji}.
\end{equation}

\paragraph{Stage H4: spinful motif HM coefficients.}
In the fourth stage, the \emph{spinful} motif HM coefficients $\tilde{t}_{(A,i,\sigma_1),(B,j,\sigma_2)}$ with $A, B \in \mathcal{N}^{n_0}_{\set{O}}$are obtained via \cref{app:eq:spinful_HMCs},
\begin{equation}
    \label{app:eq:spinful_motif_HMCs}
    t_{(A,i,\sigma_1),(B,j,\sigma_2)} = \begin{cases}
       \overline{T}^{AB}_{ij} & \text{if } \sigma_1 = \sigma_2\\
       0 & \text{otherwise}
    \end{cases}.
\end{equation}

\paragraph{(optional) Stage H4b: spinful lattice HM coefficients.}
In this optional stage, all the filtered spinful HM coefficients over the lattice $\mathcal{G}$ are obtained from \cref{app:eq:spinful_motif_HMCs} by exploiting the discrete translational invariance:
\begin{equation}
    t_{(A',i,\sigma_1),(B',j,\sigma_2)} = t_{(O,i,\sigma_1),(B,j,\sigma_2)} \quad \text{with } P'=P+\bm{R},\ \forall \bm{R}\in\mathcal{B}. 
\end{equation}

\subsubsection{Coulomb tensor coefficients}
\label{app:sec:CT_pipeline_CS}

Here we describe the steps to compute the Coulomb integrals defined in \cref{app:eq:bare_CTCs} via MC integration for a motif of order $n_{\mathrm{int}}$. Since these are $6-$dimensional integrals, their evaluation is costly. It is therefore essential to evaluate as little integrals as possible. To do so, we will take advantage of both the CS inequality introduced in \cref{app:eq:CS_ineq} and the symmetry properties of the CT of \cref{app:eq:CT_identities}. 

\paragraph{Stage CT1: fundamental CT coefficients.} In the first stage, we compute via MC integration the \emph{fundamental} terms required to apply the CS inequality of \cref{app:eq:CS_ineq}, denoted by $F^{OODD}_{iill}$. These are defined as
\begin{equation}
	\label{app:eq:CTCs_fundamental}
	F^{OODD}_{iill} = \begin{cases} 		V^{OOOO}_{iill} & \forall i, l \geq i\\
		V^{OODD}_{iill} & \forall i, l \text{ and } \forall C \in \mathcal{N}^{n_{\mathrm{int}}}_{\set{O}}\backslash\set{O}\\
	\end{cases}.
\end{equation} 

To reduce the number of Coulomb integrals to be computed it is reasonable to assume that the CT coefficients whose absolute value is smaller than a pre-determined threshold do not contribute to the properties of the material. Importantly, due to \cref{app:eq:CS_ineq}, the CT coefficient with the largest absolute value will be one of the fundamental coefficients above. Hence, we can fix a threshold as $t_{\mathrm{int}}=\tau_{\mathrm{int}} \times \mathrm{max}(|F^{OODD}_{iill}|)$ with, e.g., $\tau_{\mathrm{int}} \sim 10^{-2}$, and define the \emph{filtered} fundamental CT coefficients as
\begin{equation}
    \overline{F}^{OODD}_{i} = 
    \begin{cases}
        F^{OODD}_{iill} & \text{if } |F^{OODD}_{ill}| \geq t_{\mathrm{int}}\\
        0 &\text{otherwise}
    \end{cases}.
\end{equation}

\paragraph{Stage CT2: filtered unique CT coefficients.} In the second stage, we exploit \cref{app:eq:CS_ineq} to compute via MC integration only those \emph{unique} CT coefficients, defined as $U^{OBCD}_{ijkl} = V^{OBCD}_{ijkl}$, $\forall(O,B,C,D) \in \mathcal{C}^{OBCD}$ and $\forall (i,j,k,l) \in \mathcal{S}^{OBCD}$, such that $\overline{F}^{OODD}_{iill}\overline{F}^{OO(B-C)(B-C)}_{kkjj} \geq t_{\mathrm{int}}$. Here, $\mathcal{C}^{OBCD}$ and $\mathcal{S}^{OBCD}$ denote the minimal set of lattice sites and the unique orbital configurations required in the evaluation of terms with site structure $(O,B,C,D)$, respectively. In particular, the latter is defined as the quotient set of all the possible orbital configurations $\mathcal{M}$ induced by the equivalence relation $R^{OBCD}$, i.e., $\mathcal{S}^{OBCD} = \mathcal{M}/R^{OBCD} = \set{[(i,j,k,l)]^{OBCD}|(i,j,k,l)\in\mathcal{M}}$, with $[(i,j,k,l)]^{OBCD}$ the equivalence class associated with the orbital configuration $(i,j,k,l)$ with respect to $R^{OBCD}$. In turn, for each lattice site configuration $(O, B, C, D)$, $R^{OBCD}$ is determined by the symmetry properties of the CT outlined in \cref{app:eq:CT_identities}. For example, looking at the site structure $(O,O,O,O)$, the configurations \begin{equation*}
    (i,j,k,l), (l,k,j,i), (j,i,l,k), (l,j,k,i), (i,k,j,l), (k,l,i,j), (k,i,l,j), (j,l,i,k),
\end{equation*}
$\forall i,j,k,l \in \set{1,...,N_{\mathrm{orbitals/cell}}}$, 
are equivalent according to $R^{OOOO}$ since $V^{OOOO}_{ijkl} = V^{OOOO}_{lkji} = ... = V^{OOOO}_{jlik}$. Hence, 
\begin{equation*}
    [(i,j,l,k)]^{OOOO} = \set{(i,j,k,l), (l,k,j,i), (j,i,l,k), (l,j,k,i), (i,k,j,l), (k,l,i,j), (k,i,l,j), (j,l,i,k)}
\end{equation*}
represents the equivalence class of a given orbital configuration $(i,j,k,l)\in\mathcal{M}$. On the other hand, if we consider the site structure $(O,B,B,O)$, only the configurations
\begin{equation*}
    (i,j,k,l), (l,j,k,i), (i,k,j,l), (l,k,j,i),
\end{equation*}
$\forall i,j,k,l \in \set{1,...,N_{\mathrm{orbitals/cell}}}$, are equivalent according to $R^{OBBO}$. Indeed, for instance, $V^{OBBO}_{ijkl} \neq  V^{BOOB}_{jilk}$ and, therefore,  in contrast with the previous case, $(i,j,k,l)$ and $(j,i,l,k)$ are not equivalent. For the site structure $(O,B,B,O)$, the equivalence class of a given orbital configuration $(i,j,k,l)\in\mathcal{M}$ is therefore
\begin{equation*}
    [(i,j,k,l)]^{OOBB} = \set{(i,j,k,l), (l,j,k,i), (i,k,j,l), (l,k,j,i)}.
\end{equation*}
Note that, in this case, in order to retrieve all the possible CT coefficients (i.e., with $B\in\mathcal{N}^{n_{\mathrm{int}}}_{\set{O}}$ and $\forall (i,j,k,l)\in\mathcal{M}$) we need to compute the CT coefficients for the following minimal set of lattice site configurations 
\begin{equation*}
    \mathcal{C}^{OBBO} = \set{(O,B,B,O)| B\in\mathcal{N}^n_{\set{O}}, B \neq O}.
\end{equation*}
Following similar steps, it is possible to determine the orbital equivalence classes $[(i,j,k,l)]^{OBCD}$ and the minimal set of required lattice site configurations $\mathcal{C}^{OBCD}$ for an arbitrary lattice site structure $(O,B,C,D)$. In particular, we have
\begin{itemize}
	\item $\mathcal{C}^{OOOO} = \set{(O,O,O,O)}$ and
	\begin{equation*}
	 	[(i,j,k,l)]^{OOOO} = \set{(i,j,k,l), (l,k,j,i), (j,i,l,k), (l,j,k,i), (i,k,j,l), (k,l,i,j), (k,i,l,j), (j,l,i,k)}.
	\end{equation*} 
	\item $\mathcal{C}^{OBBO} = \set{(O,B,B,O)| B\in\mathcal{N}^n_{\set{O}}, B \neq O}$ and
	\begin{equation*}
		[(i,j,k,l)]^{OBBO} = \set{(i,j,k,l), (l,j,k,i), (i,k,j,l), (l,k,j,i)}.
	\end{equation*} 
	\item $\mathcal{C}^{OOBB} = \set{(O,O,B,B)| B\in\mathcal{N}^n_{\set{O}}, B \neq O}$ and
	\begin{equation*}
		[(i,j,k,l)]^{OOBB} = \set{(i,j,k,l), (j,i,l,k)}.
	\end{equation*} 
	\item $\mathcal{C}^{OOOB} = \set{(O,O,O,B)| B\in\mathcal{N}^n_{\set{O}}, B \neq O}$ and 
	\begin{equation*}
		[(i,j,k,l)]^{OOOB} = \set{(i,j,k,l), (i,k,j,l)}.
	\end{equation*} 
	\item $\mathcal{C}^{OBCO} = \set{(O,B,C,O)| B, C\in\mathcal{N}^n_{\set{O,B,C}}, B \neq O, C>B}$ and
	\begin{equation*}
		[(i,j,k,l)]^{OBCO} = \set{(i,j,k,l), (l,j,k,i)}.
	\end{equation*} 
	Note that coefficients with $C<B$ can be retrieved via $V^{OBCO}_{ijkl} = V^{OCBO}_{ikjl}$.
	\item $\mathcal{C}^{OOBC} = \set{(O,O,B,C)| B, C\in\mathcal{N}^n_{\set{O,B,C}}, B \neq O, C>B}$ and 
	\begin{equation*}
		[(i,j,k,l)]^{OOBC} = \set{(i,j,k,l)}, \quad\text{i.e., }\mathcal{S}^{OOBC} = \mathcal{M}.
	\end{equation*} 
	Note that coefficients with $C<B$ can be retrieved via $V^{OOBC}_{ijkl} = V^{OOCB}_{jilk}$.
	\item $\mathcal{C}^{OBBC} = \set{(O,B,B,C)| B, C\in\mathcal{N}^n_{\set{O,B,C}}, B \neq O, C \notin \set{O,B}}$ and
	\begin{equation*}
		[(i,j,k,l)]^{OBBC} = \set{(i,j,k,l), (i,k,j,l)}.
	\end{equation*} 
	\item $\mathcal{C}^{OBCB} = \set{(O,B,C,B)| B, C\in\mathcal{N}^n_{\set{O,B,C}}, B \neq O, C \notin \set{O,B}}$ and
	\begin{equation*}
		[(i,j,k,l)]^{OBCB} = \set{(i,j,k,l)}, \quad\text{i.e., }\mathcal{S}^{OBCB} = \mathcal{M}.
	\end{equation*} 
	\item $\mathcal{C}^{OBCD} = \set{(O,B,C,D)| B, C, D\in\mathcal{N}^n_{\set{O,B,C,D}}, B \neq O, C > B, D \notin \set{O,B,C}}$ and 
	\begin{equation*}
		[(i,j,k,l)]^{OBCD} = \set{(i,j,k,l)}, \quad\text{i.e., }\mathcal{S}^{OBCD} = \mathcal{M}.
	\end{equation*} 
	Note that coefficients with $C<B$ can be retrieved via $V^{OBCD}_{ijkl} = V^{OCBD}_{ikjl}$.
\end{itemize}
It can be shown that the choices above allow us to minimize the number of CT coefficients to be computed. Finally, the \emph{filtered} unique coefficients are introduced as
\begin{equation}
    \tilde{U}^{OBCD}_{i} = 
    \begin{cases}
        \overline{U}^{OBCD}_{iill} & \text{if } |U^{OBCD}_{ill}| \geq t_{\mathrm{int}}\\
        0 &\text{otherwise}
    \end{cases}.
\end{equation}

\paragraph{Stage CT3: filtered motif CT coefficients.} In the third stage, the filtered \emph{motif} CT coefficients entering \cref{app:eq:H0_Hint_motif}, $V^{m, ABCD}_{ijkl}$, $\forall A,B,C,D \in \mathcal{N}^n_{\set{O,A,B,C,D}}$ and $\forall (i,j,k,l)\in\mathcal{M}$, are obtained from the filtered unique ones. First, by exploiting the equivalence classes introduced in the previous stage, we can obtain the coefficients $V^{m, OBCD}_{ijkl}$, $\forall (O,B,C,D) \in \mathcal{C}^{OBCD}$ and $\forall(i,j,k,l)\in\mathcal{M}$. The latter can then be extended to any $ A,B,C,D \in \mathcal{N}^n_{\set{O,A,B,C,D}}$ (and at least one among $A,B,C,D$ equals to $O$) via the following identities
\begin{itemize}
	\item $V^{OBBO}_{ijkl} = V^{BOOB}_{jilk}$,
	\item $V^{OOBB}_{ijkl} = V^{OBOB}_{ikjl} = V^{BOBO}_{ljki} = V^{BBOO}_{lkji}$,
	\item $V^{OOOB}_{ijkl} = V^{OOBO}_{jilk} = V^{OBOO}_{jlik} = V^{BOOO}_{ljki}$,
	\item $V^{OBCO}_{ijkl} = V^{BOOC}_{jilk}$,
	\item $V^{OOBC}_{ijkl} = V^{OBOC}_{ikjl} = V^{CBOO}_{lkji} = V^{COBO}_{ljkl}$,
	\item $V^{OBBC}_{ijkl} = V^{CBBO}_{ljki} = V^{BOCB}_{jilk} = V^{BCOB}_{jlik}$,
	\item $V^{OBCB}_{ijkl} = V^{BBCO}_{ljki} = V^{OCBB}_{ikjl} = V^{BCBO}_{lkji} = V^{BOBC}_{jilk} = V^{BBOC}_{jlik} = V^{COBB}_{kilj} = V^{CBOB}_{klij}$,
	\item $V^{OBCD}_{ijkl} = V^{BDOC}_{jlik} = V^{BODC}_{jilk} = V^{DBCO}_{ljki}$. 
\end{itemize}

\paragraph{Stage CT4: spinful motif CT coefficients.} The filtered \emph{spinful} motif CT coefficients of \cref{app:eq:Hint_motif}, $ V^{m}_{(A,i,\sigma_1),(B,j,\sigma_2),(C,k,\sigma_3),(D,l,\sigma_4)}$ with $A,B,C,D \in \mathcal{N}^n_{\set{O, A,B,C,D}}$ and $(i,j,k,l)\in\mathcal{M}$, can be obtained via \cref{app:eq:spinful_CTCs} as
\begin{equation}
	\label{app:eq:spinful_motif_CTCs}
	V^m_{(A,i,\sigma_1),(B,j,\sigma_2),(C,k,\sigma_3),(D,l,\sigma_4)} = 
	\begin{cases}
	    V^{m,s,ABCD}_{ijkl} & \text{if } \sigma_1 = \sigma_2 = \sigma_3 = \sigma_4\\
		\frac{1}{2}V^{m,ABCD}_{ijkl} & \text{if } \sigma_1 = \sigma_3 \text{ and } \sigma_2 = \sigma_4\ (\sigma_1 \neq \sigma_2)\\
		-\frac{1}{2}V^{m,BACD}_{jikl} & \text{if } \sigma_1 = \sigma_4 \text{ and } \sigma_2 = \sigma_3 \ (\sigma_1 \neq \sigma_2)\\
		0 & \text{otherwise}
	\end{cases},
\end{equation}
with $V^{m,s,ABCD}_{ijkl} = (V^{m,ABCD}_{ijkl} - V^{m,BACD}_{jikl})/2$.

\paragraph{(optional) Stage CT4b: spinful lattice CT coefficients.} 
In this optional stage, all the filtered spinful coefficients over the lattice $\mathcal{G}$ are obtained from motif coefficients by exploiting the discrete translational invariance of the lattice: 
\begin{equation}
	V_{(A',i, \sigma_1), (B',j, \sigma_2), (C',k, \sigma_3), (D',l, \sigma_4)} = V^{m}_{(A,i, \sigma_1), (B,j, \sigma_2), (C,k, \sigma_3), (D,l, \sigma_4)},  
\end{equation}
with $P'=P+\bm{R}$, $\forall{P}\in\mathcal{N}^n_{\set{O,A,B,C,D}}$ and $\forall\bm{R}\in\mathcal{B}$.

\paragraph{Stage CT5: consistently filtered CT coefficients.} In general, the NN approximation of order $n_{\mathrm{int}}$ we performed in the previous stages is not guaranteed to be consistent, i.e., there may be CT coefficients corresponding to a higher order NN approximation whose absolute value is actually larger than the threshold $t_{\mathrm{int}}$. To avoid such a situation and obtain a consistent approximation of the CT, in this fifth stage we can
\begin{itemize}
    \item repeat the procedure above for $n_\mathrm{int}+1$ using the same value of $t_{\mathrm{int}}$;
    \item set a new threshold $t'_{\mathrm{int}}$ as the largest of the absolute values of the coefficients of the CT obtained within a NN approximation of order $n_\mathrm{int}+1$ which are not contained in the CT resulting from the NN approximation of order $n_\mathrm{int}$.
    \item filter the CT obtained within the NN approximation of order $n_\mathrm{int}$ with the new threshold $t'_{\mathrm{int}}$. 
\end{itemize}
Note that, in principle, to be sure that the approximation of the CT is consistent, one should repeat the procedure above for all NN orders $>n_{\mathrm{int}}$. However, taking into account the localization of the MLWFs and due the growing computational cost, in this work we limit our analysis to the case with $n_\mathrm{int}+1$ only. 
 
\subsubsection{Single-index HM and CT coefficients and Majorana Hamiltonian}
\label{app:sec:Majorana_H}

To obtained the Majorana form of the motif Hamiltonian, we first map each composite site-mode-spin indices triplet $(A_i,i,\sigma_i)$ to a single index $\alpha_i \in \set{1,..., M}$, with $M = 2 N_{\mathrm{orbitals/cell}} N_{\mathrm{cells/motif}}$ the total number of complex fermion modes. Here, $N_{\mathrm{cells/motif}}$ is determined as $N_{\mathrm{cells/motif}} = \mathrm{max}(N^{n_0}_O, N^{n_\mathrm{int}}_O)$ with $N^{n_0}_O$ and $N^{n_\mathrm{int}}_O$ the number of cells forming the motives used in the calculation of the HM and CT coefficients, respectively. If $N^{n_0}_O > N^{n_\mathrm{int}}_O$ ($N^{n_0}_O < N^{n_\mathrm{int}}_O$) all the CT coefficients with single indices corresponding to sites of $\mathcal{N}^{n_0}_O$ ($\mathcal{N}^{n_{\mathrm{int}}}_O$) not included in $\mathcal{N}^{n_\mathrm{int}}_O$ ($\mathcal{N}^{n_0}_O$) are set to zero. 
In terms of the single indices, the quadratic and quartic part of the motif Hamiltonian become
\begin{subequations}
\label{app:eq:H0_Hint_singleindex}
    \begin{align}
        H_0 &= \sum_{\alpha,\beta} t_{\alpha\beta} w^\dagger_\alpha w_\beta,\\
        H_\mathrm{int} &= \sum_{\alpha,\beta,\gamma,\delta} V_{\alpha\beta\gamma\delta} w^\dagger_\alpha w^\dagger_\beta w_\gamma w_\delta,
    \end{align}
\end{subequations}
respectively.

As discussed in \cref{sec:Wannier_basis_mapping}, the Majorana basis operators are defined by $w_\alpha = (\gamma_{\alpha}+i\bar{\gamma}_{\alpha})/2$ and $w^\dagger_\alpha = ( \gamma_{\alpha}-i\bar{\gamma}_{\alpha})/2$. In terms of the latter, the motif Hamiltonian reads
\begin{equation}
    \label{app:eq:H_Majorana}
    H_M = \sum_{k \in \{0,1\}^{2M}} \alpha_k \prod_j \gamma_j^{k_{2j}} \bar{\gamma}_j^{k_{2j+1}} \;,\; |k| \in \{2,4\}.
\end{equation}

\section{Materials analysis}

\subsection{Material properties}
In \cref{tab:mat_props} we list the properties of the materials studied throughout \cref{sec:results}. 
\begin{table}[]
    \begin{center}
        \input{Figs_Results/materials_results/material_properties}
    \end{center}
    \caption{Material properties}
    \label{tab:mat_props}
\end{table}

\subsection{Circuit depth}\label{sec:apdx_materials}
Here we present the complete data relating to the summary presented in \cref{tab:materials_analysed}. 
In particular, while that overview shows the circuit depths for each material by compiling the circuit including an fswap network for \gls{vqe}, here we compile with and without fswap networks for both \gls{vqe} and \gls{tds}. 
In \cref{tab:material_depths_apdx}, we report the circuit depth split between the onsite and nearest neighbour (Onsite/NN) terms, separately from the depth for the implementation of terms involving next nearest neighbours and beyond (NNN+).
In \cref{tab:n_terms_by_type} we list the number of terms within the corresponding Hamiltonians which are described as either onsite/NN or NNN+.
\par 

Here we treat the onsite/NN terms differently from the NNN+ terms, on the basis that the former are \textit{translatable} while the latter are \textit{nonstranslatable}.
Translatable terms may be tiled according to the strategy outlined in \cref{sec:compile_unit_cell}:
e.g. we first evaluate the depth of a group of terms which involve only the central cell; then we assume that equivalent terms from neighbouring cells can be implemented simultaneously, i.e. in parallel to the those terms from the central cell. 
As such, the depth of the onsite/NN terms is deemed to represent the depth of implementing all the translatable terms. 
In principle, translatable terms may be tiled indefinitely to achieve any desired lattice size, i.e. in order to simulate a given material in bulk, we need only increase the number of qubits, but the circuit depth will remain constant. 
This allows us to evaluate the circuit depth corresponding to onsite/NN terms using only a representative set of \textit{fundamental} terms: any term which is a translation of a fundamental term is omitted for the purpose of costing (but must be restored in parallel when actual circuits are built, such that the Hamiltonian is faithfully executed).
This is not true in the case of nontranslatable terms, so we must fix lattice sizes in order to retrieve meaningful circuit costs -- these lattice sizes are listed in \cref{tab:mat_props}.
To effectively cost the NNN+ terms, we must explicitly include all interactions extending beyond next-nearest-neighbour cells, e.g. for a sample of $SrVO_3$ with lattice dimensions $3 \times 3 \times 3$, where each lattice point corresponds to a cell with six modes, there are 288 NNN+ terms in the Hamiltonian, while there are 3159 onsite/NN terms. 
We see that in most cases the depth is dominated by the long range NNN+ terms, owing to the inclusion of many terms when the lattice of the fermionic encoding is tiled.
\par 

The compiler terminates by producing instructions for a quantum circuit: the instructions consist of circuit layers, where each layer is either an interaction layer or fswap layer.
Interaction layers are a set of Hamiltonian terms which the compiler has determined may be implemented in parallel, according to the decomposition rules of the given algorithm. The interaction layers implement all quadratic or quartic Hamiltonian terms in the circuit.
Fswap layers are then a set of mode-pairs, indicating which modes should be interchanged in order to facilitate subsequent interaction layers, implemented via \cref{eqn:fswap_decomposition}.  
\par 

The compiler does \textit{not} account for hardware constraints, such as available gates or qubit connectivity.
The final stage is then to iterate over the circuit layers and ensure that, as well as simultaneous with respect to the compiler's strategy, they are simultaneous on the target hardware. 
In cases where terms within a layer clash, they are placed on separate sublayers -- this is achieved using graph coloring, where terms are encoded as nodes and incompatibilities are encoded as edges. 
This mechanism is versatile and allows for arbitrary compatability rules to represent various hardware, e.g. devices where ISWAP gates are available facilitate cheaper fswaps than those without ISWAP. 
The default rule is simply that terms acting on any overlapping qubits must be in separate sublayers. 

\begin{table}
    \begin{center}
    \scalebox{0.85}{
        \input{Figs_Results/materials_results/appdx_all_data}
    }
    \caption{
        Resource requirements for each of the examined materials for a number of compiled algorithms. 
        Algorithms denoted without an asterisk are compiled with the use of an fswap network, while those with an asterisk omit the fswap network.
        Properties of the listed materials are given in \cref{tab:mat_props}, and the number of terms of each Hamiltonian type are listed in \cref{tab:n_terms_by_type}.
    }
    \label{tab:material_depths_apdx}
    \end{center}
\end{table}

\begin{table}[]
    \centering
    \input{Figs_Results/materials_results/n_terms_by_type}
    \caption{Number of terms in the materials' Hamiltonians, listed by whether they involve at most one nearest neighbour (onsite/NN) or extend to next nearest neighbour and beyond (NNN+).}
    \label{tab:n_terms_by_type}
\end{table}

\subsection{Measurement layers}
In \cref{tab:msmt_layers} we provide the data used in the analysis in \cref{sec:measurements}. 
In particular, the measurement layers presented in \cref{fig:msmt_layers} are repeated in \cref{tab:msmt_layers} for ease of interpretation.

\begin{table}[]
    \begin{center}
    \input{Figs_Results/materials_results/msmt}
    \end{center}
    \caption{
        Measurement layers required, corresponding to those represented in \cref{fig:msmt_layers}.
        Number of measurement layers are reported by the type of terms involved in the layer. 
        Terms involving only single qubit Z measurements are extracted to a first measurement layer in each case; 
        then terms are grouped by whether they are onsite/NN or NNN+, and the given strategies applied.
    }
    \label{tab:msmt_layers}
\end{table}

%% file: Figs_Results/materials_results/material_properties.tex
\begin{tabular}{lrrrrrr}
\toprule
{} &     Spacegroup &  Bands & Dimension &  Size &  Modes &  Qubits \\
Material      &                &        &           &       &        &         \\
\midrule
GaAs          &   F$\bar{4}$3m &      4 &     5x5x5 &   125 &   1000 &    1120 \\
H$_3$S        &   Im$\bar{3}$m &      7 &     5x5x5 &   125 &   1750 &    1870 \\
Li$_2$CuO$_2$ &             P1 &     11 &     5x3x3 &    45 &    990 &    1024 \\
Si            &    Fd\text{3}m &      4 &     5x5x5 &   125 &   1000 &    1120 \\
SrVO$_3$      &   Pm$\bar{3}$m &      3 &     3x3x3 &    27 &    162 &     180 \\
\bottomrule
\end{tabular}

%% file: Figs_Results/materials_results/appdx_all_data.tex
\begin{tabular}{lllllrrr}
\toprule
         &                       &     &     &  &  Onsite/NN &   NNN+ &  Total \\
Material & Dimension & Modes & Qubits & Algorithm &            &        &        \\
\midrule
\multirow{4}{*}{GaAs} & \multirow{4}{*}{$5 \times 5 \times 5$} & \multirow{4}{*}{1000} & \multirow{4}{*}{1120} & TDS &        664 &   9305 &   9969 \\
         &                       &     &     & TDS$^{*}$ &       1090 &   9305 &  10395 \\
         &                       &     &     & VQE &        662 &   7191 &   7853 \\
         &                       &     &     & VQE$^{*}$ &       1072 &   7191 &   8263 \\
\cline{1-8}
\cline{2-8}
\cline{3-8}
\cline{4-8}
\multirow{4}{*}{H$_3$S} & \multirow{4}{*}{$5 \times 5 \times 5$} & \multirow{4}{*}{1750} & \multirow{4}{*}{1870} & TDS &       1217 &  38228 &  39445 \\
         &                       &     &     & TDS$^{*}$ &       7024 &  38228 &  45252 \\
         &                       &     &     & VQE &       1214 &  36126 &  37340 \\
         &                       &     &     & VQE$^{*}$ &       7000 &  36126 &  43126 \\
\cline{1-8}
\cline{2-8}
\cline{3-8}
\cline{4-8}
\multirow{4}{*}{Li$_2$CuO$_2$} & \multirow{4}{*}{$5 \times 3 \times 3$} & \multirow{4}{*}{990} & \multirow{4}{*}{1024} & TDS &       1675 &   6885 &   8560 \\
         &                       &     &     & TDS$^{*}$ &       3385 &   6885 &  10270 \\
         &                       &     &     & VQE &       1667 &   6710 &   8377 \\
         &                       &     &     & VQE$^{*}$ &       3341 &   6710 &  10051 \\
\cline{1-8}
\cline{2-8}
\cline{3-8}
\cline{4-8}
\multirow{4}{*}{Si} & \multirow{4}{*}{$5 \times 5 \times 5$} & \multirow{4}{*}{1000} & \multirow{4}{*}{1120} & TDS &        706 &   8904 &   9610 \\
         &                       &     &     & TDS$^{*}$ &       1746 &   8904 &  10650 \\
         &                       &     &     & VQE &        704 &   7857 &   8561 \\
         &                       &     &     & VQE$^{*}$ &       1733 &   7857 &   9590 \\
\cline{1-8}
\cline{2-8}
\cline{3-8}
\cline{4-8}
\multirow{4}{*}{SrVO$_3$} & \multirow{4}{*}{$3 \times 3 \times 3$} & \multirow{4}{*}{162} & \multirow{4}{*}{180} & TDS &        408 &    700 &   1108 \\
         &                       &     &     & TDS$^{*}$ &        552 &    700 &   1252 \\
         &                       &     &     & VQE &        408 &    476 &    884 \\
         &                       &     &     & VQE$^{*}$ &        555 &    476 &   1031 \\
\bottomrule
\end{tabular}

%% file: Figs_Results/materials_results/n_terms_by_type.tex
\begin{tabular}{lrrr}
\toprule
     Material &  Onsite/NN &  NNN+ &  Total \\
\midrule
         GaAs &      27500 & 22080 &  49580 \\
       H$_3$S &      52925 & 46376 &  99301 \\
Li$_2$CuO$_2$ &      13953 &  3344 &  17297 \\
           Si &      30300 & 22880 &  53180 \\
     SrVO$_3$ &       3159 &   288 &   3447 \\
\bottomrule
\end{tabular}

%% file: Figs_Results/materials_results/msmt.tex
\begin{tabular}{llrrrr}
\toprule
         &  &  Z &  \ \ Onsite/NN &  \ \ \ \ \ \ \  NNN+ &  Total \\
 & Strategy &    &            &       &        \\
\midrule
\multirow{3}{*}{GaAs} & Commutatitvity &  1 &         15 &    22 &     38 \\
         & Noncrossing &  1 &         56 &   475 &    532 \\
         & Qubitwise Commutativity &  1 &         85 &   475 &    561 \\
\cline{1-6}
\multirow{3}{*}{H$_3$S} & Commutatitvity &  1 &         36 &    54 &     91 \\
         & Noncrossing &  1 &        161 &  2274 &   2436 \\
         & Qubitwise Commutativity &  1 &        463 &  2274 &   2738 \\
\cline{1-6}
\multirow{3}{*}{Li$_2$CuO$_2$} & Commutatitvity &  1 &         22 &    19 &     42 \\
         & Noncrossing &  1 &         88 &   539 &    628 \\
         & Qubitwise Commutativity &  1 &        248 &   539 &    788 \\
\cline{1-6}
\multirow{3}{*}{Si} & Commutatitvity &  1 &         21 &    26 &     48 \\
         & Noncrossing &  1 &         84 &   510 &    595 \\
         & Qubitwise Commutativity &  1 &        142 &   510 &    653 \\
\cline{1-6}
\multirow{3}{*}{SrVO$_3$} & Commutatitvity &  1 &          6 &     4 &     11 \\
         & Noncrossing &  1 &         36 &    46 &     83 \\
         & Qubitwise Commutativity &  1 &         41 &    46 &     88 \\
\bottomrule
\end{tabular}